\documentclass[aps,superscriptaddress,floats,twocolumn,epsf,prr]{revtex4-2}
\usepackage{graphicx}
\usepackage{epstopdf}
\usepackage{color}
\usepackage{colortbl}
\usepackage{xcolor}
\usepackage{setspace}
\usepackage[utf8]{inputenc}
\usepackage{amsmath}
\usepackage{amssymb}
\usepackage{placeins}
\usepackage{mathtools}
\usepackage{changes}
\usepackage{tikz-feynman}
\usepackage{comment}
\usepackage[export]{adjustbox}
\usepackage{physics}
\usepackage{dsfont}
\usepackage[normalem]{ulem}
\usepackage{etoolbox}
\usepackage{enumitem}
\usepackage{overpic}
\usepackage{diagbox}

\AtEndEnvironment{subequations}{\par\nobreak\noindent}
\usepackage[colorlinks=true,urlcolor=blue,citecolor=blue,linkcolor=blue,breaklinks=true]{hyperref}

\tikzset{
        linePlain/.style={draw=black, thick},
        lineWithArrowCenter/.style={draw=black, thick, postaction={decorate},decoration={markings,mark=at position .6 with {\arrow[scale=1.2]{latex}}}},
        lineArrow/.style={draw=black, thick, postaction={decorate},decoration={markings,mark=at position .7 with {\arrow[scale=1.2]{latex}}}},
        lineArrowBis/.style={draw=black, thick, postaction={decorate},decoration={markings,mark=at position .4 with {\arrow[scale=1.2]{latex}}}},
        lineWithArrowCenterEnd/.style={draw=black, thick, postaction={decorate},decoration={markings,mark=at position .85 with {\arrow[scale=1.2]{latex}}}},
        lineWithArrowCenterStart/.style={draw=black, thick, postaction={decorate},decoration={markings,mark=at position .35 with {\arrow[scale=1.2]{latex}}}},
        bosonLine/.style={draw=black, thick, decorate, decoration={snake, segment length=2mm, amplitude=0.6mm}},
}

\newcommand{\tikzm}[2]{
        \tikz[baseline=-0.65ex]{#2}
}

\newcommand{\arrowslefthalf}[2]{
   \def\shift{0.3};
   \coordinate (center) at (#1,#2);
   \draw[lineWithArrowCenterEnd] (center)  -- ($(center) + (-\shift,-\shift)$);
   \draw[lineWithArrowCenterEnd] ($(center)     + (-\shift,+\shift)$) -- (center);
}

\newcommand{\arrowsrighthalf}[2]{
   \def\shift{0.3};
   \coordinate (center) at (#1,#2);
   \draw[lineWithArrowCenterEnd] ($(center) + (+\shift,-\shift)$) -- (center);
   \draw[lineWithArrowCenterEnd] (center)    -- ($(center)   + (+\shift,+\shift)$);
}

\newcommand{\arrowslefthalffull}[3]{
   \def\shift{0.3};
   \def\shiftbox{0.3*#3};
   \coordinate (center) at (#1,#2);
   \coordinate (bottomleft)  at ($(center) + (-\shiftbox,-\shiftbox)$);
   \coordinate (topleft)     at ($(center) + (-\shiftbox,+\shiftbox)$);
   \draw[lineWithArrowCenterEnd] (bottomleft)  -- ($(bottomleft) + (-\shift,-\shift)$);
   \draw[lineWithArrowCenterEnd] ($(topleft)     + (-\shift,+\shift)$) -- (topleft);
}

\newcommand{\arrowsrighthalffull}[3]{
   \def\shift{0.3};
   \def\shiftbox{0.3*#3};
   \coordinate (center) at (#1,#2);
   \coordinate (bottomright) at ($(center) + (+\shiftbox,-\shiftbox)$);
   \coordinate (topright)    at ($(center) + (+\shiftbox,+\shiftbox)$);
   \draw[lineWithArrowCenterEnd] ($(bottomright) + (+\shift,-\shift)$) -- (bottomright);
   \draw[lineWithArrowCenterEnd] (topright)    -- ($(topright)   + (+\shift,+\shift)$);
}

\newcommand{\arrowslowerhalffull}[3]{
   \def\shift{0.3};
   \def\shiftbox{0.3*#3};
   \coordinate (center) at (#1,#2);
   \coordinate (bottomleft)  at ($(center) + (-\shiftbox,-\shiftbox)$);
   \coordinate (bottomright) at ($(center) + (+\shiftbox,-\shiftbox)$);
   \draw[lineWithArrowCenterEnd] (bottomleft)  -- ($(bottomleft) + (-\shift,-\shift)$);
   \draw[lineWithArrowCenterEnd] ($(bottomright) + (+\shift,-\shift)$) -- (bottomright);
}

\newcommand{\arrowsupperhalffull}[3]{
   \def\shift{0.3};
   \def\shiftbox{0.3*#3};
   \coordinate (center) at (#1,#2);
   \coordinate (topleft)     at ($(center) + (-\shiftbox,+\shiftbox)$);
   \coordinate (topright)    at ($(center) + (+\shiftbox,+\shiftbox)$);
   \draw[lineWithArrowCenterEnd] ($(topleft)     + (-\shift,+\shift)$) -- (topleft);
   \draw[lineWithArrowCenterEnd] (topright)    -- ($(topright)   + (+\shift,+\shift)$);
}

\newcommand{\barevertexwithlegs}[2]{
   \fill (#1,#2) circle (2pt) coordinate (center);
   \arrowslefthalf{#1}{#2}
   \arrowsrighthalf{#1}{#2}
}

\newcommand{\barevertex}[2]{
   \fill (#1,#2) circle (2pt);
}

\newcommand{\arrowsallfull}[3]{
   \arrowslefthalffull{#1}{#2}{#3}
   \arrowsrighthalffull{#1}{#2}{#3}
}

\newcommand{\phxbubble}[3]{
   \draw[lineWithArrowCenter] (#1,#2) to [out=45, in=135] (#1+1.2*#3,#2);
   \draw[lineWithArrowCenter] (#1+1.2*#3,#2) to [out=225, in=315] (#1,#2);
}

\newcommand{\phbubble}[3]{
   \draw[lineWithArrowCenter] (#1,#2) to [out=225, in=135] (#1,#2-1.2*#3);
   \draw[lineWithArrowCenter] (#1,#2-1.2*#3) to [out=45, in=315] (#1,#2);
}

\newcommand{\ppbubble}[3]{
   \draw[lineWithArrowCenter] (#1+1.2*#3,#2) to [out=45, in=135] (#1,#2);
   \draw[lineWithArrowCenter] (#1+1.2*#3,#2) to [out=225, in=315] (#1,#2);
}

\newcommand{\arrowslowerhalf}[2]{
   \def\shift{0.3};
   \coordinate (center) at (#1,#2);
   \draw[lineWithArrowCenterEnd] (center)  -- ($(center) + (-\shift,-\shift)$);
   \draw[lineWithArrowCenterEnd] ($(center) + (+\shift,-\shift)$) -- (center);
}

\newcommand{\arrowsupperhalf}[2]{
   \def\shift{0.3};
   \coordinate (center) at (#1,#2);
   \draw[lineWithArrowCenterEnd] ($(center)     + (-\shift,+\shift)$) -- (center);
   \draw[lineWithArrowCenterEnd] (center)    -- ($(center)   + (+\shift,+\shift)$);
}

\newcommand{\arrowslefthalfp}[3]{
   \def\shift{0.3};
   \coordinate (center) at (#1,#2);
   \draw[lineWithArrowCenterEnd] (center) -- ($(center) + (-\shift,-\shift)$);
   \draw[lineWithArrowCenterEnd] (center) to [out=45, in=180] ($(center) + (1.2*#3+\shift,2*\shift)$);
}

\newcommand{\arrowsrighthalfp}[3]{
   \def\shift{0.3};
   \coordinate (center) at (#1,#2);
   \draw[lineWithArrowCenterEnd] ($(center) + (+\shift,-\shift)$) -- (center);
   \draw[lineWithArrowCenterStart] ($(center) + (-1.2*#3-\shift,2*\shift)$) to [out=0, in=135] (center);
}

\newcommand{\selfenergy}[4]{
   \def\shift{0.3*#4};
   \coordinate (center) at (#2,#3);
   \draw[linePlain, fill=specialgray] (center) circle [radius=\shift];
   \node at (center) {#1};
}

\newcommand{\selfenergywithlegs}[4]{
   \selfenergy{#1}{#2}{#3}{#4}
   \draw[linePlain] (#2-0.3*#4,#3) -- (#2-0.3*#4-0.15,#3);
   \draw[linePlain] (#2+0.3*#4+0.15,#3) -- (#2+0.3*#4,#3);
}

\newcommand{\fullvertex}[4]{   
   \def\shiftbox{0.3*#4};
   \coordinate (center) at (#2,#3);
   \coordinate (bottomleft)  at ($(center) + (-\shiftbox,-\shiftbox)$);
   \coordinate (topleft)     at ($(center) + (-\shiftbox,+\shiftbox)$);
   \coordinate (bottomright) at ($(center) + (+\shiftbox,-\shiftbox)$);
   \coordinate (topright)    at ($(center) + (+\shiftbox,+\shiftbox)$);
   \draw[linePlain, fill=specialgray] (bottomleft) rectangle (topright);
   \node at (center) {#1};
}

\newcommand{\lineslefthalffull}[3]{
   \def\shift{0.1};
   \def\shiftbox{0.3*#3};
   \coordinate (center) at (#1,#2);
   \coordinate (bottomleft)  at ($(center) + (-\shiftbox,-\shiftbox)$);
   \coordinate (topleft)     at ($(center) + (-\shiftbox,+\shiftbox)$);
   \draw[linePlain] (bottomleft)  -- ($(bottomleft) + (-\shift,-\shift)$);
   \draw[linePlain] ($(topleft)     + (-\shift,+\shift)$) -- (topleft);
}

\newcommand{\linesrighthalffull}[3]{
   \def\shift{0.1};
   \def\shiftbox{0.3*#3};
   \coordinate (center) at (#1,#2);
   \coordinate (bottomright) at ($(center) + (+\shiftbox,-\shiftbox)$);
   \coordinate (topright)    at ($(center) + (+\shiftbox,+\shiftbox)$);
   \draw[linePlain] ($(bottomright) + (+\shift,-\shift)$) -- (bottomright);
   \draw[linePlain] (topright)    -- ($(topright)   + (+\shift,+\shift)$);
}

\newcommand{\lineslowerhalffull}[3]{
   \def\shift{0.1};
   \def\shiftbox{0.3*#3};
   \coordinate (center) at (#1,#2);
   \coordinate (bottomleft)  at ($(center) + (-\shiftbox,-\shiftbox)$);
   \coordinate (bottomright) at ($(center) + (+\shiftbox,-\shiftbox)$);
   \draw[linePlain] (bottomleft)  -- ($(bottomleft) + (-\shift,-\shift)$);
   \draw[linePlain] ($(bottomright) + (+\shift,-\shift)$) -- (bottomright);
}

\newcommand{\linesupperhalffull}[3]{
   \def\shift{0.1};
   \def\shiftbox{0.3*#3};
   \coordinate (center) at (#1,#2);
   \coordinate (topleft)  at ($(center) + (-\shiftbox,+\shiftbox)$);
   \coordinate (topright) at ($(center) + (+\shiftbox,+\shiftbox)$);
   \draw[linePlain] (topleft)  -- ($(topleft) + (-\shift,+\shift)$);
   \draw[linePlain] ($(topright) + (+\shift,+\shift)$) -- (topright);
}

\newcommand{\linesbottomleftcornerfull}[3]{
   \def\shift{0.1};
   \def\shiftbox{0.3*#3};
   \coordinate (center) at (#1,#2);
   \coordinate (bottomleft)  at ($(center) + (-\shiftbox,-\shiftbox)$);
   \draw[linePlain] (bottomleft)  -- ($(bottomleft) + (-\shift,-\shift)$);
}

\newcommand{\linesbottomrightcornerfull}[3]{
   \def\shift{0.1};
   \def\shiftbox{0.3*#3};
   \coordinate (center) at (#1,#2);
   \coordinate (bottomright) at ($(center) + (+\shiftbox,-\shiftbox)$);
   \draw[linePlain] ($(bottomright) + (+\shift,-\shift)$) -- (bottomright);
}

\newcommand{\linestopleftcornerfull}[3]{
   \def\shift{0.1};
   \def\shiftbox{0.3*#3};
   \coordinate (center) at (#1,#2);
   \coordinate (topleft)  at ($(center) + (-\shiftbox,+\shiftbox)$);
   \draw[linePlain] (topleft)  -- ($(topleft) + (-\shift,+\shift)$);
}

\newcommand{\loopfullvertextop}[4]{
   \def\shift{0.3*#4};
   \draw[#1] ($(#2,#3) + (\shift,\shift)$) .. controls ++(45:0.4) and ++(0:0.4) .. ($(#2,#3) + (0,0.6+0.3*#4)$) .. controls ++(180:0.4) and ++(135:0.4) .. ($(#2,#3) + (-\shift,\shift)$);
}

\newcommand{\loopfullvertexright}[4]{
   \def\shift{0.3*#4};
   \draw[#1] ($(#2,#3) + (\shift,-\shift)$) .. controls ++(315:0.4) and ++(270:0.4) .. ($(#2,#3) + (0.6+0.3*#4,0)$) .. controls ++(90:0.4) and ++(45:0.4) .. ($(#2,#3) + (\shift,\shift)$);
}

\newcommand{\fcirc}{\mathbin{\vcenter{\hbox{\scalebox{0.65}{$\bullet$}}}}}

\newcounter{subfigure}[figure]

\definecolor{specialgray}{RGB}{240,240,240}

\definecolor{darkgreen}{rgb}{0.0, 0.7, 0.0}

\newcommand{\D}{\mathrm{D}}
\newcommand{\SC}{\mathrm{SC}}
\newcommand{\M}{\mathrm{M}}
\newcommand{\X}{\mathrm{X}}

\usepackage{hyperref}
\hypersetup{colorlinks=true,breaklinks,linkcolor=blue,urlcolor=blue,citecolor=blue}


\begin{document}

\title{Functional renormalization with interaction flows: A single-boson exchange perspective and application to electron-phonon systems}

\author{Aiman Al-Eryani}
\email{aiman.al-eryani@rub.de}
\affiliation{Institute for Theoretical Physics III, Ruhr-Universit\"at Bochum, 44801 Bochum, Germany}

\author{Marcel Gievers}
\email{marcel.gievers@tuwien.ac.at}
\affiliation{Institute for Solid State Physics, TU Wien, 1040 Vienna, Austria}

\author{Kilian Fraboulet}
\email{k.fraboulet@fkf.mpg.de}
\affiliation{Max-Planck-Institut f{\"u}r Festk{\"o}rperforschung, Heisenbergstra{\ss}e 1, 70569 Stuttgart, Germany}
\affiliation{Institute of Information Systems Engineering, TU Wien, 1040 Vienna, Austria}
\affiliation{Institute for Theoretical Physics and Center for Quantum Science, Universit\"at T\"ubingen, Auf der Morgenstelle 14, 72076 T\"ubingen, Germany}

\renewcommand{\i}{{\mathrm{i}}}
\date{ \today }

\begin{abstract}
The functional renormalization group (fRG) is acknowledged as a powerful tool in quantum many-body physics and beyond. On the technical side, conventional implementations of the fRG rely on regulators for bare propagators only. Starting from Schwinger--Dyson and Bethe--Salpeter equations, we develop here an fRG formulation where both bare propagators and bare interactions can be dressed with regulators. The approach thus obtained is an extension of the multiloop fRG recently introduced for many-fermion systems. Using the single-boson exchange decomposition, we show that the underlying flow equations are simply interpreted as adding a regulator to the bosonic propagator and that such an extension scarcely changes the original structure of the flow equations. Overall, we provide a framework for implementing approaches that cannot be realized with conventional fRG methods, such as temperature flows for models with retarded interactions. For concrete applications, we analyze the loop convergence of our formulation against a conventional cutoff scheme for the Anderson impurity model. Finally, we devise a new temperature-flow scheme that implements a cutoff in both the propagator and the bare interaction, and demonstrate its validity on a model of an Anderson impurity coupled to a phonon.
\end{abstract}
\maketitle

\section{Introduction}
\label{sec:Introduction}

The renormalization group (RG) plays a significant role for the understanding of many areas of physics, especially through its implementation within the functional renormalization group (fRG), which allows for designing powerful and diverse approximation schemes to study a wide variety of problems~\cite{Berges2002,Pawlowski2007,Kopietz2010,Delamotte2012,Metzner2012,Dupuis2021}. Most fRG approaches are designed from an exact flow equation known as the Wetterich equation~\cite{Wetterich1993,Ellwanger1994,Morris1994}. In recent years, a different philosophy has emerged, deriving flow equations instead from self-consistent equations, such as the Bethe--Salpeter equations. This gave rise to the multiloop fRG for many-fermion systems~\cite{Kugler2018a,Kugler2018b,Kugler2018c} (see also Refs.~\cite{Blaizot2011,Blaizot2021} for related method developments for a bosonic model), which led to many insightful applications in condensed matter theory, notably for quantum spin systems~\cite{Thoenniss2020,Kiese2022,Ritter2022}, the Anderson impurity model~\cite{Chalupa2020,Ge2024,Ritz2024} and the two-dimensional (2D) Hubbard model~\cite{TagliaviniHille2019,Hille2020a,Hille2020b,Heinzelmann2023,HillePhDThesis}. In particular, this approach has been shown to provide quantitative approximations that go beyond the widespread one-loop ($1\ell$) truncation of the fermionic fRG derived from a vertex expansion of the Wetterich equation~\cite{Morris1994,Metzner2012} (referred to as $1\ell$ fRG). The multiloop-fRG formulation keeps its ability to describe competing orders in an unbiased manner while respecting key physical many-body principles ranging from sum rules~\cite{Chalupa2020} to the Hohenberg--Mermin--Wagner theorem due to its equivalence to the parquet formalism~\cite{Bickers1992,AlEryani_2_2025}.

The fRG flow can be pictured as an interpolation between a solvable system (which gives the initial conditions for the flow) and the more complicated system under consideration. This evolution is controlled by the RG scale (also called flow parameter). The latter can be chosen to correspond to a momentum scale, in such a way that the fRG can be viewed as a modern version of the Wilsonian momentum-shell RG~\footnote{In particular, Ref.~\cite{Delamotte2012} provides pedagogical explanations on the connection between the fRG and the Wilsonian RG.}. In the fRG, the regulator (also called cutoff function) plays a central role. It is the function that introduces the RG scale and implements the aforementioned interpolation between exactly solvable initial conditions and the studied correlated system. In practice, it is also used in particular to control divergences that might occur throughout the RG flow. Most fRG approaches, including those based on the Wetterich equation or previous multiloop fRG implementations, rely on a regulator inserted only in the bare propagator(s) of the theory. In this work, we develop an fRG scheme where the bare propagator and the bare interaction of the studied model are both dressed with a regulator. This provides extra flexibility that can be used to tackle problems in ways not achievable by other more conventional fRG schemes.

The introduction of a regulator in the bare interaction would notably give access to more correlated starting points for the fRG flow. For example, one can naturally design an fRG flow starting from dynamical mean-field theory (DMFT) results~\cite{Metzner1989,Georges1996} to treat a given lattice problem: this approach is known as DMF$^2$RG~\cite{Taranto2014,Vilardi2019,Bonetti2022}. To extend this scheme to models with non-local interactions with a starting point for the fRG flow set by, e.g., extended DMFT~\cite{Si1996,Smith2000,Chitra2000,Chitra2001,Sun2002}, one would need an fRG scheme involving a regulator both in a bare propagator and a bare interaction, as was discussed in Ref.~\cite{Katanin2019}. Furthermore, the treatment of retarded interactions with temperature flows~\cite{Honerkamp2001}, which use the temperature as RG scale, requires a regulator that dresses a bare propagator and a bare interaction. This strategy will be demonstrated in the present study.

Introducing a flow parameter in the bare interaction at the level of the Wetterich equation leads in general to very complicated flow equations, not suited to design efficient approximation schemes \footnote{We illustrate this point in more detail in App.~\ref{sec:WetterichVertexExpUflow} for generic models with quartic interactions.}. In this work, we thus consider a different approach. In the spirit of the multiloop fRG, we derive flow equations from the Schwinger--Dyson and the Bethe--Salpeter equations with a regulator both in the bare propagator and in the bare interaction for generic fermionic models with quartic interactions.

It should be noted that interaction flows, which involve regulators in bare interactions, have also been developed in the two-particle-irreducible (2PI) fRG~\cite{Dupuis2014} and in the two-particle-point-irreducible fRG~\cite{Polonyi2002,Schwenk2004,Yokota2021}, which rely on flow equations for Luttinger--Ward functionals~\cite{EnssPhDthesis} and density functionals (as encountered in density functional theory), respectively. Although these approaches are less widespread than the fRG based on the Wetterich equation (which relies on the one-particle-irreducible (1PI) effective action), they have led to several successful applications to models of interest in condensed matter physics and beyond~\cite{Rentrop2015,Rentrop2016,Fraboulet2024,Katanin2019,Kemler2013,Rentrop2015,Kemler2016,Liang2018,Yokota2019,Yokota2019_2,Yokota2019_3,Yokota2021_2,Yokota2021_3,Fraboulet2024,Yokota2025}.

In the present study, we formulate a new interaction flow for generic fermionic models. We derive flow equations for correlation functions defined in the 1PI effective action framework within a bosonization approach defined by the single-boson exchange (SBE) decomposition of the two-particle vertex $V$~\cite{Krien2019}. The SBE decomposition amounts to describing the complicated frequency and momentum dependencies of $V$ in terms of fermion-boson vertices and bosonic propagators, which give direct access to susceptibilities, the relevant objects for studying competing orders. In particular, the computational and interpretative advantages of the SBE approach have been demonstrated in the context of the $1\ell$ fRG~\cite{Bonetti2022,Fraboulet2022}, and notably within the DMF$^2$RG~\cite{Bonetti2022}.

Hence, the merging of the SBE decomposition with the multiloop fRG, which we will refer to as multiloop SBE fRG, defines a fruitful framework to construct efficient and accurate descriptions of many-electron systems that go beyond the conventional $1\ell$ fRG~\cite{Fraboulet2025}. The flow equations underlying the multiloop SBE fRG were derived in Ref.~\cite{Gievers2022} for generic fermionic models, and applications to the 2D Hubbard model were carried out recently~\cite{Fraboulet2025}, but all this was achieved considering only a regulator in the bare propagator. In the present study, we generalize the multiloop SBE fRG by introducing a regulator in both the bare propagator and the bare interaction, thus making this fRG scheme more amenable to the treatment of non-local and retarded interactions, as explained previously.

This article is structured as follows. In Sec.~\ref{sec:Formalism}, we derive the flow equations underlying our extension of the multiloop SBE fRG scheme based on an interaction flow, after reviewing the main ingredients of the SBE approach. As a next step, we demonstrate the validity of these flow equations in two concrete models: the Anderson impurity model (AIM)~\cite{Anderson1961} and the Anderson--Holstein impurity model (AHIM)~\cite{ELagos2001,Hewson2002}. In Sec.~\ref{sec:ModelsNumericalImplementationConvergence}, we present a numerical application with a simple flow scheme featuring a cutoff only in the bare interaction. We verify its loop convergence, and compare it to conventional flow schemes that utilize a cutoff only in the propagator. As a final application, in Sec.~\ref{sec:TflowRetardedInteractions}, we devise a temperature-flow scheme applicable to electron-phonon systems, where we supplement the conventional scheme with a flow in the bare interaction. A numerical application of this scheme is subsequently given for the Anderson--Holstein impurity model. Concluding remarks and future prospects are given in Sec.~\ref{sec:Conclusion}.

\section{Formalism}
\label{sec:Formalism}

In this section, we first review the SBE decomposition of the two-particle vertex and define our conventions. As a next step, we derive an extended form of the multiloop SBE fRG flow equations by including a regulator in both the bare propagator and the bare interaction. This new approach defines our \emph{interaction flow} and is the main outcome of this work. We will derive these flow equations for translationally-invariant fermionic systems relying on energy conservation. In App.~\ref{sec:SU2symmetry}, we show separately how these equations can be efficiently rewritten when $SU(2)$ spin symmetry is fulfilled. Readers familiar with the parquet and SBE decompositions may directly move to Sec.~\ref{sec:B-reducibility}, where we introduce the concept of $\mathcal{B}$-reducibility, a key ingredient for the treatment of non-local interactions, and to Sec.~\ref{sec:SBEmfRGUflow} where the multiloop SBE equations including the interaction flow are presented, the main result of this paper.

\subsection{Setting the stage}

To present the SBE approach and then the multiloop SBE fRG formalism with interaction flow, we consider a fermionic system with the bare action
\begin{equation}
S[\overline{c},c] = -  \overline{c}_{1^\prime} G_{0;1^\prime|1}^{-1} c_{1} -  \tfrac{1}{4} U_{1^\prime 2^\prime|12} \overline{c}_{1^\prime} \overline{c}_{2^\prime} c_{2} c_{1},
\label{eq:ClassicalActionS}
\end{equation}
where $G_0$ and $U$ are the bare propagator and the bare interaction, respectively. Note that $U$ is a crossing-symmetric vertex, i.e., it satisfies $U_{1^\prime 2^\prime|12} = -U_{2^\prime 1^\prime|12} = -U_{1^\prime 2^\prime|21} = U_{2^\prime 1^\prime|21}$. Furthermore, the general index $j$ encompasses all indices labeling the Grassmann fields $c_j$ (and their conjugates $\overline{c}_j$). In this paper, it includes a spin index $\sigma$, a Matsubara frequency $\nu$, and a spatial momentum $\mathbf{k}$, i.e., $j = (\sigma_j,i\nu_j,\mathbf{k}_j)$ or $j=(\sigma_j,k_j)$ with $k_j=(i\nu_j,\mathbf{k}_j)$. Repeated indices are implicitly integrated or summed over. Energy conservation and translational invariance impose frequency and momentum conservation laws, which imply the following relations for $G_0$ and $U$:
\begin{subequations}
\begin{align}
    G_{0;1^\prime|1} & = G_{0;\sigma_{1^\prime}|\sigma_1}(k_{1^\prime}|k_1) \nonumber \\
    & = \delta_{k_{1^\prime},k_1}  G_{0;\sigma_{1^\prime}|\sigma_1}(k_1) , \label{eq:FreqMomConservationLawsG0} \\
    U_{1^\prime 2^\prime|12} & = U_{\sigma_{1^\prime}\sigma_{2^\prime}|\sigma_1\sigma_2}(k_{1^\prime},k_{2^\prime}|k_1, k_2) \nonumber \\
    & = \delta_{k_{1^\prime}+k_{2^\prime},k_1+k_2} U_{\sigma_{1^\prime}\sigma_{2^\prime}|\sigma_1\sigma_2}(Q_r,k_r,k^\prime_{r}) . \label{eq:FreqMomConservationLawsU}
\end{align}
\label{eq:FreqMomConservationLawsG0U}
\end{subequations}
The full propagator $G$ and the two-particle vertex $V$ (also referred to as full vertex) obey equations identical to Eqs.~\eqref{eq:FreqMomConservationLawsG0} and~\eqref{eq:FreqMomConservationLawsU}, respectively. The frequency and momentum dependencies of two-particle objects (e.g., the bare interaction $U$ or the two-particle vertex $V$) are parametrized in terms of bosonic momenta $Q_r=(i\Omega_r,\textbf{Q}_r)$ and fermionic momenta $k^{(\prime)}_r=(i\nu^{(\prime)}_r,\textbf{k}^{(\prime)}_r)$, which are defined differently depending on the diagrammatic channel $r$ under consideration, i.e., the particle-hole ($ph$), the particle-hole crossed ($\overline{ph}$) or the particle-particle ($pp$) channel. Our conventions for those momenta are provided in Tab.~\ref{fig:FrequencyMomentumParametrization} for $r=ph,\overline{ph},pp$.

\begin{table}[t!]
\centering
\renewcommand{\arraystretch}{1.6}

\begin{tabular}{|c|c|c|c|}
\hline
\diagbox[width=1.6cm,height=1.2cm]{\scriptsize Param.}{$r$\ \ } & \textbf{$ph$} & \textbf{$\overline{ph}$} & \textbf{$pp$} \\
\hline

$\mathbf{Q}_r$ 
& $\mathbf{k}_{1} - \mathbf{k}_{1'}$
& $\mathbf{k}_1 - \mathbf{k}_{2'}$
& $\mathbf{k}_1 + \mathbf{k}_2$ \\

$\Omega_r$
& $\nu_1 - \nu_{1'}$
& $\nu_1 - \nu_{1'}$
& $\nu_1 + \nu_2$ \\
\hline

$\mathbf{k}_r$ 
& $\mathbf{k}_2$
& $\mathbf{k}_2$
& $\mathbf{k}_2$ \\

$\nu_r$
& $\nu_2 + \lfloor \Omega_{ph}/2 \rfloor$
& $\nu_2 + \lfloor \Omega_{\overline{ph}}/2\rfloor$
& $\nu_2 - \lceil \Omega_{pp}/2 \rceil$ \\
\hline

$\mathbf{k}_r'$ 
& $\mathbf{k}_{1'}$
& $\mathbf{k}_{2'}$
& $\mathbf{k}_{2'}$ \\

$\nu_r'$
& $\nu_{1'}+ \lfloor \Omega_{ph}/2 \rfloor$
& $\nu_{2'} + \lfloor \Omega_{\overline{ph}}/2 \rfloor$
& $\nu_{2'} - \lceil \Omega_{pp}/2 \rceil$ \\
\hline
\end{tabular}

\caption{Frequency and momentum conventions for the two-particle vertex $V$ in each diagrammatic channel $r$, with $r=ph,\overline{ph},pp$. Note that $\lceil ... \rceil$ and $\lfloor ... \rfloor$ round their argument up and down to the nearest bosonic Matsubara frequency, respectively.}
\label{fig:FrequencyMomentumParametrization}
\end{table}

\begin{figure}[t!]
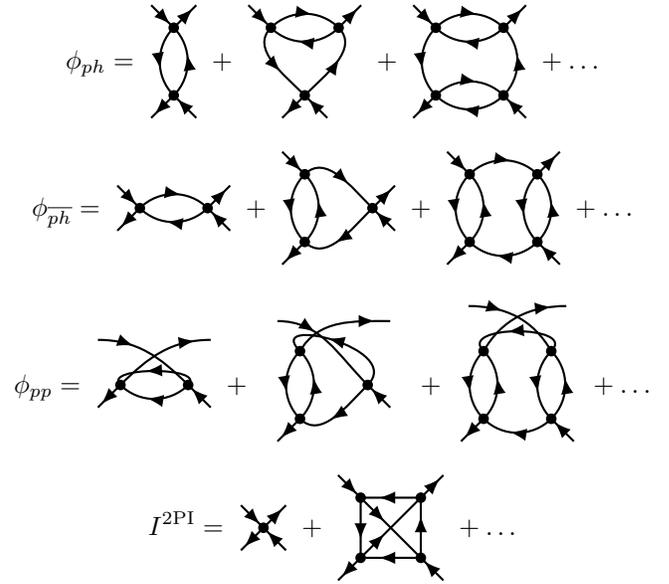

    \begin{gather*}
		\phi_{ph} = \ \tikzm{Diag_phi_ph_1}{
				\arrowsupperhalf{0}{0.45}
				\arrowslowerhalf{0}{-0.45}
				\phbubble{0}{0.45}{0.75}
				\barevertex{0}{0.45}
				\barevertex{0}{-0.45}
			} \ + \ \tikzm{Diag_phi_ph_2}{
				\arrowsupperhalffull{0}{0}{1.5}
				\phxbubble{-0.45}{0.45}{0.75}
				\arrowslowerhalf{0}{-0.45}
				\draw[lineWithArrowCenter] (-0.45,0.45) to [out=225, in=135] (0,-0.45);
				\draw[lineWithArrowCenter] (0,-0.45) to [out=45, in=315] (0.45,0.45);
				\barevertex{-0.45}{0.45}
				\barevertex{0.45}{0.45}
				\barevertex{0}{-0.45}
			} \ + \ \tikzm{Diag_phi_ph_3}{
			\arrowsupperhalffull{0}{0}{1.5}
			\arrowslowerhalffull{0}{0}{1.5}
			\phxbubble{-0.45}{0.45}{0.75}
			\phxbubble{-0.45}{-0.45}{0.75}
			\draw[lineWithArrowCenter] (-0.45,0.45) to [out=225, in=135] (-0.45,-0.45);
			\draw[lineWithArrowCenter] (0.45,-0.45) to [out=45, in=315] (0.45,0.45);
			\barevertex{0.45}{0.45}
			\barevertex{0.45}{-0.45}
			\barevertex{-0.45}{0.45}
			\barevertex{-0.45}{-0.45}
		} \ + \dots \\[8pt]
        \phi_{\overline{ph}} = \ \tikzm{Diag_phi_phx_1}{
				\arrowslefthalf{0}{0}
				\phxbubble{0}{0}{0.75}
				\arrowsrighthalf{0.9}{0}
				\barevertex{0}{0}
				\barevertex{0.9}{0}
			} \ + \ \tikzm{Diag_phi_phx_2}{
				\arrowslefthalffull{0.45}{0}{1.5}
				\phbubble{0}{0.45}{0.75}
				\draw[lineWithArrowCenter] (0,0.45) to [out=45, in=135] (0.9,0);
				\draw[lineWithArrowCenter] (0.9,0) to [out=225, in=315] (0,-0.45);
				\arrowsrighthalf{0.9}{0};
				\barevertex{0}{0.45}
				\barevertex{0}{-0.45}
				\barevertex{0.9}{0}
			} \ + \ \tikzm{Diag_phi_phx_3}{
			\phbubble{-0.45}{0.45}{0.75}
			\phbubble{0.45}{0.45}{0.75}
			\draw[lineWithArrowCenter] (-0.45,0.45) to [out=45, in=135] (0.45,0.45);
			\draw[lineWithArrowCenter] (0.45,-0.45) to [out=225, in=315] (-0.45,-0.45);
			\arrowsallfull{0}{0}{1.5}
			\barevertex{0.45}{0.45}
			\barevertex{0.45}{-0.45}
			\barevertex{-0.45}{0.45}
			\barevertex{-0.45}{-0.45}
		} \ + \dots \\[8pt]
		\phi_{pp} = \ \tikzm{Diag_phi_pp_1}{
				\arrowslefthalfp{0}{0}{0.75}
				\arrowsrighthalfp{0.9}{0}{0.75}
				\node at (0,0.6) {};
				\ppbubble{0}{0}{0.75}
				\node at (0.9,0.6) {};
				\barevertex{0}{0}
				\barevertex{0.9}{0}
			} \ + \ \tikzm{Diag_phi_pp_2}{
				\draw[lineWithArrowCenterEnd] (1.2,-0.3) -- (0.9,0);
				\draw[lineWithArrowCenterEnd] (0,-0.45) -- (-0.3,-0.75);
				\draw[lineWithArrowCenterEnd] (0,0.45) to [out=60, in=180] (1.2,0.85);
				\draw[lineWithArrowCenterStart] (-0.3,0.85) to [out=0, in=135] (0.9,0);
				\phbubble{0}{0.45}{0.75}
				\draw[lineWithArrowCenter] (0.9,0) .. controls ++(45:0.7) and ++(135:0.7) .. (0,0.45);
				\draw[lineWithArrowCenter] (0.9,0) to [out=225, in=315] (0,-0.45);
				\barevertex{0}{0.45}
				\barevertex{0}{-0.45}
				\barevertex{0.9}{0}
			} \ + \ \tikzm{Diag_phi_pp_3}{
			\arrowslowerhalffull{0}{0}{1.5}
			\draw[lineWithArrowCenterEnd] (-0.45,0.45) to [out=60, in=180] (0.65,1.05);
			\node at (0.65,1.05) {};
			\draw[lineWithArrowCenterStart] (-0.65,1.05) to [out=0, in=120] (0.45,0.45);
			\node at (-0.65,1.05) {};
			\phbubble{-0.45}{0.45}{0.75}
			\phbubble{0.45}{0.45}{0.75}
			\draw[lineWithArrowCenter] (0.45,0.45) .. controls ++(45:0.5) and ++(135:0.5) .. (-0.45,0.45);
			\draw[lineWithArrowCenter] (0.45,-0.45) to [out=225, in=315] (-0.45,-0.45);
			\barevertex{0.45}{0.45}
			\barevertex{0.45}{-0.45}
			\barevertex{-0.45}{0.45}
			\barevertex{-0.45}{-0.45}
		} \ + \dots \\[8pt]
            I^{\text{2PI}} = \ \tikzm{I2PIbarevertex}{
			\barevertexwithlegs{0}{0}
		} \ + \ \tikzm{I2PInontrivialdiagram}{
			\draw[lineArrowBis] (-0.4,0.4) -- (0.4,-0.4);
			\draw[lineArrowBis] (-0.4,-0.4) -- (0.4,0.4);
			\draw[lineArrow] (0.4,0.4) -- (-0.4,0.4);
			\draw[lineArrow] (0.4,-0.4) -- (-0.4,-0.4);
			\draw[lineArrow] (0.4,-0.4) -- (0.4,0.4);
			\draw[lineArrow] (-0.4,0.4) -- (-0.4,-0.4);
			\barevertex{-0.4}{0.4}
			\barevertex{-0.4}{-0.4}
			\barevertex{0.4}{0.4}
			\barevertex{0.4}{-0.4}
			\arrowsallfull{0}{0}{4./3.}
		} \ + \dots
    \end{gather*}
    \caption{Examples of diagrams contributing to the parquet decomposition~\eqref{eq:parquetDecomposition}. The vertices $\phi_r$ contain all diagrams that are 2PR in channel $r$, whereas the diagrams that are 2PI (i.e., not 2PR in any channel) are included in $I^{\text{2PI}}$. By definition, the diagrams contributing to $\phi_r$ can be split into two disconnected parts by cutting two propagator lines that form the $\Pi_r$ bubbles defined by Eqs.~\eqref{eq:Bubbles}. The bubbles $\Pi_{ph}$, $\Pi_{\overline{ph}}$ and $\Pi_{pp}$ are respectively made of two transverse antiparallel, two antiparallel and two parallel lines.}
    \label{fig:ParquetDecomposition}
\end{figure}

\begin{figure}[t!]
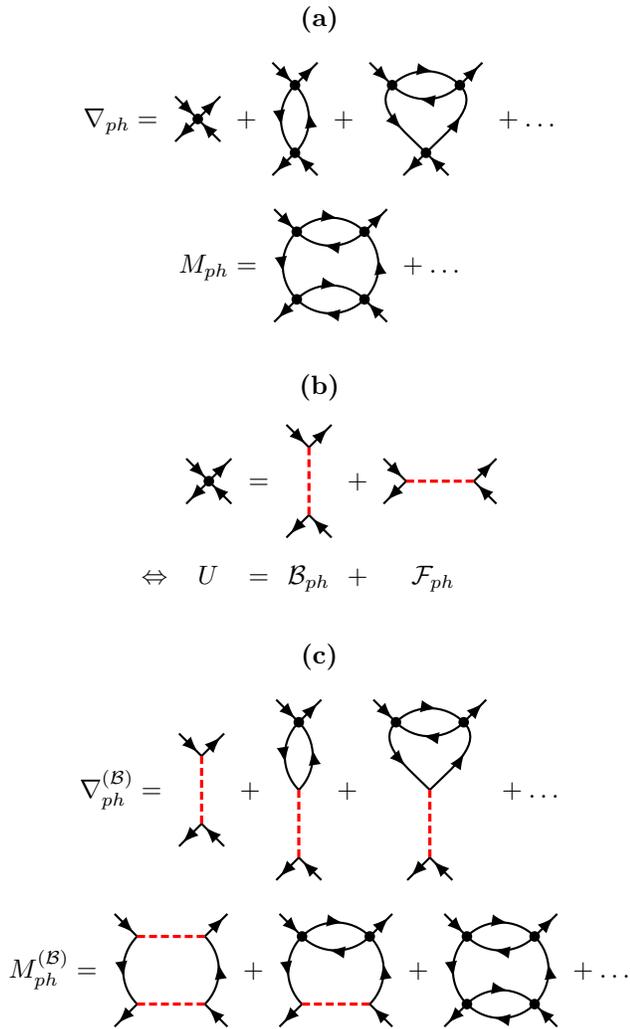

\centering
\refstepcounter{subfigure}\label{fig:UvsBreducibility_a}
  \textbf{(\alph{subfigure})}
	\begin{gather*}
		\nabla_{ph} = \ \tikzm{nablaphbarevertex}{
			\barevertexwithlegs{0}{0}
		} \ + \ \tikzm{Diag_nabla_ph_1}{
				\arrowsupperhalf{0}{0.45}
				\arrowslowerhalf{0}{-0.45}
				\phbubble{0}{0.45}{0.75}
				\barevertex{0}{0.45}
				\barevertex{0}{-0.45}
			} \ + \ \tikzm{Diag_nabla_ph_2}{
				\arrowsupperhalffull{0}{0}{1.5}
				\phxbubble{-0.45}{0.45}{0.75}
				\arrowslowerhalf{0}{-0.45}
				\draw[lineWithArrowCenter] (-0.45,0.45) to [out=225, in=135] (0,-0.45);
				\draw[lineWithArrowCenter] (0,-0.45) to [out=45, in=315] (0.45,0.45);
				\barevertex{-0.45}{0.45}
				\barevertex{0.45}{0.45}
				\barevertex{0}{-0.45}
			} \ + \dots \\[8pt]
            M_{ph} = \ \tikzm{Diag_M_ph}{
			\arrowsupperhalffull{0}{0}{1.5}
			\arrowslowerhalffull{0}{0}{1.5}
			\phxbubble{-0.45}{0.45}{0.75}
			\phxbubble{-0.45}{-0.45}{0.75}
			\draw[lineWithArrowCenter] (-0.45,0.45) to [out=225, in=135] (-0.45,-0.45);
			\draw[lineWithArrowCenter] (0.45,-0.45) to [out=45, in=315] (0.45,0.45);
			\barevertex{0.45}{0.45}
			\barevertex{0.45}{-0.45}
			\barevertex{-0.45}{0.45}
			\barevertex{-0.45}{-0.45}
		} \ + \dots
	\end{gather*}
    
\vspace{0.3cm}

\refstepcounter{subfigure}\label{fig:UvsBreducibility_b}
\textbf{(\alph{subfigure})}

\vspace{-0.4cm}

	\begin{alignat*}{3}
		\tikzm{barevertexBFsplitting}{
			\barevertexwithlegs{0}{0}
		} \ &= \ \tikzm{Diag_Br}{
				\arrowsupperhalf{0}{0.45}
				\arrowslowerhalf{0}{-0.45}
                    \draw[red, very thick, dash pattern=on 3pt off 1.5pt] (0,0.45) -- (0,-0.45);
			} \ &&+ \ \tikzm{Diag_Fr}{
				\arrowslefthalf{0}{0}
				\arrowsrighthalf{0.9}{0}
                    \draw[red, very thick, dash pattern=on 3pt off 1.5pt] (0,0) -- (0.9,0);
			} \\[5pt]
        \Leftrightarrow \hspace{0.3cm} U \hspace{0.3cm} &= \hspace{0.15cm} \mathcal{B}_{ph} &&+ \hspace{0.5cm} \mathcal{F}_{ph}
	\end{alignat*}

\vspace{0.3cm}

\refstepcounter{subfigure}\label{fig:UvsBreducibility_c}
  \textbf{(\alph{subfigure})}
	\begin{gather*}
		\nabla_{ph}^{(\mathcal{B})} = \ \tikzm{Diag_BrnablaB}{
				\arrowsupperhalf{0}{0.45}
				\arrowslowerhalf{0}{-0.45}
                    \draw[red, very thick, dash pattern=on 3pt off 1.5pt] (0,0.45) -- (0,-0.45);
			} \ + \ \tikzm{Diag_nablaB_ph_1}{
				\arrowsupperhalf{0}{0.9}
				\arrowslowerhalf{0}{-0.9}
				\phbubble{0}{0.9}{0.75}
                    \draw[red, very thick, dash pattern=on 3pt off 1.5pt] (0,0) -- (0,-0.9);
				\barevertex{0}{0.9}
			} \ + \ \tikzm{Diag_nablaB_ph_2}{
				\arrowsupperhalffull{0}{0+0.45}{1.5}
				\phxbubble{-0.45}{0.45+0.45}{0.75}
				\arrowslowerhalf{0}{-0.45-0.45}
				\draw[lineWithArrowCenter] (-0.45,0.45+0.45) to [out=225, in=135] (0,-0.45+0.45);
				\draw[lineWithArrowCenter] (0,-0.45+0.45) to [out=45, in=315] (0.45,0.45+0.45);
                    \draw[red, very thick, dash pattern=on 3pt off 1.5pt] (0,0) -- (0,-0.9);
				\barevertex{-0.45}{0.45+0.45}
				\barevertex{0.45}{0.45+0.45}
			} \ + \dots \\[8pt]
            M_{ph}^{(\mathcal{B})} = \ \tikzm{Diag_MB_ph_1}{
			\arrowsupperhalffull{0}{0}{1.5}
			\arrowslowerhalffull{0}{0}{1.5}
			\draw[lineWithArrowCenter] (-0.45,0.45) to [out=225, in=135] (-0.45,-0.45);
			\draw[lineWithArrowCenter] (0.45,-0.45) to [out=45, in=315] (0.45,0.45);
                \draw[red, very thick, dash pattern=on 3pt off 1.5pt] (-0.45,-0.45) -- (0.45,-0.45);
                \draw[red, very thick, dash pattern=on 3pt off 1.5pt] (-0.45,0.45) -- (0.45,0.45);
		} \ + \ \tikzm{Diag_MB_ph_2}{
			\arrowsupperhalffull{0}{0}{1.5}
			\arrowslowerhalffull{0}{0}{1.5}
			\phxbubble{-0.45}{0.45}{0.75}
			\draw[lineWithArrowCenter] (-0.45,0.45) to [out=225, in=135] (-0.45,-0.45);
			\draw[lineWithArrowCenter] (0.45,-0.45) to [out=45, in=315] (0.45,0.45);
                \draw[red, very thick, dash pattern=on 3pt off 1.5pt] (-0.45,-0.45) -- (0.45,-0.45);
			\barevertex{0.45}{0.45}
			\barevertex{-0.45}{0.45}
		} \ + \ \tikzm{Diag_MB_ph_3}{
			\arrowsupperhalffull{0}{0}{1.5}
			\arrowslowerhalffull{0}{0}{1.5}
			\phxbubble{-0.45}{0.45}{0.75}
			\phxbubble{-0.45}{-0.45}{0.75}
			\draw[lineWithArrowCenter] (-0.45,0.45) to [out=225, in=135] (-0.45,-0.45);
			\draw[lineWithArrowCenter] (0.45,-0.45) to [out=45, in=315] (0.45,0.45);
			\barevertex{0.45}{0.45}
			\barevertex{0.45}{-0.45}
			\barevertex{-0.45}{0.45}
			\barevertex{-0.45}{-0.45}
		} \ + \dots
	\end{gather*}
    \caption{Diagrammatic classification underlying the SBE decomposition of the two-particle vertex $V$ in the $ph$ channel. \textbf{(a)} Examples of diagrams of the $U$-reducible vertex $\nabla_{ph}$ and the SBE rest function $M_{ph}$ introduced in the SBE decomposition based on $U$-reducibility, i.e., Eq.~\eqref{eq:SBEDecomposition}. The vertex $\nabla_{ph}$ is the sum of all diagrams of $V$ which are 2PR and $U$-reducible in the $ph$ channel whereas the SBE rest function $M_{ph}$ contains all 2PR diagrams in the $ph$ channel that are not $U$-reducible. \textbf{(b)} Splitting of the bare interaction $U$ into a bosonic part $\mathcal{B}_{ph}$ and a fermionic part $\mathcal{F}_{ph}$, which corresponds to Eqs.~\eqref{eq:UrBrFr} and~\eqref{eq:UrBrFrfull} for $r=ph$. \textbf{(c)} Examples of diagrams contributing to the $\mathcal{B}_{ph}$-reducible vertex $\nabla^{(\mathcal{B})}_{ph}$ and the corresponding SBE rest function $M^{(\mathcal{B})}_{ph}$ introduced in the SBE decomposition based on $\mathcal{B}$-reducibility, i.e., Eq.~\eqref{eq:SBEDecompositionBreducibility}. The vertices $\nabla^{(\mathcal{B})}_{ph}$ and $M^{(\mathcal{B})}_{ph}$ include all $\mathcal{B}_{ph}$-reducible and 2PR $\mathcal{B}_{ph}$-irreducible diagrams in the $ph$ channel, respectively.}
    \label{fig:UvsBreducibility}
\end{figure}

\subsection{Parquet decomposition}

The parquet decomposition of the two-particle vertex $V$, which is now extensively used in quantum many-body techniques (e.g., in extensions of DMFT~\cite{BickersSelfConsistent2004,Rohringer2018}), also relies on the aforementioned diagrammatic channels. This decomposition is defined as
\begin{equation}
    V = \phi_{ph} + \phi_{\overline{ph}} + \phi_{pp} + I^{\text{2PI}},
    \label{eq:parquetDecomposition}
\end{equation}
where $\phi_r$ contains all \emph{two-particle-reducible} (2PR) diagrams contributing to $V$ in the diagrammatic channel $r$. By definition, 2PR diagrams can be split into two disconnected parts by cutting two propagator lines, usually termed as \emph{bubble} [see the definitions of Eqs.~\eqref{eq:Bubbles}]: a 2PR diagram $\phi_r$ is then assigned to a given diagrammatic channel $r$, depending on the nature of that bubble. With this in mind, the irreducible vertex $I^{\text{2PI}}$ is by construction the sum of all diagrams of $V$ that are not 2PR in any channel $r$, i.e., 2PI. We refer to Fig.~\ref{fig:ParquetDecomposition} for concrete examples regarding this diagrammatic classification.

The 2PR vertices $\phi_r$ fulfill the Bethe--Salpeter equations~\cite{Bickers1991,BickersSelfConsistent2004,Kugler2018b},
\begin{equation}
\phi_r = I_r \circ \Pi_r \circ V = V \circ \Pi_r \circ I_r,
\label{eq:BetheSalpeterEquations}
\end{equation}
where the vertex $I_r$ contains all diagrams that are not 2PR in channel $r$ (or, equivalently, that are 2PI in channel $r$), i.e.,
\begin{equation}
I_r = V - \phi_r,
\label{eq:DefinitionIr}
\end{equation}
and the bubbles $\Pi_r$ are given as
\begin{subequations}
\begin{align}
	\Pi_{ph;1^\prime 2^\prime|12} &= - G_{1^\prime |2}G_{2^\prime |1} , \\
	\Pi_{\overline{ph};1^\prime 2^\prime|12} &= G_{1^\prime |1}G_{2^\prime |2} , \\
	\Pi_{pp;1^\prime 2^\prime|12} &= \tfrac{1}{2} G_{1^\prime |1}G_{2^\prime |2} .
\end{align}
\label{eq:Bubbles}
\end{subequations}
The definition of the $\circ$ product depends on the channel $r$ under consideration. Explicitly, we have
\begin{subequations}
    \begin{align}
       ph~:\quad [A\circ B]_{1^\prime 2^\prime|12} &= A_{42^\prime |32}B_{1^\prime 3|14}, \\
       \overline{ph}~:\quad [A\circ B]_{1^\prime 2^\prime|12} &= A_{1^\prime 4|32}B_{32^\prime |14}, \\
        pp~:\quad [A\circ B]_{1^\prime 2^\prime|12} &= A_{1^\prime 2^\prime|34}B_{34|12}, \label{eq:circproductppchannel}
    \end{align}
\label{eq:Definitioncircproduct}
\end{subequations}
where $A$ and $B$ are both arbitrary four-point functions, such as $\phi_r$, $I_r$, $\Pi_r$ or $V$~\cite{Gievers2022}.

\subsection{SBE decomposition}

The SBE formalism provides an alternative decomposition of the two-particle vertex $V$. It is based on another diagrammatic criterion, called $U$-reducibility, used to split the set of 2PR diagrams of $\phi_r$ into two subclasses of diagrams~\cite{Krien2019,Gievers2022}, namely \emph{$U\!$-reducible} and \emph{$U\!$-irreducible} diagrams. We review the original SBE formulation in Sec.~\ref{sec:SBE-original}. For non-local interactions, an extension by the notion of \textit{$\mathcal{B}$-reducibility} is needed, which is discussed in Sec.~\ref{sec:B-reducibility}.

\subsubsection{Original formulation based on $U\!$-reducibility}
\label{sec:SBE-original}

We first define the meaning of $U$-reducibility. A 2PR diagram $\phi_r$ is called $U$-reducible if cutting any of the contained bare interaction vertices, $U=\tikzm{barevertexText}{\barevertexwithlegs{0}{0}}$, splits it into two disconnected parts. In the opposite case, it is referred to as $U$-irreducible. More specifically, the $U$-reducible diagrams of $\phi_r$ are termed $U$-reducible in channel $r$. Therefore, all $U$-reducible diagrams are also 2PR in the same channel $r$, which reflects the two-particle nature of the bare vertex $U$. The trivial exception is the bare interaction $U$ itself, which is indeed 2PI but $U$-reducible in all channels. Concrete examples for the diagrammatic classification underlying the SBE decomposition are given in Fig.~\ref{fig:UvsBreducibility}(\ref{fig:UvsBreducibility_a}) for the $ph$ channel.

Splitting the 2PR vertex $\phi_r$ into $U$-reducible and $U$-irreducible contributions yields
\begin{equation}
    \phi_r = \nabla_r + M_r -U,
\label{eq:phinablaM}
\end{equation}
where $\nabla_r$ contains all $U$-reducible diagrams in channel $r$, which includes the diagram for the bare interaction $U$. As mentioned previously, the bare interaction is 2PI itself and should therefore not be contained in $\phi_r$, hence the $-U$ correction. As a result, $M_r$ collects all 2PR diagrams in channel $r$ that are $U$-irreducible.

We then infer the SBE decomposition by inserting Eq.~\eqref{eq:phinablaM} into the parquet decomposition~\eqref{eq:parquetDecomposition}, which yields~\cite{Krien2019}
\begin{align}
    V &= \sum_{r=ph,\overline{ph},pp}(\nabla_r+M_r-U)+I^\mathrm{2PI} \nonumber \\
    &\equiv \sum_r \nabla_r + \mathcal{I}^{U\text{irr}}-2U,
    \label{eq:SBEDecomposition}
\end{align}
where $\mathcal{I}^{U\text{irr}}$ is the sum of all $U$-irreducible diagrams of $V$ and can be expressed as
\begin{equation}
    \mathcal{I}^{U\text{irr}} = \sum_r M_r + I^{\text{2PI}} - U.
    \label{eq:ExpressionIUirr}
\end{equation}
The central idea of the SBE decomposition amounts to rewriting the vertices $\nabla_r$ in terms of bosonic vertices, namely bosonic propagators $w_r$ and fermion-boson vertices $\lambda_r$ (also called Yukawa couplings or Hedin vertices). Owing to the $U$-reducible nature of the diagrams contributing to $\nabla_r$, one can indeed use the \emph{exact} parametrization
\begin{equation}
    \nabla_r (Q_r,k_r,k^{\prime}_r) = \overline{\lambda}_r(Q_r,k_r) \fcirc w_r(Q_r) \fcirc \lambda_r(Q_r,k^{\prime}_r),
    \label{eq:nablaSBE}
\end{equation}
with the $\fcirc$ product defined in the same way as the $\circ$ product with Eq.~\eqref{eq:Definitioncircproduct}, but excluding summation over frequencies or momenta (i.e., the $\fcirc$ product is only defined for spin indices)~\cite{Gievers2022}.

It can be shown that the objects thus introduced by the SBE decomposition satisfy the following self-consistent equations~\cite{Gievers2022}:
\begin{subequations}
\begin{align}
	w_r &= U + w_r \fcirc \lambda_r \circ \Pi_r \fcirc U \\
        &= U + U \fcirc \Pi_r \circ \overline{\lambda}_r \fcirc w_r , \\
	\lambda_r & = \mathbf{1}_r + \mathbf{1}_r \circ \Pi_r \circ \mathcal{I}_r, \\
        \overline{\lambda}_r & = \mathbf{1}_r + \mathcal{I}_r \circ \Pi_r \circ \mathbf{1}_r, \\
	M_r &= (\mathcal{I}_r - M_r ) \circ \Pi_r \circ \mathcal{I}_r \label{eq:SBEeqM1} \\
        &= \mathcal{I}_r \circ \Pi_r \circ (\mathcal{I}_r - M_r), \label{eq:SBEeqM2}
\end{align}
\label{eq:selfconsistentSBEequations}
\end{subequations}
where the irreducible vertex $\mathcal{I}_r$ collects all diagrams that are $U$-irreducible in channel $r$, namely,
\begin{equation}
    \mathcal{I}_r = V - \nabla_r.
    \label{eq:DefinitionmathcalIr}
\end{equation}
Equations~\eqref{eq:selfconsistentSBEequations}, which can be seen as the SBE versions of the Bethe--Salpeter equations~\eqref{eq:BetheSalpeterEquations}, are referred to as \emph{SBE equations}. They also rely on the identity vertices $\mathbf{1}_{r}$ defined only in spin space \cite{GieversPhDthesis}:
\begin{subequations}
	\begin{align}
		\mathbf{1}_{ph;\sigma_{1^{\prime}}\sigma_{2^{\prime}}|\sigma_1 \sigma_2} &= \delta_{\sigma_{1^{\prime}},\sigma_2} \phantom{,} \delta_{\sigma_{2^{\prime}},\sigma_1},\\
		\mathbf{1}_{\overline{ph};\sigma_{1^{\prime}}\sigma_{2^{\prime}}|\sigma_1 \sigma_2} &= \delta_{\sigma_{1^{\prime}},\sigma_1} \phantom{,} \delta_{\sigma_{2^{\prime}},\sigma_2},\\
	\mathbf{1}_{pp;\sigma_{1^{\prime}} \sigma_{2^{\prime}}|\sigma_1 \sigma_2} &= \delta_{\sigma_{1^{\prime}},\sigma_1} \phantom{,} \delta_{\sigma_{2^{\prime}},\sigma_2},
	\end{align}
    \label{eq:Identityfcirc}
\end{subequations}
which are not to be confused with the identity vertices $\mathds{1}_{r}$ defined with respect to all indices, that will also be involved in our forthcoming derivations. The latter read
\begin{subequations}
	\begin{align}
		\mathds{1}_{ph;1^{\prime}2^{\prime}|12} &= \delta_{1^{\prime},2} \phantom{,} \delta_{2^{\prime},1},\\
		\mathds{1}_{\overline{ph};1^{\prime}2^{\prime}|12} &= \delta_{1^{\prime},1} \phantom{,} \delta_{2^{\prime},2},\\
		\mathds{1}_{pp;1^{\prime}2^{\prime}|12} &= \delta_{1^{\prime},1} \phantom{,} \delta_{2^{\prime},2},
	\end{align}
    \label{eq:Identitycirc}
\end{subequations}
with $\delta_{i,j} = \delta_{k_i,k_j} \phantom{,} \delta_{\sigma_i,\sigma_j}$. The definitions of the identity vertices $\mathbf{1}_r$ and $\mathds{1}_r$, given by Eqs.~\eqref{eq:Identityfcirc} and~\eqref{eq:Identitycirc}, are determined from the conditions $V = \mathbf{1}_r \fcirc V = V \fcirc \mathbf{1}_r$ and $V = \mathds{1}_r \circ V = V \circ \mathds{1}_r$, respectively~\footnote{We refer to App. C.2 of Ref.~\cite{GieversPhDthesis} for more details on the construction of the identity vertices $\mathbf{1}_r$ and $\mathds{1}_r$.}. Those identity vertices are also used to define inverses. More specifically, we have
\begin{equation}
A^{-1} \fcirc A = A \fcirc A^{-1} = \mathbf{1}_r ,
\end{equation}
with $A$ a four-point object with respect to spin indices, and
\begin{equation}
B^{-1} \circ B = B \circ B^{-1} = \mathds{1}_r ,
\end{equation}
with $B$ a four-point object with respect to all indices (spin indices, frequencies and momenta). In particular, the bosonic propagators $w_r=w_{r;\sigma_{1^{\prime}}\sigma_{2^{\prime}}|\sigma_1 \sigma_2}(Q_r)$ and the fermion-boson vertices $\lambda_r=\lambda_{r;\sigma_{1^{\prime}}\sigma_{2^{\prime}}|\sigma_1 \sigma_2}(Q_r,k_r)$ introduced in Eq.~\eqref{eq:nablaSBE} are four-point vertices with respect to spin indices, but not with respect to frequencies or momenta.

We have introduced the SBE decomposition and the related equations. Let us focus on the merits of this approach. With the parametrization of Eq.~\eqref{eq:nablaSBE}, we efficiently exploit the frequency and momentum conservation laws [cf.\ Eqs.~\eqref{eq:FreqMomConservationLawsG0U}] by replacing the two-particle vertex $V(Q_r,k_r,k^{\prime}_r)$, with its complex frequency and momentum dependencies, by simpler objects, namely $\lambda_r(Q_r,k_r)$ and $w_r(Q_r)$, which depend on fewer frequencies and momenta. This particularly simple frequency and momentum parametrization of the $U$-reducible vertices $\nabla_r$ thus makes the SBE approach very amenable to efficient and computationally tractable approximations, while also providing deeper physical insight.

Indeed, competing orders in many-fermion systems are naturally described by introducing bosonic degrees of freedom, and this is exactly what the SBE approach does: the parametrization~\eqref{eq:nablaSBE} implicitly (i.e., without using a Hubbard--Stratonovich transformation to include a bosonic field in the theory) introduces bosons in our framework through their bosonic propagators $w_r$ and fermion-boson vertices $\lambda_r$. Furthermore, with the parametrization of Eq.~\eqref{eq:nablaSBE}, all $U$-reducible diagrams contributing to $\nabla_r$ are interpreted as the exchange of a single boson with momentum $Q_r$, which explains the name of the SBE decomposition. One can also note that the $U$-irreducible diagrams contributing to the SBE rest functions $M_r$ can be regarded as multi-boson exchange processes.

\subsubsection{Extension to non-local interactions based on $\mathcal{B}$-reducibility}
\label{sec:B-reducibility}

One advantage of our interaction flow is the possibility of introducing a temperature flow for systems with retarded interactions (cf.\ Sec.~\ref{sec:TflowRetardedInteractions}). For this, it is necessary to extend the SBE formalism to the case of a non-local bare interaction, where $U_{\sigma_{1^\prime}\sigma_{2^\prime}|\sigma_1\sigma_2}(Q_r,k_r,k^\prime_{r})$, introduced in Eq.~\eqref{eq:FreqMomConservationLawsU}, does depend on $Q_r$, $k_r$ and $k^\prime_{r}$. In this situation, the bosonic propagators and the fermion-boson vertices also depend, in principle, on those three momenta, namely $w_r=w_r(Q_r,k_r,k^\prime_{r})$ and $\lambda_r=\lambda_r(Q_r,k_r,k^\prime_{r})$. In particular, this can be directly seen for the bosonic propagators at first order in the bare interaction since $w_r(Q_r,k_r,k^\prime_{r})=U(Q_r,k_r,k^\prime_{r})+\mathcal{O}(U^2)$. It should, however, be emphasized that the parametrization $\nabla_r = \overline{\lambda}_r \fcirc w_r \fcirc \lambda_r$ is still valid for non-local interactions, whereas its formulation in terms of simplified frequency and momentum dependencies for $w_r$ and $\lambda_r$ as in Eq.~\eqref{eq:nablaSBE} is restricted to local interactions within the SBE formalism based on $U$-reducibility, as originally developed in Ref.~\cite{Krien2019}.

Importantly, the computational advantage of the SBE approach is completely spoiled if both $w_r$ and $\lambda_r$ depend on $Q_r$, $k_r$ and $k^\prime_{r}$. However, the SBE decomposition can still be efficiently generalized to non-local interactions by splitting the bare interaction $U(Q_r,k_r,k^\prime_{r})$ into a bosonic part $\mathcal{B}_r$ and a fermionic part $\mathcal{F}_r$, where the functions $\mathcal{F}_r$ encompass the whole fermionic momentum dependence of $U(Q_r,k_r,k^\prime_{r})$~\cite{AlEryani2024,AlEryani2025}. In other words, this amounts to setting
\begin{equation}
    U(Q_r,k_r,k^\prime_{r}) = \mathcal{B}_r(Q_r) + \mathcal{F}_r(Q_r,k_r,k^\prime_{r}),
    \label{eq:UrBrFr}
\end{equation}
where $U$, $\mathcal{B}_r$, and $\mathcal{F}_r$ are all four-point vertices with respect to spin indices. Similarly to $U$, both $\mathcal{B}_r$ and $\mathcal{F}_r$ satisfy relations inherent to energy conservation and translational invariance, namely,
\begin{subequations}
\begin{align}
\mathcal{B}_{r;1^\prime 2^\prime|12} & = \mathcal{B}_{r;\sigma_{1^\prime}\sigma_{2^\prime}|\sigma_1\sigma_2}(k_{1^\prime},k_{2^\prime}|k_1, k_2) \nonumber \\
& = \mathcal{B}_{r;\sigma_{1^\prime}\sigma_{2^\prime}|\sigma_1\sigma_2}(Q_r) \delta_{k_{1^\prime}+k_{2^\prime},k_1+k_2}, \\
\mathcal{F}_{r;1^\prime 2^\prime|12} & = \mathcal{F}_{r;\sigma_{1^\prime}\sigma_{2^\prime}|\sigma_1\sigma_2}(k_{1^\prime},k_{2^\prime}|k_1, k_2) \nonumber \\
& = \mathcal{F}_{r;\sigma_{1^\prime}\sigma_{2^\prime}|\sigma_1\sigma_2}(Q_r,k_r,k^\prime_{r}) \delta_{k_{1^\prime}+k_{2^\prime},k_1+k_2},
\end{align}
\label{eq:MomFreqConservationLawsBrFr}
\end{subequations}
and thus, according to Eqs.~\eqref{eq:FreqMomConservationLawsU},~\eqref{eq:UrBrFr}, and~\eqref{eq:MomFreqConservationLawsBrFr}, we also have
\begin{equation}
U_{1^\prime 2^\prime|12} = \mathcal{B}_{r;1^\prime 2^\prime|12} + \mathcal{F}_{r;1^\prime 2^\prime|12}.
\label{eq:UrBrFrfull}
\end{equation}

We then introduce the diagrammatic criterion of \emph{$\mathcal{B}$-reducibility} by defining the vertex $\nabla^{(\mathcal{B})}_r$ as the sum of all diagrams that are reducible with respect to the bosonic part $\mathcal{B}_r$ instead of the whole bare interaction $U$. We thus refer to the latter diagrams as $\mathcal{B}_r$-reducible and, most importantly, the vertex $\nabla^{(\mathcal{B})}_r$ can be parametrized as
\begin{equation}
    \nabla^{(\mathcal{B})}_r(Q_r,k_r,k^\prime_{r}) = \overline{\lambda}^{(\mathcal{B})}_r(Q_r,k_r) \fcirc w^{(\mathcal{B})}_r(Q_r) \fcirc \lambda^{(\mathcal{B})}_r(Q_r,k^{\prime}_r),
    \label{eq:nablaBreducibility}
\end{equation}
which still allows us to work with objects whose frequency and momentum dependencies are much simpler than those of the full vertex $V$, namely $w^{(\mathcal{B})}_r(Q_r)$ and $\lambda^{(\mathcal{B})}_r(Q_r,k_r)$. The SBE decomposition of $V$ then becomes
\begin{equation}
    V = \sum_r \left(\nabla^{(\mathcal{B})}_r + M^{(\mathcal{B})}_r - \mathcal{B}_r\right) + I^{\text{2PI}},
    \label{eq:SBEDecompositionBreducibility}
\end{equation}
which is analogous to its original version formulated by Eq.~\eqref{eq:SBEDecomposition}. Note in particular that the double-counting correction now involves $\mathcal{B}_r$ instead of $U$.

Concrete examples of diagrams contributing to $\nabla^{(\mathcal{B})}_{ph}$ and $M^{(\mathcal{B})}_{ph}$ are given in Fig.~\ref{fig:UvsBreducibility} with a comparison to the corresponding vertices $\nabla_{ph}$ and $M_{ph}$ introduced in the original SBE decomposition based on $U$-reducibility. Essentially, the original $U$-reducible diagrams in channel $r$ are now subdivided into $\mathcal{B}_r$- and $\mathcal{F}_r$-reducible parts:
\begin{align}
    \nabla_r=\nabla_r^{(\mathcal{B})}+\nabla_r^{(\mathcal{F})},
\end{align}
where $\nabla_r^{(\mathcal{F})}$ is $\mathcal{B}_r$-irreducible and is thus part of the new multi-boson vertex $M_r^{(\mathcal{B})}$,
\begin{align}
    M_r^{(\mathcal{B})}=M_r+\nabla_r^{(\mathcal{F})}-\mathcal{F}_r,
\end{align}
which is 2PR in channel $r$ but $\mathcal{B}_r$-irreducible~\footnote{The subtraction of $\mathcal{F}_r$ from $\nabla_r^{(\mathcal{F})}$ is needed because $M_r^{(\mathcal{B})}$ only exhibits two-particle-reducible diagrams.}.

For completeness, we review here the derivation of the SBE equations within this formulation based on $\mathcal{B}$-reducibility, following analogous steps as in Ref.~\cite{Gievers2022} that focuses on $U$-reducibility~\footnote{The derivation of the original SBE equations~\eqref{eq:selfconsistentSBEequations}, as outlined in Ref.~\cite{Gievers2022}, can then be directly recovered from our developments of Sec.~\ref{sec:B-reducibility} by setting $\mathcal{F}_r=0$ (and therefore $\mathcal{B}_r=U$)}.

We start by defining the $\mathcal{B}_r$-irreducible vertices,
\begin{equation}
    \mathcal{I}_r^{(\mathcal{B})} = V - \nabla_r^{(\mathcal{B})},
    \label{eq:DefinitionmathcalIrBreducibility}
\end{equation}
which are the counterparts of $\mathcal{I}_r$ defined by Eq.~\eqref{eq:DefinitionmathcalIr} for the original formalism based on $U$-reducibility. We can therefore replace the full vertex $V$ by $\nabla_r^{(\mathcal{B})}+\mathcal{I}_r^{(\mathcal{B})}$ in the Bethe--Salpeter equations~\eqref{eq:BetheSalpeterEquations}, which gives
\begin{align}
    \phi_r &= I_r \circ \Pi_r \circ V \nonumber \\
    &= I_r \circ \Pi_r \circ \left(\nabla_r^{(\mathcal{B})}+\mathcal{I}_r^{(\mathcal{B})}\right) \nonumber \\
    &= I_r \circ \Pi_r \circ \nabla_r^{(\mathcal{B})} + \mathcal{B}_r \circ \Pi_r \circ \mathcal{I}_r^{(\mathcal{B})} \nonumber \\
    &\phantom{=} + \left(I_r - \mathcal{B}_r\right) \circ \Pi_r \circ \mathcal{I}_r^{(\mathcal{B})}.
\end{align}
In that way, we have split the 2PR vertex $\phi_r$ into its $\mathcal{B}_r$-reducible and $\mathcal{B}_r$-irreducible parts, i.e., $I_r \circ \Pi_r \circ \nabla_r^{(\mathcal{B})} + \mathcal{B}_r \circ \Pi_r \circ \mathcal{I}_r^{(\mathcal{B})}$ and $\left(I_r - \mathcal{B}_r\right) \circ \Pi_r \circ \mathcal{I}_r^{(\mathcal{B})}$, respectively. Furthermore, we can also express $\phi_r$ as (see Eq.~\eqref{eq:phinablaM} for comparison)
\begin{equation}
    \phi_r = \nabla^{(\mathcal{B})}_r + M^{(\mathcal{B})}_r -\mathcal{B}_r,
    \label{eq:phirBreducibility}
\end{equation}
where $\nabla^{(\mathcal{B})}_r-\mathcal{B}_r$ and $M^{(\mathcal{B})}_r$ are the $\mathcal{B}_r$-reducible and $\mathcal{B}_r$-irreducible terms, respectively. By identification, we obtain
\begin{subequations}
    \begin{align}
        \nabla^{(\mathcal{B})}_r-\mathcal{B}_r &= I_r \circ \Pi_r \circ \nabla_r^{(\mathcal{B})} + \mathcal{B}_r \circ \Pi_r \circ \mathcal{I}_r^{(\mathcal{B})}, \label{eq:nablaMinusBexpression} \\
        M^{(\mathcal{B})}_r &= \left(I_r - \mathcal{B}_r\right) \circ \Pi_r \circ \mathcal{I}_r^{(\mathcal{B})}. \label{eq:MBexpression}
    \end{align}
\end{subequations}
Isolating $\nabla^{(\mathcal{B})}_r$ on the left-hand side, Eq.~\eqref{eq:nablaMinusBexpression} can be rewritten as
\begin{equation}
    \nabla^{(\mathcal{B})}_r = \left(\mathds{1}_{r} - I_r \circ \Pi_r\right)^{-1} \circ \left(\mathcal{B}_r + \mathcal{B}_r \circ \Pi_r \circ \mathcal{I}_r^{(\mathcal{B})}\right).
    \label{eq:nablaInverse}
\end{equation}
An insertion of the so-called extended Bethe--Salpeter equations $(\mathds{1}_r-I_r\circ\Pi_r)^{-1}=\mathds{1}_r+V\circ\Pi_r$~\cite{Kugler2018b,Gievers2022} yields
\begin{equation}
    \nabla^{(\mathcal{B})}_r = \left(\mathds{1}_r + V \circ \Pi_r\right) \circ \left(\mathcal{B}_r + \mathcal{B}_r \circ \Pi_r \circ \mathcal{I}_r^{(\mathcal{B})}\right).
    \label{eq:nablaBrVBrIBr}
\end{equation}
Using relations such as $\mathds{1}_r \circ \mathcal{B}_r = \mathcal{B}_r = \mathbf{1}_r \fcirc \mathcal{B}_r$ and $\mathcal{B}_r \circ \mathds{1}_r = \mathcal{B}_r = \mathcal{B}_r \fcirc \mathbf{1}_r$, as well as $V=\nabla_r^{(\mathcal{B})}+\mathcal{I}_r^{(\mathcal{B})}$, we show from Eq.~\eqref{eq:nablaBrVBrIBr} that
\begin{align}
    \nabla_r & = \left(\mathbf{1}_r + V \circ \Pi_r \circ \mathbf{1}_r\right) \fcirc \mathcal{B}_r \fcirc \left(\mathbf{1}_r + \mathbf{1}_r \circ \Pi_r \circ \mathcal{I}_r^{(\mathcal{B})}\right) \nonumber \\
    & = \left(\mathbf{1}_r + \mathcal{I}_r^{(\mathcal{B})} \circ \Pi_r \circ \mathbf{1}_r + \nabla_r^{(\mathcal{B})} \circ \Pi_r \circ \mathbf{1}_r\right) \fcirc \mathcal{B}_r \nonumber \\
    & \phantom{=} \fcirc \left(\mathbf{1}_r + \mathbf{1}_r \circ \Pi_r \circ \mathcal{I}_r^{(\mathcal{B})}\right) \nonumber \\
    & = \overline{\lambda}^{(\mathcal{B})}_r \fcirc \left(\mathbf{1}_r + w^{(\mathcal{B})}_r \fcirc \lambda^{(\mathcal{B})}_r \circ \Pi_r \circ \mathbf{1}_r \right) \fcirc \mathcal{B}_r \fcirc \lambda^{(\mathcal{B})}_r, \label{eq:nablarmathbf1}
\end{align}
where the last line was obtained by replacing $\nabla_r^{(\mathcal{B})}$ with its SBE parametrization $\overline{\lambda}^{(\mathcal{B})}_r \fcirc w^{(\mathcal{B})}_r \fcirc \lambda^{(\mathcal{B})}_r$, Eq.~\eqref{eq:nablaBreducibility}, and by using the SBE equations for the fermion-boson vertices
\begin{subequations}
    \begin{align}
        \lambda_r^{(\mathcal{B})} & = \mathbf{1}_r + \mathbf{1}_r \circ \Pi_r \circ \mathcal{I}_r^{(\mathcal{B})}, \label{eq:lambdaBexpression} \\
        \overline{\lambda}_r^{(\mathcal{B})} & = \mathbf{1}_r + \mathcal{I}_r^{(\mathcal{B})} \circ \Pi_r \circ \mathbf{1}_r. \label{eq:lambdabarBexpression}
    \end{align}
    \label{eq:lambdaBlambdabarBexpressions}
\end{subequations}
Comparing our result~\eqref{eq:nablarmathbf1} with $\nabla_r^{(\mathcal{B})}=\overline{\lambda}^{(\mathcal{B})}_r \fcirc w^{(\mathcal{B})}_r \fcirc \lambda^{(\mathcal{B})}_r$, we directly infer the following expression for $w_r^{(\mathcal{B})}$:
\begin{equation}
    w_r^{(\mathcal{B})} = \mathcal{B}_r + w_r^{(\mathcal{B})} \fcirc \lambda_r^{(\mathcal{B})} \circ \Pi_r \circ \mathcal{B}_r.
    \label{eq:wBexpression}
\end{equation}
Note that, by following the reasoning just outlined from Eq.~\eqref{eq:DefinitionmathcalIrBreducibility}, but starting from the Bethe--Salpeter equations in the form $\phi_r = V \circ \Pi_r \circ I_r$ instead of $\phi_r = I_r \circ \Pi_r \circ V$, one can also derive the equivalent expression
\begin{equation}
    w_r^{(\mathcal{B})} = \mathcal{B}_r + \mathcal{B}_r \circ \Pi_r \circ \mathcal{B}_r \circ \overline{\lambda}_r^{(\mathcal{B})} \fcirc w_r^{(\mathcal{B})},
    \label{eq:OtherwBexpression}
\end{equation}
as well as the counterpart of Eq.~\eqref{eq:MBexpression}, namely,
\begin{equation}
    M^{(\mathcal{B})}_r = \mathcal{I}_r^{(\mathcal{B})} \circ \Pi_r \circ \left(I_r - \mathcal{B}_r\right).
    \label{eq:OtherMBexpression}
\end{equation}

Collecting the results thus obtained, i.e., Eqs.~\eqref{eq:MBexpression}, \eqref{eq:lambdaBlambdabarBexpressions}--\eqref{eq:OtherMBexpression}, we can therefore conclude that, in the language of $\mathcal{B}$-reducibility, the SBE equations read
\begin{subequations}
    \label{eq:selfconsistentSBEequationsBreducibility}
    \begin{align}
	    w_r^{(\mathcal{B})} &= \mathcal{B}_r + w_r^{(\mathcal{B})} \fcirc \lambda_r^{(\mathcal{B})} \circ \Pi_r \fcirc \mathcal{B}_r \label{eq:SBEeqwB1} \\
        &= \mathcal{B}_r + \mathcal{B}_r \fcirc \Pi_r \circ \overline{\lambda}_r^{(\mathcal{B})} \fcirc w_r^{(\mathcal{B})} , \label{eq:SBEeqwB2} \\
	    \lambda_r^{(\mathcal{B})} & = \mathbf{1}_r + \mathbf{1}_r \circ \Pi_r \circ \mathcal{I}_r^{(\mathcal{B})}, \label{eq:SBEeqlambdaB} \\
        \overline{\lambda}_r^{(\mathcal{B})} & = \mathbf{1}_r + \mathcal{I}_r^{(\mathcal{B})} \circ \Pi_r \circ \mathbf{1}_r, \label{eq:SBEeqlambdabarB} \\
	    M_r^{(\mathcal{B})} &= \left(I_r - \mathcal{B}_r\right) \circ \Pi_r \circ \mathcal{I}_r^{(\mathcal{B})} \label{eq:SBEeqMB1} \\
        &= \mathcal{I}_r^{(\mathcal{B})} \circ \Pi_r \circ \left(I_r - \mathcal{B}_r\right).
        \label{eq:SBEeqMB2}
    \end{align}
\end{subequations}
The bubbles $\Pi_r$ and the identity vertices $\mathbf{1}_r$ are still given by Eqs.~\eqref{eq:Bubbles} and~\eqref{eq:Identityfcirc}, respectively. Our results~\eqref{eq:SBEeqMB1}--\eqref{eq:SBEeqMB2} for $M_r^{(\mathcal{B})}$ can also be brought to a form analogous to Eqs.~\eqref{eq:SBEeqM1}--\eqref{eq:SBEeqM2}. With $\nabla_r=\nabla_r^{(\mathcal{B})}+\nabla_r^{(\mathcal{F})}$, $M_r^{(\mathcal{B})}=M_r+\nabla_r^{(\mathcal{F})}-\mathcal{F}_r$, and $V = I_r+\phi_r=\mathcal{I}_r+\nabla_r=\mathcal{I}_r^{(\mathcal{B})}+\nabla_r^{(\mathcal{B})}$, we indeed obtain
\begin{align}
    I_r &= V-\phi_r \nonumber \\
    &= \mathcal{I}_r^{(\mathcal{B})}+\nabla_r^{(\mathcal{B})}-\left(M_r+\nabla_r^{(\mathcal{B})}-\mathcal{B}_r+\nabla_r^{(\mathcal{F})}-\mathcal{F}_r\right) \nonumber \\
    &=\mathcal{I}_r^{(\mathcal{B})}-M_r+\mathcal{B}_r-\nabla_r^{(\mathcal{F})}+\mathcal{F}_r \nonumber \\
    &=\mathcal{I}_r^{(\mathcal{B})}-M_r^{(\mathcal{B})}+\mathcal{B}_r,
    \label{eq:IrBrIrBMrB} \\
    \Rightarrow M_r^{(\mathcal{B})}&=\left(\mathcal{I}_r^{(\mathcal{B})}-M_r^{(\mathcal{B})}\right)\circ\Pi_r\circ\mathcal{I}_r^{(\mathcal{B})} \nonumber \\
    &=\mathcal{I}_r^{(\mathcal{B})}\circ\Pi_r\circ(\mathcal{I}_r^{(\mathcal{B})}-M_r^{(\mathcal{B})}).
    \label{eq:MrB_new}
\end{align}

In what follows, the ``$(\mathcal{B})$'' superscript will be dropped for conciseness of notation, i.e., the SBE vertices $\nabla_r$ will only contain $\mathcal{B}$-reducible contributions while $M_r$ will collect $\mathcal{B}$-irreducible diagrams (including $\mathcal{F}$-reducible ones). The relations derived in Sec.~\ref{sec:SBEmfRGUflow} are formulated within this generalized SBE framework for non-local interactions, where the original formulation based on $U$-reducibility is directly recovered by setting $\mathcal{F}_r=0$ (and therefore $\mathcal{B}_r=U$). This generalization of the SBE formalism for non-local interactions was already exploited in the context of the $1\ell$ fRG~\cite{AlEryani2024,AlEryani2025} to investigate the extended Hubbard model~\cite{Zhang1989,Terletska2017,Terletska2018,Paki2019} and the Hubbard--Holstein model~\cite{Holstein1959,Freericks1995,Capone2004}. It will also play an essential role in Sec.~\ref{sec:TflowRetardedInteractions} where we consider a model of an Anderson impurity coupled to a phonon, for which the fermionic bare interaction $\mathcal{F}_r$ is non-zero.

\subsection{Multiloop fRG extended by the interaction flow}
\label{sec:SBEmfRGUflow}

Let us now come to our main result, namely a derivation of the flow equations including a regulator dependence of the bare interaction. For this, we start from the SBE equations~\eqref{eq:selfconsistentSBEequationsBreducibility} that we turn into flow equations by introducing an \emph{RG scale} $\Lambda$ in both the bare propagator $G_0$ and the bare interaction $U$. This is achieved by performing the substitutions $G_0\rightarrow G_0^\Lambda$ and $U\rightarrow U^\Lambda$ (with both $\mathcal{B}_r\rightarrow \mathcal{B}_r^\Lambda$ and $\mathcal{F}_r\rightarrow \mathcal{F}_r^\Lambda$ in principle). We stress again that most fRG approaches (including the multiloop fRG~\cite{Kugler2018a,Kugler2018b,Kugler2018c,Gievers2022}) are formulated with a regulator inserted in $G_0$ only~\footnote{The regulators (or cutoff functions) are typically introduced as $G_0^\Lambda=\Theta^\Lambda G_0$ or $\left(G_0^\Lambda\right)^{-1}=G_0^{-1}+R^\Lambda$. With these definitions, $\Theta^\Lambda$ and $R^\Lambda$ are referred to as multiplicative and additive regulators, respectively. The regulators used in the present study are defined in Sec.~III.}, i.e., using $G_0\rightarrow G_0^\Lambda$ while keeping bare interactions such as $U$ $\Lambda$-independent. The interaction flow generated by $U\rightarrow U^\Lambda$ is thus a novel feature of our forthcoming derivations. In this section, we determine the underlying flow equations for a generic fermionic model given by Eq.~\eqref{eq:ClassicalActionS} with energy conservation and translational invariance [see Eqs.~\eqref{eq:FreqMomConservationLawsG0U}] whereas the addition of the $SU(2)$ spin symmetry to this setup is treated specifically in App.~\ref{sec:SU2symmetry}.

We derive flow equations involving a flowing bare interaction starting from the SBE equations in Sec.~\ref{sec:U-flow_SBE} and parquet equations in Sec.~\ref{sec:U-flow_parquet}. Further, we give an interpretation of the flow equation for the bosonic propagator in Sec.~\ref{sec:Katanin} and present the flow equation for the self-energy in Sec.~\ref{sec:SelfEnergyFlow}.

\subsubsection{Flow equations derived from the SBE equations}
\label{sec:U-flow_SBE}

In what follows, we will repeatedly use the shorthand notation $\dot{X}\equiv\partial_\Lambda X$ for any vertex $X$. Our present aim thus consists in deriving expressions for the derivatives $\dot{w}_r$, $\dot{\lambda}_r$, $\dot{\overline{\lambda}}_r$ and $\dot{M}_r$, which will constitute our flow equations. In contrast to the derivation in Ref.~\cite{Gievers2022}, we perform this after introducing the RG scale $\Lambda$ via $G_0\rightarrow G_0^\Lambda$ and, importantly, $U\rightarrow U^\Lambda$. All objects in the SBE equations~\eqref{eq:selfconsistentSBEequationsBreducibility} depend on $\Lambda$, with the exception of the identity vertices $\mathds{1}_r$. By differentiating then both sides of the SBE equations with respect to $\Lambda$, expressions for $\dot{w}_r$, $\dot{\lambda}_r$, $\dot{\overline{\lambda}}_r$ and $\dot{M}_r$ directly follow. For the SBE equation~\eqref{eq:SBEeqMB1} for $M_r$, this yields
\begin{align}
\dot{M}_r &= \left(\dot{I}_r - \dot{\mathcal{B}}_r\right) \circ \Pi_r \circ \mathcal{I}_r + \left(I_r - \mathcal{B}_r\right) \circ \dot{\Pi}_r \circ \mathcal{I}_r \nonumber \\
&\phantom{=} +\left(I_r - \mathcal{B}_r\right) \circ \Pi_r \circ \left(\dot{M}_r + \dot{I}_r -\dot{\mathcal{B}}_r\right), \label{eq:MrdotDerivation}
\end{align}
where the last term is obtained by replacing $\dot{\mathcal{I}}_r$ according to Eq.~\eqref{eq:IrBrIrBMrB}. Isolating $\dot{M}_r$ on the left-hand side, Eq.~\eqref{eq:MrdotDerivation} becomes
\begin{align}
    \dot{M}_r &= \left(\mathds{1}_r - (I_r - \mathcal{B}_r) \circ \Pi_r\right)^{-1} \nonumber \\
    &\phantom{=} \circ \bigg[\left(\dot{I}_r - \dot{\mathcal{B}}_r\right) \circ \Pi_r \circ \mathcal{I}_r + \left(I_r - \mathcal{B}_r\right) \circ \dot{\Pi}_r \circ \mathcal{I}_r \nonumber \\
    &\phantom{= \circ \bigg[} + \left(I_r - \mathcal{B}_r\right) \circ \Pi_r \circ \left(\dot{I}_r -\dot{\mathcal{B}}_r\right) \bigg]. \label{eq:MrdotIntermediaryStep}
\end{align}
The inverse $\left(\mathds{1}_r - (I_r - \mathcal{B}_r) \circ \Pi_r\right)^{-1}$ can also be reexpressed with the help of an extended Bethe--Salpeter equation. Indeed, Eq.~\eqref{eq:MrB_new} has the same structure as the original Bethe--Salpeter equation:
\begin{subequations}
    \begin{alignat}{3}
        &V &&= I_r+\phi_r &&=I_r + V\circ \Pi_r \circ I_r,\\
        &\mathcal{I}_r &&= \left(\mathcal{I}_r-M_r\right)+M_r &&= \left(I_r-\mathcal{B}_r\right) +\mathcal{I}_r\circ\Pi_r\circ\left(I_r-\mathcal{B}_r\right).
    \end{alignat}
\end{subequations}
The corresponding extended Bethe--Salpeter equations thus take the following form:
\begin{subequations}
    \begin{align}
        \mathds{1}_r+V\circ\Pi_r &= (\mathds{1}_r-I_r\circ\Pi_r)^{-1},\\
        \mathds{1}_r+\mathcal{I}_r\circ\Pi_r&=(\mathds{1}_r-(I_r-\mathcal{B}_r)\circ\Pi_r)^{-1}. \label{eq:ExtendedBSEIrBr}
    \end{align}
\end{subequations}
From Eq.~\eqref{eq:ExtendedBSEIrBr}, the expression~\eqref{eq:MrdotIntermediaryStep} for $\dot{M}_r$ can be rewritten as
\begin{align}
    \dot{M}_r &= (\mathds{1}_r + \mathcal{I}_r \circ \Pi_r) \nonumber \\
    &\phantom{=} \circ \bigg[\left(\dot{I}_r - \dot{\mathcal{B}}_r\right) \circ \Pi_r \circ \mathcal{I}_r + \left(I_r - \mathcal{B}_r\right) \circ \dot{\Pi}_r \circ \mathcal{I}_r \nonumber \\
    &\phantom{= \circ \bigg[} + \left(I_r - \mathcal{B}_r\right) \circ \Pi_r \circ \left(\dot{I}_r -\dot{\mathcal{B}}_r\right) \bigg].
\end{align}
Rearranging the terms, we obtain
\begin{align}
    \dot{M}_r &= \mathcal{I}_r \circ \dot{\Pi}_r \circ \mathcal{I}_r + \mathcal{I}_r \circ \Pi_r \circ \left( \dot{I}_r - \dot{\mathcal{B}}_r \right) \nonumber \\
    &\phantom{=} + \left( \dot{I}_r - \dot{\mathcal{B}}_r \right) \circ \Pi_r \circ \mathcal{I}_r \nonumber \\
    &\phantom{=} + \mathcal{I}_r \circ \Pi_r \circ \left(\dot{I}_r - \dot{\mathcal{B}}_r\right) \circ \Pi_r \circ \mathcal{I}_r, \label{eq:FlowEquationMr}
\end{align}
where we have used the relation
\begin{equation}
    \mathcal{I}_r = (I_r - \mathcal{B}_r) + \mathcal{I}_r \circ \Pi_r \circ (I_r - \mathcal{B}_r),
\end{equation}
which follows from the SBE equation~\eqref{eq:SBEeqMB2} and Eq.~\eqref{eq:IrBrIrBMrB}.

We then focus on the flow equations for the fermion-boson vertices. After performing the substitutions $G_0\rightarrow G_0^\Lambda$ and $U\rightarrow U^\Lambda$, differentiating both sides of the SBE equations~\eqref{eq:SBEeqlambdaB} and~\eqref{eq:SBEeqlambdabarB} with respect to $\Lambda$ leads to
\begin{subequations}
    \begin{align}
        \dot{\lambda}_r & = \mathbf{1}_r \circ \dot{\Pi}_r \circ \mathcal{I}_r + \mathbf{1}_r \circ \Pi_r \circ \dot{\mathcal{I}}_r, \\
        \dot{\overline{\lambda}}_r & = \dot{\mathcal{I}}_r \circ \Pi_r \circ \mathbf{1}_r + \mathcal{I}_r \circ \dot{\Pi}_r \circ \mathbf{1}_r,
    \end{align}
\end{subequations}
where $\dot{\mathcal{I}}_r$ can again be replaced using Eq.~\eqref{eq:IrBrIrBMrB}, thus yielding
\begin{subequations}
    \begin{align}
        \dot{\lambda}_r & = \mathbf{1}_r \circ \dot{\Pi}_r \circ \mathcal{I}_r + \mathbf{1}_r \circ \Pi_r \circ \left(\dot{M}_r + \dot{I}_r - \dot{\mathcal{B}}_r\right), \\
        \dot{\overline{\lambda}}_r & = \left(\dot{M}_r + \dot{I}_r - \dot{\mathcal{B}}_r\right) \circ \Pi_r \circ \mathbf{1}_r + \mathcal{I}_r \circ \dot{\Pi}_r \circ \mathbf{1}_r.
    \end{align}
    \label{eq:lambdardotIntermediaryStep}
\end{subequations}
Inserting our expression~\eqref{eq:FlowEquationMr} for $\dot{M}_r$ in the right-hand sides of Eqs.~\eqref{eq:lambdardotIntermediaryStep} and then simplifying the equations thus obtained with the SBE equations for the fermion-boson vertices [i.e., Eqs.~\eqref{eq:SBEeqlambdaB} and~\eqref{eq:SBEeqlambdabarB}] yields
\begin{subequations}
    \begin{align}
        \dot{\lambda}_r & = \lambda_r \circ \dot{\Pi}_r \circ \mathcal{I}_r + \lambda_r \circ \Pi_r \circ \left(\dot{I}_r - \dot{\mathcal{B}}_r\right) \nonumber \\
        &\phantom{=} + \lambda_r \circ \Pi_r \circ \left(\dot{I}_r - \dot{\mathcal{B}}_r\right) \circ \Pi_r \circ \mathcal{I}_r, \label{eq:FlowEquationlambdar} \\
        \dot{\overline{\lambda}}_r & = \mathcal{I}_r \circ \dot{\Pi}_r \circ \overline{\lambda}_r + \left(\dot{I}_r - \dot{\mathcal{B}}_r\right) \circ \Pi_r \circ \overline{\lambda}_r \nonumber \\
        &\phantom{=} + \mathcal{I}_r \circ \Pi_r \circ \left(\dot{I}_r - \dot{\mathcal{B}}_r\right) \circ \Pi_r \circ \overline{\lambda}_r. \label{eq:FlowEquationlambdabarr}
    \end{align}
    \label{eq:FlowEquationlambdalambdabarr}
\end{subequations}

Finally, we derive the flow equation for the bosonic propagators $w_r$. To that end, we first rewrite the SBE equation~\eqref{eq:SBEeqwB1} by isolating $w_r$ on the left-hand side, namely,
\begin{equation}
    w_r = \mathcal{B}_r \fcirc (\mathbf{1}_r - \lambda_r \circ \Pi_r \circ \mathcal{B}_r)^{-1} .
    \label{eq:rewriteSBEeqwB1}
\end{equation}
Differentiating both sides of this equation with respect to the RG scale $\Lambda$ after introducing the latter via $G_0 \rightarrow G_0^\Lambda$ and $U \rightarrow U^\Lambda$, we find
\begin{align}
    \dot{w}_r &= \dot{\mathcal{B}}_r \fcirc (\mathbf{1}_r - \lambda_r \circ \Pi_r \circ \mathcal{B}_r)^{-1} \nonumber \\
    &\phantom{=} + w_r \fcirc \left(\dot{\lambda}_r \circ \Pi_r + \lambda_r \circ \dot{\Pi}_r\right) \fcirc w_r \nonumber \\
    &\phantom{=} + w_r \fcirc \lambda_r \circ \Pi_r \circ \dot{\mathcal{B}}_r \fcirc (\mathbf{1}_r - \lambda_r \circ \Pi_r \circ \mathcal{B}_r)^{-1},
\end{align}
where $w_r$ was inserted on the right-hand side using Eq.~\eqref{eq:rewriteSBEeqwB1}. We then replace $\dot{\lambda}_r$ with its expression~\eqref{eq:FlowEquationlambdar} and simplify the equation thus obtained by introducing $\overline{\lambda}_r$ through its SBE equation~\eqref{eq:SBEeqlambdabarB}. This results in
\begin{align}
    \dot{w}_r &= \dot{\mathcal{B}}_r \fcirc (\mathbf{1}_r - \lambda_r \circ \Pi_r \circ \mathcal{B}_r)^{-1} + w_r \fcirc \lambda_r \circ \dot{\Pi}_r \circ \overline{\lambda}_r \fcirc w_r \nonumber \\
    &\phantom{=} + w_r \fcirc \lambda_r \circ \Pi_r \circ \left( \dot{I}_r - \dot{\mathcal{B}}_r \right) \circ \Pi_r \circ \overline{\lambda}_r \fcirc w_r \nonumber \\
    &\phantom{=} + w_r \fcirc \lambda_r \circ \Pi_r \circ \dot{\mathcal{B}}_r \fcirc (\mathbf{1}_r - \lambda_r \circ \Pi_r \circ \mathcal{B}_r)^{-1}. \label{eq:wdotIntermediaryStep}
\end{align}
Furthermore, by comparing Eqs.~\eqref{eq:SBEeqwB2} and~\eqref{eq:rewriteSBEeqwB1}, one can directly infer that
\begin{equation}
(\mathbf{1}_r - \lambda_r \circ \Pi_r \circ \mathcal{B}_r)^{-1} = \mathbf{1}_r + \mathbf{1}_r \circ \Pi_r \circ \overline{\lambda}_r \fcirc w_r.
\end{equation}
Inserting the last equality into Eq.~\eqref{eq:wdotIntermediaryStep} leads to
\begin{align}
    \dot{w}_r &= w_r \fcirc \lambda_r \circ \dot{\Pi}_r \circ \overline{\lambda}_r \fcirc w_r \nonumber \\
    &\phantom{=} + w_r \fcirc \lambda_r \circ \Pi_r \circ \dot{I}_r \circ \Pi_r \circ \overline{\lambda}_r \fcirc w_r + \dot{\mathcal{B}}_r \nonumber \\
    &\phantom{=} + w_r \fcirc \lambda_r \circ \Pi_r \circ \dot{\mathcal{B}}_r + \dot{\mathcal{B}}_r \circ \Pi_r \circ \overline{\lambda}_r \fcirc w_r. \label{eq:FlowEquationwr}
\end{align}

Eqs.~\eqref{eq:FlowEquationMr},~\eqref{eq:FlowEquationlambdar},~\eqref{eq:FlowEquationlambdabarr} and~\eqref{eq:FlowEquationwr} give our final expressions for $\dot{M}_r$, $\dot{\lambda}_r$, $\dot{\overline{\lambda}}_r$ and $\dot{w}_r$, respectively. We collect them below for clarity
\begin{subequations}
\begin{align}
    \dot{w}_r &= w_r \fcirc \lambda_r \circ \dot{\Pi}_r \circ \overline{\lambda}_r \fcirc w_r \nonumber \\
    &\phantom{=} + w_r \fcirc \lambda_r \circ \Pi_r \circ \dot{I}_r \circ \Pi_r \circ \overline{\lambda}_r \fcirc w_r + \dot{\mathcal{B}}_r \nonumber \\
    &\phantom{=} + w_r \fcirc \lambda_r \circ \Pi_r \circ \dot{\mathcal{B}}_r + \dot{\mathcal{B}}_r \circ \Pi_r \circ \overline{\lambda}_r \fcirc w_r, \label{eq:FinalFlowEquationwr} \\
    \dot{\lambda}_r & = \lambda_r \circ \dot{\Pi}_r \circ \mathcal{I}_r + \lambda_r \circ \Pi_r \circ \left(\dot{I}_r - \dot{\mathcal{B}}_r\right) \nonumber \\
    &\phantom{=} + \lambda_r \circ \Pi_r \circ \left(\dot{I}_r - \dot{\mathcal{B}}_r\right) \circ \Pi_r \circ \mathcal{I}_r, \label{eq:FinalFlowEquationlambdar} \\
    \dot{\overline{\lambda}}_r & = \mathcal{I}_r \circ \dot{\Pi}_r \circ \overline{\lambda}_r + \left(\dot{I}_r - \dot{\mathcal{B}}_r\right) \circ \Pi_r \circ \overline{\lambda}_r \nonumber \\
    &\phantom{=} + \mathcal{I}_r \circ \Pi_r \circ \left(\dot{I}_r - \dot{\mathcal{B}}_r\right) \circ \Pi_r \circ \overline{\lambda}_r, \label{eq:FinalFlowEquationlambdabarr} \\
    \dot{M}_r &= \mathcal{I}_r \circ \dot{\Pi}_r \circ \mathcal{I}_r + \mathcal{I}_r \circ \Pi_r \circ \left( \dot{I}_r - \dot{\mathcal{B}}_r \right) \nonumber \\
    &\phantom{=} + \left( \dot{I}_r - \dot{\mathcal{B}}_r \right) \circ \Pi_r \circ \mathcal{I}_r \nonumber \\
    &\phantom{=} + \mathcal{I}_r \circ \Pi_r \circ \left(\dot{I}_r - \dot{\mathcal{B}}_r\right) \circ \Pi_r \circ \mathcal{I}_r. \label{eq:FinalFlowEquationMr}
\end{align}
\label{eq:FinalFlowEquationswrlambdarMr}
\end{subequations}
The original flow equations presented in Ref.~\cite{Gievers2022}, where $\dot U = 0$ was assumed, are straightforwardly recovered by setting $\dot{\mathcal{B}}_r\to 0$. In the flow equations for $\overline{\lambda}_r, \lambda_r$, and $M_r$ such terms always come with $\dot I_r-\dot{\mathcal{B}}_r$. Only in the flow equation of $\dot w_r$, there are three additional terms with a new structure.

All these flow equations are \emph{exact}, i.e., no approximation has been made so far. Moreover, the right-hand sides of those equations still involve $\dot{w}_r$, $\dot{\lambda}_r$, $\dot{\overline{\lambda}}_r$ and $\dot{M}_r$ via the derivatives $\dot{I}_r$, which reflects the self-consistent character of the SBE equations from which those flow equations are derived. In practice, it is more convenient for the numerical implementation to reorganize this structure based on the $\dot{I}_r$ vertex by performing a loop expansion of the flow equations. For this, the left-hand sides of Eqs.~\eqref{eq:FinalFlowEquationswrlambdarMr} are expanded as
\begin{align}
        \dot{w}_r  &= \sum_{\ell=1}^\infty \dot{w}_r^{(\ell)}, \quad
        \dot{\lambda}_r  = \sum_{\ell=1}^\infty \dot{\lambda}_r^{(\ell)}, \nonumber \\
        \dot{\overline{\lambda}}_r  &= \sum_{\ell=1}^\infty \dot{\overline{\lambda}}_r^{(\ell)},\quad 
        \dot{M}_r  = \sum_{\ell=1}^\infty \dot{M}_r^{(\ell)}, \label{eq:LoopExpansionwrlambdarMr}
\end{align}
whereas $\dot{I}_r$ is rewritten on the right-hand sides according to
\begin{equation}
    \dot{I}_r = \sum_{r^\prime \neq r} \dot{\phi}_{r^\prime} + \dot{I}^{\text{2PI}} = \sum_{r^\prime \neq r} \sum_{\ell=1}^\infty \dot{\phi}_{r^\prime}^{(\ell)} + \dot{I}^{\text{2PI}}.
    \label{eq:dotIrwithI2PI}
\end{equation}

In our numerical computations, we impose the \emph{only} approximation
\begin{equation}
    \dot{I}^{\text{2PI}} = \dot{U} = \dot{\mathcal{B}}_r + \dot{\mathcal{F}}_r,
    \label{eq:I2PIapprox}
\end{equation}
which is analogous to the condition $\dot{I}^{\text{2PI}} = 0$ used in the original version of the multiloop fRG where $\dot{U}=0$~\cite{Kugler2018a,Kugler2018b,Kugler2018c}. Note that the latter approach with the approximation $\dot{I}^{\text{2PI}} = 0$ yields the results of the well-known \emph{parquet approximation}~\cite{Kugler2018a} obtained by iteratively solving the Bethe--Salpeter equations~\eqref{eq:BetheSalpeterEquations} together with the Schwinger--Dyson equation and the condition $I^{\text{2PI}} = U$. The parquet approximation neglects totally two-particle irreducible diagrams from fourth order $\mathcal{O}(U^4)$ on. At the same time, it yields full self-consistency at the two-particle level.

Furthermore, Eq.~\eqref{eq:I2PIapprox} enables us to rewrite Eq.~\eqref{eq:dotIrwithI2PI} as
\begin{equation}
    \dot{I}_r = \sum_{\ell=1}^\infty \dot{\phi}^{(\ell)}_{\overline{r}} + \dot{U}, \quad \text{with} \quad
    \dot{\phi}^{(\ell)}_{\overline{r}} = \sum_{r^\prime \neq r} \dot{\phi}_{r^\prime}^{(\ell)},
    \label{eq:dotIrwithoutI2PI}
\end{equation}
and the derivatives $\dot{\phi}_r^{(\ell)}$ satisfy
\begin{equation}
    \dot{\phi}_{r}^{(\ell)} = \dot{\overline{\lambda}}_r^{(\ell)} \fcirc w_r \fcirc \lambda_r + \overline{\lambda}_r \fcirc \dot{w}_r^{(\ell)} \fcirc \lambda_r + \overline{\lambda}_r \fcirc w_r \fcirc \dot{\lambda}_r^{(\ell)} + \dot{M}_r^{(\ell)} - \dot{\mathcal{B}}_r \delta_{\ell,1}.
    \label{eq:phirdotell}
\end{equation}
After expanding both sides of the flow equations~\eqref{eq:FinalFlowEquationswrlambdarMr} in that way, we can infer the following relations: For $\ell = 1$,
\begin{subequations}
    \begin{align}
    \dot{w}_r^{(1)} & = w_r \fcirc \lambda_r \circ \dot{\Pi}_r \circ \overline{\lambda}_r \fcirc w_r \nonumber \\
    &\phantom{=} + w_r \fcirc \lambda_r \circ \Pi_r \circ \dot{U} \circ \Pi_r \circ \overline{\lambda}_r \fcirc w_r + \dot{\mathcal{B}}_r \nonumber \\
    &\phantom{=} + w_r \fcirc \lambda_r \circ \Pi_r \circ \dot{\mathcal{B}}_r + \dot{\mathcal{B}}_r \circ \Pi_r \circ \overline{\lambda}_r \fcirc w_r, \label{eq:UflowmultiloopSBEfRGEq1l_wr} \\
    \dot{\lambda}_r^{(1)} & = \lambda_r \circ \dot{\Pi}_r \circ \mathcal{I}_r + \lambda_r \circ \Pi_r \circ \dot{\mathcal{F}}_r \nonumber \\
    &\phantom{=} + \lambda_r \circ \Pi_r \circ \dot{\mathcal{F}}_r \circ \Pi_r \circ \mathcal{I}_r, \label{eq:UflowmultiloopSBEfRGEq1l_lambdar} \\
    \dot{\overline{\lambda}}_r^{(1)} & = \mathcal{I}_r \circ \dot{\Pi}_r \circ \overline{\lambda}_r + \dot{\mathcal{F}}_r \circ \Pi_r \circ \overline{\lambda}_r \nonumber \\
    &\phantom{=} + \mathcal{I}_r \circ \Pi_r \circ \dot{\mathcal{F}}_r \circ \Pi_r \circ \overline{\lambda}_r, \label{eq:UflowmultiloopSBEfRGEq1l_lambdabarr} \\
    \dot{M}_r^{(1)} & = \mathcal{I}_r \circ \dot{\Pi}_r \circ \mathcal{I}_r + \mathcal{I}_r \circ \Pi_r \circ \dot{\mathcal{F}}_r \nonumber \\
    &\phantom{=} + \dot{\mathcal{F}}_r \circ \Pi_r \circ \mathcal{I}_r + \mathcal{I}_r \circ \Pi_r \circ \dot{\mathcal{F}}_r \circ \Pi_r \circ \mathcal{I}_r, \label{eq:UflowmultiloopSBEfRGEq1l_Mr}
    \end{align}
\label{eq:UflowmultiloopSBEfRGEq1l}
\end{subequations}
for $\ell = 2$,
\begin{subequations}
    \begin{align}
    \dot{w}_r^{(2)} & = 0, \label{eq:UflowmultiloopSBEfRGEq2l_wr} \\
    \dot{\lambda}_r^{(2)} & = \lambda_r \circ \Pi_r \circ \dot{\phi}_{\overline{r}}^{(1)}, \label{eq:UflowmultiloopSBEfRGEq2l_lambdar} \\
    \dot{\overline{\lambda}}_r^{(2)} & = \dot{\phi}_{\overline{r}}^{(1)} \circ \Pi_r \circ \overline{\lambda}_r , \label{eq:UflowmultiloopSBEfRGEq2l_lambdabarr} \\
    \dot{M}_r^{(2)} & = \dot{\phi}_{\overline{r}}^{(1)} \circ \Pi_r \circ \mathcal{I}_r + \mathcal{I}_r \circ \Pi_r \circ \dot{\phi}_{\overline{r}}^{(1)} , \label{eq:UflowmultiloopSBEfRGEq2l_Mr}
    \end{align}
\label{eq:UflowmultiloopSBEfRGEq2l}
\end{subequations}
and, for $\ell\geq 3$,
\begin{subequations}
    \begin{align}
    \dot{w}_r^{(\ell)} & = w_r \fcirc \lambda_r \circ \Pi_r \circ \dot{\phi}_{\overline{r}}^{(\ell-2)} \circ \Pi_r \circ \overline{\lambda}_r \fcirc w_r , \label{eq:UflowmultiloopSBEfRGEqnl_wr} \\
    \dot{\lambda}_r^{(\ell)} & = \lambda_r \circ \Pi_r \circ \dot{\phi}_{\overline{r}}^{(\ell-1)} \nonumber \\
    & \phantom{=} + \lambda_r \circ \Pi_r \circ \dot{\phi}_{\overline{r}}^{(\ell-2)} \circ \Pi_r \circ \mathcal{I}_r , \label{eq:UflowmultiloopSBEfRGEqnl_lambdar} \\
    \dot{\overline{\lambda}}_r^{(\ell)} & = \dot{\phi}_{\overline{r}}^{(\ell-1)} \circ \Pi_r \circ \overline{\lambda}_r \nonumber \\
    & \phantom{=} + \mathcal{I}_r \circ \Pi_r \circ \dot{\phi}_{\overline{r}}^{(\ell-2)} \circ \Pi_r \circ \overline{\lambda}_r , \label{eq:UflowmultiloopSBEfRGEqnl_lambdabarr} \\
    \dot{M}_r^{(\ell)} & = \dot{\phi}_{\overline{r}}^{(\ell-1)} \circ \Pi_r \circ \mathcal{I}_r + \mathcal{I}_r \circ \Pi_r \circ \dot{\phi}_{\overline{r}}^{(\ell-1)} \nonumber \\
    & \phantom{=} + \mathcal{I}_r \circ \Pi_r \circ \dot{\phi}_{\overline{r}}^{(\ell-2)} \circ \Pi_r \circ \mathcal{I}_r . \label{eq:UflowmultiloopSBEfRGEqnl_Mr}
    \end{align}
\label{eq:UflowmultiloopSBEfRGEqnl}
\end{subequations}
The derivatives $\dot{\mathcal{F}}_r$ in the $1\ell$ flow equations~\eqref{eq:UflowmultiloopSBEfRGEq1l} originate from the expansion of the terms $\dot{I}_r-\dot{\mathcal{B}}_r$ on the right-hand sides of Eqs.~\eqref{eq:FinalFlowEquationswrlambdarMr}. From Eq.~\eqref{eq:dotIrwithoutI2PI} [and therefore from our approximation imposed by Eq.~\eqref{eq:I2PIapprox}], it indeed follows that
\begin{equation}
    \dot{I}_r - \dot{\mathcal{B}}_r = \sum_{\ell=1}^\infty \dot{\phi}^{(\ell)}_{\overline{r}} + \dot{\mathcal{F}}_r.
\end{equation}
Hence, almost all extra terms generated by the $\Lambda$-dependence of the bare interaction $U$ are contained in the $1\ell$ flow equations. More precisely, the flow equations reported above at $2\ell$ and at higher loop orders by Eqs.~\eqref{eq:UflowmultiloopSBEfRGEq2l}--\eqref{eq:UflowmultiloopSBEfRGEqnl} coincide exactly with the conventional SBE multiloop fRG equations for which $\dot{U}=\dot{\mathcal{B}}_r=\dot{\mathcal{F}}_r=0$~\cite{Gievers2022}, with the only exception that $\dot{\phi}_{\overline{r}}^{(1)}$ contains a contribution from $\dot{\mathcal{B}}_r$, as can be seen from Eq.~\eqref{eq:phirdotell}. Note however that $\dot{\phi}_{\overline{r}}^{(1)}$ is only involved in the $2\ell$ and $3\ell$ flow equations, namely Eqs.~\eqref{eq:UflowmultiloopSBEfRGEq2l}--\eqref{eq:UflowmultiloopSBEfRGEqnl} with $\ell=3$.

To summarize the procedure that led to the multiloop flow equations~\eqref{eq:UflowmultiloopSBEfRGEq1l}--\eqref{eq:UflowmultiloopSBEfRGEqnl}, we have reshuffled the $\dot{\phi}_{\overline{r}}^{(\ell)}$ terms introduced by the expansion of $\dot{I}_r$ across all loop orders of $\dot{w}_r$, $\dot{\lambda}_r$, $\dot{\overline{\lambda}}_r$ and $\dot{M}_r$, whereas the remaining terms involving either $\dot{\Pi}_r$, $\dot{\mathcal{B}}_r$ or $\dot{\mathcal{F}}_r$ define their $1\ell$ contributions in Eqs.~\eqref{eq:UflowmultiloopSBEfRGEq1l}. In conclusion, our results~\eqref{eq:UflowmultiloopSBEfRGEq1l}--\eqref{eq:UflowmultiloopSBEfRGEqnl}, which are exact apart from the condition~\eqref{eq:I2PIapprox} imposed to the 2PI vertex $I^{\text{2PI}}$, constitute the \emph{multiloop SBE fRG} flow equations within our \emph{interaction flow} scheme (where the bare propagator and the bare interaction can both depend on the RG scale $\Lambda$), for a generic model with an action of the form~\eqref{eq:ClassicalActionS} relying on both energy conservation and translational invariance.

In the limit of $\dot{U}=0$, these results reduce to the original \emph{multiloop SBE fRG} flow equations derived in Ref.~\cite{Gievers2022}. Furthermore, \emph{spin rotation invariance}, i.e., \emph{$SU(2)$ spin symmetry}, is always preserved for our analysis of the Anderson impurity model and the Anderson--Holstein impurity model presented in Secs.~\ref{sec:ModelsNumericalImplementationConvergence} and \ref{sec:TflowRetardedInteractions}. Following Ref.~\cite{Fraboulet2025}, the flow equations derived in this section for our interaction flow scheme can be efficiently rewritten in terms of \emph{physical channels} assuming that $SU(2)$ spin symmetry is respected. This is shown in detail in App.~\ref{sec:SU2symmetry}, especially for Eqs.~\eqref{eq:UflowmultiloopSBEfRGEq1l}--\eqref{eq:UflowmultiloopSBEfRGEqnl}.

\subsubsection{Flow equations derived from the parquet decomposition and the Bethe--Salpeter equations}
\label{sec:U-flow_parquet}

We note that our previous derivation of the flow equations underlying our interaction flow scheme is not unique. Alternatively, Eqs.~\eqref{eq:UflowmultiloopSBEfRGEq1l}--\eqref{eq:UflowmultiloopSBEfRGEqnl} can be obtained starting from the parquet decomposition with the Bethe--Salpeter equations~\eqref{eq:BetheSalpeterEquations}, instead of the SBE equations~\eqref{eq:selfconsistentSBEequationsBreducibility}. Indeed, after introducing the RG scale $\Lambda$ as previously via $G_0\rightarrow G_0^\Lambda$ and $U\rightarrow U^\Lambda$, one can also differentiate the Bethe--Salpeter equations~\eqref{eq:BetheSalpeterEquations}, leading to the relation
\begin{equation}
    \dot{\phi}_r = \dot{I}_r \circ \Pi_r \circ V + I_r \circ \dot{\Pi}_r \circ V + I_r \circ \Pi_r \circ \dot{V},
    \label{eq:DifferentiateBSE}
\end{equation}
which, after exploiting the definition~\eqref{eq:DefinitionIr} of the irreducible vertex $I_r$, can be rewritten as
\begin{align}
    \dot{\phi}_r &= V \circ \dot{\Pi}_r \circ V + \dot{I}_r \circ \Pi_r \circ V + V \circ \Pi_r \circ \dot{I}_r \nonumber \\
    &\phantom{=} + V \circ \Pi_r \circ \dot{\phi}_r - \phi_r \circ \dot{\Pi}_r \circ V - \phi_r \circ \Pi_r \circ \dot{V}. \label{eq:DifferentiateBSE2}
\end{align}
Inserting $\phi_r=V \circ \Pi_r \circ I_r$ and replacing $\dot{\phi}_r$ by its expression~\eqref{eq:DifferentiateBSE} in the last line of Eq.~\eqref{eq:DifferentiateBSE2}, we arrive at
\begin{align}
    \dot{\phi}_r &= V \circ \dot{\Pi}_r \circ V + \dot{I}_r \circ \Pi_r \circ V + V \circ \Pi_r \circ \dot{I}_r \nonumber \\
    &\phantom{=} + V \circ \Pi_r \circ \dot{I}_r \circ \Pi_r \circ V. \label{eq:DifferentiateBSE3}
\end{align}
This equation is identical to that underlying the original implementation of the multiloop fRG~\cite{Kugler2018a,Kugler2018b,Kugler2018c} since the $\Lambda$-dependence of the bare interaction $U$ does not add any explicit new terms to those generated by $G_0\rightarrow G_0^\Lambda$~\footnote{In particular, Eq.~\eqref{eq:DifferentiateBSE3} is identical to Eq.~(10) of Ref.~\cite{Kugler2018c}.}. We then expand $\dot{I}_r$ in the right-hand side of Eq.~\eqref{eq:DifferentiateBSE3} with the relation~\eqref{eq:dotIrwithoutI2PI} [which implements the approximation $\dot{I}^{\text{2PI}} = \dot{U}$ set by Eq.~\eqref{eq:I2PIapprox}] and $\dot{\phi}_{r}$ in its left-hand side according to
\begin{equation}
    \dot{\phi}_{r} = \sum_{\ell=1}^\infty \dot{\phi}^{(\ell)}_{r}.
\end{equation}
This yields the multiloop fRG flow equations based on the parquet decomposition for our interaction flow scheme:
\begin{subequations}
\begin{align}
\dot{\phi}^{(1)}_r &= V \circ \dot{\Pi}_r \circ V + V \circ \Pi_r \circ \dot{U} + \dot{U} \circ \Pi_r \circ V \nonumber \\
& \phantom{=} + V \circ \Pi_r \circ \dot{U} \circ \Pi_r \circ V, \label{eq:UflowmultiloopParquetfRGEq1l} \\
\dot{\phi}^{(2)}_r &= \dot{\phi}^{(1)}_{\overline{r}} \circ \Pi_r \circ V + V \circ \Pi_r \circ \dot{\phi}^{(1)}_{\overline{r}}, \label{eq:UflowmultiloopParquetfRGEq2l} \\
\dot{\phi}^{(\ell+2)}_r &= \dot{\phi}^{(\ell+1)}_{\overline{r}} \circ \Pi_r \circ V + V \circ \Pi_r \circ \dot{\phi}^{(\ell+1)}_{\overline{r}} \nonumber \\
&\phantom{=} + V \circ \Pi_r \circ \dot{\phi}^{(\ell)}_{\overline{r}} \circ \Pi_r \circ V. \label{eq:UflowmultiloopParquetfRGEqnl}
\end{align}
\label{eq:UflowmultiloopParquetfRGEqs}
\end{subequations}

In comparison to the original multiloop flow equations~\cite{Kugler2018a,Kugler2018b,Kugler2018c}, a term including $\dot{U}$ only explicitly enters in the $1\ell$ flow equation.
By inserting the SBE decomposition $V = \lambda_r \fcirc w_r \fcirc \overline{\lambda}_r + \mathcal{I}_r$ [which follows from Eqs.~\eqref{eq:nablaBreducibility} and~\eqref{eq:DefinitionmathcalIrBreducibility}] on the right-hand sides of Eqs.~\eqref{eq:UflowmultiloopParquetfRGEqs} and rewriting their left-hand sides using Eq.~\eqref{eq:phirdotell}, we can infer all expressions of $\dot{w}_r^{(\ell)}$, $\dot{\lambda}_r^{(\ell)}$, $\dot{\overline{\lambda}}_r^{(\ell)}$ and $\dot{M}_r^{(\ell)}$ as given by Eqs.~\eqref{eq:UflowmultiloopSBEfRGEq1l}--\eqref{eq:UflowmultiloopSBEfRGEqnl}. This consistency between the derivation of the multiloop SBE fRG flow equations obtained from the SBE equations or from the Bethe--Salpeter equations based on the parquet decomposition was already found in Ref.~\cite{Gievers2022}, but in the conventional framework where $\dot{U}=\dot{\mathcal{B}}_r=\dot{\mathcal{F}}_r=0$.

\subsubsection{Katanin substitution for bosonic propagators}
\label{sec:Katanin}

We stress that the SBE framework itself allows for a very compact and elegant formulation of our interaction flow derived from self-consistent equations. To see this more clearly, we introduce the polarization $P_r$ (or bosonic self-energy~\cite{Krien2021,Gievers2022}) defined as
\begin{equation}
    P_r = \lambda_r \circ \Pi_r \circ \mathbf{1}_r = \mathbf{1}_r \circ \Pi_r \circ \overline{\lambda}_r.
\label{eq:DefinitionPr}
\end{equation}
As a next step, we derive a flow equation for $P_r$ within our interaction flow scheme. Differentiating the first equality with respect to the RG scale yields
\begin{equation}
    \dot{P}_r = \dot{\lambda}_r \circ \Pi_r \circ \mathbf{1}_r + \lambda_r \circ \dot{\Pi}_r \circ \mathbf{1}_r.
\end{equation}
Replacing $\dot{\lambda}_r$ by its expression~\eqref{eq:FinalFlowEquationlambdar} and then simplifying the terms obtained with the SBE equation~\eqref{eq:SBEeqlambdabarB} leads to
\begin{equation}
    \dot{P}_r = \lambda_r \circ \dot{\Pi}_r \circ \overline{\lambda}_r + \lambda_r \circ \Pi_r \circ \left(\dot{I}_r - \dot{\mathcal{B}}_r\right) \circ \Pi_r \circ \overline{\lambda}_r.
\label{eq:FlowEquationPr}
\end{equation}

With Eq.~\eqref{eq:FlowEquationPr}, we can rewrite the flow equation~\eqref{eq:FinalFlowEquationwr} in a more compact form as follows
\begin{align}
    \dot{w}_r &= w_r \fcirc \dot{P}_r \fcirc w_r \nonumber \\
    &\phantom{=} + w_r \fcirc \lambda_r \circ \Pi_r \circ \dot{\mathcal{B}}_r \circ \Pi_r \circ \overline{\lambda}_r \fcirc w_r + \dot{\mathcal{B}}_r \nonumber \\
    &\phantom{=} + w_r \fcirc \lambda_r \circ \Pi_r \circ \dot{\mathcal{B}}_r + \dot{\mathcal{B}}_r \circ \Pi_r \circ \overline{\lambda}_r \fcirc w_r. \label{eq:FlowEqwrPr}
\end{align}
The terms in the last two lines of Eq.~\eqref{eq:FlowEqwrPr} can also be further simplified with the help of the SBE equations~\eqref{eq:SBEeqwB1} and~\eqref{eq:SBEeqwB2}, which are equivalent to
\begin{subequations}
\begin{align}
    w_r \fcirc \lambda_r \circ \Pi_r \circ \mathbf{1}_r = w_r \fcirc \mathcal{B}_r^{-1} - \mathbf{1}_r, \\
    \mathbf{1}_r \circ \Pi_r \circ \overline{\lambda}_r \fcirc w_r = \mathcal{B}_r^{-1} \fcirc w_r - \mathbf{1}_r.
\end{align}    
\end{subequations}
Combining these two relations with Eq.~\eqref{eq:FlowEqwrPr}, we obtain a particularly simple expression for $\dot{w}_r$:
\begin{equation}
    \dot{w}_r = w_r \fcirc \dot{P}_r \fcirc w_r + w_r \fcirc \mathcal{B}_r^{-1} \fcirc \dot{\mathcal{B}}_r \fcirc \mathcal{B}_r^{-1} \fcirc w_r.
\end{equation}
Since $\partial_\Lambda\!\left(w_r^{-1}\right) = -w_r^{-1} \fcirc \dot{w}_r \fcirc w_r^{-1}$ and $\partial_\Lambda\!\left(\mathcal{B}_r^{-1}\right) = -\mathcal{B}_r^{-1} \fcirc \dot{\mathcal{B}}_r \fcirc \mathcal{B}_r^{-1}$, this can be rearranged as
\begin{equation}
    \partial_\Lambda\!\left(w_r^{-1}\right) = \partial_\Lambda\!\left(\mathcal{B}_r^{-1}\right) - \dot{P}_r.
\end{equation}
This is consistent with the fact that $w_r$ satisfies a Dyson equation where $P_r$ plays the role of the self-energy, namely,
\begin{equation}
    w_r^{-1} = \mathcal{B}_r^{-1} - P_r,
    \label{eq:DysonEqwr}
\end{equation}
which is equivalent to the SBE equations~\eqref{eq:SBEeqwB1}--\eqref{eq:SBEeqwB2}.

Therefore, all additional terms generated by the $\Lambda$-dependence of the bare interaction $U$ (i.e., all terms depending explicitly on $\dot{\mathcal{B}}_r$ or $\dot{\mathcal{F}}_r$) in the SBE flow equations~\eqref{eq:FinalFlowEquationswrlambdarMr} can be obtained via the substitutions $\dot{I}_r\rightarrow \dot{I}_r - \dot{\mathcal{B}}_r$ in each of those flow equations and $\dot{w}_r\rightarrow \dot{w}_r - w_r \fcirc \partial_\Lambda\!\left(\mathcal{B}_r^{-1}\right)\!\fcirc w_r$ in the flow equation expressing $\dot{w}_r$. We also stress that introducing a regulator for the bare interaction, and especially for $\mathcal{B}_r$, gives a direct control on bosonic fluctuations, as it regularizes bosonic propagators through Eq.~\eqref{eq:DysonEqwr}.

An interesting connection can be observed between the last substitution applied to $\dot{w}_r$ and the Katanin substitution~\cite{Katanin2004}, which is applied to the single-scale propagator $S \equiv \left.\dot{G}\right|_{\dot{\Sigma}=0}$ (usually considered in the $1\ell$-fRG approach, see App.~\ref{sec:WetterichVertexExpUflow}) as
\begin{equation}
S \rightarrow \dot{G}=S + G \, \dot{\Sigma} \, G,
\label{eq:KataninSubstitution}
\end{equation}
with all indices left implicit. The expression of $\dot{G}$ in the substitution~\eqref{eq:KataninSubstitution} follows directly from the Dyson equation $G^{-1}=G_0^{-1}-\Sigma$, namely, $\dot{G}=-G\!\left(\partial_\Lambda G^{-1}\right)\! G = S + G \, \dot{\Sigma} \, G$ where $S=-G\!\left(\partial_\Lambda G_0^{-1}\right)\! G$. Returning to the case of the bosonic propagator $w_r$, the substitution $\dot{w}_r\rightarrow \dot{w}_r - w_r \fcirc \partial_\Lambda\!\left(\mathcal{B}_r^{-1}\right)\!\fcirc w_r$, which follows from the interaction flow, i.e., $\dot U \neq 0$, can be seen as a Katanin substitution for the bosonic propagator:
\begin{subequations}
    \begin{align}
        \dot w_r &= S_{w_r} + w_r\fcirc \dot P_r\fcirc w_r,\\
        S_{w_r} &= - w_r \fcirc \partial_\Lambda\!\left(\mathcal{B}_r^{-1}\right)\!\fcirc w_r = \left. \dot{w}_r \right|_{\dot{P}_r = 0},
    \end{align}
\end{subequations}
where $S_{w_r}$ is the bosonic counterpart of the single-scale propagator $S$. The latter is added to the flow of $\dot w_r$ retroactively in contrast to the fermionic Katanin substitution where the term $G\dot\Sigma G$ (i.e., the counterpart of $w_r\fcirc \dot P_r\fcirc w_r$) is added.

The identification of the single-scale propagator $S_{w_r}$ in the flow equation for the bosonic propagator underlines the fact that a flowing interaction, i.e., $\dot U$ (or $\dot{\mathcal{B}}_r$) comes along with a direct regularization of bosonic fluctuations. The SBE formalism is equivalent to a Hubbard--Stratonovich theory where $U$ takes the role of the bare bosonic propagator~\cite{Bonetti2022,GieversPhDthesis}. Whereas previous fRG approaches with both fermionic and bosonic regulators were sensitive to the cutoff scheme~\cite{Floerchinger2009Exact,Friederich2010Fourpoint,Homenda2024Generalized}, our formalism guarantees regulator independence at multiloop convergence.

\subsubsection{Self-energy flow equations}
\label{sec:SelfEnergyFlow}

The flow equations~\eqref{eq:FinalFlowEquationswrlambdarMr} or~\eqref{eq:UflowmultiloopSBEfRGEq1l}--\eqref{eq:UflowmultiloopSBEfRGEqnl} determine the flow of the two-particle vertex $V$. They should be coupled with a flow equation for the self-energy $\Sigma$, which directly contributes to Eqs.~\eqref{eq:UflowmultiloopSBEfRGEq1l}--\eqref{eq:UflowmultiloopSBEfRGEqnl} via the bubbles $\Pi_r$, defined from the full propagator $G$ in Eqs.~\eqref{eq:Bubbles}. In the conventional fRG where the regulator is introduced in the bare propagator only (i.e., with $\dot{U}=\dot{\mathcal{B}}_r=\dot{\mathcal{F}}_r=0$), this self-energy flow equation is determined from the Wetterich equation~\cite{Wetterich1993,Ellwanger1994,Morris1994}. When starting from the Wetterich equation, however, adding a regulator in the bare interaction, and thus allowing the derivative $\dot{U}$ to be non-zero, drastically increases the complexity, which is demonstrated in App.~\ref{sec:WetterichVertexExpUflow}. 

Another self-energy flow equation was formulated by differentiating the Schwinger--Dyson equation~\cite{Kugler2018c,Hille2020a,Hille2020b,HillePhDThesis}, which echoes the derivation of the multiloop fRG equations for the two-particle vertex from the Bethe--Salpeter equations. This version was actually shown to be advantageous as compared to the conventional self-energy flow equation in multiloop fRG studies of the pseudogap physics in the 2D Hubbard model~\cite{Hille2020a,HillePhDThesis}, when treating momentum dependencies within the truncated-unity fRG approach~\cite{Husemann2009,Wang2012,Lichtenstein2017} which is now very widespread in condensed-matter applications~\cite{Schober2018,Eckhardt2020,Profe2022,Gneist2022,Bonetti2022,Fraboulet2022,Beyer2023,Profe2024}. We will thus follow this route and derive a self-energy flow equation for our interaction flow scheme by differentiating the Schwinger--Dyson equation~\cite{BickersSelfConsistent2004,KuglerPhDthesis},
\begin{equation}
    \Sigma_{1^\prime|1} = - U_{1^\prime 2^\prime|12} G_{2|2^\prime} - \tfrac{1}{2} U_{1^\prime 2^\prime|4 3} G_{2|2^\prime} G_{3|3^\prime} G_{4|4^\prime} V_{4^\prime 3^\prime|1 2},
\label{eq:SchwingerDysonEquation}
\end{equation}
which can be rewritten in terms of the bubbles $\Pi_r$ defined by Eqs.~\eqref{eq:Bubbles},
\begin{subequations}
\label{eq:SigmaPirSDE}
    \begin{align}
    \Sigma &= G \cdot \left( U + \tfrac{1}{2} \; U \circ \Pi_{ph} \circ V \right), \\
    \Sigma &= - \left( U + \tfrac{1}{2} \; U \circ \Pi_{\overline{ph}} \circ V \right) \cdot G, \\
    \Sigma &= G \cdot \left( U + U \circ \Pi_{pp} \circ V \right).
\end{align}
\end{subequations}
Here, we have introduced the $\cdot$ product defined as follows:
\begin{subequations}
    \begin{align}
    \left[ A \cdot G \right]_{1^\prime|1} &= A_{1^\prime 2^\prime | 1 2} G_{2|2^\prime} \\
    &= - G_{2|2^\prime} A_{2^\prime 1^\prime | 1 2} 
    = - \left[ G \cdot A \right]_{1^\prime|1},
\end{align}
\end{subequations}
with $A$ an arbitrary four-point object~\cite{Patricolo2025, GieversPhDthesis}.

Introducing the RG scale in the bare propagator and the bare interaction through $G_0\rightarrow G_0^\Lambda$ and $U\rightarrow U^\Lambda$ as before, Eqs.~\eqref{eq:SigmaPirSDE} can be turned into the following flow equations:
\begin{subequations}
\begin{align}
    \dot{\Sigma} &= \dot{G} \cdot \left( U + \tfrac{1}{2} \; U \circ \Pi_{ph} \circ V \right) \nonumber \\
    &\phantom{=} + \tfrac{1}{2} \: G \cdot \left( U \circ \dot{\Pi}_{ph} \circ V + U \circ \Pi_{ph} \circ \dot{V} \right) \nonumber \\
    &\phantom{=} + G \cdot \left( \dot{U} + \tfrac{1}{2} \; \dot{U} \circ \Pi_{ph} \circ V \right), \label{eq:SigmadotPirSDEph} \\
    \dot{\Sigma} &= - \left( U + \tfrac{1}{2} \; U \circ \Pi_{\overline{ph}} \circ V \right) \cdot \dot{G} \nonumber \\
    &\phantom{=} - \tfrac{1}{2} \left( U \circ \dot{\Pi}_{\overline{ph}} \circ V + U \circ \Pi_{\overline{ph}} \circ \dot{V} \right) \cdot G \nonumber \\
    &\phantom{=} - \left( \dot{U} + \tfrac{1}{2} \; \dot{U} \circ \Pi_{\overline{ph}} \circ V \right) \cdot G, \\
    \dot{\Sigma} &= \dot{G} \cdot \left( U + U \circ \Pi_{pp} \circ V \right) \nonumber \\
    &\phantom{=} + G \cdot \left( U \circ \dot{\Pi}_{pp} \circ V + U \circ \Pi_{pp} \circ \dot{V} \right) \nonumber \\
    &\phantom{=} + G \cdot \left( \dot{U} + \dot{U} \circ \Pi_{pp} \circ V \right).
\end{align}
\label{eq:SigmadotPirSDE}
\end{subequations}

Equations~\eqref{eq:SigmadotPirSDE} are all \emph{exact}. By adding one of these three self-energy flow equations to the set of differential equations formed by Eqs.~\eqref{eq:UflowmultiloopSBEfRGEq1l}--\eqref{eq:UflowmultiloopSBEfRGEqnl} and imposing a condition for the 2PI vertex, e.g., the parquet approximation $\dot{I}^{\text{2PI}} = \dot{U}$, Eq.~\eqref{eq:I2PIapprox}, we obtain a \emph{closed system} of equations. It is also worth noting that, in the SBE framework based on $U$-reducibility, the Schwinger--Dyson equation can be simplified efficiently by combining Eqs.~\eqref{eq:SigmaPirSDE} with the relation $w_r \fcirc \lambda_r = U + U \circ \Pi_r \circ V$~\cite{Patricolo2025}. This SBE formulation of the self-energy flow equations~\eqref{eq:SigmadotPirSDE} is presented in App.~\ref{sec:SU2symmetry} for $SU(2)$-spin-symmetric systems, and will be used in our study for the Anderson impurity model with a local interaction (i.e., with $\mathcal{F}_r=0$). In the SBE formalism based on $\mathcal{B}$-reducibility, we have instead the relation $w_r \fcirc \lambda_r = \mathcal{B}_r + \mathcal{B}_r \circ \Pi_r \circ V$, which does not enable us to replace the full terms $U \circ \Pi_r \circ V = (\mathcal{B}_r+\mathcal{F}_r) \circ \Pi_r \circ V$ in Eqs.~\eqref{eq:SigmaPirSDE} by the simpler structures $w_r \fcirc \lambda_r - \mathcal{B}_r$ if $\mathcal{F}_r$ is finite. In the case of the Anderson--Holstein impurity model where the fermionic bare interactions $\mathcal{F}_r$ are non-trivial, we will therefore employ a self-energy flow equation from Eqs.~\eqref{eq:SigmadotPirSDE}.

\section{Numerical results for loop convergence with interaction regulators}
\label{sec:ModelsNumericalImplementationConvergence}

In this section, we demonstrate by a numerical analysis that our flow equations can be solved with similar convergence behavior to that of a conventional flow scheme. In particular, we illustrate numerically the loop convergence of our equations by applying them to the Anderson impurity model. To rule out the role of a propagator regulator in convergence, we consider a simple multiplicative regulator introduced exclusively in the bare interaction that interpolates between the non-interacting and interacting models, which we refer to as ``$U$-flow (2025)''. This can be seen as the simplest implementation of our interaction-flow scheme, which only amounts to switching on adiabatically the bare interaction throughout the flow, starting from the non-interacting theory.

\begin{figure*}[t!]
    \centering
    \scalebox{1}[1.0]{\includegraphics[width=1.0\linewidth, keepaspectratio=false]{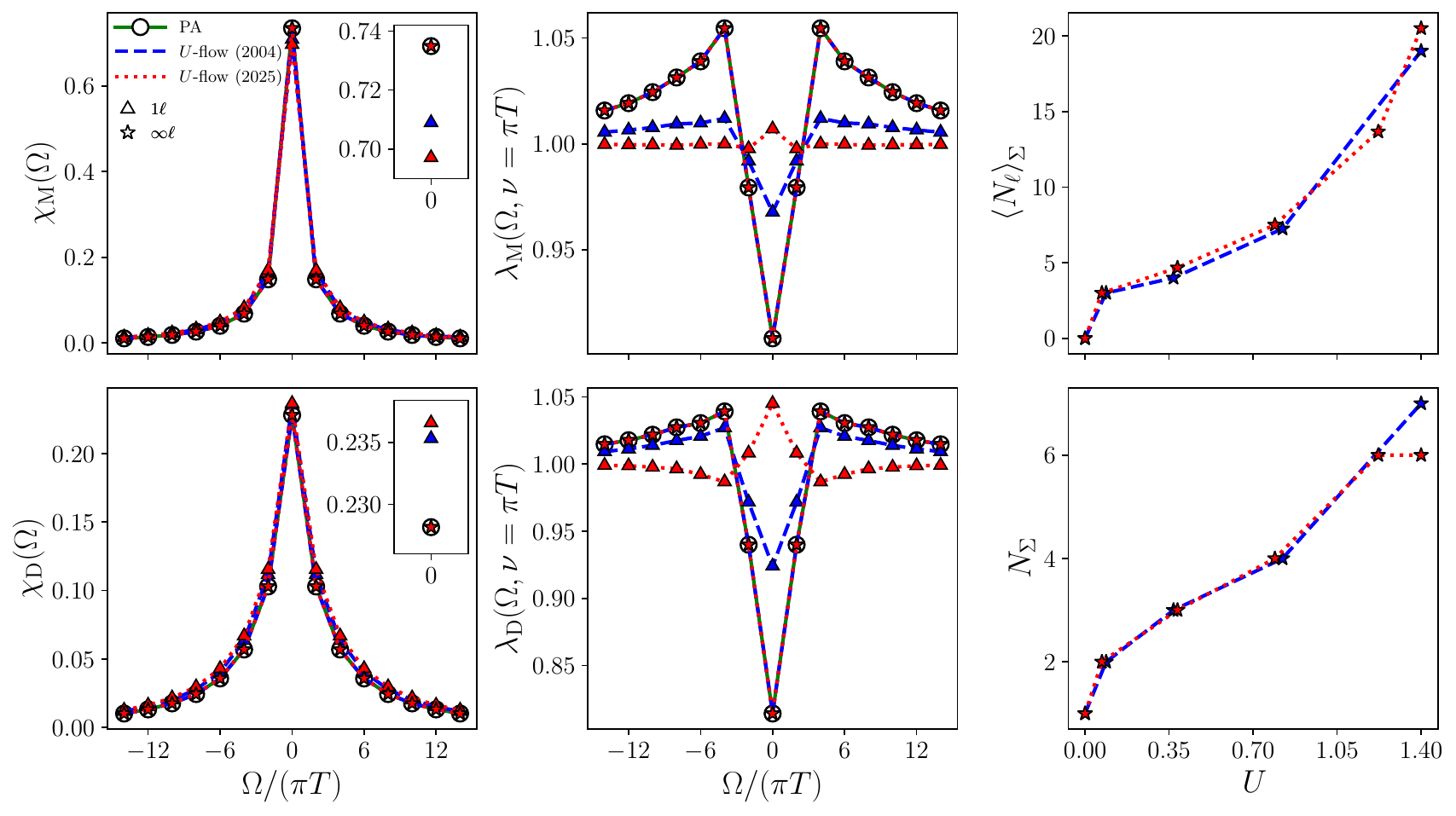}}
    \caption{Bosonic frequency dependence of susceptibilities and fermion-boson vertices for the AIM in the magnetic and density channels at $\beta = 5$, $U = 1.4$, $\Delta_0 = \pi/5$ and at half filling. The results are shown at $1\ell$ and $\infty \ell$ (fully converged) multiloop order for the two flow schemes specified in the text. Note that the $\D$ and $\SC$ channels are degenerate, i.e., $\chi_\D=\chi_\SC$ and $\lambda_\D=\lambda_\SC$. The insets in the left-hand panels show the same susceptibilities as the corresponding main panels, but magnified around $\Omega=0$. It is observed that, like the $U$-flow (2004), the $U$-flow (2025) converges to the PA solution, which, by construction, does not depend on the choice of a cutoff scheme. The panels on the right-hand side show the average number of loop corrections per self-energy iteration $\left\langle N_{\ell} \right\rangle_{\Sigma}$ and the corresponding number of self-energy iterations $N_{\Sigma}$ at each step of the flow.}
    \label{fig:aim_loopconvergence}
\end{figure*}

We benchmark its performance against self-consistent SBE calculations (that result from solving equations \eqref{eq:selfconsistentSBEequations} self-consistently), and against another analogous flow scheme based on a regulator in the bare propagator only within the multiloop fRG. To highlight the novel features of the new flow equations, we focus on an impurity model without momentum dependence. While momentum dependence generally increases the complexity of the problem, this complication is generic to all fRG schemes and is not specific to our new scheme. In particular, applying our approach to momentum-dependent models does not introduce additional regulator-induced or conceptual complications beyond those inherent to conventional fRG schemes. Further benchmarks on finite loop truncations and on another model, the Hubbard atom~\cite{Hubbard1963}, are presented in App.~\ref{sec:hubbard_atom}.

\subsection{Model definition and flow schemes}
\label{sec:ModelDefinitionsCutoffSchemes}

\par
\emph{The Anderson impurity model (AIM).} The AIM exhibits full $SU(2)$ spin symmetry, and by frequency conservation, one has $G_{0;\sigma_1|\sigma_2}(i\nu_1|i\nu_2) = \delta_{\sigma_1,\sigma_2}\delta_{i\nu_1, i\nu_2} G_0(i\nu)$. It is then also convenient to work in physical channels (see App.~\ref{sec:SU2symmetry} for details).

The AIM consists of an impurity coupled to a fermionic bath. It is relevant in the self-consistency cycle of DMFT and has become an important benchmark model for $1\ell$ fRG \cite{Karrasch2008Finite,Jakobs2010Nonequilibrium} and multiloop fRG \cite{Chalupa2020,Ge2024,Ritz2024} computations. The bath degrees of freedom can be integrated out to yield a bare propagator given by
\begin{align}
G_{0}(i\nu) = \frac{1}{i\nu - \mu - \Delta(\nu)}, \label{eq:anderson_impurity_propagator}
\end{align}
where $\Delta(\nu)$ is the hybridization function. We consider a bath with a box-shaped density of states with half bandwidth $D$ and an isotropic hybridization strength $\mathcal{V}$. The corresponding hybridization function takes the form
\begin{align}
\Delta(\nu) = i\frac{\mathcal{V}^2}{D} \arctan(\frac{D}{\nu}).
\end{align}
This can be simplified in the so-called wide-band limit $D \rightarrow \infty$ so that the hybridization function reduces to 
\begin{align}
\lim_{D \rightarrow \infty}\Delta(\nu) \equiv i\, \text{sgn}(\nu)\, \Delta_0,
\end{align}
where $\mathcal{V}$ is scaled in such a way that $\Delta_0$ is kept finite. As the model fulfills $SU(2)$ spin symmetry, we here use physical channels $\mathrm{X}=\D,\M,\SC$, which are defined in App.~\ref{sec:SU2symmetry}. In these channels, the bare interaction is given by
\begin{align}
U_\D = U_\SC = -U_\M = -U,
\label{eq:hubbard_interaction_physical_channels}
\end{align}
where $U$ is a Hubbard interaction parameter (not to be confused with the bare vertex from Sec.~\ref{sec:Formalism}). In terms of the $\mathcal{B}+\mathcal{F}$-splitting, the situation corresponds to $\mathcal{B}_\X = U_\X$ and $\mathcal{F}_\X = 0$.

In this model, the criterion for weak coupling, where the parquet approximation (PA), Eq.~\eqref{eq:I2PIapprox}, becomes quantitatively accurate, is governed by the requirement that
\begin{align}
\max_{\X}\abs{\Pi_\X(i\Omega = 0) U_\X} \approx  \frac{U}{\pi\Delta_0}< 1.\label{eq:PA_valid_copndition_AIM}
\end{align}
 When satisfied, we can expect good quantitative accuracy in our multiloop scheme up to $\mathcal{O}([\beta U/4]^4)$.

\par
\emph{Flow schemes.} To properly benchmark our flow equations, we utilize a simple multiplicative regulator~\footnote{We remark that although we use the term ``regulator'', this flow scheme is not really regularizing since it does not handle the singularities of the bare propagator in Eq.~\eqref{eq:anderson_impurity_propagator}.} \emph{exclusively} in the bare interaction.
\begin{enumerate}[leftmargin=*]
    \item[$\bullet$] \textbf{$U$-flow (2025).}  We consider a scheme specified by making the bare model dependent on a flow parameter $\Lambda$ in the following way:
\begin{subequations}
    \begin{align}
    U_\X^\Lambda &\equiv \Lambda U_\X \quad (\mathcal{B}_\X^\Lambda \equiv \Lambda \mathcal{B}_\X, \  \mathcal{F}_\X^\Lambda \equiv\Lambda\mathcal{F}_\X),\\
    G_0^\Lambda &\equiv G_0,
\end{align}
\end{subequations}
where $\Lambda$ flows from $\Lambda_{\text{init}} = 0$ to $\Lambda_{\text{final}} = 1$. We stress here that the bare propagator does not flow. At the beginning of the flow, we have $U_\X^{\Lambda_{\text{init}}} = 0$, the non-interacting system. The intermediate steps in this flow clearly correspond to solutions of the family of models with the respective bare interaction  $\Lambda U$. 
\end{enumerate}

An analogous flow scheme based on introducing a regulator only in the propagator was developed in previous works.

\begin{enumerate}[leftmargin=*]
\item[$\bullet$] \textbf{$U$-flow (2004).} Introduced in Ref.~\cite{Honerkamp2004}, it is specified by a multiplicative regulator in the bare propagator as    
    \begin{align}
    &G_0^{\Lambda} \equiv \Lambda G_0,
    \label{eq:interaction_flow_2004}
    \end{align} 
    where $\Lambda$ flows from $\Lambda_{\text{init}} = 0$ to $\Lambda_{\text{final}} = 1$. This regulator is a result of multiplying the quartic interaction part of the action by $\Lambda^2$, and then absorbing it away from the quartic part through a redefinition of the fields $\overline{c}_i, c_i \rightarrow \sqrt{\Lambda}\overline{c}_i, \sqrt{\Lambda}c_i$ leading to $\Lambda$ appearing next to the bare propagator $G_0$ in Eq.~\eqref{eq:interaction_flow_2004}. We stress that the bare interaction $U$ in this scheme does not flow. Nonetheless, intermediate steps of the flow can be interpreted to correspond to an interacting theory at various bare interaction values after rescaling the vertex functions by appropriate powers of $\Lambda$. In particular, the rescaled vertex $\Lambda^2V^\Lambda$ at step $\Lambda$ corresponds to a system with bare interaction $\Lambda^2 U$. This is in contrast to $U$-flow (2025), where the vertices at intermediate steps of the flow need \emph{not} be rescaled.
\end{enumerate}
Intuitively, while both give flows of increasing interaction strength, the difference between the $U$-flow (2004) and $U$-flow (2025) schemes can be thought of as follows. In the $U$-flow (2025), the propagator (i.e., the kinetic energy) is fixed while the Hubbard interaction (i.e., the potential energy) is increased during the flow. On the other hand, the $U$-flow (2004) does it by keeping the potential energy fixed, but by reducing the kinetic energy, so that the relative interaction strength increases along the flow. As we will see, when truncating the multiloop series, these two flow schemes become distinct.

\subsection{Numerical results}
\label{sec:numerical_results}

In principle, for a complete multiloop fRG calculation, one must calculate and sum up each term in the multiloop series for the vertex derivatives, Eqs.~\eqref{eq:LoopExpansionwrlambdarMr}. This series, however, can be terminated at a finite loop order where the remaining terms are smaller than a chosen error. The correctness of this relies on the assumption that the multiloop series converges -- a requirement made plausible by the fact that $m\ell$ and $(m+1)\ell$ terms differ by a product of a bubble and a vertex. Moreover, the flow equations exhibit coupling between the self-energy derivative $\dot{\Sigma}$ and the vertex derivatives, which needs to be resolved iteratively at each step of the flow. When the multiloop expansion is truncated at a fixed loop order $n$, we refer to this as an $n\ell$ calculation. In a similar vein, one may fix a maximum number of self-energy iterations. Technical details and frequency parametrization in the implementation are given in App.~\ref{sec:technical_aspects_and_freq_param}.

We performed calculations for the AIM at $\beta = 5$, $\Delta_0 = \pi/5$ and $U=1.4$. This choice of parameters satisfies the convergence criterion, Eq.~\eqref{eq:PA_valid_copndition_AIM}. For the chemical potential, we take $\mu = -U/2$, which corresponds to half filling.

\begin{figure}[t]
    \centering
    \includegraphics[width=0.9\linewidth]{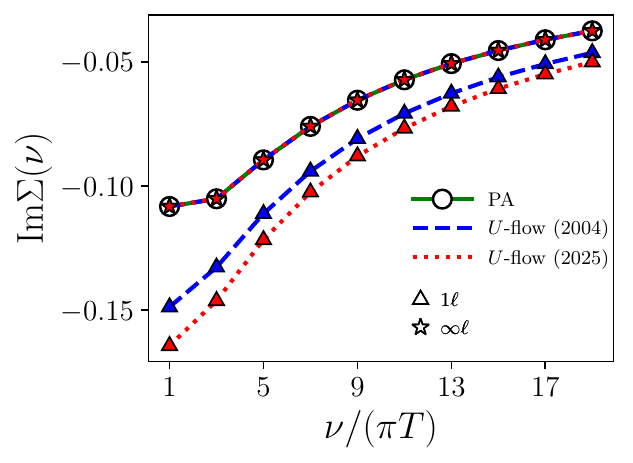}
    \caption{Imaginary part of the self-energy from the $1\ell$ fRG and the converged multiloop fRG ($\infty\ell$) compared to the self-consistent result in the PA. The results are obtained for the AIM with the same parameters as in Fig.~\ref{fig:aim_loopconvergence} for two flow schemes.}
    \label{fig:selfenergy_1l_vs_inftyl}
\end{figure}

In Fig.~\ref{fig:aim_loopconvergence}, we show results for the susceptibilities $\chi_\X$ and the fermion-boson vertices $\lambda_\X$ of fully converged multiloop calculations, marked as ``$\infty \ell$'', for the $U$-flow (2025), and benchmark them against fully converged $U$-flow (2004) calculations, as well as self-consistent results of the SBE equations~\eqref{eq:selfconsistentSBEequations} in the parquet approximation, i.e., $I^{2\mathrm{PI}}=U$,  Eq.~\eqref{eq:I2PIapprox}, which we abbreviate as PA. We also show the $1\ell$ results for the two flow schemes (with a fixed number of $N_{\Sigma}=2$ self-energy iterations). Note that at half filling, there is a degeneracy between the superconducting $\X = \SC$ and density $\X = \D$ channels, so that $\chi_\D = \chi_\SC$ and $\lambda_\D = \lambda_\SC$. Additionally, Fig.~\ref{fig:selfenergy_1l_vs_inftyl} shows the imaginary part of electronic self-energies. Overall, we find excellent quantitative agreement with the self-consistent solution of the SBE equations~\eqref{eq:selfconsistentSBEequations} in the PA, demonstrating the convergence of our flow equations~\eqref{eq:UflowmultiloopSBEfRGEq1l}--\eqref{eq:UflowmultiloopSBEfRGEqnl}.

In Fig.~\ref{fig:aim_loopconvergence}, we also show the average number of multiloop corrections (averaged over the self-energy iterations and denoted by $\left\langle N_{\ell} \right\rangle_{\Sigma}$) and the number of self-energy iterations (denoted by $N_\Sigma$) that were required at each step of the flow to reach multiloop convergence. We find these numbers for the $U$-flow (2025) to be nearly identical to those of the $U$-flow (2004). As discussed in the previous section, although both schemes implement the same RG flows, the two flow schemes are nonetheless distinct. This can be seen in the results: the two flow schemes yield different results at truncated loop order, as can be seen from the finite $1\ell$ results. Thus, we have demonstrated that our flow equations also converge if there is no regulator in the bare propagator but exclusively in the bare interaction. Furthermore, the convergence rate of our flow equations is similar to that in the conventional multiloop fRG $U$-flow (2004), which relies on a regulator in the bare propagator only.

\section{Beyond propagator cutoffs: Temperature flows for retarded interactions}
\label{sec:TflowRetardedInteractions}

In Sec.~\ref{sec:ModelsNumericalImplementationConvergence}, our interaction flow was applied as a simple example of a flow scheme with a regulator in the bare interaction to straightforwardly implement the idea of a flow in the microscopic interaction of the system [see $U$-flow (2025)]. The flexibility to deform both the propagator and bare interaction, however, goes beyond this simple application, as it provides full control over the flow of the bare couplings of the interacting theory. To demonstrate this, we here present a first practical application that requires taking full advantage of this flexibility. To this end, we address the problem of developing a flow scheme wherein the RG flow corresponds to lowering the temperature of the system. Such a scheme aligns well with the intuition that an RG scale can be loosely interpreted as a temperature scale, and rather makes this idea concrete. For Hubbard models, this problem was addressed in the temperature-flow scheme first introduced in Ref.~\cite{Honerkamp2001}. Recently, this approach was adapted for spin systems~\cite{Schneider2023}. The convenience of this scheme has already been exploited to efficiently perform temperature scans for phase diagrams \cite{Gneist2022,Braun2025,Gneist_2022_2,Schneider2023,Igoshev2023} and to naturally calculate various thermodynamic variables expressed as derivatives of temperature throughout the flow. As we show below, the original temperature-flow scheme (referred to as ``$T$-flow (2001)'') is limited to systems with only static interactions, and is therefore not applicable to models featuring retarded bare interactions, such as those in electron-phonon systems after phonons have been integrated out. To remedy this limitation, one must allow the bare interaction to flow as we show below.

\subsection{Temperature flow for instantaneous interactions}
We briefly review the idea behind the temperature-flow scheme of Ref.~\cite{Honerkamp2001}, which we will refer to as ``$T$-flow (2001)''. Our starting point is an action of the form Eq.~\eqref{eq:ClassicalActionS}, which we rewrite here focusing on frequency dependence and making the Matsubara frequency summations and their normalization factors explicit: 
\begin{align}
S[\overline{c},c] &= - T\sum_{i\nu_1}\overline{c}_{i\nu_1} G_{0;i\nu_1|i\nu_1}^{-1} c_{i\nu_1} \nonumber \\
&\phantom{=} - T^3\sum_{\mathclap{i\nu_1,i\nu_2,i\nu_1^\prime}} \tfrac{1}{4}U_{i\nu_1^\prime i\nu_2^\prime|i\nu_1i\nu_2} \overline{c}_{i\nu_1^\prime} \overline{c}_{i\nu_2^\prime} c_{i\nu_2} c_{i\nu_1},
\end{align}
where we have invoked energy conservation with $i\nu_2^\prime = i\nu_1 + i\nu_2 - i\nu_1^\prime$ in the second line. The main step is to switch to rescaled fields 
\begin{align}
\tilde{c}(i\tilde{\nu}) = T^{3/4}c(i\nu),\quad \overline{\tilde{c}}(i\tilde{\nu}) = T^{3/4}\overline{c}(i\nu), \label{eq:rescaling_T_flow}
\end{align}
where the new variable $\tilde{\nu} \equiv \nu/T$ is a dimensionless Matsubara frequency. This brings the action to the form:
\begin{align}
    S[\overline{\tilde{c}},\tilde{c}] &= - \sum_{i\tilde{\nu}_1}\overline{\tilde{c}}_{i\tilde{\nu}_1} (T^{1/2}G_{0;i{\nu}_1|i{\nu}_1})^{-1} \tilde{c}_{i\tilde{\nu}_1} \nonumber \\
    &\phantom{=} - \sum_{\mathclap{i\tilde{\nu}_1,i\tilde{\nu}_2,i\tilde{\nu}_1^\prime}} \tfrac{1}{4}U_{i{\nu}_1^\prime i{\nu}_2^\prime|i{\nu}_1i{\nu}_2} \overline{\tilde{c}}_{i\tilde{\nu}_1^\prime} \overline{\tilde{c}}_{i\tilde{\nu}_2^\prime} \tilde{c}_{i\tilde{\nu}_2} \tilde{c}_{i\tilde{\nu}_1}.
    \label{eq:rescaled_tflow_action}
\end{align}
In this way, if $U$ is, e.g., a Hubbard interaction, all $T$-dependence has been shifted to the bare propagator. Regarding $T$ as an RG flow parameter in the flow-dependent propagator, one arrives at flow equations, which at intermediate steps can be interpreted as the physical state of the system at the corresponding temperature:
\begin{enumerate}
\item[$\bullet$] \textbf{$T$-flow (2001).} This scheme is implemented by
\begin{align}
    G^T_0(i\nu) \equiv T^{1/2}G_0(i\nu),
    \label{eq:T_flow_2001}
\end{align}
where the flow parameter starts at $T^{\text{init}} = \infty$ and ends at the desired $T^{\text{final}}$. Note that this regulator is not purely multiplicative since the temperature dependence enters also the Matsubara frequency $i\nu = {i \pi T (2n + 1)}$. Furthermore, the physical self-energy at temperature $T$ is $\Sigma = \sqrt{T}\Sigma^T$, where $\Sigma^T$ is the self-energy calculated throughout the flow.
\end{enumerate}
Clearly, this construction only works as long as the bare interaction function itself does not depend on temperature. A situation where this is not the case is for retarded interactions, in which the bare interaction depends on Matsubara frequencies. A physically relevant example of this is a result of integrating out phonons that couple linearly to the fermions yields, and hence $U \rightarrow U(i\Omega)$ depends implicitly on temperature through the transfer Matsubara frequency $i\Omega = i2\pi nT$. Another example is in a high-energy fermionic system where retardation effects of the Coulomb interaction due to the finite speed of light must be included.

\subsection{Temperature flow for retarded interactions}

\begin{figure*}[t!]
    \centering
    \scalebox{1}[1.0]{\includegraphics[width=1.0\linewidth, keepaspectratio=false]{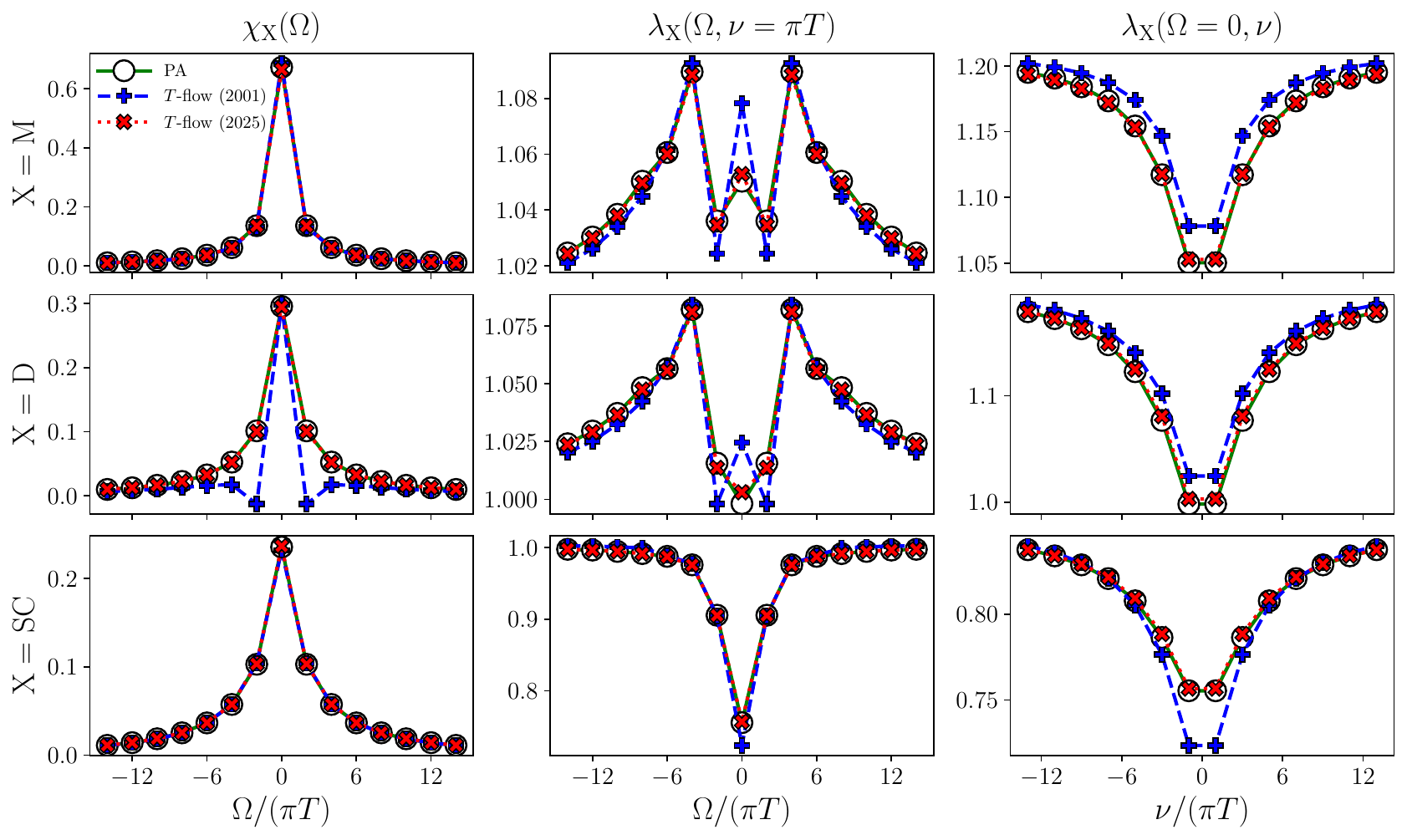}}
    \caption{Susceptibilities and fermion-boson vertices for the AHIM at inverse temperature $\beta = 5$, hybridization parameter $\Delta_0 = \pi/5$, Hubbard interaction $U = 1.4$, phonon frequency $\omega_0 = 1$, electron-phonon coupling $g = 0.5$, at half filling from the fully converged multiloop equations ($\infty\ell$). The $T$-flow (2001) that does not implement the necessary flow of the interaction fails qualitatively whereas the $T$-flow (2025) yields results that lie on top of the self-consistent solution of the SBE equations in the PA.}
    \label{fig:new_T_flow_vs old_T_flow_convergence}
\end{figure*}

The freedom of using a regulator in the interaction allows to develop an RG flow in temperatures even for systems with retarded interactions. To this end, we consider a paradigmatic model, in which phonons couple linearly to the fermion density.

\par
\emph{The Anderson--Holstein impurity model (AHIM).} This model describes an Anderson impurity where additionally the mechanical vibration of the impurity site from its equilibrium position is taken into account~\cite{Hewson2002}. Previously, it has been analyzed in the $1\ell$-fRG approach~\cite{Laakso2014Functional}. In this model, the electron density at the impurity is coupled to a phonon, which we take to have a fixed frequency $\omega_0$,  with electron-phonon coupling $g$. The bare impurity-electron propagator is still given by Eq.~\eqref{eq:anderson_impurity_propagator}. Integrating out the phonon degrees of freedom, one ends up with a density-density interaction of the form
\begin{align}
U_{1^\prime2^\prime|12} &= -\delta_{\nu_1^\prime+\nu_2^\prime, \nu_1 + \nu_2}(\delta_{\sigma_1^\prime,\sigma_1} \delta_{\sigma_2^\prime,\sigma_2} - \delta_{\sigma_1^\prime,\sigma_2} \delta_{\sigma_2^\prime,\sigma_1})\nonumber\\
&\phantom{=}\quad \times U(i\nu_1 - i\nu_1^\prime) ,
\end{align}
with the coupling function
\begin{align}
U(i\Omega) = U - V_H\frac{\omega_0^2}{\Omega^2 + \omega_0^2},
\end{align}
where $U$ is the Hubbard repulsion and the ratio $V_H \equiv 2g^2/\omega_0$ is the effective phonon-induced electron-electron interaction strength. The ratio $V_H$ is independent of the impurity mass $M_{\text{ion}}$ whereas $\omega_0 \sim 1/\sqrt{M_{\text{ion}}}$. Thus, exchanging one of the chemical constituents of the impurity with its isotope changes only $\omega_0$. Relevant to our discussion is the anti-adiabatic limit $M_{\text{ion}} \rightarrow \infty$, for which the coupling 
$U(i\Omega) \rightarrow U_{\text{eff}}$ becomes instantaneous. Here, $U_{\text{eff}}  \equiv U - V_H$ represents the effective instantaneous part of the interaction. The quantity $\omega_0$ thus also controls retardation; the smaller $\omega_0$, the more the retarded nature of the interaction becomes apparent. 
\par
\emph{$\mathcal{B}+\mathcal{F}$-splitting.} For a treatment of this model within the SBE formalism, which keeps the pure bosonic nature of the bosonic propagators intact, a $\mathcal{B}+\mathcal{F}$-splitting is necessary. Note that, in general, this splitting is not unique. Following Ref.~\cite{AlEryani2025}, we set
\begin{align}
\mathcal{B}_\X \equiv \lim_{i\nu \rightarrow 0,i\nu' \rightarrow 0} U_\X(i\Omega, i\nu, i\nu^\prime),
\end{align}
which ensures that $\mathcal{B}_\X$ contains the instantaneous part of the interaction. The following expressions are found in the physical channels:
\begin{align}
\mathcal{B}_\D(i\Omega) &= -U_{\text{eff}} - 2V_H \frac{\Omega^2}{\Omega^2 + \omega_0^2},\nonumber\\
\mathcal{B}_\SC &= -\mathcal{B}_\M = -U_{\text{eff}},  \label{eq:B_part}
\end{align}
for the bosonic parts and 
\begin{align}
\mathcal{F}_\D(i\nu,i\nu') &= \mathcal{F}_\M = -\mathcal{F}_\SC =2V_H \frac{(\nu - \nu')^2}{(\nu - \nu')^2 + \omega_0^2},\label{eq:F_part}
\end{align}
for the fermionic parts.

\par
\emph{Temperature flow.} As discussed previously, a flow that uses the temperature as a flow parameter, as implemented in the $T$-flow (2001) scheme, is not valid for our system, since the bare interaction contains non-trivial temperature dependence through the dependence on Matsubara frequencies. However, this issue can be remedied in a natural and straightforward way by interpreting the residual $T$-dependence of our bare interaction, after the rescaling in Eq.~\eqref{eq:rescaling_T_flow}, as dependence on the RG scale. As a result, we no longer have $\dot{\mathcal{B}}_\X = \dot{\mathcal{F}}_\X = 0$; instead we have the following new scheme.

\begin{enumerate}[leftmargin=*]
\item[$\bullet$] \textbf{$T$-flow (2025).}  Similarly to the $T$-flow (2001), a regulator in the propagator $G_0^T$ is implemented by Eq.~\eqref{eq:T_flow_2001}, with the temperature $T$ playing the role of the flow parameter starting at $T_{\text{init}} = \infty$ and ending at a chosen final temperature. The bare interactions Eqs.~\eqref{eq:B_part}--\eqref{eq:F_part} depend implicitly on the temperature through the dependence on the Matsubara frequencies $i\Omega = i 2\pi Tn$ or $i\nu = i \pi T (2n +1)$ with $n\in\mathbb{Z}$ . The scale derivatives for the bare interaction are given by
\begin{align}
\dot{\mathcal{B}}_\D(i\Omega) = -2V_H \frac{\mathrm{d}}{\mathrm{d}T}K^T(i\Omega), \quad \dot{\mathcal{B}}_{\SC} = \dot{\mathcal{B}}_{\M} = 0,
\end{align}
for the bosonic parts and
\begin{align}
\dot{\mathcal{F}}^T_\D(i\nu, i\nu^\prime) = \dot{\mathcal{F}}^T_\M = -\dot{\mathcal{F}}^T_\SC = 2V_H \frac{\mathrm{d}}{\mathrm{d}T}K^T(i\nu - i\nu^\prime),
\end{align}
for the fermionic part, where
\begin{align}
\frac{\mathrm{d}}{\mathrm{d}T}K^T(i2\pi T n ) = -\frac{8\pi^2 T n^2  \omega_0^2 }{[(2\pi T n)^2 + \omega_0^2]^2}.
\end{align} 
Throughout the flow, the vertex $V^T$ and the scaled self-energy $\sqrt{T}\Sigma^T$ correspond to the vertex and the self-energy of the model at temperature $T$. 
\end{enumerate}
We note that the conventional $T$-flow (2001) can be recovered from the $T$-flow (2025) if we set $\mathrm{d}K^T/{\mathrm{d}T}= 0$. At high temperatures, this term scales as ${\mathrm{d}}K^T/{\mathrm{d}T}\sim 1/T^3$, and therefore, the limitations of the $T$-flow (2001) scheme are expected to become most apparent as the temperature decreases.

\begin{figure}[t!]
    \centering
    \scalebox{1}{\includegraphics[width=0.9\linewidth, keepaspectratio=false]{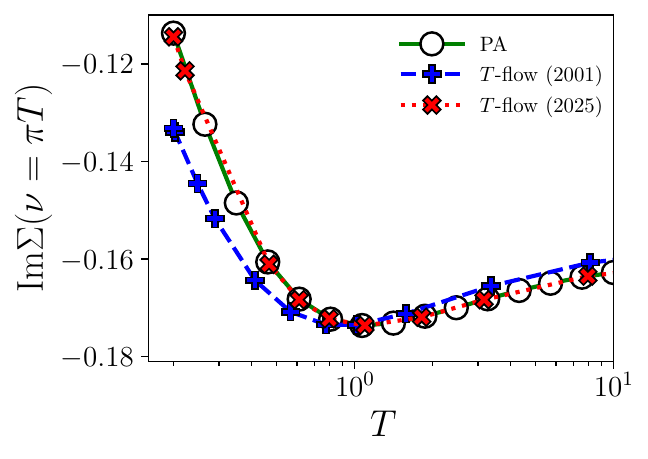}}
    \caption{Imaginary part of the self-energy obtained for different temperatures within the $T$-flow (2001) and $T$-flow (2025) calculations against 15 self-consistent PA calculations of temperatures between $T = 10$ and $T = 0.2$ for the AHIM with the same parameters as in Fig.~\ref{fig:new_T_flow_vs old_T_flow_convergence}. The $T$-flow (2001) exhibits a clear discrepancy with the PA results, especially for $T < 1$, whereas the results extracted from the $T$-flow (2025) calculation agree quantitatively with the PA calculations across the entire temperature range.}
    \label{fig:T_dependence_Im_Sig}
\end{figure}

\begin{figure}
    \centering
    \scalebox{1}[1.0]{\includegraphics[width=0.9\linewidth, keepaspectratio=false]{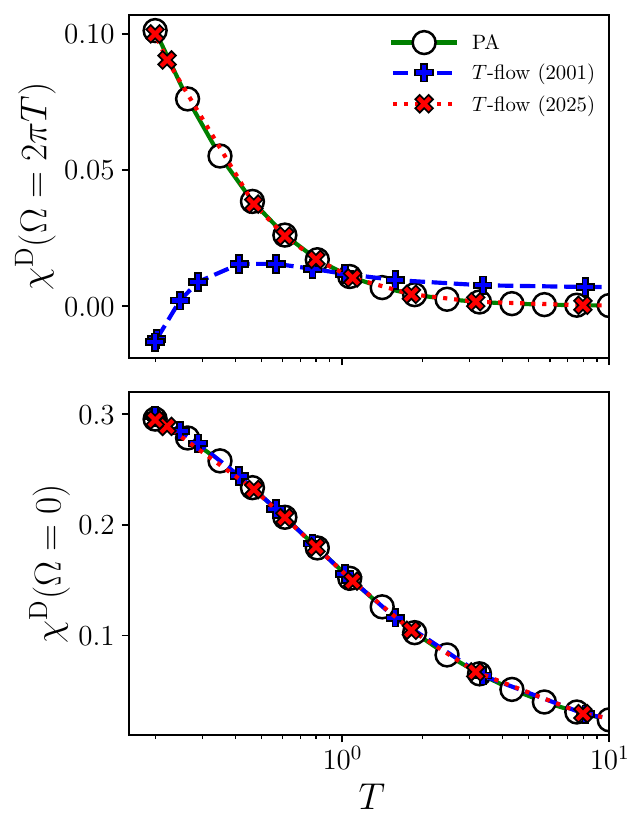}}
    \caption{Density susceptibility evaluated at its zeroth and first bosonic Matsubara frequency and obtained for different temperatures within the $T$-flow (2001) and the $T$-flow (2025), corresponding to the results shown in Fig.~\ref{fig:T_dependence_Im_Sig}. The first Matsubara frequency of the density susceptibility in the $T$-flow (2001) in general deviates from the self-consistent PA result, and starts to do so drastically for $T < 1$ and shows a nonphysical non-monotonic behavior with $T$. On the other hand, the $T$-flow (2025) yields excellent agreement with the PA results.}
    \label{fig:T_dependence_suscs}
\end{figure}

\subsection{Numerical results: Application to the Anderson--Holstein impurity model}
We consider the AHIM with effective hybridization $\Delta_0 = \pi/5$, half filling, i.e., $\mu = -U/2$, and a similar parameter set to that considered for the AIM in Sec.~\ref{sec:ModelsNumericalImplementationConvergence}. Additionally, here, the impurity is coupled to a phonon mode of frequency $\omega_0 = 1$ with electron-phonon coupling $g = 0.5$.

To begin, we performed calculations comparing fully converged multiloop calculations in the $T$-flow (2025) and the $T$-flow (2001) schemes with $T_{\text{final}} = 1/\beta$ and $\beta = 5$. As a benchmark, we also compare against the result of the self-consistent PA calculation for the same parameters at inverse temperature $\beta = 5$.

The results are shown in Fig.~\ref{fig:new_T_flow_vs old_T_flow_convergence} for the susceptibilities and fermion-boson vertices. Note that due to the non-zero $\mathcal{F}_\X$ in this model, $\lambda_\X$ does not in general decay asymptotically to $1$ when $i\nu \rightarrow \pm i\infty$, see also Ref.~\cite{AlEryani2024}. Unsurprisingly, the $T$-flow (2001) does not converge to the PA solution, as expected in dealing with a model with retarded interactions. Within the $T$-flow (2025) scheme on the other hand, which includes the necessary flow of the bare interaction, we find excellent agreement.

The results show that the finite frequency phonon, which couples preferentially to the density of electrons, leads to the breaking of the pseudospin symmetry of the AIM  responsible for the degeneracy between the density $\D$ and superconducting $\SC$ channels. While in the AIM the $\D$ and $\SC$ channels are degenerate (cf.\ Fig.~\ref{fig:aim_loopconvergence}), in the AHIM, the renormalized susceptibility $\chi_{\D}$ and the fermion-boson vertex $\lambda_{\D}$ at $\Omega=0$ are increased compared to $\chi_{\SC}$ and $\lambda_{\SC}$. This underlines enhanced phonon-driven density fluctuations relative to the superconducting ones. 

Additionally, we perform 15 separate PA calculations at logarithmically spaced inverse temperatures between $T = 10$ and $T = 0.2$. The results are then compared to single $T$-flow (2001) and $T$-flow (2025) calculations that are stopped at $T_{\text{final}} = 0.2$ and are displayed in Fig.~\ref{fig:T_dependence_Im_Sig} for the self-energy, and in Fig.~\ref{fig:T_dependence_suscs} for the density susceptibility at the zero and first Matsubara frequencies.

Whereas the static susceptibilities, obtained from the two $T$-flow schemes, are still in good agreement, which we can attribute to the vanishing of $\tfrac{\mathrm{d}}{\mathrm{d}T}K^T(i\Omega)$ at $i\Omega = 0$, the susceptibility at finite frequencies displays non-monotonicity with temperature in the $T$-flow (2001) scheme. As this non-monotonicity is absent in the $T$-flow (2025) as well as the self-consistent PA, we interpret it as nonphysical. We mention in passing the recent finding that connects the finite Matsubara-frequency components of the susceptibility to the entanglement of the system~\cite{Hauke2016Measuring,Bippus2025}.

As expected, we find that the $T$-flow (2001) starts to fail drastically for $T < 1$, whereas the $T$-flow (2025) scheme converges squarely to the self-consistent PA results for all $T$.

\section{Conclusion}
\label{sec:Conclusion}

In this work, we have introduced an extension of the multiloop fRG that incorporates regulators in both the bare propagator and the bare interaction, derived from the Bethe--Salpeter and Schwinger--Dyson equations. In the SBE formalism, these equations take a particularly transparent form, where they can be interpreted as a Katanin-type substitution for the SBE bosonic propagators. The resulting flow equations remain compact and computationally feasible. As a result, one gains complete control over the bare theory and thus virtually full control over how the RG flow explores theory space.

In a numerical application to the Anderson impurity model, we demonstrated loop convergence when a flow parameter is introduced in the bare interaction, comparing it to the conventional multiloop fRG when only the bare propagator is flowing. Furthermore, the enlarged flexibility of our equations allows to design a temperature-flow scheme with an RG flow interpreted as lowering the temperature of the system for models with retarded interactions, a possibility that has been absent in schemes that only involve regulators in bare propagators. A demonstration on an Anderson impurity model coupled to phonons, coined Anderson--Holstein impurity model~\cite{ELagos2001,Hewson2002}, confirms that this temperature-flow scheme yields physically consistent results, contrasting it with the original application of temperature-flow scheme that only utilizes a regulator for the bare propagator~\cite{Honerkamp2001}.

For the future, we see a large number of potential applications. The unprecedented control of the RG procedure in our flow equations opens the door to many interesting possibilities. For example, RG flows that start from a reference system such as the DMF$^2$RG find a new perspective to be used even for systems with non-local interactions, by turning on the non-local part of the interaction throughout the flow. In terms of new potential approximations or improvements, note that, since in the fermionic action, the bare interaction couples to \emph{four} fields whereas the bare propagator only couples to \emph{two} fields, the interaction flow implicitly relates two-point with four-point correlation functions. This might be beneficial for exploring flow schemes that respect Ward--Takahashi identities~\cite{EnssPhDthesis,Chalupa2020} and other exact functional relations to systematically improve RG flows beyond what is possible in standard fRG approaches.

Another family of applications of our equations is in problems concerning optimal control of fermionic quantum systems. There, one seeks an optimal sequence of steps to adjust a number of knobs, which may involve changing parameters in both the propagator and the interaction, to arrive at a final state while minimizing a certain cost, such as the response of the system.

\acknowledgments 
The authors thank S.\ Andergassen, S.\ Heinzelmann, A.\ Katanin, M.\ Krämer, M.\ Patricolo, J.\ Profe, N.\ Ritz, B.\ Schneider, M.\ Scherer, D.\ Vilardi and J.\ von Delft for valuable discussions. K.F.\ acknowledges financial support from the Austrian Science Fund (FWF) 10.55776/I6946 and from the Deutsche Forschungsgemeinschaft (DFG, German Research Foundation) within the research unit FOR 5413/1 (Grant No. 465199066). A.A.-E.\ acknowledges funding from the DFG under Project No.~277146847 (SFB 1238, project C02). Computational resources provided by the Paderborn Center for Parallel Computing (PC2), specifically the Noctua 2 high-performance computing system, was used for this research.

\appendix
\section*{Appendices}
\section{Interaction flow and conventional fRG formalism based on the Wetterich equation}
\label{sec:WetterichVertexExpUflow}

\begin{figure*}
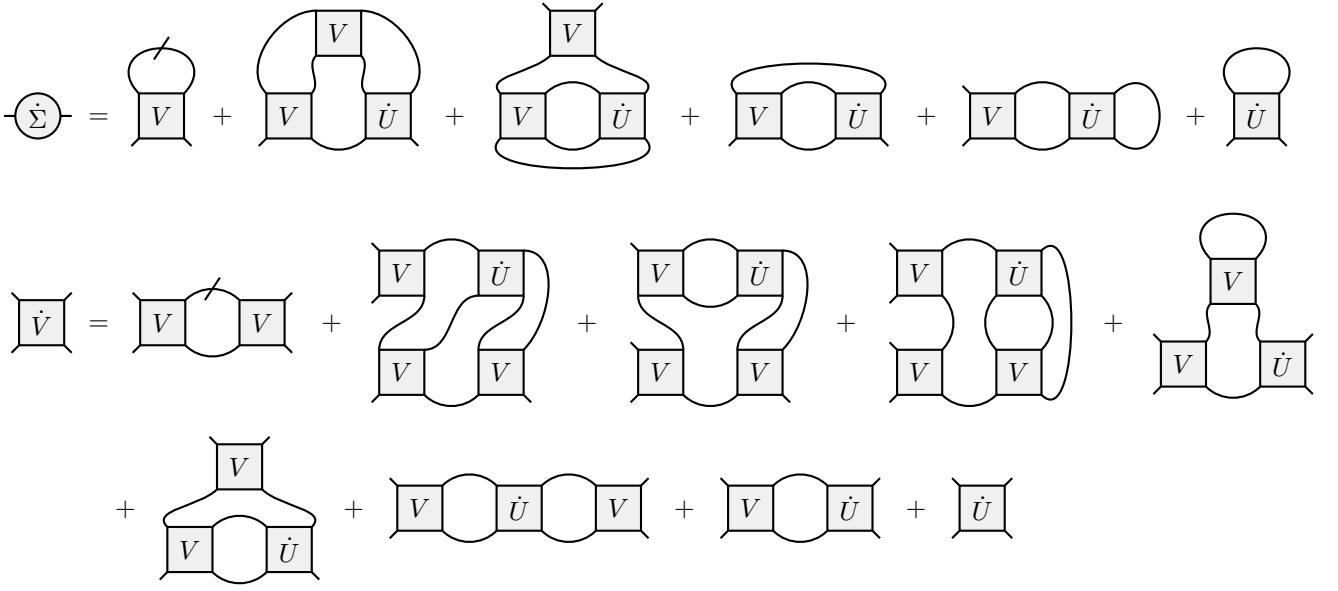

\begin{align*}
	\begin{array}{rl}
			\tikzm{VertexExpWetterichEqSigmadot}{
                \selfenergywithlegs{$\dot{\Sigma}$}{0}{0}{1}
                }
			&=
			\tikzm{VertexExpWetterichEqSigma1}{
			\fullvertex{$V$}{0}{0}{1}
			\lineslowerhalffull{0}{0}{1}
			\loopfullvertextop{linePlain}{0}{0}{1}
			\draw[linePlain] (-0.1,0.75) -- (0.1,1.05);
			}
			+
                \tikzm{VertexExpWetterichEqSigma2}{
			\fullvertex{$V$}{0}{0}{1}
			\fullvertex{$V$}{0.66}{1.1}{1}
			\fullvertex{$\dot{U}$}{1.32}{0}{1}
			\linesbottomleftcornerfull{0}{0}{1}
			\linesbottomrightcornerfull{1.32}{0}{1}
			\draw[linePlain] (1.02,-0.3) to [out=225, in=315] (0.3,-0.3);
			\draw[linePlain] (-0.3,0.3) to [out=135, in=180] (0.36,1.4);
			\draw[linePlain] (0.3,0.3) to [out=45, in=225] (0.36,0.8);
			\draw[linePlain] (1.62,0.3) to [out=45, in=0] (0.96,1.4);
			\draw[linePlain] (1.02,0.3) to [out=135, in=315] (0.96,0.8);
			}
                +\hspace{-0.15cm}
                \tikzm{VertexExpWetterichEqSigma3}{
			\fullvertex{$V$}{0}{0}{1}
			\fullvertex{$V$}{0.66}{1.1}{1}
			\fullvertex{$\dot{U}$}{1.32}{0}{1}
			\linesupperhalffull{0.66}{1.1}{1}
			\draw[linePlain] (0.3,0.3) to [out=45, in=135] (1.02,0.3);
			\draw[linePlain] (1.02,-0.3) to [out=225, in=315] (0.3,-0.3);
			\draw[linePlain] (-0.3,-0.3) to [out=225, in=315] (1.62,-0.3);
			\draw[linePlain] (-0.3,0.3) to [out=135, in=225] (0.36,0.8);
			\draw[linePlain] (1.62,0.3) to [out=45, in=315] (0.96,0.8);
			}
                \hspace{-0.15cm}+\hspace{-0.15cm}
                \tikzm{VertexExpWetterichEqSigma4}{
			\fullvertex{$V$}{0}{0}{1}
			\fullvertex{$\dot{U}$}{1.32}{0}{1}
			\linesbottomleftcornerfull{0}{0}{1}
			\linesbottomrightcornerfull{1.32}{0}{1}
			\draw[linePlain] (0.3,0.3) to [out=45, in=135] (1.02,0.3);
			\draw[linePlain] (1.02,-0.3) to [out=225, in=315] (0.3,-0.3);
			\draw[linePlain] (-0.3,0.3) to [out=135, in=45] (1.62,0.3);
			}
			\hspace{-0.15cm}+\hspace{0.25cm}
                \tikzm{VertexExpWetterichEqSigma5}{
			\fullvertex{$V$}{0}{0}{1}
			\fullvertex{$\dot{U}$}{1.32}{0}{1}
			\lineslefthalffull{0}{0}{1}
			\loopfullvertexright{linePlain}{1.32}{0}{1}
			\draw[linePlain] (0.3,0.3) to [out=45, in=135] (1.02,0.3);
			\draw[linePlain] (1.02,-0.3) to [out=225, in=315] (0.3,-0.3);
			}
                \hspace{0.25cm}+
                \tikzm{VertexExpWetterichEqSigma6}{
			\fullvertex{$\dot{U}$}{0}{0}{1}
			\lineslowerhalffull{0}{0}{1}
			\loopfullvertextop{linePlain}{0}{0}{1}
			}
			\\
			\\
			\tikzm{VertexExpWetterichEqVdot}{
			\fullvertex{$\dot{V}$}{0}{0}{1}
			\lineslefthalffull{0}{0}{1}
			\linesrighthalffull{0}{0}{1}
			}
			&=\hspace{0.2cm}
			\tikzm{VertexExpWetterichEqV1}{
			\fullvertex{$V$}{0}{0}{1}
			\fullvertex{$V$}{1.32}{0}{1}
			\lineslefthalffull{0}{0}{1}
			\linesrighthalffull{1.32}{0}{1}
			\draw[linePlain] (0.3,0.3) to [out=45, in=135] (1.02,0.3);
			\draw[linePlain] (1.02,-0.3) to [out=225, in=315] (0.3,-0.3);
			\draw[linePlain] (0.56,0.3) -- (0.76,0.6);
			}
			\hspace{0.3cm}+\hspace{0.3cm}
                \tikzm{VertexExpWetterichEqV2}{
			\fullvertex{$V$}{0}{0-0.66}{1}
			\fullvertex{$V$}{0}{1.32-0.66}{1}
			\fullvertex{$V$}{1.32}{0-0.66}{1}
			\fullvertex{$\dot{U}$}{1.32}{1.32-0.66}{1}
			\linesbottomleftcornerfull{0}{0-0.66}{1}
			\linesbottomrightcornerfull{1.32}{0-0.66}{1}
			\lineslefthalffull{0}{1.32-0.66}{1}
			\draw[linePlain] (1.02,-0.3-0.66) to [out=225, in=315] (0.3,-0.3-0.66);
			\draw[linePlain] (0.3,1.62-0.66) to [out=45, in=135] (1.02,1.62-0.66);
			\draw[linePlain] (-0.3,0.3-0.66) to [out=90, in=270] (0.3,1.02-0.66);
			\draw[linePlain] (1.02,0.3-0.66) to [out=90, in=270] (1.62,1.02-0.66);
			\draw[linePlain] (0.3,0.3-0.66) to [out=0, in=180] (1.02,1.02-0.66);
			\draw[linePlain] (1.62,0.3-0.66) to [out=45, in=0] (1.62,1.62-0.66);
			}
			\hspace{0.1cm}+\hspace{0.35cm}
                \tikzm{VertexExpWetterichEqV3}{
			\fullvertex{$V$}{0}{0-0.66}{1}
			\fullvertex{$V$}{0}{1.32-0.66}{1}
			\fullvertex{$V$}{1.32}{0-0.66}{1}
			\fullvertex{$\dot{U}$}{1.32}{1.32-0.66}{1}
			\linestopleftcornerfull{0}{1.32-0.66}{1}
			\linesbottomrightcornerfull{1.32}{0-0.66}{1}
			\lineslefthalffull{0}{0-0.66}{1}
			\draw[linePlain] (1.02,-0.3-0.66) to [out=225, in=315] (0.3,-0.3-0.66);
			\draw[linePlain] (0.3,0.3-0.66) to [out=90, in=270] (-0.3,1.02-0.66);
			\draw[linePlain] (1.02,1.02-0.66) to [out=225, in=315] (0.3,1.02-0.66);
			\draw[linePlain] (0.3,1.62-0.66) to [out=45, in=135] (1.02,1.62-0.66);
			\draw[linePlain] (1.02,0.3-0.66) to [out=90, in=270] (1.62,1.02-0.66);
			\draw[linePlain] (1.62,0.3-0.66) to [out=45, in=0] (1.62,1.62-0.66);
			}
			\hspace{0.1cm}+\hspace{0.35cm}
                \tikzm{VertexExpWetterichEqV4}{
			\fullvertex{$V$}{0}{0-0.66}{1}
			\fullvertex{$V$}{0}{1.32-0.66}{1}
			\fullvertex{$V$}{1.32}{0-0.66}{1}
			\fullvertex{$\dot{U}$}{1.32}{1.32-0.66}{1}
			\lineslefthalffull{0}{0-0.66}{1}
			\lineslefthalffull{0}{1.32-0.66}{1}
			\draw[linePlain] (1.02,-0.3-0.66) to [out=225, in=315] (0.3,-0.3-0.66);
			\draw[linePlain] (0.3,1.62-0.66) to [out=45, in=135] (1.02,1.62-0.66);
			\draw[linePlain] (0.3,0.3-0.66) to [out=45, in=315] (0.3,1.02-0.66);
			\draw[linePlain] (1.02,0.3-0.66) to [out=135, in=225] (1.02,1.02-0.66);
			\draw[linePlain] (1.62,0.3-0.66) to [out=45, in=315] (1.62,1.02-0.66);
			\draw[linePlain] (1.62,-0.3-0.66) to [out=315, in=45] (1.62,1.62-0.66);
			}
			\hspace{0.2cm}+\hspace{0.3cm}
                \tikzm{VertexExpWetterichEqV5}{
			\fullvertex{$V$}{0}{0-0.55}{1}
			\fullvertex{$V$}{0.66}{1.1-0.55}{1}
			\fullvertex{$\dot{U}$}{1.32}{0-0.55}{1}
			\lineslefthalffull{0}{0-0.55}{1}
			\linesrighthalffull{1.32}{0-0.55}{1}
			\loopfullvertextop{linePlain}{0.66}{1.1-0.55}{1}
			\draw[linePlain] (1.02,-0.3-0.55) to [out=225, in=315] (0.3,-0.3-0.55);
			\draw[linePlain] (0.3,0.3-0.55) to [out=45, in=225] (0.36,0.8-0.55);
			\draw[linePlain] (1.02,0.3-0.55) to [out=135, in=315] (0.96,0.8-0.55);
			} \\
			& \phantom{=} +\hspace{0.1cm}
                \tikzm{VertexExpWetterichEqV6}{
			\fullvertex{$V$}{0}{0-0.55}{1}
			\fullvertex{$V$}{0.66}{1.1-0.55}{1}
			\fullvertex{$\dot{U}$}{1.32}{0-0.55}{1}
			\linesbottomleftcornerfull{0}{0-0.55}{1}
			\linesbottomrightcornerfull{1.32}{0-0.55}{1}
			\linesupperhalffull{0.66}{1.1-0.55}{1}
			\draw[linePlain] (0.3,0.3-0.55) to [out=45, in=135] (1.02,0.3-0.55);
			\draw[linePlain] (1.02,-0.3-0.55) to [out=225, in=315] (0.3,-0.3-0.55);
			\draw[linePlain] (-0.3,0.3-0.55) to [out=135, in=225] (0.36,0.8-0.55);
			\draw[linePlain] (1.62,0.3-0.55) to [out=45, in=315] (0.96,0.8-0.55);
			}
                \hspace{0.1cm}+\hspace{0.25cm}
                \tikzm{VertexExpWetterichEqV7}{
			\fullvertex{$V$}{0}{0}{1}
			\fullvertex{$\dot{U}$}{1.32}{0}{1}
			\fullvertex{$V$}{2.64}{0}{1}
			\lineslefthalffull{0}{0}{1}
			\linesrighthalffull{2.64}{0}{1}
			\draw[linePlain] (0.3,0.3) to [out=45, in=135] (1.02,0.3);
			\draw[linePlain] (1.02,-0.3) to [out=225, in=315] (0.3,-0.3);
			\draw[linePlain] (1.62,0.3) to [out=45, in=135] (2.34,0.3);
			\draw[linePlain] (2.34,-0.3) to [out=225, in=315] (1.62,-0.3);
			}
			\hspace{0.25cm}+\hspace{0.25cm}
                \tikzm{VertexExpWetterichEqV8}{
			\fullvertex{$V$}{0}{0}{1}
			\fullvertex{$\dot{U}$}{1.32}{0}{1}
			\lineslefthalffull{0}{0}{1}
			\linesrighthalffull{1.32}{0}{1}
			\draw[linePlain] (0.3,0.3) to [out=45, in=135] (1.02,0.3);
			\draw[linePlain] (1.02,-0.3) to [out=225, in=315] (0.3,-0.3);
			}
			\hspace{0.25cm}+\hspace{0.25cm}
                \tikzm{VertexExpWetterichEqV9}{
			\fullvertex{$\dot{U}$}{0}{0}{1}
			\lineslefthalffull{0}{0}{1}
			\linesrighthalffull{0}{0}{1}
			}
	\end{array}
\end{align*}
\caption{Diagrammatic representation of the flow equations resulting from the vertex expansion of the Wetterich equation~\cite{Morris1994,Kopietz2010,Metzner2012} generalized by introducing a regulator in both the bare propagator $G_0$ and the bare interaction $U$, i.e., by introducing the RG scale $\Lambda$ via the substitutions $G_0\rightarrow G_0^\Lambda$ and $U\rightarrow U^\Lambda$ in the classical action~\eqref{eq:ClassicalActionS}. The equations shown in this figure are obtained from the level-2 truncation, which means that all diagrams involving the 3-particle vertex or higher-order vertices are ignored. Solid lines with a dash represent the single-scale propagator $S \equiv \left.\dot{G}\right|_{\dot{\Sigma}=0}$ whereas the other solid lines correspond to the full propagator $G$.}
\label{fig:UflowVertexExpansionWetterichEquation}
\end{figure*}

In Sec.~\ref{sec:Formalism}, we have formulated an fRG approach based on an interaction flow starting from self-consistent equations [either the Bethe--Salpeter equations~\eqref{eq:BetheSalpeterEquations} or the SBE equations~\eqref{eq:selfconsistentSBEequationsBreducibility}], which is not the predominant way to derive flow equations in the fRG community. In most applications~\cite{Dupuis2021}, fRG approaches are designed from the \emph{Wetterich equation}~\cite{Wetterich1993,Ellwanger1994,Morris1994}, which can be derived from the generating functional
\begin{equation}
    Z\!\left[\overline{J},J\right] = \int\! \mathcal{D}c\,\mathcal{D}\overline{c} \ e^{-S[\overline{c},c]+\overline{J}_{1}c_{1}+\overline{c}_{1^\prime}J_{1^\prime}},
    \label{eq:GeneratingFunctionalZ}
\end{equation}
for a model with classical action $S[\overline{c},c]$, e.g., Eq.~\eqref{eq:ClassicalActionS}. The 1PI effective action $\Gamma\!\left[\overline{\psi},\psi\right]$ is defined as the Legendre transform of $W\!\left[\overline{J},J\right]\equiv\ln\!\left(Z\!\left[\overline{J},J\right]\right)$ with respect to the source fields $\overline{J}$ and $J$, with $\overline{\psi}=-\frac{\delta W[\overline{J},J]}{\delta J}$ and $\psi=\frac{\delta W[\overline{J},J]}{\delta \overline{J}}$. By introducing the regulator via $G_0\rightarrow G_0^\Lambda$ in the generating functional~\eqref{eq:GeneratingFunctionalZ} and slightly modifying the Legendre transform defining $\Gamma\!\left[\overline{\psi},\psi\right]$, one can then derive from this modified Legendre transform an \emph{exact} $1\ell$ equation expressing the derivative $\partial_{\Lambda}\Gamma\!\left[\overline{\psi},\psi\right]\equiv\dot{\Gamma}\!\left[\overline{\psi},\psi\right]$, known as the \emph{Wetterich equation}. In practice, this equation cannot be solved exactly (unless for very simple specific cases) and one must design an approximation scheme to extract information from it. The choice of this approximation scheme crucially depends on the problem under consideration. To study competing orders in condensed matter theory, the most widespread approach is the \emph{vertex expansion}~\cite{Morris1994,Kopietz2010,Metzner2012}, which consists in expanding both sides of the Wetterich equation with respect to the fields $\overline{\psi}$ and $\psi$~\footnote{Since $\overline{\psi}$ and $\psi$ are Grassmann fields in the present case, this expansion must be performed about $\overline{\psi}=\psi=0$}. From this expansion, one can then extract an infinite hierarchy of exact flow equations, that take the form of expressions for the derivatives
\begin{equation}
\partial_{\Lambda}\Gamma_{1^\prime ... n^\prime|1 ... n}^{(2n)} \equiv \partial_{\Lambda}\!\left(\left.\frac{\delta^{2n} \Gamma\!\left[\overline{\psi},\psi\right]}{\delta \overline{\psi}_{1^\prime}...\delta \overline{\psi}_{n^\prime} \delta \psi_{n} ... \delta\psi_1}\right|_{\overline{\psi},\psi=0}\right).
\end{equation}
With this setup, the approximation that is typically implemented consists in retaining only the flow equation of the self-energy $\Sigma\sim \Gamma^{(2)}$ and that of the two-particle vertex $V\sim \Gamma^{(4)}$, while ignoring the contribution of higher-order vertices (i.e., $\Gamma^{(m)}=0 \:\text{ for~} m>4$) to close the set of differential equations thus obtained. This defines the so-called \emph{level-2 or $1\ell$ truncation} of the fRG. It has led to numerous insightful studies of many-fermion systems~\cite{Metzner2012,Dupuis2021}. For more technical details on the treatment of the Wetterich equation based on the vertex expansion, we refer to the review~\cite{Metzner2012} and the textbook~\cite{Kopietz2010}.

The question that we want to address at the present stage is how this widespread approach, i.e., the level-2 truncation obtained from the vertex expansion of the Wetterich equation, is modified when a regulator is added in the bare interaction $U$. More precisely, we consider once again the substitutions $G_0\rightarrow G_0^\Lambda$ and $U\rightarrow U^\Lambda$ in the classical action $S[\overline{c},c]$ of the form~\eqref{eq:ClassicalActionS}, but this time we follow the aforementioned procedure of the vertex expansion starting from the generating functional $Z\!\left[\overline{J},J\right]$ of Eq.~\eqref{eq:GeneratingFunctionalZ} to determine a closed set of flow equations for the self-energy $\Sigma$ and the two-particle vertex $V$. The differential equations thus obtained are depicted in Fig.~\ref{fig:UflowVertexExpansionWetterichEquation}. Even though we are still considering the fermionic model of Eq.~\eqref{eq:ClassicalActionS} with complex Grassmann fields, we do not represent propagator lines with arrows in this figure to make the diagrammatic expressions less cluttered. We only want to focus here on the general structure of the diagrams involved in those equations, which will be sufficient to make our point. Most of the diagrams shown in Fig.~\ref{fig:UflowVertexExpansionWetterichEquation} involve the derivative $\dot{U}$ and are thus generated by the substitution $U\rightarrow U^\Lambda$. In fact, setting $\dot{U}=0$ in the flow equations of Fig.~\ref{fig:UflowVertexExpansionWetterichEquation}, the latter reduce to the much simpler and well-known $1\ell$-fRG equations, namely,
\begin{subequations}
\begin{align}
    \tikzm{1lfRGSigmadot}{
                \selfenergywithlegs{$\dot{\Sigma}$}{0}{0}{1}
                }
			&=
			\tikzm{1lfRGSigmadotRHS}{
			\fullvertex{$V$}{0}{0}{1}
			\lineslowerhalffull{0}{0}{1}
			\loopfullvertextop{linePlain}{0}{0}{1}
			\draw[linePlain] (-0.1,0.75) -- (0.1,1.05);
			}, \label{eq:1lfRGSigma} \\
    \tikzm{1lfRGVdot}{
			\fullvertex{$\dot{V}$}{0}{0}{1}
			\lineslefthalffull{0}{0}{1}
			\linesrighthalffull{0}{0}{1}
			}
			&=\hspace{0.2cm}
			\tikzm{1lfRGVdotRHS}{
			\fullvertex{$V$}{0}{0}{1}
			\fullvertex{$V$}{1.32}{0}{1}
			\lineslefthalffull{0}{0}{1}
			\linesrighthalffull{1.32}{0}{1}
			\draw[linePlain] (0.3,0.3) to [out=45, in=135] (1.02,0.3);
			\draw[linePlain] (1.02,-0.3) to [out=225, in=315] (0.3,-0.3);
			\draw[linePlain] (0.56,0.3) -- (0.76,0.6);
			} \hspace{0.2cm} . \label{eq:1lfRGV}
\end{align}
\label{eq:1lfRG}
\end{subequations}
We can see with Fig.~\ref{fig:UflowVertexExpansionWetterichEquation} that most of the $\dot{U}$-dependent diagrams generated by the interaction flow have a more complicated nested loop structure, which would significantly increase the cost of the underlying numerical treatment. For that reason, we conclude that formulating the interaction flow within the conventional fRG approach of the vertex expansion combined with the level-2 truncation does not constitute an efficient approximation scheme to study many-electron systems, at least for fermionic models with a quartic interaction [such as the studied model based on Eq.~\eqref{eq:ClassicalActionS}]~\footnote{By ``does not constitute an efficient approximation scheme'', we exclude here the interaction flow scheme of Ref.~\cite{Honerkamp2004}, which was originally developed within the conventional fRG framework of the vertex expansion for fermionic models with a quartic interaction. However, as explained in Sec.~\ref{sec:ModelDefinitionsCutoffSchemes}, the dependence of the bare interaction with respect to the flow parameter is entirely reshuffled in the bare propagator within this approach, so that the corresponding flow equations reduce to Eqs.~\eqref{eq:1lfRG} within the level-2 truncation.}. One might try to design further approximations to treat the flow equations of Fig.~\ref{fig:UflowVertexExpansionWetterichEquation}, but we can also point out that, within the interaction flow formulated within the multiloop SBE fRG framework in Sec.~\ref{sec:SBEmfRGUflow}, the terms depending explicitly on $\dot{U}$, more precisely on $\dot{\mathcal{B}}_r$ or $\dot{\mathcal{F}}_r$, exhibit a $1\ell$ or a $2\ell$ structure at most (e.g., $\dot{\mathcal{F}}_r \circ \Pi_r \circ \mathcal{I}_r$ and $\mathcal{I}_r \circ \Pi_r \circ \dot{\mathcal{F}}_r \circ \Pi_r \circ \mathcal{I}_r$ respectively have a $1\ell$ and a $2\ell$ structure). Hence, although the substitution $U\rightarrow U^\Lambda$ also generates additional terms in this framework, it does not noticeably reduce the efficiency of the multiloop SBE fRG approach.

It should be stressed, however, that this feature of our interaction flow is not inherent to the SBE framework, but is rather due to the fact that we started our derivation from self-consistent equations, thus circumventing the expansions at the origin of the diagrammatic expressions of Fig.~\ref{fig:UflowVertexExpansionWetterichEquation}. This can be seen notably from the multiloop fRG equations underlying our interaction flow scheme within the parquet decomposition, i.e., Eqs.~\eqref{eq:UflowmultiloopParquetfRGEqs}. Indeed, the diagrammatic structure of the $\dot{U}$-dependent terms in the latter equations is also of $1\ell$ or $2\ell$ nature at most. In fact, it is known that, without any interaction flow ($\dot{U}=0$), the $1\ell$ flow equation~\eqref{eq:1lfRGV} combined with the parquet decomposition of $V$ coincides with Eq.~\eqref{eq:UflowmultiloopParquetfRGEq1l} (i.e., $\dot{\phi}^{(1)}_r = V \circ \dot{\Pi}_r \circ V$ for $\dot{U}=0$) if the single-scale propagator $S \equiv \left.\dot{G}\right|_{\dot{\Sigma}=0}$ [represented by a solid line with a dash in Eq.~\eqref{eq:1lfRGV}] is replaced by the derivative $\dot{G}$~\cite{Kugler2018b}, which is known as the Katanin substitution~\cite{Katanin2004}. Hence, the interaction flow approach developed in this study from the Bethe--Salpeter or the SBE equations can also be directly related to the conventional $1\ell$-fRG scheme and can be used to improve or go beyond the latter. We note in passing that, even though our discussion was fully focused on the two-particle vertex $V$, or vertices introduced via its decomposition ($\phi_r$, $w_r$, $\lambda_r$, $\overline{\lambda}_r$, $M_r$), similar remarks can be made for the self-energy flow equations~\eqref{eq:SigmadotPirSDE}. Each of these equations involves $\dot{U}$-dependent terms with a $1\ell$ or a $2\ell$ structure, which are of the same complexity as the other terms like $\dot{G} \cdot \left( U \circ \Pi_r \circ V \right)$ already present when $\dot{U}=\dot{\mathcal{B}}_r=\dot{\mathcal{F}}_r=0$~$\forall r$. This stands in stark contrast to the expression of $\dot{\Sigma}$ shown in Fig.~\ref{fig:UflowVertexExpansionWetterichEquation}.

\section{$SU(2)$ spin symmetry and physical channels}
\label{sec:SU2symmetry}

In this appendix, we rewrite the multiloop SBE fRG flow equations~\eqref{eq:UflowmultiloopSBEfRGEq1l},~\eqref{eq:UflowmultiloopSBEfRGEq2l} and~\eqref{eq:UflowmultiloopSBEfRGEqnl} for $SU(2)$-spin-symmetric systems in a compact manner by introducing \emph{physical channels}. We note that this was already done in Ref.~\cite{Fraboulet2025} for the multiloop SBE fRG flow equations without interaction flow (i.e., for Eqs.~\eqref{eq:UflowmultiloopSBEfRGEq1l},~\eqref{eq:UflowmultiloopSBEfRGEq2l} and~\eqref{eq:UflowmultiloopSBEfRGEqnl} with $\dot{U}=\dot{\mathcal{B}}_r=\dot{\mathcal{F}}_r=0$), so we also refer to this work for more details on our forthcoming derivations. It can be shown that $SU(2)$ spin symmetry imposes the following relations for the full propagator $G$ and the full vertex $V$~\cite{Salmhofer2001}:
\begin{subequations}
\begin{align}
    G_{1^\prime|1}  & = G_{\sigma_{1^\prime}|\sigma_1}(k_{1^\prime}|k_1) \\
    & = \delta_{\sigma_{1^\prime},\sigma_1} G(k_{1^\prime}|k_1) \label{eq:FullPropagatorSU2symmetry} \\
    V_{1^\prime 2^\prime|12} & = V_{\sigma_{1^\prime}\sigma_{2^\prime}|\sigma_1\sigma_2}(k_{1^\prime},k_{2^\prime}|k_1, k_2) \nonumber \\
    & = \delta_{\sigma_{1^\prime},\sigma_1} \delta_{\sigma_{2^\prime},\sigma_2} V^{\uparrow\downarrow}(k_{1^\prime},k_{2^\prime}|k_1, k_2)  \nonumber \\
    & \phantom{=} - \delta_{\sigma_{1^\prime},\sigma_2} \delta_{\sigma_{2^\prime},\sigma_1} V^{\uparrow\downarrow}(k_{2^\prime},k_{1^\prime}|k_1, k_2) , \label{eq:FullVertexSU2symmetry}
\end{align}
\label{eq:GVSU2symmetry}
\end{subequations}
with $V^{\uparrow\downarrow}=V_{\uparrow\downarrow|\uparrow\downarrow}$. In subsequent equations, we will repeatedly use the shorthand notations
\begin{align}
       A^{\sigma\sigma}=A_{\sigma\sigma|\sigma\sigma}, \quad
       A^{\sigma\overline{\sigma}}=A_{\sigma\overline{\sigma}|\sigma\overline{\sigma}}, \quad A^{\widehat{\sigma\overline{\sigma}}}=A_{\sigma\overline{\sigma}|\overline{\sigma}\sigma},
\end{align}
where $A$ denotes an arbitrary four-point object with respect to spin indices. Note that $\mathord{\overline{\uparrow}} = \mathord{\downarrow}$, $\mathord{\overline{\downarrow}} = \mathord{\uparrow}$. We can also infer from Eq.~\eqref{eq:FullVertexSU2symmetry} that the $SU(2)$-spin-symmetric two-particle vertex $V$ only has 6 non-zero spin components, which reduce to 3 distinct components according to the relations
\begin{align}
       V^{\uparrow\uparrow} = V^{\downarrow\downarrow}, \quad V^{\uparrow\downarrow} = V^{\downarrow\uparrow}, \quad V^{\widehat{\uparrow\downarrow}} = V^{\widehat{\downarrow\uparrow}}.
\end{align}
Similarly, for $SU(2)$-spin-symmetric systems, the objects introduced in the SBE formalism, i.e., $w_r$, $\lambda_r$ and $M_r$ are also fully determined by their $\uparrow\uparrow$, $\uparrow\downarrow$ and $\widehat{\uparrow\downarrow}$ spin components~\cite{Gievers2025Subleading,GieversPhDthesis}.

As a next step, we introduce the \emph{physical channels}, originally introduced to diagonalize the Bethe--Salpeter equations with respect to their spin indices for $SU(2)$-spin-symmetric systems~\cite{BickersSelfConsistent2004}. In what follows, physical channels will be referred to with the letter $\text{X}$ (as opposed to $r$ for diagrammatic channels). More precisely, we define three physical channels called magnetic~(M), density~(D) and superconducting~(SC) channels which satisfy~\footnote{Like the full vertex $V$~\cite{Salmhofer2001}, the vertex $A_{ph}$ can be expressed as $A_{ph}^{\uparrow\uparrow}=A_{ph}^{\uparrow\downarrow}+A_{ph}^{\widehat{\uparrow\downarrow}}$ whereas its crossing symmetry implies that $A_{ph}^{\widehat{\uparrow\downarrow}}=-A_{\overline{ph}}^{\uparrow\downarrow}$. According to these two relations, one can derive the equality $A^{\uparrow\uparrow}_{ph} - A^{\uparrow\downarrow}_{ph} = - A^{\uparrow\downarrow}_{\overline{ph}}$ used in Eq.~\eqref{eq:DefinitionMchannel}.}
\begin{subequations}
    \begin{align}
       A_{\text{M}} & = A^{\uparrow\uparrow}_{ph} - A^{\uparrow\downarrow}_{ph} = - A^{\uparrow\downarrow}_{\overline{ph}}, \label{eq:DefinitionMchannel} \\
       A_{\text{D}} & = A^{\uparrow\uparrow}_{ph} + A^{\uparrow\downarrow}_{ph}, \label{eq:DefinitionDchannel} \\
       A_{\text{SC}} & = A^{\uparrow\downarrow}_{pp}, \label{eq:DefinitionSCchannel}
    \end{align}
    \label{eq:DefinitionsPhysicalChannels}
\end{subequations}
where the vertices $A_r$ on the right-hand sides can be any four-point object with respect to spin indices, with the exception of the fermion-boson vertices $\lambda_r$ and the bubbles $\Pi_r$. The physical channels for the fermion-boson vertices are defined as
\begin{subequations}
    \begin{align}
       \lambda_{\text{M}} & = \lambda^{\uparrow\uparrow}_{ph} - \lambda^{\uparrow\downarrow}_{ph} = +\lambda^{\uparrow\downarrow}_{\overline{ph}}, \label{eq:DefinitionlambdaMchannel} \\
       \lambda_{\text{D}} & = \lambda^{\uparrow\uparrow}_{ph} + \lambda^{\uparrow\downarrow}_{ph}, \label{eq:DefinitionlambdaDchannel} \\
       \lambda_{\text{SC}} & = \lambda_{pp}^{\uparrow\downarrow} - \lambda_{pp}^{\widehat{\uparrow\downarrow}},\label{eq:DefinitionlambdaSCchannel}
    \end{align}
    \label{eq:DefinitionslambdasPhysicalChannels}
\end{subequations}
whereas the physical channels for the bubbles will be defined later with Eqs.~\eqref{eq:BubblesPiX}. The definitions~\eqref{eq:DefinitionslambdasPhysicalChannels} are chosen such that the vertices $\nabla_{\text{X}}$ satisfy
\begin{equation}
    \nabla_{\text{X}} (Q,k,k^{\prime}) = \overline{\lambda}_{\text{X}}(Q,k) w_{\text{X}}(Q) \lambda_{\text{X}}(Q,k^{\prime}),
    \label{eq:nablaSBEphysicalChannel}
\end{equation}
for $\text{X}=\text{M},\text{D},\text{SC}$. In the case of the $\text{SC}$ channel for example, one can use the definition of the $\fcirc$ product in the $pp$ channel (i.e., Eq.~\eqref{eq:circproductppchannel} for spin indices only) to obtain
\begin{align}
   \nabla_{\text{SC}} & = \nabla_{pp}^{\uparrow\downarrow} \nonumber \\
   & = \left[\overline{\lambda}_{pp} \fcirc w_{pp} \fcirc \lambda_{pp}\right]^{\uparrow\downarrow} \nonumber \\
   & = \overline{\lambda}_{pp;\uparrow\downarrow|\sigma_1 \sigma_{1^\prime}} w_{pp;\sigma_1 \sigma_{1^\prime}|\sigma_2 \sigma_{2^\prime}} \lambda_{pp;\sigma_2 \sigma_{2^\prime}|\uparrow\downarrow} \nonumber \\
   & = \overline{\lambda}_{pp}^{\uparrow\downarrow} w_{pp}^{\uparrow\downarrow} \lambda_{pp}^{\uparrow\downarrow} + \overline{\lambda}_{pp}^{\widehat{\uparrow\downarrow}} w_{pp}^{\uparrow\downarrow} \lambda_{pp}^{\widehat{\uparrow\downarrow}} + \overline{\lambda}_{pp}^{\widehat{\uparrow\downarrow}} w_{pp}^{\widehat{\uparrow\downarrow}} \lambda_{pp}^{\uparrow\downarrow} + \overline{\lambda}_{pp}^{\uparrow\downarrow} w_{pp}^{\widehat{\uparrow\downarrow}} \lambda_{pp}^{\widehat{\uparrow\downarrow}} \nonumber \\
   & = \left( \overline{\lambda}_{pp}^{\uparrow\downarrow} - \overline{\lambda}_{pp}^{\widehat{\uparrow\downarrow}} \right) w^{\uparrow\downarrow}_{pp} \left( \lambda_{pp}^{\uparrow\downarrow} - \lambda_{pp}^{\widehat{\uparrow\downarrow}} \right) ,
\end{align}
where the last line follows from the crossing symmetry of $w_{pp}(Q)$ (in the form of $w_{pp}^{\widehat{\uparrow\downarrow}}(Q)=-w_{pp}^{\uparrow\downarrow}(Q)$). By setting $w_{\text{SC}} = w_{pp}^{\uparrow\downarrow}$ (in accordance with Eq.~\eqref{eq:DefinitionSCchannel}), $\lambda_{\text{SC}} = \lambda_{pp}^{\uparrow\downarrow} - \lambda_{pp}^{\widehat{\uparrow\downarrow}}$ (in accordance with Eq.~\eqref{eq:DefinitionlambdaSCchannel}) and $\overline{\lambda}_{\text{SC}} = \overline{\lambda}_{pp}^{\uparrow\downarrow} - \overline{\lambda}_{pp}^{\widehat{\uparrow\downarrow}}$, Eq.~\eqref{eq:nablaSBEphysicalChannel} is directly recovered for $\text{X}=\text{SC}$. This is how our definition~\eqref{eq:DefinitionlambdaSCchannel} for $\lambda_{\text{SC}}$ is derived and the analogous relations for the other two channels (i.e., Eqs.~\eqref{eq:DefinitionlambdaMchannel} and~\eqref{eq:DefinitionlambdaDchannel}) can be inferred in the same way. Furthermore, we note that $\lambda^{\uparrow\uparrow}_{ph} + \lambda^{\uparrow\downarrow}_{ph} = +\lambda^{\uparrow\downarrow}_{\overline{ph}}$ in the definition of $\lambda_{\text{M}}$ given by Eq.~\eqref{eq:DefinitionlambdaMchannel} differs by a minus sign from $A^{\uparrow\uparrow}_{ph} + A^{\uparrow\downarrow}_{ph} = -A^{\uparrow\downarrow}_{\overline{ph}}$ in Eq.~\eqref{eq:DefinitionMchannel}. This originates from the fact that the fermion-boson vertices obey different crossing symmetry relations as compared to other vertices like $w_r$ or $M_r$ (i.e., $A_r$ in Eq.~\eqref{eq:DefinitionMchannel})~\footnote{We have $\lambda^{\widehat{\uparrow\downarrow}}_{ph}=+\lambda^{\uparrow\downarrow}_{\overline{ph}}$ whereas $A^{\widehat{\uparrow\downarrow}}_{ph}=-A^{\uparrow\downarrow}_{\overline{ph}}$ for Eq.~\eqref{eq:DefinitionMchannel} (see notably Eq.~(4.31a) and related discussions in Ref.~\cite{Gievers2022}). The definition $\lambda_{\text{M}}=+\lambda^{\uparrow\downarrow}_{\overline{ph}}$ in Eq.~\eqref{eq:DefinitionlambdaMchannel} is then obtained by combining the relation $\lambda^{\widehat{\uparrow\downarrow}}_{ph}=+\lambda^{\uparrow\downarrow}_{\overline{ph}}$ with $\lambda_{ph}^{\uparrow\uparrow}=\lambda_{ph}^{\uparrow\downarrow}+\lambda_{ph}^{\widehat{\uparrow\downarrow}}$.}. Finally, we can also make another remark on the conjugates $\overline{\lambda}_{\text{X}}$, which are defined analogously to Eqs.~\eqref{eq:DefinitionslambdasPhysicalChannels}. It can be shown that, under time-reversal and crossing symmetry, we have~\cite{Bonetti2022,GieversPhDthesis}
\begin{equation}
    \overline{\lambda}_{\text{X}}=\lambda_{\text{X}},
    \label{eq:lambdabarXtimereversal}
\end{equation}
for $\text{X}=\text{M},\text{D},\text{SC}$. This property will be used in the rest of this appendix, therefore assuming that time-reversal symmetry is also respected.

In order to rewrite the multiloop SBE fRG flow equations underlying our interaction flow scheme, i.e., Eqs.~\eqref{eq:UflowmultiloopSBEfRGEq1l},~\eqref{eq:UflowmultiloopSBEfRGEq2l} and~\eqref{eq:UflowmultiloopSBEfRGEqnl}, for $SU(2)$-spin-symmetric systems, we will also evaluate and simplify sums over frequencies and momenta which take the form
\begin{equation}
        A \circ \Pi_r \circ B,
\end{equation}
with $\Pi_r$ replaced by its derivative $\dot{\Pi}_r$ in some of the terms of the $1\ell$ flow equations~\eqref{eq:UflowmultiloopSBEfRGEq1l}. Here, $A$ and $B$ are four-point objects with respect to frequencies and momenta that satisfy the relations
\begin{subequations}
        \begin{align}
            A(k_{1^\prime},k_{2^\prime}|k_1, k_2) & = A(Q_r,k_r,k^\prime_{r}) \delta_{k_{1^\prime}+k_{2^\prime},k_1+k_2}, \\
            B(k_{1^\prime},k_{2^\prime}|k_1, k_2) & = B(Q_r,k_r,k^\prime_{r}) \delta_{k_{1^\prime}+k_{2^\prime},k_1+k_2},
        \end{align}
        \label{eq:AandBFreqMomentum}
\end{subequations}
which also apply to the whole term $A \circ \Pi_r \circ B$, i.e.,
\begin{align}
        [A \circ \Pi_r \circ B](k_{1^\prime},k_{2^\prime}|k_1, k_2) & = [A \circ \Pi_r \circ B](Q_r,k_r,k^\prime_{r}) \nonumber \\
        & \phantom{=} \times \delta_{k_{1^\prime}+k_{2^\prime},k_1+k_2}. \label{eq:ApiBFreqMomentum}
\end{align}
Moreover, frequency and momentum conservation laws for the full propagator yield
\begin{equation}
        G(k_{1^\prime}|k_1)=G(k_1)\delta_{k_{1^\prime},k_1}.
        \label{eq:GMomentum}
\end{equation}
Combining the latter equality with the definitions~\eqref{eq:Bubbles} for the bubbles leads to
\begin{subequations}
\begin{align}
	\Pi_{ph}(k_{1^\prime},k_{2^\prime}|k_1, k_2) &= - G(k_2) G(k_1) \delta_{k_{1^\prime},k_2} \delta_{k_{2^\prime},k_1} , \\
	\Pi_{\overline{ph}}(k_{1^\prime},k_{2^\prime}|k_1, k_2) &= G(k_1) G(k_2) \delta_{k_{1^\prime},k_1} \delta_{k_{2^\prime},k_2} , \\
	\Pi_{pp}(k_{1^\prime},k_{2^\prime}|k_1, k_2) &= \tfrac{1}{2} G(k_1) G(k_2) \delta_{k_{1^\prime},k_1} \delta_{k_{2^\prime},k_2} .
\end{align}
\label{eq:BubblesFreqMomentum}
\end{subequations}
From Eqs.~\eqref{eq:AandBFreqMomentum},~\eqref{eq:ApiBFreqMomentum}, and~\eqref{eq:BubblesFreqMomentum}, it follows that
\begin{widetext}
\begin{subequations}
    \begin{align}
        [A\circ\Pi_{ph}\circ B](Q_{ph},k_{ph},k^{\prime}_{ph}) &= \sum_{k^{\prime\prime}_{ph}}A(Q_{ph},k_{ph},k^{\prime\prime}_{ph}) \fcirc \Pi_{ph}(Q_{ph},k^{\prime\prime}_{ph}) \fcirc B(Q_{ph},k^{\prime\prime}_{ph},k^{\prime}_{ph}), \label{eq:EvaluateSumsFreqMomentaphchannel} \\
        [A\circ\Pi_{\overline{ph}}\circ B](Q_{\overline{ph}},k_{\overline{ph}},k^{\prime}_{\overline{ph}}) &= \sum_{k^{\prime\prime}_{\overline{ph}}}A(Q_{\overline{ph}},k_{\overline{ph}},k^{\prime\prime}_{\overline{ph}}) \fcirc \Pi_{\overline{ph}}(Q_{\overline{ph}},k^{\prime\prime}_{\overline{ph}}) \fcirc B(Q_{\overline{ph}},k^{\prime\prime}_{\overline{ph}},k^{\prime}_{\overline{ph}}), \label{eq:EvaluateSumsFreqMomentaphxchannel} \\
        [A\circ\Pi_{pp}\circ B](Q_{pp},k_{pp},k^{\prime}_{pp}) &= \sum_{k^{\prime\prime}_{pp}}A(Q_{pp},k^{\prime\prime}_{pp},k^{\prime}_{pp}) \fcirc \Pi_{pp}(Q_{pp},k^{\prime\prime}_{pp}) \fcirc B(Q_{pp},k_{pp},k^{\prime\prime}_{pp}), \label{eq:EvaluateSumsFreqMomentappchannel}
        \end{align}
    \label{eq:EvaluateSumsFreqMomenta}
\end{subequations}
where
\begin{subequations}
    \begin{align}
    \Pi_{ph;\sigma_{1^\prime} \sigma_{2^\prime}|\sigma_1 \sigma_2}(Q_{ph},k_{ph}) & = - G_{\sigma_{2^\prime}|\sigma_1}\Bigg(\nu_{ph}+\left\lceil\frac{\Omega_{ph}}{2}\right\rceil,\mathbf{k}_{ph}+\mathbf{Q}_{ph}\Bigg)G_{\sigma_{1^\prime}|\sigma_2}\Bigg(\nu_{ph}-\left\lfloor\frac{\Omega_{ph}}{2}\right\rfloor,\mathbf{k}_{ph}\Bigg), \\
    \Pi_{\overline{ph};\sigma_{1^\prime} \sigma_{2^\prime}|\sigma_1 \sigma_2}(Q_{\overline{ph}},k_{\overline{ph}}) & = G_{\sigma_{1^\prime}|\sigma_1}\Bigg(\nu_{\overline{ph}}+\left\lceil\frac{\Omega_{\overline{ph}}}{2}\right\rceil,\mathbf{k}_{\overline{ph}}+\mathbf{Q}_{\overline{ph}}\Bigg)G_{\sigma_{2^\prime}|\sigma_2}\Bigg(\nu_{\overline{ph}}-\left\lfloor\frac{\Omega_{\overline{ph}}}{2}\right\rfloor,\mathbf{k}_{\overline{ph}}\Bigg), \\
    \Pi_{pp;\sigma_{1^\prime} \sigma_{2^\prime}|\sigma_1 \sigma_2}(Q_{pp},k_{pp}) & = \frac{1}{2} G_{\sigma_{1^\prime}|\sigma_1}\Bigg(\left\lfloor\frac{\Omega_{pp}}{2}\right\rfloor-\nu_{pp},\mathbf{Q}_{pp}-\mathbf{k}_{pp}\Bigg)G_{\sigma_{2^\prime}|\sigma_2}\Bigg(\nu_{pp}+\left\lceil\frac{\Omega_{pp}}{2}\right\rceil,\mathbf{k}_{pp}\Bigg).
    \end{align}
    \label{eq:BubblesEvaluateSumsFreqMomenta}
\end{subequations}
\end{widetext}
Note that Eqs.~\eqref{eq:EvaluateSumsFreqMomenta} are still valid if the bubbles $\Pi_r$ are substituted with their derivatives $\dot{\Pi}_r$.

We then turn to the evaluation of the sums over spin indices in the flow equations~\eqref{eq:UflowmultiloopSBEfRGEq1l}--\eqref{eq:UflowmultiloopSBEfRGEqnl} for the $SU(2)$-spin-symmetric case. For this, we recall our definition of the $\fcirc$ product [in accordance with that of the $\circ$ product in Eqs.~\eqref{eq:Definitioncircproduct}] between two arbitrary four-point objects with respect to spin indices $A$ and $B$:
\begin{subequations}
\begin{align}
       ph~:\quad [A\fcirc B]_{\sigma_{1^\prime} \sigma_{2^\prime}|\sigma_{1} \sigma_{2}} &= A_{\sigma_{4} \sigma_{2^\prime} |\sigma_{3} \sigma_{2}}B_{\sigma_{1^\prime} \sigma_{3}|\sigma_{1} \sigma_{4}}, \\
       \overline{ph}~:\quad [A\fcirc B]_{\sigma_{1^\prime} \sigma_{2^\prime}|\sigma_{1} \sigma_{2}} &= A_{\sigma_{1^\prime} \sigma_{4}|\sigma_{3} \sigma_{2}}B_{\sigma_{3} \sigma_{2^\prime} |\sigma_{1} \sigma_{4}}, \\
        pp~:\quad [A\fcirc B]_{\sigma_{1^\prime} \sigma_{2^\prime}|\sigma_{1} \sigma_{2}} &= A_{\sigma_{1^\prime} \sigma_{2^\prime}|\sigma_{3} \sigma_{4}}B_{\sigma_{3}\sigma_{4}|\sigma_{1}\sigma_{2}}.
    \end{align}
\label{eq:Definitionfcircproduct}
\end{subequations}
Repeated spin indices [$\sigma_3$ and $\sigma_4$ in Eqs.~\eqref{eq:Definitionfcircproduct}] are summed over their two configurations $\uparrow$ and $\downarrow$. When $SU(2)$ spin symmetry is preserved, the four-point objects manipulated in our formalism ($A = w_r, \lambda_r, M_r, ...$) only have 6 non-zero spin components ($A^{\uparrow\uparrow}$, $A^{\downarrow\downarrow}$, $A^{\uparrow\downarrow}$, $A^{\downarrow\uparrow}$, $A^{\widehat{\uparrow\downarrow}}$ and $A^{\widehat{\downarrow\uparrow}}$). Using this property together with the definition~\eqref{eq:Definitionfcircproduct} for the $\fcirc$ product, we find
\begin{subequations}
    \begin{align}
       ph~:\quad [A\fcirc B]^{\widehat{\uparrow\downarrow}} &= A^{\widehat{\uparrow\downarrow}} B^{\widehat{\uparrow\downarrow}}, \nonumber \\
       [A\fcirc B]^{\uparrow\downarrow} &= A^{\uparrow\downarrow}B^{\uparrow\uparrow}+A^{\downarrow\downarrow}B^{\uparrow\downarrow}, \nonumber \\
       [A\fcirc B]^{\uparrow\uparrow} &= A^{\uparrow\uparrow}B^{\uparrow\uparrow}+A^{\downarrow\uparrow}B^{\uparrow\downarrow}, \\
       \overline{ph}~:\quad [A\fcirc B]^{\uparrow\downarrow} &= A^{\uparrow\downarrow} B^{\uparrow\downarrow}, \nonumber \\
       [A\fcirc B]^{\uparrow\uparrow} &= A^{\uparrow\uparrow}B^{\uparrow\uparrow}+A^{\widehat{\uparrow\downarrow}}B^{\widehat{\downarrow\uparrow}}, \nonumber \\
       [A\fcirc B]^{\widehat{\uparrow\downarrow}} &= A^{\uparrow\uparrow}B^{\widehat{\uparrow\downarrow}}+A^{\widehat{\uparrow\downarrow}}B^{\downarrow\downarrow}, \\
        pp~:\quad [A\fcirc B]^{\uparrow\uparrow} &= A^{\uparrow\uparrow} B^{\uparrow\uparrow}, \nonumber \\
       [A\fcirc B]^{\widehat{\uparrow\downarrow}} &= A^{\uparrow\downarrow}B^{\widehat{\uparrow\downarrow}}+A^{\widehat{\uparrow\downarrow}}B^{\downarrow\uparrow}, \nonumber \\
       [A\fcirc B]^{\uparrow\downarrow} &= A^{\uparrow\downarrow}B^{\uparrow\downarrow}+A^{\widehat{\uparrow\downarrow}}B^{\widehat{\downarrow\uparrow}},
    \end{align}
\label{eq:ABFormulasfcircproduct}
\end{subequations}
which can be extended to products between three vertices, thus yielding
\begin{subequations}
    \begin{align}
       ph~:\quad [A\fcirc B\fcirc C]^{\uparrow\uparrow} &= A^{\uparrow\uparrow} B^{\uparrow\uparrow} C^{\uparrow\uparrow} + A^{\uparrow\uparrow} B^{\downarrow\uparrow} C^{\uparrow\downarrow} \nonumber \\
       &\phantom{=} + A^{\downarrow\uparrow} B^{\uparrow\downarrow} C^{\uparrow\uparrow} + A^{\downarrow\uparrow} B^{\downarrow\downarrow} C^{\uparrow\downarrow}, \nonumber \\
       \quad [A\fcirc B\fcirc C]^{\uparrow\downarrow} &= A^{\uparrow\downarrow} B^{\uparrow\uparrow} C^{\uparrow\uparrow} + A^{\uparrow\downarrow} B^{\downarrow\uparrow} C^{\uparrow\downarrow} \nonumber \\
       &\phantom{=} + A^{\downarrow\downarrow} B^{\uparrow\downarrow} C^{\uparrow\uparrow} + A^{\downarrow\downarrow} B^{\downarrow\downarrow} C^{\uparrow\downarrow}, \\
       \overline{ph}~:\quad [A\fcirc B\fcirc C]^{\uparrow\downarrow} &= A^{\uparrow\downarrow} B^{\uparrow\downarrow} C^{\uparrow\downarrow}, \\
        pp~:\quad [A\fcirc B\fcirc C]^{\uparrow\downarrow} &= A^{\uparrow\downarrow} B^{\uparrow\downarrow} C^{\uparrow\downarrow} + A^{\widehat{\uparrow\downarrow}} B^{\widehat{\downarrow\uparrow}} C^{\uparrow\downarrow} \nonumber \\
        &\phantom{=} + A^{\uparrow\downarrow} B^{\widehat{\uparrow\downarrow}} C^{\widehat{\downarrow\uparrow}} + A^{\widehat{\uparrow\downarrow}} B^{\downarrow\uparrow} C^{\widehat{\downarrow\uparrow}}, \nonumber \\
       \quad [A\fcirc B\fcirc C]^{\widehat{\uparrow\downarrow}} &= A^{\uparrow\downarrow} B^{\uparrow\downarrow} C^{\widehat{\uparrow\downarrow}} + A^{\widehat{\uparrow\downarrow}} B^{\widehat{\downarrow\uparrow}} C^{\widehat{\uparrow\downarrow}} \nonumber \\
       &\phantom{=} + A^{\uparrow\downarrow} B^{\widehat{\uparrow\downarrow}} C^{\downarrow\uparrow} + A^{\widehat{\uparrow\downarrow}} B^{\downarrow\uparrow} C^{\downarrow\uparrow},
    \end{align}
\label{eq:ABCFormulasfcircproduct}
\end{subequations}
where $C$ is also an arbitrary four-point object with respect to spin indices~\footnote{As explained in Refs.~\cite{Patricolo2025,GieversPhDthesis}, it is also possible to reformulate the product~\eqref{eq:Definitionfcircproduct} in such a way that Eqs.~\eqref{eq:ABFormulasfcircproduct} and~\eqref{eq:ABCFormulasfcircproduct} can be directly derived from standard matrix products}. When rewriting the generic flow equations~\eqref{eq:UflowmultiloopSBEfRGEq1l},~\eqref{eq:UflowmultiloopSBEfRGEq2l} and~\eqref{eq:UflowmultiloopSBEfRGEqnl}, Eqs.~\eqref{eq:ABCFormulasfcircproduct} will be often used with $B=\Pi_r$. The $SU(2)$ spin symmetry also implies that $\Pi_{ph}^{\uparrow\downarrow}=\Pi_{ph}^{\downarrow\uparrow}=0$ and $\Pi_{pp}^{\widehat{\uparrow\downarrow}}=\Pi_{pp}^{\widehat{\downarrow\uparrow}}=0$, as can be seen by combining Eq.~\eqref{eq:FullPropagatorSU2symmetry} with the definitions of the bubbles given by Eqs.~\eqref{eq:Bubbles}. In that case, Eqs.~\eqref{eq:ABCFormulasfcircproduct} can be rewritten as
\begin{subequations}
    \begin{align}
       ph~:\quad [A\fcirc \Pi_{ph}\fcirc B]^{\uparrow\uparrow} &= A^{\uparrow\uparrow} \Pi_{ph}^{\uparrow\uparrow} B^{\uparrow\uparrow} + A^{\downarrow\uparrow} \Pi_{ph}^{\downarrow\downarrow} B^{\uparrow\downarrow}, \nonumber \\
       \quad [A\fcirc \Pi_{ph}\fcirc B]^{\uparrow\downarrow} &= A^{\uparrow\downarrow} \Pi_{ph}^{\uparrow\uparrow} B^{\uparrow\uparrow} + A^{\downarrow\downarrow} \Pi_{ph}^{\downarrow\downarrow} B^{\uparrow\downarrow}, \\
       \overline{ph}~:\quad [A\fcirc \Pi_{\overline{ph}}\fcirc B]^{\uparrow\downarrow} &= A^{\uparrow\downarrow} \Pi_{\overline{ph}}^{\uparrow\downarrow} B^{\uparrow\downarrow}, \\
        pp~:\quad [A\fcirc \Pi_{pp}\fcirc B]^{\uparrow\downarrow} &= A^{\uparrow\downarrow} \Pi_{pp}^{\uparrow\downarrow} B^{\uparrow\downarrow} + A^{\widehat{\uparrow\downarrow}} \Pi_{pp}^{\downarrow\uparrow} B^{\widehat{\downarrow\uparrow}}, \nonumber \\
       \quad [A\fcirc \Pi_{pp}\fcirc B]^{\widehat{\uparrow\downarrow}} &= A^{\uparrow\downarrow} \Pi_{pp}^{\uparrow\downarrow} B^{\widehat{\uparrow\downarrow}} + A^{\widehat{\uparrow\downarrow}} \Pi_{pp}^{\downarrow\uparrow} B^{\downarrow\uparrow}.
    \end{align}
\label{eq:APiBFormulasfcircproduct}
\end{subequations}
These equations also hold if each bubble $\Pi_r$ is substituted with its derivative $\dot{\Pi}_r$.

We have thus introduced a set of relations that we can use to rewrite efficiently the generic flow equations~\eqref{eq:UflowmultiloopSBEfRGEq1l},~\eqref{eq:UflowmultiloopSBEfRGEq2l} and~\eqref{eq:UflowmultiloopSBEfRGEqnl} for $SU(2)$-spin-symmetric systems. We focus on the $1\ell$ flow equations for the bosonic propagators given by Eq.~\eqref{eq:UflowmultiloopSBEfRGEq1l_wr}. According to the definitions of our physical channels set by Eqs.~\eqref{eq:DefinitionsPhysicalChannels}, the $1\ell$ corrections for the bosonic propagators for the physical channels satisfy
\begin{subequations}
    \begin{align}
       \dot{w}^{(1\ell)}_{\text{M}} & = - \dot{w}^{(1\ell)\uparrow\downarrow}_{\overline{ph}}, \\
       \dot{w}^{(1\ell)}_{\text{D}} & = \dot{w}^{(1\ell)\uparrow\uparrow}_{ph} + \dot{w}^{(1\ell)\uparrow\downarrow}_{ph}, \\
       \dot{w}^{(1\ell)}_{\text{SC}} & = \dot{w}^{(1\ell)\uparrow\downarrow}_{pp}.
    \end{align}
    \label{eq:DerivationwX1loopStep1}
\end{subequations}
One can then insert the expressions of $\dot{w}_r^{(1\ell)}$ for $r=ph,\overline{ph},pp$ given by Eq.~\eqref{eq:UflowmultiloopSBEfRGEq1l_wr} in the right-hand sides of Eqs.~\eqref{eq:DerivationwX1loopStep1} which, after simplification using some of the relations that we have just introduced (Eqs.~\eqref{eq:EvaluateSumsFreqMomenta},~\eqref{eq:ABCFormulasfcircproduct} and~\eqref{eq:APiBFormulasfcircproduct} in particular), would lead to the $1\ell$ SBE fRG equations in physical channels for our interaction flow scheme. Let us illustrate how this is done in detail for the term
\begin{equation}
    \dot{\widetilde{w}}_r^{(1\ell)} = w_r \fcirc \lambda_r \circ \Pi_r \circ \dot{U} \circ \Pi_r \circ \lambda_r \fcirc w_r,
\end{equation}
which is involved in the expression of $\dot{w}_r^{(1\ell)}$ in Eq.~\eqref{eq:UflowmultiloopSBEfRGEq1l_wr}. The contributions of $\dot{\widetilde{w}}_r^{(1\ell)}$ to $\dot{w}_{\text{M}}^{(1\ell)}$, $\dot{w}_{\text{D}}^{(1\ell)}$ and $\dot{w}_{\text{SC}}^{(1\ell)}$ in Eqs.~\eqref{eq:DerivationwX1loopStep1} read
\begin{subequations}
    \begin{align}
       \dot{\widetilde{w}}^{(1\ell)}_{\text{M}} &= - \dot{\widetilde{w}}^{(1\ell)\uparrow\downarrow}_{\overline{ph}} \nonumber \\
       &= - \left[w_{\overline{ph}} \fcirc \lambda_{\overline{ph}} \circ \Pi_{\overline{ph}} \circ \dot{U} \circ \Pi_{\overline{ph}} \circ \lambda_{\overline{ph}} \fcirc w_{\overline{ph}}\right]^{\uparrow\downarrow}, \\
       \dot{\widetilde{w}}^{(1\ell)}_{\text{D}}&= \dot{\widetilde{w}}^{(1\ell)\uparrow\uparrow}_{ph} + \dot{\widetilde{w}}^{(1\ell)\uparrow\downarrow}_{ph} \nonumber \\
       &= \left[w_{ph} \fcirc \lambda_{ph} \circ \Pi_{ph} \circ \dot{U} \circ \Pi_{ph} \circ \lambda_{ph} \fcirc w_{ph}\right]^{\uparrow\uparrow} \nonumber \\
       & \phantom{=} + \left[w_{ph} \fcirc \lambda_{ph} \circ \Pi_{ph} \circ \dot{U} \circ \Pi_{ph} \circ \lambda_{ph} \fcirc w_{ph}\right]^{\uparrow\downarrow}, \\
       \dot{\widetilde{w}}^{(1\ell)}_{\text{SC}} &= \dot{\widetilde{w}}^{(1\ell)\uparrow\downarrow}_{pp} \nonumber \\
       &= \left[w_{pp} \fcirc \lambda_{pp} \circ \Pi_{pp} \circ \dot{U} \circ \Pi_{pp} \circ \lambda_{pp} \fcirc w_{pp}\right]^{\uparrow\downarrow}.
    \end{align}
    \label{eq:DerivationwX1loopStep2}
\end{subequations}
Eqs.~\eqref{eq:ABCFormulasfcircproduct} can then be used to evaluate some of the sums over spin indices with $A=C=w_r$ and $B=\lambda_r \circ \Pi_r \circ \dot{U} \circ \Pi_r \circ \lambda_r$ for $r=ph,\overline{ph},pp$, which gives
\begin{subequations}
    \begin{align}
       \dot{\widetilde{w}}^{(1\ell)}_{\text{M}} & = -\left(w_{\text{M}}\right)^2\left[\lambda_{\overline{ph}} \circ \Pi_{\overline{ph}} \circ \dot{U} \circ \Pi_{\overline{ph}} \circ \lambda_{\overline{ph}}\right]^{\uparrow\downarrow}, \\
       \dot{\widetilde{w}}^{(1\ell)}_{\text{D}} & = \left(w_{\text{D}}\right)^2 \bigg(\left[\lambda_{ph} \circ \Pi_{ph} \circ \dot{U} \circ \Pi_{ph} \circ \lambda_{ph}\right]^{\uparrow\uparrow} \nonumber \\
       & \phantom{= \left(w_{\text{D}}\right)^2 \bigg(} + \left[\lambda_{ph} \circ \Pi_{ph} \circ \dot{U} \circ \Pi_{ph} \circ \lambda_{ph}\right]^{\uparrow\downarrow}\bigg), \\
       \dot{\widetilde{w}}^{(1\ell)}_{\text{SC}} & = 2\left(w_{\text{SC}}\right)^2 \bigg(\left[\lambda_{pp} \circ \Pi_{pp} \circ \dot{U} \circ \Pi_{pp} \circ \lambda_{pp}\right]^{\uparrow\downarrow} \nonumber \\
       & \phantom{= 2\left(w_{\text{SC}}\right)^2 \bigg(} - \left[\lambda_{pp} \circ \Pi_{pp} \circ \dot{U} \circ \Pi_{pp} \circ \lambda_{pp}\right]^{\widehat{\uparrow\downarrow}}\bigg), \label{eq:DerivationwSC1loopStep3}
    \end{align}
    \label{eq:DerivationwX1loopStep3}
\end{subequations}
with $w_{\text{M}}=-w_{\overline{ph}}^{\uparrow\downarrow}$, $w_{\text{D}}=w_{ph}^{\uparrow\uparrow}+w_{ph}^{\uparrow\downarrow}$, $w_{\text{SC}}=w^{\uparrow\downarrow}_{pp}$, and Eq.~\eqref{eq:DerivationwSC1loopStep3} also follows from the relation $w_{pp}^{\widehat{\uparrow\downarrow}}=-w_{pp}^{\uparrow\downarrow}$, which reflects the crossing symmetry of $w_{pp}$. As a next step, we perform some of the sums over momenta and frequencies in Eqs.~\eqref{eq:DerivationwX1loopStep3} by considering Eqs.~\eqref{eq:EvaluateSumsFreqMomenta} with $A=\lambda_r$ and $B=\dot{U} \circ \Pi_r \circ \lambda_r$ for $r=ph,\overline{ph},pp$. This yields
\begin{widetext}
\begin{subequations}
    \begin{align}
       \dot{\widetilde{w}}^{(1\ell)}_{\text{M}}(Q) & = -\left(w_{\text{M}}(Q)\right)^2 \sum_k \left[\lambda_{\overline{ph}}(Q,k) \fcirc \Pi_{\overline{ph}}(Q,k) \fcirc \left[\dot{U} \circ \Pi_{\overline{ph}} \circ \lambda_{\overline{ph}}\right]\!(Q,k)\right]^{\uparrow\downarrow}, \\
       \dot{\widetilde{w}}^{(1\ell)}_{\text{D}}(Q) & = \left(w_{\text{D}}(Q)\right)^2 \sum_k \bigg(\left[\lambda_{ph}(Q,k) \fcirc \Pi_{ph}(Q,k) \fcirc \left[\dot{U} \circ \Pi_{ph} \circ \lambda_{ph}\right]\!(Q,k)\right]^{\uparrow\uparrow} \nonumber \\
       & \phantom{= \left(w_{\text{D}}(Q)\right)^2 \sum_k \bigg(} + \left[\lambda_{ph}(Q,k) \fcirc \Pi_{ph}(Q,k) \fcirc \left[\dot{U} \circ \Pi_{ph} \circ \lambda_{ph}\right]\!(Q,k)\right]^{\uparrow\downarrow}\bigg), \\
       \dot{\widetilde{w}}^{(1\ell)}_{\text{SC}}(Q) & = 2\left(w_{\text{SC}}(Q)\right)^2 \sum_k \bigg(\left[\lambda_{pp}(Q,k) \fcirc \Pi_{pp}(Q,k) \fcirc \left[\dot{U} \circ \Pi_{pp} \circ \lambda_{pp}\right]\!(Q,k)\right]^{\uparrow\downarrow} \nonumber \\
       & \phantom{=2\left(w_{\text{SC}}(Q)\right)^2 \sum_k \bigg(} - \left[\lambda_{pp}(Q,k) \fcirc \Pi_{pp}(Q,k) \fcirc \left[\dot{U} \circ \Pi_{pp} \circ \lambda_{pp}\right]\!(Q,k)\right]^{\widehat{\uparrow\downarrow}}\bigg),
    \end{align}
    \label{eq:DerivationwX1loopStep4}
\end{subequations}
We can then exploit Eqs.~\eqref{eq:APiBFormulasfcircproduct} to evaluate the $\fcirc$ products in Eqs.~\eqref{eq:DerivationwX1loopStep4}, which leads to
\begin{subequations}
    \begin{align}
       \dot{\widetilde{w}}^{(1\ell)}_{\text{M}}(Q) & = \left(w_{\text{M}}(Q)\right)^2 \sum_k \lambda_{\text{M}}(Q,k) \Pi_{\text{M}}(Q,k) \left[\dot{U} \circ \Pi_{\overline{ph}} \circ \lambda_{\overline{ph}}\right]^{\uparrow\downarrow}\!(Q,k), \\
       \dot{\widetilde{w}}^{(1\ell)}_{\text{D}}(Q) & = \left(w_{\text{D}}(Q)\right)^2 \sum_k \lambda_{\text{D}}(Q,k) \Pi_{\text{D}}(Q,k) \bigg(\left[\dot{U} \circ \Pi_{ph} \circ \lambda_{ph}\right]^{\uparrow\uparrow}\!(Q,k) + \left[\dot{U} \circ \Pi_{ph} \circ \lambda_{ph}\right]^{\uparrow\downarrow}\!(Q,k)\bigg), \\
       \dot{\widetilde{w}}^{(1\ell)}_{\text{SC}}(Q) & = \left(w_{\text{SC}}(Q)\right)^2 \sum_k \lambda_{\text{SC}}(Q,k) \Pi_{\text{SC}}(Q,k) \bigg(\left[\dot{U} \circ \Pi_{pp} \circ \lambda_{pp}\right]^{\uparrow\downarrow}\!(Q,k) - \left[\dot{U} \circ \Pi_{pp} \circ \lambda_{pp}\right]^{\widehat{\uparrow\downarrow}}\!(Q,k)\bigg).
    \end{align}
    \label{eq:DerivationwX1loopStep5}
\end{subequations}
We have used here the definitions of the physical channels for the fermion-boson vertices $\lambda_{\text{X}}$, i.e., Eqs.~\eqref{eq:DefinitionslambdasPhysicalChannels}, and for the bubbles $\Pi_{\text{X}}$, i.e.,
\begin{subequations}
        \begin{align}
        \Pi_{\text{M}}(Q,k) & = \Pi_{\text{D}}(Q,k) = -\Pi_{\overline{ph}}^{\uparrow\downarrow}(Q,k) = \Pi_{ph}^{\uparrow\uparrow}(Q,k) = -G\bigg(\nu+\left\lceil\frac{\Omega}{2}\right\rceil,\mathbf{k}+\mathbf{Q}\bigg)G\bigg(\nu-\left\lfloor\frac{\Omega}{2}\right\rfloor,\mathbf{k}\bigg), \\
        \Pi_{\text{SC}}(Q,k) & = 2\Pi_{pp}^{\uparrow\downarrow}(Q,k) = G\bigg(\left\lfloor\frac{\Omega}{2}\right\rfloor-\nu,\mathbf{Q}-\mathbf{k}\bigg)G\bigg(\nu+\left\lceil\frac{\Omega}{2}\right\rceil,\mathbf{k}\bigg),
        \end{align}
        \label{eq:BubblesPiX}
    \end{subequations}
where $G(\nu,\mathbf{k})=G(k)$ was already introduced in the right-hand sides of Eqs.~\eqref{eq:GMomentum} and~\eqref{eq:BubblesFreqMomentum}. Finally, one can proceed with the same reasoning to treat the sums involved in the terms $\dot{U}\circ \Pi_r \circ \lambda_r$ in Eq.~\eqref{eq:DerivationwX1loopStep5}, namely, by using Eqs.~\eqref{eq:EvaluateSumsFreqMomenta} and then Eqs.~\eqref{eq:APiBFormulasfcircproduct} with $A=\dot{U}$ and $B=\lambda_r$ for $r=ph,\overline{ph},pp$. In that way, we obtain
\begin{equation}
\dot{\widetilde{w}}^{(1\ell)}_{\text{X}}(Q) = \left(w_{\text{X}}(Q)\right)^2 \sum_{k,k^\prime} \lambda_{\text{X}}(Q,k) \Pi_{\text{X}}(Q,k) \dot{U}_{\text{X}}(Q,k,k^\prime) \Pi_{\text{X}}(Q,k^\prime) \lambda_{\text{X}}(Q,k^\prime),
\label{eq:DerivationwX1loopStep6}
\end{equation}
for $\text{X}=\text{M},\text{D},\text{SC}$, and $U_{\text{X}}$ is defined following Eq.~\eqref{eq:DefinitionsPhysicalChannels} as
\begin{subequations}
    \begin{align}
       U_{\text{M}} & = U^{\uparrow\uparrow} - U^{\uparrow\downarrow} = - U^{\uparrow\downarrow}, \\
       U_{\text{D}} & = U^{\uparrow\uparrow} + U^{\uparrow\downarrow}, \\
       U_{\text{SC}} & = U^{\uparrow\downarrow}.
    \end{align}
\end{subequations}
To derive Eq.~\eqref{eq:DerivationwX1loopStep6} for $\text{X}=\text{SC}$, we also used the relation
\begin{equation}
\sum_{k^\prime} \left(\dot{U}^{\uparrow\downarrow}(Q,k^\prime,k)-\dot{U}^{\widehat{\uparrow\downarrow}}(Q,k^\prime,k)\right) \Pi_{pp}^{\uparrow\downarrow}(Q,k^\prime) \lambda_{\text{SC}}(Q,k^\prime) = 2 \sum_{k^\prime} \dot{U}^{\uparrow\downarrow}(Q,k^\prime,k) \Pi_{pp}^{\uparrow\downarrow}(Q,k^\prime) \lambda_{\text{SC}}(Q,k^\prime),
\end{equation}
that follows from the crossing symmetry of $\dot{U}$, $\Pi_{pp}$ and $\lambda_{pp}$. Each term involved in the right-hand sides of the generic flow equations~\eqref{eq:UflowmultiloopSBEfRGEq1l},~\eqref{eq:UflowmultiloopSBEfRGEq2l} and~\eqref{eq:UflowmultiloopSBEfRGEqnl} can be treated in the same way as $\dot{\widetilde{w}}_r^{(1\ell)}$, still by exploiting Eqs.~\eqref{eq:EvaluateSumsFreqMomenta},~\eqref{eq:ABCFormulasfcircproduct} and~\eqref{eq:APiBFormulasfcircproduct} to obtain a final contribution in the form of Eq.~\eqref{eq:DerivationwX1loopStep6}. This procedure yields the following flow equations:
\begin{subequations}
    \begin{align}
        \dot{w}_{\text{X}}^{(1)}(Q) & = \left(w_{\text{X}}(Q)\right)^2 \sum_k \lambda_{\text{X}}(Q,k) \dot{\Pi}_{\text{X}}(Q,k) \lambda_{\text{X}}(Q,k) \nonumber \\
        &\phantom{=} + \left(w_{\text{X}}(Q)\right)^2 \sum_{k,k^{\prime}} \lambda_{\text{X}}(Q,k) \Pi_{\text{X}}(Q,k) \dot{U}_{\text{X}}(Q,k,k^{\prime}) \Pi_{\text{X}}(Q,k^{\prime}) \lambda_{\text{X}}(Q,k^{\prime}) + \dot{\mathcal{B}}_{\text{X}}(Q) \nonumber \\
        &\phantom{=} + w_{\text{X}}(Q)\sum_k\left(\lambda_{\text{X}}(Q,k) \Pi_{\text{X}}(Q,k) \dot{\mathcal{B}}_{\text{X}}(Q) + \dot{\mathcal{B}}_{\text{X}}(Q) \Pi_{\text{X}}(Q,k) \lambda_{\text{X}}(Q,k)\right), \\
        \dot{\lambda}_{\text{X}}^{(1)}(Q,k) & = \sum_{k^{\prime}} \lambda_{\text{X}}(Q,k^{\prime}) \dot{\Pi}_{\text{X}}(Q,k^{\prime}) \mathcal{I}_{\text{X}}(Q,k^{\prime},k) + \sum_{k^{\prime}} \lambda_{\text{X}}(Q,k^{\prime}) \Pi_{\text{X}}(Q,k^{\prime}) \dot{\mathcal{F}}_{\text{X}}(Q,k^{\prime},k) \nonumber \\
        &\phantom{=} + \sum_{k^{\prime},k^{\prime\prime}} \lambda_{\text{X}}(Q,k^{\prime}) \Pi_{\text{X}}(Q,k^{\prime}) \dot{\mathcal{F}}_{\text{X}}(Q,k^{\prime},k^{\prime\prime}) \Pi_{\text{X}}(Q,k^{\prime\prime}) \mathcal{I}_{\text{X}}(Q,k^{\prime\prime},k), \\
        \dot{M}_{\text{X}}^{(1)}(Q,k,k^{\prime}) & = \sum_{k^{\prime\prime}} \mathcal{I}_{\text{X}}(Q,k,k^{\prime\prime}) \dot{\Pi}_{\text{X}}(Q,k^{\prime\prime}) \mathcal{I}_{\text{X}}(Q,k^{\prime\prime},k) + \sum_{k^{\prime\prime}} \mathcal{I}_{\text{X}}(Q,k,k^{\prime\prime}) \Pi_{\text{X}}(Q,k^{\prime\prime}) \dot{\mathcal{F}}_{\text{X}}(Q,k^{\prime\prime},k^{\prime}) \nonumber \\
        &\phantom{=} + \sum_{k^{\prime\prime}} \dot{\mathcal{F}}_{\text{X}}(Q,k,k^{\prime\prime}) \Pi_{\text{X}}(Q,k^{\prime\prime}) \mathcal{I}_{\text{X}}(Q,k^{\prime\prime},k^{\prime}) \nonumber \\
        &\phantom{=} + \sum_{k^{\prime\prime},k^{\prime\prime\prime}} \mathcal{I}_{\text{X}}(Q,k,k^{\prime\prime}) \Pi_{\text{X}}(Q,k^{\prime\prime}) \dot{\mathcal{F}}_{\text{X}}(Q,k^{\prime\prime},k^{\prime\prime\prime}) \Pi_{\text{X}}(Q,k^{\prime\prime\prime}) \mathcal{I}_{\text{X}}(Q,k^{\prime\prime\prime},k^{\prime}),
    \end{align}
    \label{eq:mfRGequationsPhysicalChannels1l}
\end{subequations}
\begin{subequations}
    \begin{align}
        \dot{w}_{\text{X}}^{(2)}(Q) & = 0, \\
        \dot{\lambda}_{\text{X}}^{(2)}(Q,k) & = \sum_{k^{\prime}} \lambda_{\text{X}}(Q,k^{\prime}) \Pi_{\text{X}}(Q,k^{\prime}) \dot{\phi}_{\overline{\text{X}}}^{(1)}(Q,k^{\prime},k), \\
        \dot{M}_{\text{X}}^{(2)}(Q,k,k^{\prime}) & = \sum_{k^{\prime\prime}} \dot{\phi}_{\overline{\text{X}}}^{(1)}(Q,k,k^{\prime\prime}) \Pi_{\text{X}}(Q,k^{\prime\prime}) \mathcal{I}_{\text{X}}(Q,k^{\prime\prime},k^{\prime}) \nonumber\\
        & \phantom{=} + \sum_{k^{\prime\prime}} \mathcal{I}_{\text{X}}(Q,k,k^{\prime\prime}) \Pi_{\text{X}}(Q,k^{\prime\prime}) \dot{\phi}_{\overline{\text{X}}}^{(1)}(Q,k^{\prime\prime},k^{\prime}),
    \end{align}
    \label{eq:mfRGequationsPhysicalChannels2l}
\end{subequations}
and, for $\ell\geq 3$,
\begin{subequations}
    \begin{align}
        \dot{w}_{\text{X}}^{(\ell)}(Q) & = \left(w_{\text{X}}(Q)\right)^2 \sum_{k,k^{\prime}} \lambda_{\text{X}}(Q,k) \Pi_{\text{X}}(Q,k) \dot{\phi}^{(\ell-2)}_{\overline{\text{X}}}(Q,k,k^{\prime}) \Pi_{\text{X}}(Q,k^{\prime}) \lambda_{\text{X}}(Q,k^{\prime}), \\
        \dot{\lambda}_{\text{X}}^{(\ell)}(Q,k) & = \sum_{k^{\prime}} \lambda_{\text{X}}(Q,k^{\prime}) \Pi_{\text{X}}(Q,k^{\prime}) \dot{\phi}_{\overline{\text{X}}}^{(\ell-1)}(Q,k^{\prime},k) \nonumber \\
        & \phantom{=} + \sum_{k^{\prime},k^{\prime\prime}} \lambda_{\text{X}}(Q,k^{\prime}) \Pi_{\text{X}}(Q,k^{\prime}) \dot{\phi}^{(\ell-2)}_{\overline{\text{X}}}(Q,k^{\prime},k^{\prime\prime}) \Pi_{\text{X}}(Q,k^{\prime\prime}) \mathcal{I}_{\text{X}}(Q,k^{\prime\prime},k), \\
        \dot{M}_{\text{X}}^{(\ell)}(Q,k,k^{\prime}) & = \sum_{k^{\prime\prime}} \dot{\phi}_{\overline{\text{X}}}^{(\ell-1)}(Q,k,k^{\prime\prime}) \Pi_{\text{X}}(Q,k^{\prime\prime}) \mathcal{I}_{\text{X}}(Q,k^{\prime\prime},k^{\prime}) \nonumber \\
        & \phantom{=} + \sum_{k^{\prime\prime}} \mathcal{I}_{\text{X}}(Q,k,k^{\prime\prime}) \Pi_{\text{X}}(Q,k^{\prime\prime}) \dot{\phi}_{\overline{\text{X}}}^{(\ell-1)}(Q,k^{\prime\prime},k^{\prime}) \nonumber \\
        & \phantom{=} + \sum_{k^{\prime\prime},k^{\prime\prime\prime}} \mathcal{I}_{\text{X}}(Q,k,k^{\prime\prime}) \Pi_{\text{X}}(Q,k^{\prime\prime}) \dot{\phi}^{(\ell-2)}_{\overline{\text{X}}}(Q,k^{\prime\prime},k^{\prime\prime\prime}) \Pi_{\text{X}}(Q,k^{\prime\prime\prime}) \mathcal{I}_{\text{X}}(Q,k^{\prime\prime\prime},k^{\prime}),
    \end{align}
    \label{eq:mfRGequationsPhysicalChannelsnl}
\end{subequations}
where $\text{X}=\text{M},\text{D},\text{SC}$. We recall that all objects defined in physical channels (i.e., labeled by the letter $\text{X}$) are defined by Eqs.~\eqref{eq:DefinitionsPhysicalChannels}, with the exception of the fermion-boson vertices $\lambda_{\text{X}}$ and the bubbles $\Pi_{\text{X}}$ given by Eqs.~\eqref{eq:DefinitionslambdasPhysicalChannels} and~\eqref{eq:BubblesPiX}, respectively. We also have the following expressions for $\mathcal{I}_{\text{X}}$:
\begin{subequations}
\begin{align}
    \mathcal{I}_{\text{M}} & = M_{\text{M}} + \frac{1}{2} P^{ph\rightarrow \overline{ph}}\left(\phi_{\text{M}}-\phi_{\text{D}}\right) - P^{pp\rightarrow \overline{ph}}\phi_{\text{SC}} + \mathcal{F}_{\text{M}}, \\
    \mathcal{I}_{\text{D}} & = M_{\text{D}} - 2 P^{\overline{ph} \rightarrow ph}\phi_{\text{M}} + 2 P^{pp\rightarrow ph}\phi_{\text{SC}} - P^{pp\rightarrow \overline{ph}}\phi_{\text{SC}} + \frac{1}{2} P^{ph\rightarrow \overline{ph}}\left(\phi_{\text{M}}-\phi_{\text{D}}\right) + \mathcal{F}_{\text{D}}, \\
    \mathcal{I}_{\text{SC}} & = M_{\text{SC}} + \frac{1}{2} P^{ph\rightarrow pp}\left(\phi_{\text{D}} - \phi_{\text{M}}\right) - P^{\overline{ph}\rightarrow pp}\phi_{\text{M}} + \mathcal{F}_{\text{SC}},
\end{align}
\label{eq:ExpressionsMathcalIX}
\end{subequations}
\end{widetext}
where
\begin{equation}
\phi_{\text{X}}=\nabla_{\text{X}}+M_{\text{X}}-\mathcal{B}_{\text{X}},
\end{equation}
and the vertices $\nabla_{\text{X}}(Q,k,k^{\prime})$ still satisfy Eq.~\eqref{eq:nablaSBEphysicalChannel} [with Eq.~\eqref{eq:lambdabarXtimereversal}]. $P^{r\rightarrow r^{\prime}}$ are the projection matrices, which translate conventions between our different channel notations set by Tab.~\ref{fig:FrequencyMomentumParametrization} for frequencies and momenta. Note that, for our numerical implementation, we use the expressions of the projection matrices $P^{r\rightarrow r^{\prime}}$ given in Refs.~\cite{HillePhDThesis,HeinzelmannPhDThesis}. Furthermore, the derivatives $\dot{\phi}^{(\ell)}_{\overline{\text{X}}}(Q,k,k^{\prime})$ can be written as
\begin{subequations}
\begin{align}
    \dot{\phi}_{\overline{\text{M}}}^{(\ell)} & = \frac{1}{2} P^{ph\rightarrow \overline{ph}}\left(\dot{\phi}_{\text{M}}^{(\ell)}-\dot{\phi}_{\text{D}}^{(\ell)}\right) - P^{pp\rightarrow \overline{ph}}\dot{\phi}_{\text{SC}}^{(\ell)}, \\
    \dot{\phi}_{\overline{\text{D}}}^{(\ell)} & = - 2 P^{\overline{ph} \rightarrow ph}\dot{\phi}_{\text{M}}^{(\ell)} + 2 P^{pp\rightarrow ph}\dot{\phi}_{\text{SC}}^{(\ell)} - P^{pp\rightarrow \overline{ph}}\dot{\phi}_{\text{SC}}^{(\ell)} \nonumber \\
    &\phantom{=} + \frac{1}{2} P^{ph\rightarrow \overline{ph}}\left(\dot{\phi}_{\text{M}}^{(\ell)}-\dot{\phi}_{\text{D}}^{(\ell)}\right), \\
    \dot{\phi}_{\overline{\text{SC}}}^{(\ell)} & = \frac{1}{2} P^{ph\rightarrow pp}\left(\dot{\phi}_{\text{D}}^{(\ell)} - \dot{\phi}_{\text{M}}^{(\ell)}\right) - P^{\overline{ph}\rightarrow pp}\dot{\phi}_{\text{M}}^{(\ell)},
\end{align}
\label{eq:ExpressionsdotphilXbar}
\end{subequations}
with
\begin{equation}
\dot{\phi}^{(\ell)}_{\text{X}}=\dot{\nabla}^{(\ell)}_{\text{X}}+\dot{M}^{(\ell)}_{\text{X}}-\dot{\mathcal{B}}_{\text{X}}\delta_{\ell,1},
\end{equation}
and
\begin{align}
\dot{\nabla}^{(\ell)}_{\text{X}}(Q,k,k^{\prime})&=\dot{\lambda}^{(\ell)}_{\text{X}}(Q,k)w_{\text{X}}(Q)\lambda_{\text{X}}(Q,k^{\prime}) \nonumber \\
& \phantom{=} +\lambda_{\text{X}}(Q,k)\dot{w}^{(\ell)}_{\text{X}}(Q)\lambda_{\text{X}}(Q,k^{\prime}) \nonumber \\
& \phantom{=} +\lambda_{\text{X}}(Q,k)w_{\text{X}}(Q)\dot{\lambda}^{(\ell)}_{\text{X}}(Q,k^{\prime}).
\end{align}
The results~\eqref{eq:mfRGequationsPhysicalChannels1l}--\eqref{eq:mfRGequationsPhysicalChannelsnl}, with Eqs.~\eqref{eq:ExpressionsMathcalIX} and~\eqref{eq:ExpressionsdotphilXbar}, are the multiloop SBE fRG flow equations underlying our interaction flow scheme in physical channels. The corresponding initial conditions for the flowing objects are given by $w_\X^{\Lambda_{\text{init}}} = \mathcal{B}_\X^{\Lambda_{\text{init}}}$, $\lambda_\X^{\Lambda_{\text{init}}} = 1$, and $M_\X^{\Lambda_{\text{init}}} = 0$, where $\Lambda_{\text{init}}$ is the initial value for the RG scale $\Lambda$. They are valid for fermionic systems with a classical action consistent with Eq.~\eqref{eq:ClassicalActionS}, possessing translational-invariance and energy conservation [see Eqs.~\eqref{eq:FreqMomConservationLawsG0U}] and respecting time-reversal and $SU(2)$ spin symmetry [see Eqs.~\eqref{eq:GVSU2symmetry} and Eq.~\eqref{eq:lambdabarXtimereversal}].

\begin{figure*}[t!]
    \centering
    \scalebox{1}[1.0]{\includegraphics[width=1.0\linewidth, keepaspectratio=false]{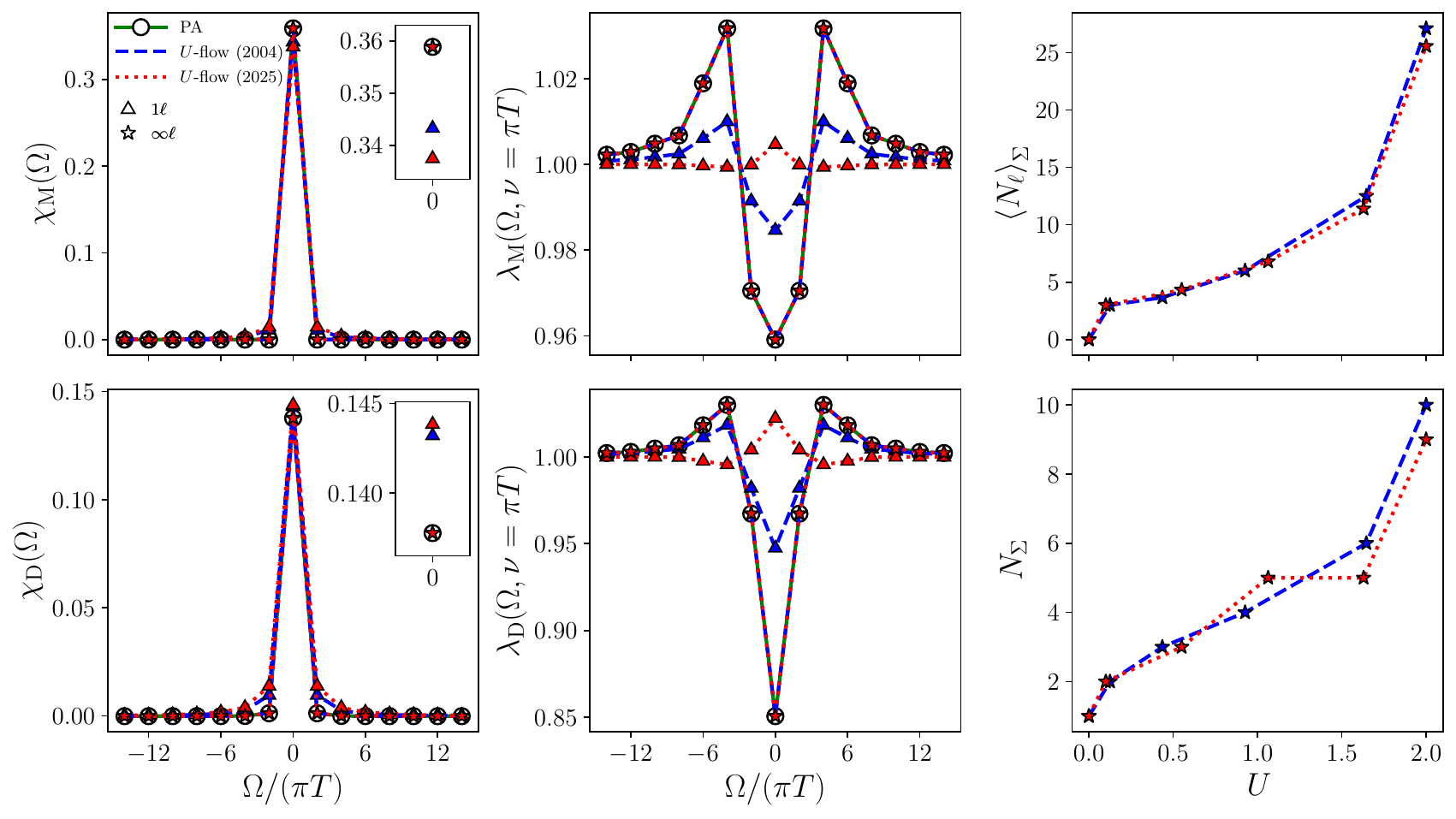}}
    \caption{Susceptibilities and fermion-boson vertices for the HA in the magnetic and density channels at $\beta = 1$, $U = 2$ and at half filling. Similarly to Fig.~\ref{fig:aim_loopconvergence}, the results are shown at $1\ell$ and $\infty \ell$ (fully converged) multiloop order for the two flow schemes defined in Sec.~\ref{sec:ModelDefinitionsCutoffSchemes}. Note that the $\D$ and $\SC$ channels are degenerate, i.e., $\chi_\D=\chi_\SC$ and $\lambda_\D=\lambda_\SC$. The insets in the left-hand panels show the same susceptibilities as the corresponding main panels, but magnified around $\Omega=0$. The panels on the right-hand side show the average number of loop corrections per self-energy iteration $\left\langle N_{\ell} \right\rangle_{\Sigma}$ and the corresponding number of self-energy iterations $N_{\Sigma}$ at each step of the flow.}
    \label{fig:hubbard_atom_loopconvergence}
\end{figure*}

Regarding the self-energy flow equations, we explained in Sec.~\ref{sec:SelfEnergyFlow} that we will notably consider the Schwinger--Dyson equation rewritten within the SBE framework~\cite{Patricolo2025} for the studied models with $\mathcal{F}_r=0$ (Hubbard atom and AIM). Under the same assumptions as those used to derive the flow equations~\eqref{eq:mfRGequationsPhysicalChannels1l}--\eqref{eq:mfRGequationsPhysicalChannelsnl}, this formulation of the Schwinger--Dyson equation reads in the magnetic channel~\cite{Patricolo2025}
\begin{equation}
    \Sigma(k) = \sum_{Q} w_{\mathrm{M}}(Q) \lambda_{\mathrm{M}}\Big(Q,\widetilde{k}\Big) G(k+Q),
    \label{eq:SigmaMEquation}
\end{equation}
where $\widetilde{k}=\left(\nu + \left\lfloor\frac{\Omega}{2}\right\rfloor,\mathbf{k}\right)$, with $Q=(\Omega, \mathbf{Q})$ and $k=(\nu, \mathbf{k})$. Introducing the RG scale $\Lambda$ via the substitutions $G_0\rightarrow G_0^\Lambda$ and $U\rightarrow U^\Lambda$, we obtain the following flow equation from Eq.~\eqref{eq:SigmaMEquation}:
\begin{align}
    \dot{\Sigma}(k) & = \sum_{Q} \dot{w}_{\mathrm{M}}(Q) \lambda_{\mathrm{M}}\Big(Q,\widetilde{k}\Big) G(k+Q) \nonumber \\
    & \phantom{=} + \sum_{Q} w_{\mathrm{M}}(Q) \dot{\lambda}_{\mathrm{M}}\Big(Q,\widetilde{k}\Big) G(k+Q) \nonumber \\
    & \phantom{=} + \sum_{Q} w_{\mathrm{M}}(Q) \lambda_{\mathrm{M}}\Big(Q,\widetilde{k}\Big) \dot{G}(k+Q). \label{eq:SigmaMdotFlowEquation}
\end{align}
Following Ref.~\cite{Patricolo2025}, relations analogous to Eqs.~\eqref{eq:SigmaMEquation} and~\eqref{eq:SigmaMdotFlowEquation} can also be derived in the $\mathrm{D}$ and $\mathrm{SC}$ channels. Since the right-hand side of Eq.~\eqref{eq:SigmaMEquation} does not explicitly involve the bare interaction $U$, the flow equation~\eqref{eq:SigmaMdotFlowEquation} does not contain any additional $\dot{U}$-dependent term compared to its counterpart used in a conventional fRG approach relying on a regulator introduced in the bare propagator only (with $\dot{U}=0$). In the main text, we consider additionally a model with non-trivial fermionic bare interactions $\mathcal{F}_r$, the AHIM. In this case, the form Eq.~\eqref{eq:SigmaMdotFlowEquation} is no longer valid (cf. Ref.~\cite{Patricolo2025} App.~B), and we must in this case work directly adapt the self-energy flow equation~\eqref{eq:SigmadotPirSDEph} to $SU(2)$-spin-symmetric systems to arrive at
\begin{widetext}
    \begin{align}
\nonumber\dot{\Sigma}(i\nu) =& - 
\sum_{i\nu', i\Omega} \left[2\dot{U}\left(i\nu' - i\nu - \left\lfloor \frac{i\Omega}{2} \right\rfloor \right) - \dot{U}(i\Omega) \right]
\Pi_\M(i\Omega - i\nu, i\nu')  V_\M\left(i\Omega - i\nu, i\nu + \left\lfloor \frac{i\Omega - i\nu'}{2}\right\rfloor, i\nu' \right)G(i\nu')\\
&-\sum_{i\nu', i\Omega} \left[2U\left(i\nu' - i\nu - \left\lfloor \frac{i\Omega}{2} \right\rfloor \right) - U(i\Omega) \right]
\dot{\Pi}_\M(i\Omega - i\nu, i\nu')  V_\M\left(i\Omega - i\nu, i\nu + \left\lfloor \frac{i\Omega - i\nu'}{2}\right\rfloor, i\nu' \right)G(i\nu')\nonumber \\
&-\sum_{i\nu', i\Omega} \left[2U\left(i\nu' - i\nu - \left\lfloor \frac{i\Omega}{2} \right\rfloor \right) - U(i\Omega) \right]
\Pi_\M(i\Omega - i\nu, i\nu')  \dot{V}_\M\left(i\Omega - i\nu, i\nu + \left\lfloor \frac{i\Omega - i\nu'}{2}\right\rfloor, i\nu' \right)G(i\nu')\nonumber \\
&-\sum_{i\nu', i\Omega} \left[2U\left(i\nu' - i\nu - \left\lfloor \frac{i\Omega}{2} \right\rfloor \right) - U(i\Omega) \right]
\Pi_\M(i\Omega - i\nu, i\nu')  V_\M\left(i\Omega - i\nu, i\nu + \left\lfloor \frac{i\Omega - i\nu'}{2}\right\rfloor, i\nu' \right)\dot{G}(i\nu')\nonumber \\
&- \sum_{i\nu'}\dot{U}(i\nu - i\nu')G(i\nu') - \sum_{i\nu'}U(i\nu - i\nu')\dot{G}(i\nu'),
\label{eq:SDEdensityflowEquation}
\end{align}
\end{widetext}
with $V_\M=V^{\uparrow\uparrow}-V^{\uparrow\downarrow}$.

\section{Results for the Hubbard atom and $2\ell$ fRG}
\label{sec:hubbard_atom}

\begin{figure*}[t!]
    \centering
    \scalebox{1}[1.0]{\includegraphics[width=1.0\linewidth, keepaspectratio=false]{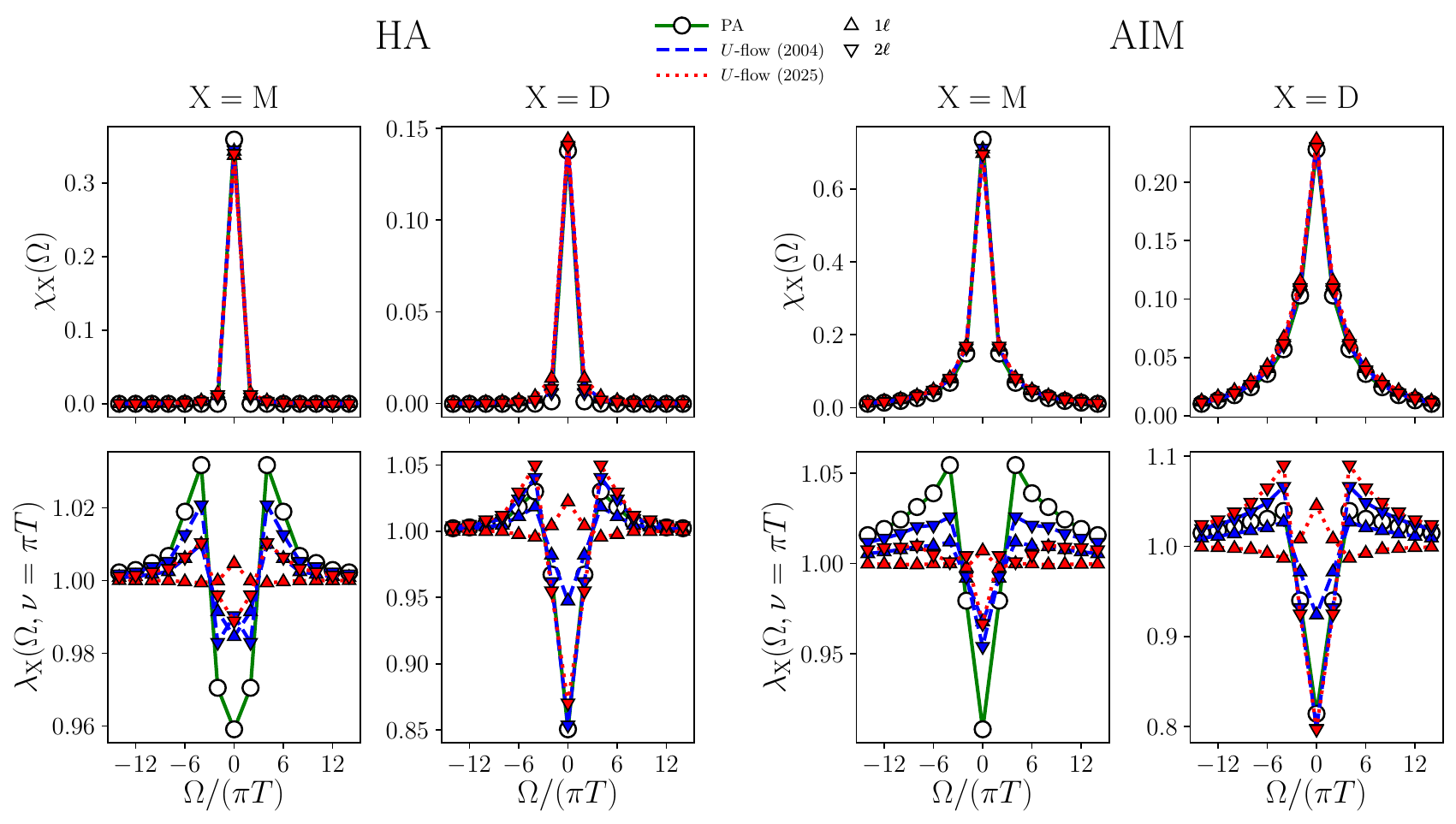}}
    \caption{Susceptibilities and fermion-boson vertices at $1\ell$ and $2\ell$ truncations for the HA (left) and the AIM (right), for the same set of parameters as Figs.~\ref{fig:aim_loopconvergence} and \ref{fig:hubbard_atom_loopconvergence}.}    
    \label{fig:HA_AIM_1l2l}
\end{figure*}

\begin{figure}[t]
    \centering
    \includegraphics[width=0.9\linewidth]{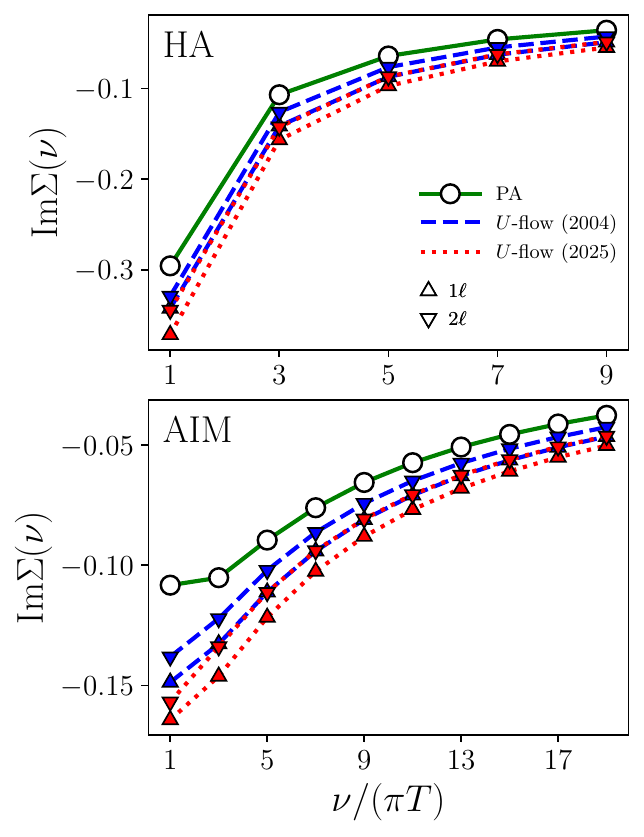}
    \caption{Imaginary part of the self-energy at the $1\ell$ and $2\ell$ truncations for the HA and the AIM with the same parameters as Figs.~\ref{fig:aim_loopconvergence} and \ref{fig:hubbard_atom_loopconvergence} for the two flow schemes defined in Sec.~\ref{sec:ModelDefinitionsCutoffSchemes}. The converged $\infty\ell$ results (not shown) lie exactly on top of the corresponding result from the self-consistent PA.}
    \label{fig:selfenergy_1l_vs_2l}
\end{figure}

Here, we provide benchmarks for the Hubbard atom (HA). Additionally, we comment briefly on the performance of the $U$-flow (2025) in the $2\ell$ fRG.

The HA describes a system consisting of a localized site, which can host up to two electrons (one with spin up and another with spin down). It is $SU(2)$-spin-symmetric and satisfies frequency conservation. The HA is an important benchmark for the Hubbard model at strong interactions, where the hopping is negligible. The bare propagator is given by
\begin{align}
G_{0}(i\nu) = \frac{1}{i\nu - \mu},
\end{align}
where $\mu$ is the chemical potential. The bare interaction is given by the on-site Hubbard-like interaction of Eq.~\eqref{eq:hubbard_interaction_physical_channels}. In this model, the criterion of weak coupling is governed by the requirement that
\begin{align}
\max_{\X}\abs{\Pi_\X(i\Omega = 0) U_\X} \approx  \frac{\beta U}{4}< 1.\label{eq:PA_valid_copndition_atom}
\end{align}

We performed calculations for the HA at $\beta = 1$ and $U = 2$, at half filling. This parameter choice satisfies Eq.~\eqref{eq:PA_valid_copndition_atom}. Once again, we benchmarked the performance of the $U$-flow (2025), similar to what is discussed in Sec.~\ref{sec:ModelsNumericalImplementationConvergence}. The corresponding results are shown in Fig.~\ref{fig:hubbard_atom_loopconvergence} for the susceptibilities and the fermion-boson vertices. The conclusions we draw are exactly the same. In particular, the $U$-flow (2025) converges to the self-consistent PA result in a very similar manner to that of the $U$-flow (2004). Thus, we have shown loop convergence of our flow equations with an interaction in the regulator only yet for another model.

Although both flow schemes reach full convergence within a comparable number of loop corrections and self-energy iterations, the results of the $1\ell$ calculations for both the $U$-flow (2025) and the $U$-flow (2004) look quite different. To investigate this further, we performed $2\ell$-fRG calculations for both the AIM and HA. The corresponding results are shown in Fig.~\ref{fig:HA_AIM_1l2l} for the susceptibilities and fermion-boson vertices and in Fig.~\ref{fig:selfenergy_1l_vs_2l} for the self-energies.

In the $1\ell$ truncation, the $U$-flow (2025) scheme actually performs worse than the $U$-flow (2004). In the $2\ell$ truncation, both results are improved, but the relative improvement from $1\ell$ is larger for the $U$-flow (2025). This can be seen particularly from the fermion-boson vertices $\lambda_\X$. Thus, it appears that the $U$-flow (2025), while it performs worse at low loop orders, converges with a higher rate for higher loop orders.

In the 2PI formalism, it has notably been observed that interaction flows similar to the $U$-flow (2025) perform worse overall than schemes involving $G_0\rightarrow G_0^\Lambda$~\cite{Fraboulet2024}. An aspect of non-regularizing flow schemes like the $U$-flow (2004) or the $U$-flow (2025) is that they consider all frequency modes at every step of the flow, an effect which leads to an overestimation of fluctuations in truncated flow equations~\cite{Honerkamp2004}. As such, we conclude that the $U$-flow (2025) scheme, which we introduced here only for its simplicity as a test ground for loop convergence, should be avoided at low loop orders.

However, there is an important caveat. At finite loops and especially at low loop orders, the results still depend on the choice of the regulators (i.e., on the functions $G_0^\Lambda$ and $U^\Lambda$). This could be improved by identifying the functions $G_0^\Lambda$ and $U^\Lambda$ that optimize the $1\ell$ (or other low-order) truncation(s) within our scheme. This substantial task is beyond the scope of our paper (see notably Ref.~\cite{DePolsi2022} for a detailed study of regulator dependence in the context of other fRG approaches). We stress that our interaction-flow scheme provides more freedom and flexibility to perform such optimization since it relies on two regulators ($G_0^\Lambda$ and $U^\Lambda$) instead of one in traditional fRG approaches (only $G_0^\Lambda$).

\section{Technical aspects and frequency parametrization}
\label{sec:technical_aspects_and_freq_param}

In this appendix, we specify the technical aspects of the numerical calculations performed in Secs.~\ref{sec:ModelsNumericalImplementationConvergence}--\ref{sec:TflowRetardedInteractions} and App.~\ref{sec:hubbard_atom}. All vertex objects are calculated inside finite frequency boxes. An algorithmic advantage of SBE-fRG formulations is that under usual circumstances (in the symmetry-unbroken phase), their complexity decreases as their importance increases. The most important objects being the bosonic propagators $w_\mathrm{X}$, depend on only one bosonic argument, whereas the multiboson/rest functions $M_\mathrm{X}$, often found to be less important~\cite{Fraboulet2022,AlEryani2024,AlEryani2025,Fraboulet2025}, depend on one bosonic and two fermionic arguments. As such, it is natural to choose bigger frequency boxes for $w_\mathrm{X}$, smaller for $\lambda_\mathrm{X}$ and smallest for the rest functions $M_\mathrm{X}$. In what follows, we specify the number of \emph{positive} Matsubara frequencies that define the boxes of frequencies calculated and stored for each object every step of the flow.
\begin{enumerate}
\item[$\bullet$] 240 fermionic frequencies for the self-energy $\Sigma$ and its derivative $\dot{\Sigma}$.
\item[$\bullet$] 1536 bosonic frequencies for the bosonic propagators $w_\mathrm{X}, \dot{w}_\mathrm{X}^{(\ell)}$, and the bubbles $\Pi_\mathrm{X}, \dot{\Pi}_\mathrm{X}$, and 1536 fermionic fermionic for the bubbles $\Pi_\mathrm{X}, \dot{\Pi}_\mathrm{X}$.
\item[$\bullet$] 48 bosonic frequencies for the fermion-boson vertices $\lambda_\mathrm{X}, \dot{\lambda}_\mathrm{X}^{(\ell)}$ as well as the rest functions $M_\mathrm{X}, \dot{M}_\mathrm{X}^{(\ell)}$ and irreducible vertex derivatives $\dot{\phi}_\mathrm{X}^{(\ell)}$.
\item[$\bullet$] 48 fermionic frequencies for the fermion-boson vertices $\lambda_\mathrm{X}, \dot{\lambda}_\mathrm{X}^{(\ell)}$, and 24 fermionic frequencies for the rest functions $M_\mathrm{X}, \dot{M}_\mathrm{X}^{(\ell)}$ and the irreducible vertices $\dot{I}_\mathrm{X}^{(\ell)}$. 
\end{enumerate}
Outside of the boxes, the objects take on constant asymptotic values. In particular, care must be taken when $\dot{U} \neq 0$ that $\dot{w}_\mathrm{X}^{(\ell)}$ satisfies
\begin{align}
\lim_{i\Omega \rightarrow \pm \infty} \dot{w}_\X^{(\ell)}(i\Omega) &= \lim_{i\Omega \rightarrow \pm \infty} \dot{\mathcal{B}}_\X(i\Omega)\delta_{\ell,1}.
\end{align}
For more details on the asymptotics of the other SBE objects even when $\mathcal{F}_\mathrm{X}\neq 0$, we refer the reader to Ref.~\cite{AlEryani2025}. 

For the multiloop convergence criterion, we choose a tolerance $\varepsilon_{\text{vtx}} = 10^{-4}$, so that we cut the multiloop series at $\ell^*$ when $\max |V^{(\ell^*)}| < \varepsilon_{\text{vtx}}$. For convergence in self-energy iterations, we choose a tolerance $\varepsilon_{\text{se}} = 10^{-3}$, i.e., we stop iterating once updating the self-energy changes the vertex by less than $\varepsilon_{\text{se}}$. For all our calculations (including those performed with the $1\ell$ and $2\ell$ truncations), we also use $\dot{\Sigma}=0$ as initial guess for the derivative of the self-energy for the first self-energy iteration. We refer the interested reader to Ref.~\cite{Fraboulet2025} for more details.

\newpage

\bibliography{main.bbl}

\begin{thebibliography}{120}%
\makeatletter
\providecommand \@ifxundefined [1]{%
 \@ifx{#1\undefined}
}%
\providecommand \@ifnum [1]{%
 \ifnum #1\expandafter \@firstoftwo
 \else \expandafter \@secondoftwo
 \fi
}%
\providecommand \@ifx [1]{%
 \ifx #1\expandafter \@firstoftwo
 \else \expandafter \@secondoftwo
 \fi
}%
\providecommand \natexlab [1]{#1}%
\providecommand \enquote  [1]{``#1''}%
\providecommand \bibnamefont  [1]{#1}%
\providecommand \bibfnamefont [1]{#1}%
\providecommand \citenamefont [1]{#1}%
\providecommand \href@noop [0]{\@secondoftwo}%
\providecommand \href [0]{\begingroup \@sanitize@url \@href}%
\providecommand \@href[1]{\@@startlink{#1}\@@href}%
\providecommand \@@href[1]{\endgroup#1\@@endlink}%
\providecommand \@sanitize@url [0]{\catcode `\\12\catcode `\$12\catcode `\&12\catcode `\#12\catcode `\^12\catcode `\_12\catcode `\%12\relax}%
\providecommand \@@startlink[1]{}%
\providecommand \@@endlink[0]{}%
\providecommand \url  [0]{\begingroup\@sanitize@url \@url }%
\providecommand \@url [1]{\endgroup\@href {#1}{\urlprefix }}%
\providecommand \urlprefix  [0]{URL }%
\providecommand \Eprint [0]{\href }%
\providecommand \doibase [0]{https://doi.org/}%
\providecommand \selectlanguage [0]{\@gobble}%
\providecommand \bibinfo  [0]{\@secondoftwo}%
\providecommand \bibfield  [0]{\@secondoftwo}%
\providecommand \translation [1]{[#1]}%
\providecommand \BibitemOpen [0]{}%
\providecommand \bibitemStop [0]{}%
\providecommand \bibitemNoStop [0]{.\EOS\space}%
\providecommand \EOS [0]{\spacefactor3000\relax}%
\providecommand \BibitemShut  [1]{\csname bibitem#1\endcsname}%
\let\auto@bib@innerbib\@empty
\bibitem [{\citenamefont {Berges}\ \emph {et~al.}(2002)\citenamefont {Berges}, \citenamefont {Tetradis},\ and\ \citenamefont {Wetterich}}]{Berges2002}%
  \BibitemOpen
  \bibfield  {author} {\bibinfo {author} {\bibfnamefont {J.}~\bibnamefont {Berges}}, \bibinfo {author} {\bibfnamefont {N.}~\bibnamefont {Tetradis}},\ and\ \bibinfo {author} {\bibfnamefont {C.}~\bibnamefont {Wetterich}},\ }\bibfield  {title} {\bibinfo {title} {{Non-perturbative renormalization flow in quantum field theory and statistical physics}},\ }\href {https://doi.org/https://doi.org/10.1016/S0370-1573(01)00098-9} {\bibfield  {journal} {\bibinfo  {journal} {Physics Reports}\ }\textbf {\bibinfo {volume} {363}},\ \bibinfo {pages} {223} (\bibinfo {year} {2002})}\BibitemShut {NoStop}%
\bibitem [{\citenamefont {Pawlowski}(2007)}]{Pawlowski2007}%
  \BibitemOpen
  \bibfield  {author} {\bibinfo {author} {\bibfnamefont {J.~M.}\ \bibnamefont {Pawlowski}},\ }\bibfield  {title} {\bibinfo {title} {{Aspects of the functional renormalisation group}},\ }\href {https://doi.org/https://doi.org/10.1016/j.aop.2007.01.007} {\bibfield  {journal} {\bibinfo  {journal} {Annals of Physics}\ }\textbf {\bibinfo {volume} {322}},\ \bibinfo {pages} {2831} (\bibinfo {year} {2007})}\BibitemShut {NoStop}%
\bibitem [{\citenamefont {Kopietz}\ \emph {et~al.}(2010)\citenamefont {Kopietz}, \citenamefont {Bartosch},\ and\ \citenamefont {Sch\"{u}tz}}]{Kopietz2010}%
  \BibitemOpen
  \bibfield  {author} {\bibinfo {author} {\bibfnamefont {P.}~\bibnamefont {Kopietz}}, \bibinfo {author} {\bibfnamefont {L.}~\bibnamefont {Bartosch}},\ and\ \bibinfo {author} {\bibfnamefont {F.}~\bibnamefont {Sch\"{u}tz}},\ }\href {https://doi.org/10.1007/978-3-642-05094-7} {\emph {\bibinfo {title} {{Introduction to the Functional Renormalization Group}}}},\ Lecture Notes in Physics\ (\bibinfo  {publisher} {Springer, Berlin},\ \bibinfo {year} {2010})\BibitemShut {NoStop}%
\bibitem [{\citenamefont {Delamotte}(2012)}]{Delamotte2012}%
  \BibitemOpen
  \bibfield  {author} {\bibinfo {author} {\bibfnamefont {B.}~\bibnamefont {Delamotte}},\ }\bibinfo {title} {An introduction to the nonperturbative renormalization group},\ in\ \href {https://doi.org/10.1007/978-3-642-27320-9_2} {\emph {\bibinfo {booktitle} {Renormalization Group and Effective Field Theory Approaches to Many-Body Systems}}},\ \bibinfo {editor} {edited by\ \bibinfo {editor} {\bibfnamefont {A.}~\bibnamefont {Schwenk}}\ and\ \bibinfo {editor} {\bibfnamefont {J.}~\bibnamefont {Polonyi}}}\ (\bibinfo  {publisher} {Springer Berlin Heidelberg},\ \bibinfo {address} {Berlin, Heidelberg},\ \bibinfo {year} {2012})\ pp.\ \bibinfo {pages} {49--132}\BibitemShut {NoStop}%
\bibitem [{\citenamefont {Metzner}\ \emph {et~al.}(2012)\citenamefont {Metzner}, \citenamefont {Salmhofer}, \citenamefont {Honerkamp}, \citenamefont {Meden},\ and\ \citenamefont {Sch\"onhammer}}]{Metzner2012}%
  \BibitemOpen
  \bibfield  {author} {\bibinfo {author} {\bibfnamefont {W.}~\bibnamefont {Metzner}}, \bibinfo {author} {\bibfnamefont {M.}~\bibnamefont {Salmhofer}}, \bibinfo {author} {\bibfnamefont {C.}~\bibnamefont {Honerkamp}}, \bibinfo {author} {\bibfnamefont {V.}~\bibnamefont {Meden}},\ and\ \bibinfo {author} {\bibfnamefont {K.}~\bibnamefont {Sch\"onhammer}},\ }\bibfield  {title} {\bibinfo {title} {{Functional renormalization group approach to correlated fermion systems}},\ }\href {https://doi.org/10.1103/RevModPhys.84.299} {\bibfield  {journal} {\bibinfo  {journal} {Rev. Mod. Phys.}\ }\textbf {\bibinfo {volume} {84}},\ \bibinfo {pages} {299} (\bibinfo {year} {2012})}\BibitemShut {NoStop}%
\bibitem [{\citenamefont {Dupuis}\ \emph {et~al.}(2021)\citenamefont {Dupuis}, \citenamefont {Canet}, \citenamefont {Eichhorn}, \citenamefont {Metzner}, \citenamefont {Pawlowski}, \citenamefont {Tissier},\ and\ \citenamefont {Wschebor}}]{Dupuis2021}%
  \BibitemOpen
  \bibfield  {author} {\bibinfo {author} {\bibfnamefont {N.}~\bibnamefont {Dupuis}}, \bibinfo {author} {\bibfnamefont {L.}~\bibnamefont {Canet}}, \bibinfo {author} {\bibfnamefont {A.}~\bibnamefont {Eichhorn}}, \bibinfo {author} {\bibfnamefont {W.}~\bibnamefont {Metzner}}, \bibinfo {author} {\bibfnamefont {J.~M.}\ \bibnamefont {Pawlowski}}, \bibinfo {author} {\bibfnamefont {M.}~\bibnamefont {Tissier}},\ and\ \bibinfo {author} {\bibfnamefont {N.}~\bibnamefont {Wschebor}},\ }\bibfield  {title} {\bibinfo {title} {{The nonperturbative functional renormalization group and its applications}},\ }\href {https://doi.org/https://doi.org/10.1016/j.physrep.2021.01.001} {\bibfield  {journal} {\bibinfo  {journal} {Physics Reports}\ }\textbf {\bibinfo {volume} {910}},\ \bibinfo {pages} {1} (\bibinfo {year} {2021})}\BibitemShut {NoStop}%
\bibitem [{\citenamefont {Wetterich}(1993)}]{Wetterich1993}%
  \BibitemOpen
  \bibfield  {author} {\bibinfo {author} {\bibfnamefont {C.}~\bibnamefont {Wetterich}},\ }\bibfield  {title} {\bibinfo {title} {{Exact evolution equation for the effective potential}},\ }\href {https://doi.org/10.1016/0370-2693(93)90726-x} {\bibfield  {journal} {\bibinfo  {journal} {Physics Letters B}\ }\textbf {\bibinfo {volume} {301}},\ \bibinfo {pages} {90} (\bibinfo {year} {1993})}\BibitemShut {NoStop}%
\bibitem [{\citenamefont {Ellwanger}(1994)}]{Ellwanger1994}%
  \BibitemOpen
  \bibfield  {author} {\bibinfo {author} {\bibfnamefont {U.}~\bibnamefont {Ellwanger}},\ }\bibfield  {title} {\bibinfo {title} {{Flow equations for $N$ point functions and bound states}},\ }\href {https://doi.org/10.1007/BF01555911} {\bibfield  {journal} {\bibinfo  {journal} {Z. Phys. C - Particles and Fields}\ }\textbf {\bibinfo {volume} {62}},\ \bibinfo {pages} {503} (\bibinfo {year} {1994})}\BibitemShut {NoStop}%
\bibitem [{\citenamefont {Morris}(1994)}]{Morris1994}%
  \BibitemOpen
  \bibfield  {author} {\bibinfo {author} {\bibfnamefont {T.~R.}\ \bibnamefont {Morris}},\ }\bibfield  {title} {\bibinfo {title} {{The Exact Renormalization Group and Approximate Solutions}},\ }\href {https://doi.org/10.1142/S0217751X94000972} {\bibfield  {journal} {\bibinfo  {journal} {International Journal of Modern Physics A}\ }\textbf {\bibinfo {volume} {09}},\ \bibinfo {pages} {2411} (\bibinfo {year} {1994})},\ \Eprint {https://arxiv.org/abs/https://doi.org/10.1142/S0217751X94000972} {https://doi.org/10.1142/S0217751X94000972} \BibitemShut {NoStop}%
\bibitem [{\citenamefont {Kugler}\ and\ \citenamefont {von Delft}(2018{\natexlab{a}})}]{Kugler2018a}%
  \BibitemOpen
  \bibfield  {author} {\bibinfo {author} {\bibfnamefont {F.~B.}\ \bibnamefont {Kugler}}\ and\ \bibinfo {author} {\bibfnamefont {J.}~\bibnamefont {von Delft}},\ }\bibfield  {title} {\bibinfo {title} {{Multiloop Functional Renormalization Group That Sums Up All Parquet Diagrams}},\ }\href {https://doi.org/10.1103/PhysRevLett.120.057403} {\bibfield  {journal} {\bibinfo  {journal} {Phys. Rev. Lett.}\ }\textbf {\bibinfo {volume} {120}},\ \bibinfo {pages} {057403} (\bibinfo {year} {2018}{\natexlab{a}})}\BibitemShut {NoStop}%
\bibitem [{\citenamefont {Kugler}\ and\ \citenamefont {von Delft}(2018{\natexlab{b}})}]{Kugler2018b}%
  \BibitemOpen
  \bibfield  {author} {\bibinfo {author} {\bibfnamefont {F.~B.}\ \bibnamefont {Kugler}}\ and\ \bibinfo {author} {\bibfnamefont {J.}~\bibnamefont {von Delft}},\ }\bibfield  {title} {\bibinfo {title} {{Multiloop functional renormalization group for general models}},\ }\href {https://doi.org/10.1103/PhysRevB.97.035162} {\bibfield  {journal} {\bibinfo  {journal} {Phys. Rev. B}\ }\textbf {\bibinfo {volume} {97}},\ \bibinfo {pages} {035162} (\bibinfo {year} {2018}{\natexlab{b}})}\BibitemShut {NoStop}%
\bibitem [{\citenamefont {Kugler}\ and\ \citenamefont {von Delft}(2018{\natexlab{c}})}]{Kugler2018c}%
  \BibitemOpen
  \bibfield  {author} {\bibinfo {author} {\bibfnamefont {F.~B.}\ \bibnamefont {Kugler}}\ and\ \bibinfo {author} {\bibfnamefont {J.}~\bibnamefont {von Delft}},\ }\bibfield  {title} {\bibinfo {title} {{Derivation of exact flow equations from the self-consistent parquet relations}},\ }\href {https://doi.org/10.1088/1367-2630/aaf65f} {\bibfield  {journal} {\bibinfo  {journal} {New Journal of Physics}\ }\textbf {\bibinfo {volume} {20}},\ \bibinfo {pages} {123029} (\bibinfo {year} {2018}{\natexlab{c}})}\BibitemShut {NoStop}%
\bibitem [{\citenamefont {Blaizot}\ \emph {et~al.}(2011)\citenamefont {Blaizot}, \citenamefont {Pawlowski},\ and\ \citenamefont {Reinosa}}]{Blaizot2011}%
  \BibitemOpen
  \bibfield  {author} {\bibinfo {author} {\bibfnamefont {J.-P.}\ \bibnamefont {Blaizot}}, \bibinfo {author} {\bibfnamefont {J.~M.}\ \bibnamefont {Pawlowski}},\ and\ \bibinfo {author} {\bibfnamefont {U.}~\bibnamefont {Reinosa}},\ }\bibfield  {title} {\bibinfo {title} {{Exact renormalization group and $\Phi$-derivable approximations}},\ }\href {https://doi.org/https://doi.org/10.1016/j.physletb.2010.12.058} {\bibfield  {journal} {\bibinfo  {journal} {Physics Letters B}\ }\textbf {\bibinfo {volume} {696}},\ \bibinfo {pages} {523} (\bibinfo {year} {2011})}\BibitemShut {NoStop}%
\bibitem [{\citenamefont {Blaizot}\ \emph {et~al.}(2021)\citenamefont {Blaizot}, \citenamefont {Pawlowski},\ and\ \citenamefont {Reinosa}}]{Blaizot2021}%
  \BibitemOpen
  \bibfield  {author} {\bibinfo {author} {\bibfnamefont {J.-P.}\ \bibnamefont {Blaizot}}, \bibinfo {author} {\bibfnamefont {J.~M.}\ \bibnamefont {Pawlowski}},\ and\ \bibinfo {author} {\bibfnamefont {U.}~\bibnamefont {Reinosa}},\ }\bibfield  {title} {\bibinfo {title} {{Functional renormalization group and 2PI effective action formalism}},\ }\href {https://doi.org/https://doi.org/10.1016/j.aop.2021.168549} {\bibfield  {journal} {\bibinfo  {journal} {Annals of Physics}\ }\textbf {\bibinfo {volume} {431}},\ \bibinfo {pages} {168549} (\bibinfo {year} {2021})}\BibitemShut {NoStop}%
\bibitem [{\citenamefont {Thoenniss}\ \emph {et~al.}(2020)\citenamefont {Thoenniss}, \citenamefont {Ritter}, \citenamefont {Kugler}, \citenamefont {von Delft},\ and\ \citenamefont {Punk}}]{Thoenniss2020}%
  \BibitemOpen
  \bibfield  {author} {\bibinfo {author} {\bibfnamefont {J.}~\bibnamefont {Thoenniss}}, \bibinfo {author} {\bibfnamefont {M.~K.}\ \bibnamefont {Ritter}}, \bibinfo {author} {\bibfnamefont {F.~B.}\ \bibnamefont {Kugler}}, \bibinfo {author} {\bibfnamefont {J.}~\bibnamefont {von Delft}},\ and\ \bibinfo {author} {\bibfnamefont {M.}~\bibnamefont {Punk}},\ }\href {https://arxiv.org/abs/2011.01268} {\bibinfo {title} {{Multiloop pseudofermion functional renormalization for quantum spin systems: Application to the spin-$\frac{1}{2}$ kagome Heisenberg model}}} (\bibinfo {year} {2020}),\ \Eprint {https://arxiv.org/abs/2011.01268} {arXiv:2011.01268 [cond-mat.str-el]} \BibitemShut {NoStop}%
\bibitem [{\citenamefont {Kiese}\ \emph {et~al.}(2022)\citenamefont {Kiese}, \citenamefont {M\"uller}, \citenamefont {Iqbal}, \citenamefont {Thomale},\ and\ \citenamefont {Trebst}}]{Kiese2022}%
  \BibitemOpen
  \bibfield  {author} {\bibinfo {author} {\bibfnamefont {D.}~\bibnamefont {Kiese}}, \bibinfo {author} {\bibfnamefont {T.}~\bibnamefont {M\"uller}}, \bibinfo {author} {\bibfnamefont {Y.}~\bibnamefont {Iqbal}}, \bibinfo {author} {\bibfnamefont {R.}~\bibnamefont {Thomale}},\ and\ \bibinfo {author} {\bibfnamefont {S.}~\bibnamefont {Trebst}},\ }\bibfield  {title} {\bibinfo {title} {{Multiloop functional renormalization group approach to quantum spin systems}},\ }\href {https://doi.org/10.1103/PhysRevResearch.4.023185} {\bibfield  {journal} {\bibinfo  {journal} {Phys. Rev. Res.}\ }\textbf {\bibinfo {volume} {4}},\ \bibinfo {pages} {023185} (\bibinfo {year} {2022})}\BibitemShut {NoStop}%
\bibitem [{\citenamefont {Ritter}\ \emph {et~al.}(2022)\citenamefont {Ritter}, \citenamefont {Kiese}, \citenamefont {Müller}, \citenamefont {Kugler}, \citenamefont {Thomale}, \citenamefont {Trebst},\ and\ \citenamefont {von Delft}}]{Ritter2022}%
  \BibitemOpen
  \bibfield  {author} {\bibinfo {author} {\bibfnamefont {M.~K.}\ \bibnamefont {Ritter}}, \bibinfo {author} {\bibfnamefont {D.}~\bibnamefont {Kiese}}, \bibinfo {author} {\bibfnamefont {T.}~\bibnamefont {Müller}}, \bibinfo {author} {\bibfnamefont {F.~B.}\ \bibnamefont {Kugler}}, \bibinfo {author} {\bibfnamefont {R.}~\bibnamefont {Thomale}}, \bibinfo {author} {\bibfnamefont {S.}~\bibnamefont {Trebst}},\ and\ \bibinfo {author} {\bibfnamefont {J.}~\bibnamefont {von Delft}},\ }\bibfield  {title} {\bibinfo {title} {{Benchmark calculations of multiloop pseudofermion fRG}},\ }\href {https://doi.org/10.1140/epjb/s10051-022-00349-2} {\bibfield  {journal} {\bibinfo  {journal} {Eur. Phys. J. B}\ }\textbf {\bibinfo {volume} {95}},\ \bibinfo {pages} {102} (\bibinfo {year} {2022})}\BibitemShut {NoStop}%
\bibitem [{\citenamefont {Chalupa-Gantner}\ \emph {et~al.}(2022)\citenamefont {Chalupa-Gantner}, \citenamefont {Kugler}, \citenamefont {Hille}, \citenamefont {von Delft}, \citenamefont {Andergassen},\ and\ \citenamefont {Toschi}}]{Chalupa2020}%
  \BibitemOpen
  \bibfield  {author} {\bibinfo {author} {\bibfnamefont {P.}~\bibnamefont {Chalupa-Gantner}}, \bibinfo {author} {\bibfnamefont {F.~B.}\ \bibnamefont {Kugler}}, \bibinfo {author} {\bibfnamefont {C.}~\bibnamefont {Hille}}, \bibinfo {author} {\bibfnamefont {J.}~\bibnamefont {von Delft}}, \bibinfo {author} {\bibfnamefont {S.}~\bibnamefont {Andergassen}},\ and\ \bibinfo {author} {\bibfnamefont {A.}~\bibnamefont {Toschi}},\ }\bibfield  {title} {\bibinfo {title} {{Fulfillment of sum rules and Ward identities in the multiloop functional renormalization group solution of the Anderson impurity model}},\ }\href {https://doi.org/10.1103/PhysRevResearch.4.023050} {\bibfield  {journal} {\bibinfo  {journal} {Phys. Rev. Res.}\ }\textbf {\bibinfo {volume} {4}},\ \bibinfo {pages} {023050} (\bibinfo {year} {2022})}\BibitemShut {NoStop}%
\bibitem [{\citenamefont {Ge}\ \emph {et~al.}(2024)\citenamefont {Ge}, \citenamefont {Ritz}, \citenamefont {Walter}, \citenamefont {Aguirre}, \citenamefont {von Delft},\ and\ \citenamefont {Kugler}}]{Ge2024}%
  \BibitemOpen
  \bibfield  {author} {\bibinfo {author} {\bibfnamefont {A.}~\bibnamefont {Ge}}, \bibinfo {author} {\bibfnamefont {N.}~\bibnamefont {Ritz}}, \bibinfo {author} {\bibfnamefont {E.}~\bibnamefont {Walter}}, \bibinfo {author} {\bibfnamefont {S.}~\bibnamefont {Aguirre}}, \bibinfo {author} {\bibfnamefont {J.}~\bibnamefont {von Delft}},\ and\ \bibinfo {author} {\bibfnamefont {F.~B.}\ \bibnamefont {Kugler}},\ }\bibfield  {title} {\bibinfo {title} {{Real-frequency quantum field theory applied to the single-impurity Anderson model}},\ }\href {https://doi.org/10.1103/PhysRevB.109.115128} {\bibfield  {journal} {\bibinfo  {journal} {Phys. Rev. B}\ }\textbf {\bibinfo {volume} {109}},\ \bibinfo {pages} {115128} (\bibinfo {year} {2024})}\BibitemShut {NoStop}%
\bibitem [{\citenamefont {Ritz}\ \emph {et~al.}(2024)\citenamefont {Ritz}, \citenamefont {Ge}, \citenamefont {Walter}, \citenamefont {Aguirre}, \citenamefont {von Delft},\ and\ \citenamefont {Kugler}}]{Ritz2024}%
  \BibitemOpen
  \bibfield  {author} {\bibinfo {author} {\bibfnamefont {N.}~\bibnamefont {Ritz}}, \bibinfo {author} {\bibfnamefont {A.}~\bibnamefont {Ge}}, \bibinfo {author} {\bibfnamefont {E.}~\bibnamefont {Walter}}, \bibinfo {author} {\bibfnamefont {S.}~\bibnamefont {Aguirre}}, \bibinfo {author} {\bibfnamefont {J.}~\bibnamefont {von Delft}},\ and\ \bibinfo {author} {\bibfnamefont {F.~B.}\ \bibnamefont {Kugler}},\ }\bibfield  {title} {\bibinfo {title} {{KeldyshQFT: A C++ codebase for real-frequency multiloop functional renormalization group and parquet computations of the single-impurity Anderson model}},\ }\href {https://doi.org/10.1063/5.0221340} {\bibfield  {journal} {\bibinfo  {journal} {The Journal of Chemical Physics}\ }\textbf {\bibinfo {volume} {161}},\ \bibinfo {pages} {054118} (\bibinfo {year} {2024})}\BibitemShut {NoStop}%
\bibitem [{\citenamefont {Tagliavini}\ \emph {et~al.}(2019)\citenamefont {Tagliavini}, \citenamefont {Hille}, \citenamefont {Kugler}, \citenamefont {Andergassen}, \citenamefont {Toschi},\ and\ \citenamefont {Honerkamp}}]{TagliaviniHille2019}%
  \BibitemOpen
  \bibfield  {author} {\bibinfo {author} {\bibfnamefont {A.}~\bibnamefont {Tagliavini}}, \bibinfo {author} {\bibfnamefont {C.}~\bibnamefont {Hille}}, \bibinfo {author} {\bibfnamefont {F.~B.}\ \bibnamefont {Kugler}}, \bibinfo {author} {\bibfnamefont {S.}~\bibnamefont {Andergassen}}, \bibinfo {author} {\bibfnamefont {A.}~\bibnamefont {Toschi}},\ and\ \bibinfo {author} {\bibfnamefont {C.}~\bibnamefont {Honerkamp}},\ }\bibfield  {title} {\bibinfo {title} {{Multiloop functional renormalization group for the two-dimensional Hubbard model: Loop convergence of the response functions}},\ }\href {https://doi.org/10.21468/SciPostPhys.6.1.009} {\bibfield  {journal} {\bibinfo  {journal} {SciPost Phys.}\ }\textbf {\bibinfo {volume} {6}},\ \bibinfo {pages} {009} (\bibinfo {year} {2019})}\BibitemShut {NoStop}%
\bibitem [{\citenamefont {Hille}\ \emph {et~al.}(2020{\natexlab{a}})\citenamefont {Hille}, \citenamefont {Rohe}, \citenamefont {Honerkamp},\ and\ \citenamefont {Andergassen}}]{Hille2020a}%
  \BibitemOpen
  \bibfield  {author} {\bibinfo {author} {\bibfnamefont {C.}~\bibnamefont {Hille}}, \bibinfo {author} {\bibfnamefont {D.}~\bibnamefont {Rohe}}, \bibinfo {author} {\bibfnamefont {C.}~\bibnamefont {Honerkamp}},\ and\ \bibinfo {author} {\bibfnamefont {S.}~\bibnamefont {Andergassen}},\ }\bibfield  {title} {\bibinfo {title} {{Pseudogap opening in the two-dimensional Hubbard model: A functional renormalization group analysis}},\ }\href {https://doi.org/10.1103/PhysRevResearch.2.033068} {\bibfield  {journal} {\bibinfo  {journal} {Phys. Rev. Res.}\ }\textbf {\bibinfo {volume} {2}},\ \bibinfo {pages} {033068} (\bibinfo {year} {2020}{\natexlab{a}})}\BibitemShut {NoStop}%
\bibitem [{\citenamefont {Hille}\ \emph {et~al.}(2020{\natexlab{b}})\citenamefont {Hille}, \citenamefont {Kugler}, \citenamefont {Eckhardt}, \citenamefont {He}, \citenamefont {Kauch}, \citenamefont {Honerkamp}, \citenamefont {Toschi},\ and\ \citenamefont {Andergassen}}]{Hille2020b}%
  \BibitemOpen
  \bibfield  {author} {\bibinfo {author} {\bibfnamefont {C.}~\bibnamefont {Hille}}, \bibinfo {author} {\bibfnamefont {F.~B.}\ \bibnamefont {Kugler}}, \bibinfo {author} {\bibfnamefont {C.~J.}\ \bibnamefont {Eckhardt}}, \bibinfo {author} {\bibfnamefont {Y.-Y.}\ \bibnamefont {He}}, \bibinfo {author} {\bibfnamefont {A.}~\bibnamefont {Kauch}}, \bibinfo {author} {\bibfnamefont {C.}~\bibnamefont {Honerkamp}}, \bibinfo {author} {\bibfnamefont {A.}~\bibnamefont {Toschi}},\ and\ \bibinfo {author} {\bibfnamefont {S.}~\bibnamefont {Andergassen}},\ }\bibfield  {title} {\bibinfo {title} {{Quantitative functional renormalization group description of the two-dimensional Hubbard model}},\ }\href {https://doi.org/10.1103/PhysRevResearch.2.033372} {\bibfield  {journal} {\bibinfo  {journal} {Phys. Rev. Res.}\ }\textbf {\bibinfo {volume} {2}},\ \bibinfo {pages} {033372} (\bibinfo {year} {2020}{\natexlab{b}})}\BibitemShut {NoStop}%
\bibitem [{\citenamefont {Heinzelmann}\ \emph {et~al.}(2023)\citenamefont {Heinzelmann}, \citenamefont {Toschi},\ and\ \citenamefont {Andergassen}}]{Heinzelmann2023}%
  \BibitemOpen
  \bibfield  {author} {\bibinfo {author} {\bibfnamefont {S.}~\bibnamefont {Heinzelmann}}, \bibinfo {author} {\bibfnamefont {A.}~\bibnamefont {Toschi}},\ and\ \bibinfo {author} {\bibfnamefont {S.}~\bibnamefont {Andergassen}},\ }\href {https://arxiv.org/abs/2308.06497} {\bibinfo {title} {{Entangled magnetic, charge, and superconducting pairing correlations in the two-dimensional Hubbard model: a functional renormalization-group analysis}}} (\bibinfo {year} {2023}),\ \Eprint {https://arxiv.org/abs/2308.06497} {arXiv:2308.06497 [cond-mat.str-el]} \BibitemShut {NoStop}%
\bibitem [{\citenamefont {Hille}(2020)}]{HillePhDThesis}%
  \BibitemOpen
  \bibfield  {author} {\bibinfo {author} {\bibfnamefont {C.}~\bibnamefont {Hille}},\ }\emph {\bibinfo {title} {{The role of the self-energy in the functional renormalization group description of interacting Fermi systems}}},\ \href {https://doi.org/10.15496/publikation-46212} {Ph.D. thesis},\ \bibinfo  {school} {Universit\"at T\"ubingen} (\bibinfo {year} {2020})\BibitemShut {NoStop}%
\bibitem [{\citenamefont {Bickers}\ and\ \citenamefont {Scalapino}(1992)}]{Bickers1992}%
  \BibitemOpen
  \bibfield  {author} {\bibinfo {author} {\bibfnamefont {N.~E.}\ \bibnamefont {Bickers}}\ and\ \bibinfo {author} {\bibfnamefont {D.~J.}\ \bibnamefont {Scalapino}},\ }\bibfield  {title} {\bibinfo {title} {{Critical behavior of electronic parquet solutions}},\ }\href {https://doi.org/10.1103/PhysRevB.46.8050} {\bibfield  {journal} {\bibinfo  {journal} {Phys. Rev. B}\ }\textbf {\bibinfo {volume} {46}},\ \bibinfo {pages} {8050} (\bibinfo {year} {1992})}\BibitemShut {NoStop}%
\bibitem [{\citenamefont {Al-Eryani}(2026)}]{AlEryani_2_2025}%
  \BibitemOpen
  \bibfield  {author} {\bibinfo {author} {\bibfnamefont {A.}~\bibnamefont {Al-Eryani}},\ }\bibfield  {title} {\bibinfo {title} {{Diagrammatic bosonization, aspects of criticality, and the Hohenberg--Mermin--Wagner theorem in parquet approaches}},\ }\href {https://doi.org/10.1140/epjb/s10051-026-01134-1} {\bibfield  {journal} {\bibinfo  {journal} {Eur. Phys. J. B}\ }\textbf {\bibinfo {volume} {99}},\ \bibinfo {pages} {19} (\bibinfo {year} {2026})}\BibitemShut {NoStop}%
\bibitem [{Note1()}]{Note1}%
  \BibitemOpen
  \bibinfo {note} {In particular, Ref.~\cite {Delamotte2012} provides pedagogical explanations on the connection between the fRG and the Wilsonian RG.}\BibitemShut {Stop}%
\bibitem [{\citenamefont {Metzner}\ and\ \citenamefont {Vollhardt}(1989)}]{Metzner1989}%
  \BibitemOpen
  \bibfield  {author} {\bibinfo {author} {\bibfnamefont {W.}~\bibnamefont {Metzner}}\ and\ \bibinfo {author} {\bibfnamefont {D.}~\bibnamefont {Vollhardt}},\ }\bibfield  {title} {\bibinfo {title} {{Correlated Lattice Fermions in $d=\ensuremath{\infty}$ Dimensions}},\ }\href {https://doi.org/10.1103/PhysRevLett.62.324} {\bibfield  {journal} {\bibinfo  {journal} {Phys. Rev. Lett.}\ }\textbf {\bibinfo {volume} {62}},\ \bibinfo {pages} {324} (\bibinfo {year} {1989})}\BibitemShut {NoStop}%
\bibitem [{\citenamefont {Georges}\ \emph {et~al.}(1996)\citenamefont {Georges}, \citenamefont {Kotliar}, \citenamefont {Krauth},\ and\ \citenamefont {Rozenberg}}]{Georges1996}%
  \BibitemOpen
  \bibfield  {author} {\bibinfo {author} {\bibfnamefont {A.}~\bibnamefont {Georges}}, \bibinfo {author} {\bibfnamefont {G.}~\bibnamefont {Kotliar}}, \bibinfo {author} {\bibfnamefont {W.}~\bibnamefont {Krauth}},\ and\ \bibinfo {author} {\bibfnamefont {M.~J.}\ \bibnamefont {Rozenberg}},\ }\bibfield  {title} {\bibinfo {title} {{Dynamical mean-field theory of strongly correlated fermion systems and the limit of infinite dimensions}},\ }\href {https://doi.org/10.1103/RevModPhys.68.13} {\bibfield  {journal} {\bibinfo  {journal} {Rev. Mod. Phys.}\ }\textbf {\bibinfo {volume} {68}},\ \bibinfo {pages} {13} (\bibinfo {year} {1996})}\BibitemShut {NoStop}%
\bibitem [{\citenamefont {Taranto}\ \emph {et~al.}(2014)\citenamefont {Taranto}, \citenamefont {Andergassen}, \citenamefont {Bauer}, \citenamefont {Held}, \citenamefont {Katanin}, \citenamefont {Metzner}, \citenamefont {Rohringer},\ and\ \citenamefont {Toschi}}]{Taranto2014}%
  \BibitemOpen
  \bibfield  {author} {\bibinfo {author} {\bibfnamefont {C.}~\bibnamefont {Taranto}}, \bibinfo {author} {\bibfnamefont {S.}~\bibnamefont {Andergassen}}, \bibinfo {author} {\bibfnamefont {J.}~\bibnamefont {Bauer}}, \bibinfo {author} {\bibfnamefont {K.}~\bibnamefont {Held}}, \bibinfo {author} {\bibfnamefont {A.}~\bibnamefont {Katanin}}, \bibinfo {author} {\bibfnamefont {W.}~\bibnamefont {Metzner}}, \bibinfo {author} {\bibfnamefont {G.}~\bibnamefont {Rohringer}},\ and\ \bibinfo {author} {\bibfnamefont {A.}~\bibnamefont {Toschi}},\ }\bibfield  {title} {\bibinfo {title} {{From Infinite to Two Dimensions through the Functional Renormalization Group}},\ }\href {https://doi.org/10.1103/PhysRevLett.112.196402} {\bibfield  {journal} {\bibinfo  {journal} {Phys. Rev. Lett.}\ }\textbf {\bibinfo {volume} {112}},\ \bibinfo {pages} {196402} (\bibinfo {year} {2014})}\BibitemShut {NoStop}%
\bibitem [{\citenamefont {Vilardi}\ \emph {et~al.}(2019)\citenamefont {Vilardi}, \citenamefont {Taranto},\ and\ \citenamefont {Metzner}}]{Vilardi2019}%
  \BibitemOpen
  \bibfield  {author} {\bibinfo {author} {\bibfnamefont {D.}~\bibnamefont {Vilardi}}, \bibinfo {author} {\bibfnamefont {C.}~\bibnamefont {Taranto}},\ and\ \bibinfo {author} {\bibfnamefont {W.}~\bibnamefont {Metzner}},\ }\bibfield  {title} {\bibinfo {title} {{Antiferromagnetic and $d$-wave pairing correlations in the strongly interacting two-dimensional Hubbard model from the functional renormalization group}},\ }\href {https://doi.org/10.1103/PhysRevB.99.104501} {\bibfield  {journal} {\bibinfo  {journal} {Phys. Rev. B}\ }\textbf {\bibinfo {volume} {99}},\ \bibinfo {pages} {104501} (\bibinfo {year} {2019})}\BibitemShut {NoStop}%
\bibitem [{\citenamefont {Bonetti}\ \emph {et~al.}(2022)\citenamefont {Bonetti}, \citenamefont {Toschi}, \citenamefont {Hille}, \citenamefont {Andergassen},\ and\ \citenamefont {Vilardi}}]{Bonetti2022}%
  \BibitemOpen
  \bibfield  {author} {\bibinfo {author} {\bibfnamefont {P.~M.}\ \bibnamefont {Bonetti}}, \bibinfo {author} {\bibfnamefont {A.}~\bibnamefont {Toschi}}, \bibinfo {author} {\bibfnamefont {C.}~\bibnamefont {Hille}}, \bibinfo {author} {\bibfnamefont {S.}~\bibnamefont {Andergassen}},\ and\ \bibinfo {author} {\bibfnamefont {D.}~\bibnamefont {Vilardi}},\ }\bibfield  {title} {\bibinfo {title} {{Single-boson exchange representation of the functional renormalization group for strongly interacting many-electron systems}},\ }\href {https://doi.org/10.1103/PhysRevResearch.4.013034} {\bibfield  {journal} {\bibinfo  {journal} {Phys. Rev. Research}\ }\textbf {\bibinfo {volume} {4}},\ \bibinfo {pages} {013034} (\bibinfo {year} {2022})}\BibitemShut {NoStop}%
\bibitem [{\citenamefont {Si}\ and\ \citenamefont {Smith}(1996)}]{Si1996}%
  \BibitemOpen
  \bibfield  {author} {\bibinfo {author} {\bibfnamefont {Q.}~\bibnamefont {Si}}\ and\ \bibinfo {author} {\bibfnamefont {J.~L.}\ \bibnamefont {Smith}},\ }\bibfield  {title} {\bibinfo {title} {{Kosterlitz-Thouless Transition and Short Range Spatial Correlations in an Extended Hubbard Model}},\ }\href {https://doi.org/10.1103/PhysRevLett.77.3391} {\bibfield  {journal} {\bibinfo  {journal} {Phys. Rev. Lett.}\ }\textbf {\bibinfo {volume} {77}},\ \bibinfo {pages} {3391} (\bibinfo {year} {1996})}\BibitemShut {NoStop}%
\bibitem [{\citenamefont {Smith}\ and\ \citenamefont {Si}(2000)}]{Smith2000}%
  \BibitemOpen
  \bibfield  {author} {\bibinfo {author} {\bibfnamefont {J.~L.}\ \bibnamefont {Smith}}\ and\ \bibinfo {author} {\bibfnamefont {Q.}~\bibnamefont {Si}},\ }\bibfield  {title} {\bibinfo {title} {{Spatial correlations in dynamical mean-field theory}},\ }\href {https://doi.org/10.1103/PhysRevB.61.5184} {\bibfield  {journal} {\bibinfo  {journal} {Phys. Rev. B}\ }\textbf {\bibinfo {volume} {61}},\ \bibinfo {pages} {5184} (\bibinfo {year} {2000})}\BibitemShut {NoStop}%
\bibitem [{\citenamefont {Chitra}\ and\ \citenamefont {Kotliar}(2000)}]{Chitra2000}%
  \BibitemOpen
  \bibfield  {author} {\bibinfo {author} {\bibfnamefont {R.}~\bibnamefont {Chitra}}\ and\ \bibinfo {author} {\bibfnamefont {G.}~\bibnamefont {Kotliar}},\ }\bibfield  {title} {\bibinfo {title} {{Effect of Long Range Coulomb Interactions on the Mott Transition}},\ }\href {https://doi.org/10.1103/PhysRevLett.84.3678} {\bibfield  {journal} {\bibinfo  {journal} {Phys. Rev. Lett.}\ }\textbf {\bibinfo {volume} {84}},\ \bibinfo {pages} {3678} (\bibinfo {year} {2000})}\BibitemShut {NoStop}%
\bibitem [{\citenamefont {Chitra}\ and\ \citenamefont {Kotliar}(2001)}]{Chitra2001}%
  \BibitemOpen
  \bibfield  {author} {\bibinfo {author} {\bibfnamefont {R.}~\bibnamefont {Chitra}}\ and\ \bibinfo {author} {\bibfnamefont {G.}~\bibnamefont {Kotliar}},\ }\bibfield  {title} {\bibinfo {title} {{Effective-action approach to strongly correlated fermion systems}},\ }\href {https://doi.org/10.1103/PhysRevB.63.115110} {\bibfield  {journal} {\bibinfo  {journal} {Phys. Rev. B}\ }\textbf {\bibinfo {volume} {63}},\ \bibinfo {pages} {115110} (\bibinfo {year} {2001})}\BibitemShut {NoStop}%
\bibitem [{\citenamefont {Sun}\ and\ \citenamefont {Kotliar}(2002)}]{Sun2002}%
  \BibitemOpen
  \bibfield  {author} {\bibinfo {author} {\bibfnamefont {P.}~\bibnamefont {Sun}}\ and\ \bibinfo {author} {\bibfnamefont {G.}~\bibnamefont {Kotliar}},\ }\bibfield  {title} {\bibinfo {title} {{Extended dynamical mean-field theory and $\mathrm{GW}$ method}},\ }\href {https://doi.org/10.1103/PhysRevB.66.085120} {\bibfield  {journal} {\bibinfo  {journal} {Phys. Rev. B}\ }\textbf {\bibinfo {volume} {66}},\ \bibinfo {pages} {085120} (\bibinfo {year} {2002})}\BibitemShut {NoStop}%
\bibitem [{\citenamefont {Katanin}(2019)}]{Katanin2019}%
  \BibitemOpen
  \bibfield  {author} {\bibinfo {author} {\bibfnamefont {A.~A.}\ \bibnamefont {Katanin}},\ }\bibfield  {title} {\bibinfo {title} {{Extended dynamical mean field theory combined with the two-particle irreducible functional renormalization-group approach as a tool to study strongly correlated systems}},\ }\href {https://doi.org/10.1103/PhysRevB.99.115112} {\bibfield  {journal} {\bibinfo  {journal} {Phys. Rev. B}\ }\textbf {\bibinfo {volume} {99}},\ \bibinfo {pages} {115112} (\bibinfo {year} {2019})}\BibitemShut {NoStop}%
\bibitem [{\citenamefont {Honerkamp}\ and\ \citenamefont {Salmhofer}(2001)}]{Honerkamp2001}%
  \BibitemOpen
  \bibfield  {author} {\bibinfo {author} {\bibfnamefont {C.}~\bibnamefont {Honerkamp}}\ and\ \bibinfo {author} {\bibfnamefont {M.}~\bibnamefont {Salmhofer}},\ }\bibfield  {title} {\bibinfo {title} {{Temperature-flow renormalization group and the competition between superconductivity and ferromagnetism}},\ }\href {https://doi.org/10.1103/PhysRevB.64.184516} {\bibfield  {journal} {\bibinfo  {journal} {Phys. Rev. B}\ }\textbf {\bibinfo {volume} {64}},\ \bibinfo {pages} {184516} (\bibinfo {year} {2001})}\BibitemShut {NoStop}%
\bibitem [{Note2()}]{Note2}%
  \BibitemOpen
  \bibinfo {note} {We illustrate this point in more detail in App.~\ref {sec:WetterichVertexExpUflow} for generic models with quartic interactions.}\BibitemShut {Stop}%
\bibitem [{\citenamefont {Dupuis}(2014)}]{Dupuis2014}%
  \BibitemOpen
  \bibfield  {author} {\bibinfo {author} {\bibfnamefont {N.}~\bibnamefont {Dupuis}},\ }\bibfield  {title} {\bibinfo {title} {{Nonperturbative renormalization-group approach to fermion systems in the two-particle-irreducible effective action formalism}},\ }\href {https://doi.org/10.1103/PhysRevB.89.035113} {\bibfield  {journal} {\bibinfo  {journal} {Phys. Rev. B}\ }\textbf {\bibinfo {volume} {89}},\ \bibinfo {pages} {035113} (\bibinfo {year} {2014})}\BibitemShut {NoStop}%
\bibitem [{\citenamefont {Polonyi}\ and\ \citenamefont {Sailer}(2002)}]{Polonyi2002}%
  \BibitemOpen
  \bibfield  {author} {\bibinfo {author} {\bibfnamefont {J.}~\bibnamefont {Polonyi}}\ and\ \bibinfo {author} {\bibfnamefont {K.}~\bibnamefont {Sailer}},\ }\bibfield  {title} {\bibinfo {title} {{Effective action and density-functional theory}},\ }\href {https://doi.org/10.1103/PhysRevB.66.155113} {\bibfield  {journal} {\bibinfo  {journal} {Phys. Rev. B}\ }\textbf {\bibinfo {volume} {66}},\ \bibinfo {pages} {155113} (\bibinfo {year} {2002})}\BibitemShut {NoStop}%
\bibitem [{\citenamefont {Schwenk}\ and\ \citenamefont {Polonyi}(2004)}]{Schwenk2004}%
  \BibitemOpen
  \bibfield  {author} {\bibinfo {author} {\bibfnamefont {A.}~\bibnamefont {Schwenk}}\ and\ \bibinfo {author} {\bibfnamefont {J.}~\bibnamefont {Polonyi}},\ }\href {https://arxiv.org/abs/nucl-th/0403011} {\bibinfo {title} {{Towards Density Functional Calculations from Nuclear Forces}}} (\bibinfo {year} {2004}),\ \Eprint {https://arxiv.org/abs/nucl-th/0403011} {arXiv:nucl-th/0403011 [nucl-th]} \BibitemShut {NoStop}%
\bibitem [{\citenamefont {Yokota}\ \emph {et~al.}(2020)\citenamefont {Yokota}, \citenamefont {Kasuya}, \citenamefont {Yoshida},\ and\ \citenamefont {Kunihiro}}]{Yokota2021}%
  \BibitemOpen
  \bibfield  {author} {\bibinfo {author} {\bibfnamefont {T.}~\bibnamefont {Yokota}}, \bibinfo {author} {\bibfnamefont {H.}~\bibnamefont {Kasuya}}, \bibinfo {author} {\bibfnamefont {K.}~\bibnamefont {Yoshida}},\ and\ \bibinfo {author} {\bibfnamefont {T.}~\bibnamefont {Kunihiro}},\ }\bibfield  {title} {\bibinfo {title} {{Microscopic derivation of density functional theory for superfluid systems based on effective action formalism}},\ }\href {https://doi.org/10.1093/ptep/ptaa173} {\bibfield  {journal} {\bibinfo  {journal} {Progress of Theoretical and Experimental Physics}\ }\textbf {\bibinfo {volume} {2021}},\ \bibinfo {pages} {013A03} (\bibinfo {year} {2020})}\BibitemShut {NoStop}%
\bibitem [{\citenamefont {Enss}(2005)}]{EnssPhDthesis}%
  \BibitemOpen
  \bibfield  {author} {\bibinfo {author} {\bibfnamefont {T.}~\bibnamefont {Enss}},\ }\emph {\bibinfo {title} {{Renormalization, Conservation Laws and Transport in Correlated Electron Systems}}},\ \href {https://doi.org/10.18419/opus-6568} {Ph.D. thesis},\ \bibinfo  {school} {Universit\"at Stuttgart} (\bibinfo {year} {2005}),\ \Eprint {https://arxiv.org/abs/cond-mat/0504703} {arXiv:cond-mat/0504703 [cond-mat.str-el]} \BibitemShut {NoStop}%
\bibitem [{\citenamefont {Rentrop}\ \emph {et~al.}(2015)\citenamefont {Rentrop}, \citenamefont {Jakobs},\ and\ \citenamefont {Meden}}]{Rentrop2015}%
  \BibitemOpen
  \bibfield  {author} {\bibinfo {author} {\bibfnamefont {J.~F.}\ \bibnamefont {Rentrop}}, \bibinfo {author} {\bibfnamefont {S.~G.}\ \bibnamefont {Jakobs}},\ and\ \bibinfo {author} {\bibfnamefont {V.}~\bibnamefont {Meden}},\ }\bibfield  {title} {\bibinfo {title} {{Two-particle irreducible functional renormalization group schemes—a comparative study}},\ }\href {https://doi.org/10.1088/1751-8113/48/14/145002} {\bibfield  {journal} {\bibinfo  {journal} {Journal of Physics A: Mathematical and Theoretical}\ }\textbf {\bibinfo {volume} {48}},\ \bibinfo {pages} {145002} (\bibinfo {year} {2015})}\BibitemShut {NoStop}%
\bibitem [{\citenamefont {Rentrop}\ \emph {et~al.}(2016)\citenamefont {Rentrop}, \citenamefont {Meden},\ and\ \citenamefont {Jakobs}}]{Rentrop2016}%
  \BibitemOpen
  \bibfield  {author} {\bibinfo {author} {\bibfnamefont {J.~F.}\ \bibnamefont {Rentrop}}, \bibinfo {author} {\bibfnamefont {V.}~\bibnamefont {Meden}},\ and\ \bibinfo {author} {\bibfnamefont {S.~G.}\ \bibnamefont {Jakobs}},\ }\bibfield  {title} {\bibinfo {title} {{Renormalization group flow of the Luttinger-Ward functional: Conserving approximations and application to the Anderson impurity model}},\ }\href {https://doi.org/10.1103/PhysRevB.93.195160} {\bibfield  {journal} {\bibinfo  {journal} {Phys. Rev. B}\ }\textbf {\bibinfo {volume} {93}},\ \bibinfo {pages} {195160} (\bibinfo {year} {2016})}\BibitemShut {NoStop}%
\bibitem [{\citenamefont {Fraboulet}\ and\ \citenamefont {Ebran}(2024)}]{Fraboulet2024}%
  \BibitemOpen
  \bibfield  {author} {\bibinfo {author} {\bibfnamefont {K.}~\bibnamefont {Fraboulet}}\ and\ \bibinfo {author} {\bibfnamefont {J.-P.}\ \bibnamefont {Ebran}},\ }\bibfield  {title} {\bibinfo {title} {{Addressing energy density functionals in the language of path-integrals II: comparative study of functional renormalization group techniques applied to the (0+0)-D O(N)-symmetric $\varphi^{4}$-theory}},\ }\href {https://doi.org/10.1140/epja/s10050-023-01069-6} {\bibfield  {journal} {\bibinfo  {journal} {Eur. Phys. J. A}\ }\textbf {\bibinfo {volume} {60}},\ \bibinfo {pages} {45} (\bibinfo {year} {2024})}\BibitemShut {NoStop}%
\bibitem [{\citenamefont {Kemler}\ and\ \citenamefont {Braun}(2013)}]{Kemler2013}%
  \BibitemOpen
  \bibfield  {author} {\bibinfo {author} {\bibfnamefont {S.}~\bibnamefont {Kemler}}\ and\ \bibinfo {author} {\bibfnamefont {J.}~\bibnamefont {Braun}},\ }\bibfield  {title} {\bibinfo {title} {{Towards a renormalization group approach to density functional theory—general formalism and case studies}},\ }\href {https://doi.org/10.1088/0954-3899/40/8/085105} {\bibfield  {journal} {\bibinfo  {journal} {Journal of Physics G: Nuclear and Particle Physics}\ }\textbf {\bibinfo {volume} {40}},\ \bibinfo {pages} {085105} (\bibinfo {year} {2013})}\BibitemShut {NoStop}%
\bibitem [{\citenamefont {Kemler}\ \emph {et~al.}(2016)\citenamefont {Kemler}, \citenamefont {Pospiech},\ and\ \citenamefont {Braun}}]{Kemler2016}%
  \BibitemOpen
  \bibfield  {author} {\bibinfo {author} {\bibfnamefont {S.}~\bibnamefont {Kemler}}, \bibinfo {author} {\bibfnamefont {M.}~\bibnamefont {Pospiech}},\ and\ \bibinfo {author} {\bibfnamefont {J.}~\bibnamefont {Braun}},\ }\bibfield  {title} {\bibinfo {title} {{Formation of selfbound states in a one-dimensional nuclear model—a renormalization group based density functional study}},\ }\href {https://doi.org/10.1088/0954-3899/44/1/015101} {\bibfield  {journal} {\bibinfo  {journal} {Journal of Physics G: Nuclear and Particle Physics}\ }\textbf {\bibinfo {volume} {44}},\ \bibinfo {pages} {015101} (\bibinfo {year} {2016})}\BibitemShut {NoStop}%
\bibitem [{\citenamefont {Liang}\ \emph {et~al.}(2018)\citenamefont {Liang}, \citenamefont {Niu},\ and\ \citenamefont {Hatsuda}}]{Liang2018}%
  \BibitemOpen
  \bibfield  {author} {\bibinfo {author} {\bibfnamefont {H.}~\bibnamefont {Liang}}, \bibinfo {author} {\bibfnamefont {Y.}~\bibnamefont {Niu}},\ and\ \bibinfo {author} {\bibfnamefont {T.}~\bibnamefont {Hatsuda}},\ }\bibfield  {title} {\bibinfo {title} {{Functional renormalization group and Kohn–Sham scheme in density functional theory}},\ }\href {https://doi.org/https://doi.org/10.1016/j.physletb.2018.02.034} {\bibfield  {journal} {\bibinfo  {journal} {Physics Letters B}\ }\textbf {\bibinfo {volume} {779}},\ \bibinfo {pages} {436} (\bibinfo {year} {2018})}\BibitemShut {NoStop}%
\bibitem [{\citenamefont {Yokota}\ \emph {et~al.}(2019{\natexlab{a}})\citenamefont {Yokota}, \citenamefont {Yoshida},\ and\ \citenamefont {Kunihiro}}]{Yokota2019}%
  \BibitemOpen
  \bibfield  {author} {\bibinfo {author} {\bibfnamefont {T.}~\bibnamefont {Yokota}}, \bibinfo {author} {\bibfnamefont {K.}~\bibnamefont {Yoshida}},\ and\ \bibinfo {author} {\bibfnamefont {T.}~\bibnamefont {Kunihiro}},\ }\bibfield  {title} {\bibinfo {title} {{Functional renormalization-group calculation of the equation of state of one-dimensional uniform matter inspired by the Hohenberg-Kohn theorem}},\ }\href {https://doi.org/10.1103/PhysRevC.99.024302} {\bibfield  {journal} {\bibinfo  {journal} {Phys. Rev. C}\ }\textbf {\bibinfo {volume} {99}},\ \bibinfo {pages} {024302} (\bibinfo {year} {2019}{\natexlab{a}})}\BibitemShut {NoStop}%
\bibitem [{\citenamefont {Yokota}\ \emph {et~al.}(2019{\natexlab{b}})\citenamefont {Yokota}, \citenamefont {Yoshida},\ and\ \citenamefont {Kunihiro}}]{Yokota2019_2}%
  \BibitemOpen
  \bibfield  {author} {\bibinfo {author} {\bibfnamefont {T.}~\bibnamefont {Yokota}}, \bibinfo {author} {\bibfnamefont {K.}~\bibnamefont {Yoshida}},\ and\ \bibinfo {author} {\bibfnamefont {T.}~\bibnamefont {Kunihiro}},\ }\bibfield  {title} {\bibinfo {title} {{Ab~initio description of excited states of 1D uniform matter with the Hohenberg–Kohn-theorem-inspired functional-renormalization-group method}},\ }\href {https://doi.org/10.1093/ptep/pty139} {\bibfield  {journal} {\bibinfo  {journal} {Progress of Theoretical and Experimental Physics}\ }\textbf {\bibinfo {volume} {2019}},\ \bibinfo {pages} {011D01} (\bibinfo {year} {2019}{\natexlab{b}})}\BibitemShut {NoStop}%
\bibitem [{\citenamefont {Yokota}\ and\ \citenamefont {Naito}(2019)}]{Yokota2019_3}%
  \BibitemOpen
  \bibfield  {author} {\bibinfo {author} {\bibfnamefont {T.}~\bibnamefont {Yokota}}\ and\ \bibinfo {author} {\bibfnamefont {T.}~\bibnamefont {Naito}},\ }\bibfield  {title} {\bibinfo {title} {{Functional-renormalization-group aided density functional analysis for the correlation energy of the two-dimensional homogeneous electron gas}},\ }\href {https://doi.org/10.1103/PhysRevB.99.115106} {\bibfield  {journal} {\bibinfo  {journal} {Phys. Rev. B}\ }\textbf {\bibinfo {volume} {99}},\ \bibinfo {pages} {115106} (\bibinfo {year} {2019})}\BibitemShut {NoStop}%
\bibitem [{\citenamefont {Yokota}\ and\ \citenamefont {Naito}(2021)}]{Yokota2021_2}%
  \BibitemOpen
  \bibfield  {author} {\bibinfo {author} {\bibfnamefont {T.}~\bibnamefont {Yokota}}\ and\ \bibinfo {author} {\bibfnamefont {T.}~\bibnamefont {Naito}},\ }\bibfield  {title} {\bibinfo {title} {{Ab initio construction of the energy density functional for electron systems with the functional-renormalization-group-aided density functional theory}},\ }\href {https://doi.org/10.1103/PhysRevResearch.3.L012015} {\bibfield  {journal} {\bibinfo  {journal} {Phys. Rev. Res.}\ }\textbf {\bibinfo {volume} {3}},\ \bibinfo {pages} {L012015} (\bibinfo {year} {2021})}\BibitemShut {NoStop}%
\bibitem [{\citenamefont {Yokota}\ \emph {et~al.}(2021)\citenamefont {Yokota}, \citenamefont {Haruyama},\ and\ \citenamefont {Sugino}}]{Yokota2021_3}%
  \BibitemOpen
  \bibfield  {author} {\bibinfo {author} {\bibfnamefont {T.}~\bibnamefont {Yokota}}, \bibinfo {author} {\bibfnamefont {J.}~\bibnamefont {Haruyama}},\ and\ \bibinfo {author} {\bibfnamefont {O.}~\bibnamefont {Sugino}},\ }\bibfield  {title} {\bibinfo {title} {{Functional-renormalization-group approach to classical liquids with short-range repulsion: A scheme without repulsive reference system}},\ }\href {https://doi.org/10.1103/PhysRevE.104.014124} {\bibfield  {journal} {\bibinfo  {journal} {Phys. Rev. E}\ }\textbf {\bibinfo {volume} {104}},\ \bibinfo {pages} {014124} (\bibinfo {year} {2021})}\BibitemShut {NoStop}%
\bibitem [{\citenamefont {Yokota}\ \emph {et~al.}(2025)\citenamefont {Yokota}, \citenamefont {Haruyama},\ and\ \citenamefont {Sugino}}]{Yokota2025}%
  \BibitemOpen
  \bibfield  {author} {\bibinfo {author} {\bibfnamefont {T.}~\bibnamefont {Yokota}}, \bibinfo {author} {\bibfnamefont {J.}~\bibnamefont {Haruyama}},\ and\ \bibinfo {author} {\bibfnamefont {O.}~\bibnamefont {Sugino}},\ }\href {https://arxiv.org/abs/2510.16710} {\bibinfo {title} {{Functional renormalization group for classical liquids without recourse to hard-core reference systems: A study of three-dimensional Lennard-Jones liquids}}} (\bibinfo {year} {2025}),\ \Eprint {https://arxiv.org/abs/2510.16710} {arXiv:2510.16710 [cond-mat.soft]} \BibitemShut {NoStop}%
\bibitem [{\citenamefont {Krien}\ \emph {et~al.}(2019)\citenamefont {Krien}, \citenamefont {Valli},\ and\ \citenamefont {Capone}}]{Krien2019}%
  \BibitemOpen
  \bibfield  {author} {\bibinfo {author} {\bibfnamefont {F.}~\bibnamefont {Krien}}, \bibinfo {author} {\bibfnamefont {A.}~\bibnamefont {Valli}},\ and\ \bibinfo {author} {\bibfnamefont {M.}~\bibnamefont {Capone}},\ }\bibfield  {title} {\bibinfo {title} {{Single-boson exchange decomposition of the vertex function}},\ }\href {https://doi.org/10.1103/PhysRevB.100.155149} {\bibfield  {journal} {\bibinfo  {journal} {Phys. Rev. B}\ }\textbf {\bibinfo {volume} {100}},\ \bibinfo {pages} {155149} (\bibinfo {year} {2019})}\BibitemShut {NoStop}%
\bibitem [{\citenamefont {Fraboulet}\ \emph {et~al.}(2022)\citenamefont {Fraboulet}, \citenamefont {Heinzelmann}, \citenamefont {Bonetti}, \citenamefont {Al-Eryani}, \citenamefont {Vilardi}, \citenamefont {Toschi},\ and\ \citenamefont {Andergassen}}]{Fraboulet2022}%
  \BibitemOpen
  \bibfield  {author} {\bibinfo {author} {\bibfnamefont {K.}~\bibnamefont {Fraboulet}}, \bibinfo {author} {\bibfnamefont {S.}~\bibnamefont {Heinzelmann}}, \bibinfo {author} {\bibfnamefont {P.~M.}\ \bibnamefont {Bonetti}}, \bibinfo {author} {\bibfnamefont {A.}~\bibnamefont {Al-Eryani}}, \bibinfo {author} {\bibfnamefont {D.}~\bibnamefont {Vilardi}}, \bibinfo {author} {\bibfnamefont {A.}~\bibnamefont {Toschi}},\ and\ \bibinfo {author} {\bibfnamefont {S.}~\bibnamefont {Andergassen}},\ }\bibfield  {title} {\bibinfo {title} {{Single-boson exchange functional renormalization group application to the two-dimensional Hubbard model at weak coupling}},\ }\href {https://doi.org/10.1140/epjb/s10051-022-00438-2} {\bibfield  {journal} {\bibinfo  {journal} {Eur. Phys. J. B}\ }\textbf {\bibinfo {volume} {95}},\ \bibinfo {pages} {202} (\bibinfo {year} {2022})}\BibitemShut {NoStop}%
\bibitem [{\citenamefont {Fraboulet}\ \emph {et~al.}(2025)\citenamefont {Fraboulet}, \citenamefont {Al-Eryani}, \citenamefont {Heinzelmann}, \citenamefont {Kauch},\ and\ \citenamefont {Andergassen}}]{Fraboulet2025}%
  \BibitemOpen
  \bibfield  {author} {\bibinfo {author} {\bibfnamefont {K.}~\bibnamefont {Fraboulet}}, \bibinfo {author} {\bibfnamefont {A.}~\bibnamefont {Al-Eryani}}, \bibinfo {author} {\bibfnamefont {S.}~\bibnamefont {Heinzelmann}}, \bibinfo {author} {\bibfnamefont {A.}~\bibnamefont {Kauch}},\ and\ \bibinfo {author} {\bibfnamefont {S.}~\bibnamefont {Andergassen}},\ }\href {https://arxiv.org/abs/2512.11190} {\bibinfo {title} {{Multiloop functional renormalization group from single bosons}}} (\bibinfo {year} {2025}),\ \Eprint {https://arxiv.org/abs/2512.11190} {arXiv:2512.11190 [cond-mat.str-el]} \BibitemShut {NoStop}%
\bibitem [{\citenamefont {Gievers}\ \emph {et~al.}(2022)\citenamefont {Gievers}, \citenamefont {Walter}, \citenamefont {Ge}, \citenamefont {von Delft},\ and\ \citenamefont {Kugler}}]{Gievers2022}%
  \BibitemOpen
  \bibfield  {author} {\bibinfo {author} {\bibfnamefont {M.}~\bibnamefont {Gievers}}, \bibinfo {author} {\bibfnamefont {E.}~\bibnamefont {Walter}}, \bibinfo {author} {\bibfnamefont {A.}~\bibnamefont {Ge}}, \bibinfo {author} {\bibfnamefont {J.}~\bibnamefont {von Delft}},\ and\ \bibinfo {author} {\bibfnamefont {F.~B.}\ \bibnamefont {Kugler}},\ }\bibfield  {title} {\bibinfo {title} {{Multiloop flow equations for single-boson exchange fRG}},\ }\href {https://doi.org/10.1140/epjb/s10051-022-00353-6} {\bibfield  {journal} {\bibinfo  {journal} {Eur. Phys. J. B}\ }\textbf {\bibinfo {volume} {95}},\ \bibinfo {pages} {108} (\bibinfo {year} {2022})}\BibitemShut {NoStop}%
\bibitem [{\citenamefont {Anderson}(1961)}]{Anderson1961}%
  \BibitemOpen
  \bibfield  {author} {\bibinfo {author} {\bibfnamefont {P.~W.}\ \bibnamefont {Anderson}},\ }\bibfield  {title} {\bibinfo {title} {{Localized Magnetic States in Metals}},\ }\href {https://doi.org/10.1103/PhysRev.124.41} {\bibfield  {journal} {\bibinfo  {journal} {Phys. Rev.}\ }\textbf {\bibinfo {volume} {124}},\ \bibinfo {pages} {41} (\bibinfo {year} {1961})}\BibitemShut {NoStop}%
\bibitem [{\citenamefont {{E. Lagos}}(2001)}]{ELagos2001}%
  \BibitemOpen
  \bibfield  {author} {\bibinfo {author} {\bibfnamefont {R.}~\bibnamefont {{E. Lagos}}},\ }\bibfield  {title} {\bibinfo {title} {{The Holstein–Anderson impurity model}},\ }\href {https://doi.org/https://doi.org/10.1016/S0304-8853(01)00106-8} {\bibfield  {journal} {\bibinfo  {journal} {Journal of Magnetism and Magnetic Materials}\ }\textbf {\bibinfo {volume} {226-230}},\ \bibinfo {pages} {103} (\bibinfo {year} {2001})},\ \bibinfo {note} {proceedings of the International Conference on Magnetism (ICM 2000)}\BibitemShut {NoStop}%
\bibitem [{\citenamefont {Hewson}\ and\ \citenamefont {Meyer}(2001)}]{Hewson2002}%
  \BibitemOpen
  \bibfield  {author} {\bibinfo {author} {\bibfnamefont {A.~C.}\ \bibnamefont {Hewson}}\ and\ \bibinfo {author} {\bibfnamefont {D.}~\bibnamefont {Meyer}},\ }\bibfield  {title} {\bibinfo {title} {{Numerical renormalization group study of the Anderson-Holstein impurity model}},\ }\href {https://doi.org/10.1088/0953-8984/14/3/312} {\bibfield  {journal} {\bibinfo  {journal} {Journal of Physics: Condensed Matter}\ }\textbf {\bibinfo {volume} {14}},\ \bibinfo {pages} {427} (\bibinfo {year} {2001})}\BibitemShut {NoStop}%
\bibitem [{\citenamefont {Bickers}(2004)}]{BickersSelfConsistent2004}%
  \BibitemOpen
  \bibfield  {author} {\bibinfo {author} {\bibfnamefont {N.~E.}\ \bibnamefont {Bickers}},\ }\bibinfo {title} {{Self-Consistent Many-Body Theory for Condensed Matter Systems}},\ in\ \href {https://doi.org/10.1007/0-387-21717-7_6} {\emph {\bibinfo {booktitle} {{Theoretical Methods for Strongly Correlated Electrons}}}},\ \bibinfo {editor} {edited by\ \bibinfo {editor} {\bibfnamefont {D.}~\bibnamefont {S{\'e}n{\'e}chal}}, \bibinfo {editor} {\bibfnamefont {A.-M.}\ \bibnamefont {Tremblay}},\ and\ \bibinfo {editor} {\bibfnamefont {C.}~\bibnamefont {Bourbonnais}}}\ (\bibinfo  {publisher} {Springer New York},\ \bibinfo {address} {New York, NY},\ \bibinfo {year} {2004})\ pp.\ \bibinfo {pages} {237--296}\BibitemShut {NoStop}%
\bibitem [{\citenamefont {Rohringer}\ \emph {et~al.}(2018)\citenamefont {Rohringer}, \citenamefont {Hafermann}, \citenamefont {Toschi}, \citenamefont {Katanin}, \citenamefont {Antipov}, \citenamefont {Katsnelson}, \citenamefont {Lichtenstein}, \citenamefont {Rubtsov},\ and\ \citenamefont {Held}}]{Rohringer2018}%
  \BibitemOpen
  \bibfield  {author} {\bibinfo {author} {\bibfnamefont {G.}~\bibnamefont {Rohringer}}, \bibinfo {author} {\bibfnamefont {H.}~\bibnamefont {Hafermann}}, \bibinfo {author} {\bibfnamefont {A.}~\bibnamefont {Toschi}}, \bibinfo {author} {\bibfnamefont {A.~A.}\ \bibnamefont {Katanin}}, \bibinfo {author} {\bibfnamefont {A.~E.}\ \bibnamefont {Antipov}}, \bibinfo {author} {\bibfnamefont {M.~I.}\ \bibnamefont {Katsnelson}}, \bibinfo {author} {\bibfnamefont {A.~I.}\ \bibnamefont {Lichtenstein}}, \bibinfo {author} {\bibfnamefont {A.~N.}\ \bibnamefont {Rubtsov}},\ and\ \bibinfo {author} {\bibfnamefont {K.}~\bibnamefont {Held}},\ }\bibfield  {title} {\bibinfo {title} {{Diagrammatic routes to nonlocal correlations beyond dynamical mean field theory}},\ }\href {https://doi.org/10.1103/RevModPhys.90.025003} {\bibfield  {journal} {\bibinfo  {journal} {Rev. Mod. Phys.}\ }\textbf {\bibinfo {volume} {90}},\ \bibinfo {pages} {025003} (\bibinfo {year} {2018})}\BibitemShut {NoStop}%
\bibitem [{\citenamefont {Bickers}\ and\ \citenamefont {White}(1991)}]{Bickers1991}%
  \BibitemOpen
  \bibfield  {author} {\bibinfo {author} {\bibfnamefont {N.~E.}\ \bibnamefont {Bickers}}\ and\ \bibinfo {author} {\bibfnamefont {S.~R.}\ \bibnamefont {White}},\ }\bibfield  {title} {\bibinfo {title} {{Conserving approximations for strongly fluctuating electron systems. II. Numerical results and parquet extension}},\ }\href {https://doi.org/10.1103/PhysRevB.43.8044} {\bibfield  {journal} {\bibinfo  {journal} {Phys. Rev. B}\ }\textbf {\bibinfo {volume} {43}},\ \bibinfo {pages} {8044} (\bibinfo {year} {1991})}\BibitemShut {NoStop}%
\bibitem [{\citenamefont {Gievers}(2025)}]{GieversPhDthesis}%
  \BibitemOpen
  \bibfield  {author} {\bibinfo {author} {\bibfnamefont {M.}~\bibnamefont {Gievers}},\ }\emph {\bibinfo {title} {{Functional approaches to Fermi polarons in cold atomic gases and solid-state systems}}},\ \href {http://nbn-resolving.de/urn:nbn:de:bvb:19-349955} {Ph.D. thesis},\ \bibinfo  {school} {Ludwig-Maximilians-Universit{\"a}t M{\"u}nchen} (\bibinfo {year} {2025})\BibitemShut {NoStop}%
\bibitem [{Note3()}]{Note3}%
  \BibitemOpen
  \bibinfo {note} {We refer to App. C.2 of Ref.~\cite {GieversPhDthesis} for more details on the construction of the identity vertices $\protect \mathbf {1}_r$ and $\protect \mathds {1}_r$.}\BibitemShut {Stop}%
\bibitem [{\citenamefont {Al-Eryani}\ \emph {et~al.}(2024)\citenamefont {Al-Eryani}, \citenamefont {Heinzelmann}, \citenamefont {Fraboulet}, \citenamefont {Krien},\ and\ \citenamefont {Andergassen}}]{AlEryani2024}%
  \BibitemOpen
  \bibfield  {author} {\bibinfo {author} {\bibfnamefont {A.}~\bibnamefont {Al-Eryani}}, \bibinfo {author} {\bibfnamefont {S.}~\bibnamefont {Heinzelmann}}, \bibinfo {author} {\bibfnamefont {K.}~\bibnamefont {Fraboulet}}, \bibinfo {author} {\bibfnamefont {F.}~\bibnamefont {Krien}},\ and\ \bibinfo {author} {\bibfnamefont {S.}~\bibnamefont {Andergassen}},\ }\href {https://arxiv.org/abs/2412.07323} {\bibinfo {title} {{Screening and effective RPA-like charge susceptibility in the extended Hubbard model}}} (\bibinfo {year} {2024}),\ \Eprint {https://arxiv.org/abs/2412.07323} {arXiv:2412.07323 [cond-mat.str-el]} \BibitemShut {NoStop}%
\bibitem [{\citenamefont {Al-Eryani}\ \emph {et~al.}(2025)\citenamefont {Al-Eryani}, \citenamefont {Andergassen},\ and\ \citenamefont {Scherer}}]{AlEryani2025}%
  \BibitemOpen
  \bibfield  {author} {\bibinfo {author} {\bibfnamefont {A.}~\bibnamefont {Al-Eryani}}, \bibinfo {author} {\bibfnamefont {S.}~\bibnamefont {Andergassen}},\ and\ \bibinfo {author} {\bibfnamefont {M.~M.}\ \bibnamefont {Scherer}},\ }\bibfield  {title} {\bibinfo {title} {{Intertwined fluctuations and isotope effects in the Hubbard-Holstein model on the square lattice from functional renormalization}},\ }\href {https://doi.org/10.1103/6vyb-h379} {\bibfield  {journal} {\bibinfo  {journal} {Phys. Rev. Res.}\ }\textbf {\bibinfo {volume} {7}},\ \bibinfo {pages} {043052} (\bibinfo {year} {2025})}\BibitemShut {NoStop}%
\bibitem [{Note4()}]{Note4}%
  \BibitemOpen
  \bibinfo {note} {The subtraction of $\protect \mathcal {F}_r$ from $\nabla _r^{(\protect \mathcal {F})}$ is needed because $M_r^{(\protect \mathcal {B})}$ only exhibits two-particle-reducible diagrams.}\BibitemShut {Stop}%
\bibitem [{Note5()}]{Note5}%
  \BibitemOpen
  \bibinfo {note} {The derivation of the original SBE equations~\protect \eqref {eq:selfconsistentSBEequations}, as outlined in Ref.~\cite {Gievers2022}, can then be directly recovered from our developments of Sec.~\ref {sec:B-reducibility} by setting $\protect \mathcal {F}_r=0$ (and therefore $\protect \mathcal {B}_r=U$)}\BibitemShut {NoStop}%
\bibitem [{\citenamefont {Zhang}\ and\ \citenamefont {Callaway}(1989)}]{Zhang1989}%
  \BibitemOpen
  \bibfield  {author} {\bibinfo {author} {\bibfnamefont {Y.}~\bibnamefont {Zhang}}\ and\ \bibinfo {author} {\bibfnamefont {J.}~\bibnamefont {Callaway}},\ }\bibfield  {title} {\bibinfo {title} {{Extended Hubbard model in two dimensions}},\ }\href {https://doi.org/10.1103/PhysRevB.39.9397} {\bibfield  {journal} {\bibinfo  {journal} {Phys. Rev. B}\ }\textbf {\bibinfo {volume} {39}},\ \bibinfo {pages} {9397} (\bibinfo {year} {1989})}\BibitemShut {NoStop}%
\bibitem [{\citenamefont {Terletska}\ \emph {et~al.}(2017)\citenamefont {Terletska}, \citenamefont {Chen},\ and\ \citenamefont {Gull}}]{Terletska2017}%
  \BibitemOpen
  \bibfield  {author} {\bibinfo {author} {\bibfnamefont {H.}~\bibnamefont {Terletska}}, \bibinfo {author} {\bibfnamefont {T.}~\bibnamefont {Chen}},\ and\ \bibinfo {author} {\bibfnamefont {E.}~\bibnamefont {Gull}},\ }\bibfield  {title} {\bibinfo {title} {{Charge ordering and correlation effects in the extended Hubbard model}},\ }\href {https://doi.org/10.1103/PhysRevB.95.115149} {\bibfield  {journal} {\bibinfo  {journal} {Phys. Rev. B}\ }\textbf {\bibinfo {volume} {95}},\ \bibinfo {pages} {115149} (\bibinfo {year} {2017})}\BibitemShut {NoStop}%
\bibitem [{\citenamefont {Terletska}\ \emph {et~al.}(2018)\citenamefont {Terletska}, \citenamefont {Chen}, \citenamefont {Paki},\ and\ \citenamefont {Gull}}]{Terletska2018}%
  \BibitemOpen
  \bibfield  {author} {\bibinfo {author} {\bibfnamefont {H.}~\bibnamefont {Terletska}}, \bibinfo {author} {\bibfnamefont {T.}~\bibnamefont {Chen}}, \bibinfo {author} {\bibfnamefont {J.}~\bibnamefont {Paki}},\ and\ \bibinfo {author} {\bibfnamefont {E.}~\bibnamefont {Gull}},\ }\bibfield  {title} {\bibinfo {title} {{Charge ordering and nonlocal correlations in the doped extended Hubbard model}},\ }\href {https://doi.org/10.1103/PhysRevB.97.115117} {\bibfield  {journal} {\bibinfo  {journal} {Phys. Rev. B}\ }\textbf {\bibinfo {volume} {97}},\ \bibinfo {pages} {115117} (\bibinfo {year} {2018})}\BibitemShut {NoStop}%
\bibitem [{\citenamefont {Paki}\ \emph {et~al.}(2019)\citenamefont {Paki}, \citenamefont {Terletska}, \citenamefont {Iskakov},\ and\ \citenamefont {Gull}}]{Paki2019}%
  \BibitemOpen
  \bibfield  {author} {\bibinfo {author} {\bibfnamefont {J.}~\bibnamefont {Paki}}, \bibinfo {author} {\bibfnamefont {H.}~\bibnamefont {Terletska}}, \bibinfo {author} {\bibfnamefont {S.}~\bibnamefont {Iskakov}},\ and\ \bibinfo {author} {\bibfnamefont {E.}~\bibnamefont {Gull}},\ }\bibfield  {title} {\bibinfo {title} {{Charge order and antiferromagnetism in the extended Hubbard model}},\ }\href {https://doi.org/10.1103/PhysRevB.99.245146} {\bibfield  {journal} {\bibinfo  {journal} {Phys. Rev. B}\ }\textbf {\bibinfo {volume} {99}},\ \bibinfo {pages} {245146} (\bibinfo {year} {2019})}\BibitemShut {NoStop}%
\bibitem [{\citenamefont {Holstein}(1959)}]{Holstein1959}%
  \BibitemOpen
  \bibfield  {author} {\bibinfo {author} {\bibfnamefont {T.}~\bibnamefont {Holstein}},\ }\bibfield  {title} {\bibinfo {title} {{Studies of polaron motion: Part I. The molecular-crystal model}},\ }\href {https://doi.org/https://doi.org/10.1016/0003-4916(59)90002-8} {\bibfield  {journal} {\bibinfo  {journal} {Annals of Physics}\ }\textbf {\bibinfo {volume} {8}},\ \bibinfo {pages} {325} (\bibinfo {year} {1959})}\BibitemShut {NoStop}%
\bibitem [{\citenamefont {Freericks}\ and\ \citenamefont {Jarrell}(1995)}]{Freericks1995}%
  \BibitemOpen
  \bibfield  {author} {\bibinfo {author} {\bibfnamefont {J.~K.}\ \bibnamefont {Freericks}}\ and\ \bibinfo {author} {\bibfnamefont {M.}~\bibnamefont {Jarrell}},\ }\bibfield  {title} {\bibinfo {title} {{Competition between Electron-Phonon Attraction and Weak Coulomb Repulsion}},\ }\href {https://doi.org/10.1103/PhysRevLett.75.2570} {\bibfield  {journal} {\bibinfo  {journal} {Phys. Rev. Lett.}\ }\textbf {\bibinfo {volume} {75}},\ \bibinfo {pages} {2570} (\bibinfo {year} {1995})}\BibitemShut {NoStop}%
\bibitem [{\citenamefont {Capone}\ \emph {et~al.}(2004)\citenamefont {Capone}, \citenamefont {Sangiovanni}, \citenamefont {Castellani}, \citenamefont {Di~Castro},\ and\ \citenamefont {Grilli}}]{Capone2004}%
  \BibitemOpen
  \bibfield  {author} {\bibinfo {author} {\bibfnamefont {M.}~\bibnamefont {Capone}}, \bibinfo {author} {\bibfnamefont {G.}~\bibnamefont {Sangiovanni}}, \bibinfo {author} {\bibfnamefont {C.}~\bibnamefont {Castellani}}, \bibinfo {author} {\bibfnamefont {C.}~\bibnamefont {Di~Castro}},\ and\ \bibinfo {author} {\bibfnamefont {M.}~\bibnamefont {Grilli}},\ }\bibfield  {title} {\bibinfo {title} {{Phase Separation Close to the Density-Driven Mott Transition in the Hubbard-Holstein Model}},\ }\href {https://doi.org/10.1103/PhysRevLett.92.106401} {\bibfield  {journal} {\bibinfo  {journal} {Phys. Rev. Lett.}\ }\textbf {\bibinfo {volume} {92}},\ \bibinfo {pages} {106401} (\bibinfo {year} {2004})}\BibitemShut {NoStop}%
\bibitem [{Note6()}]{Note6}%
  \BibitemOpen
  \bibinfo {note} {The regulators (or cutoff functions) are typically introduced as $G_0^\Lambda =\Theta ^\Lambda G_0$ or $\left (G_0^\Lambda \right )^{-1}=G_0^{-1}+R^\Lambda $. With these definitions, $\Theta ^\Lambda $ and $R^\Lambda $ are referred to as multiplicative and additive regulators, respectively. The regulators used in the present study are defined in Sec.~III.}\BibitemShut {Stop}%
\bibitem [{Note7()}]{Note7}%
  \BibitemOpen
  \bibinfo {note} {In particular, Eq.~\protect \eqref {eq:DifferentiateBSE3} is identical to Eq.~(10) of Ref.~\cite {Kugler2018c}.}\BibitemShut {Stop}%
\bibitem [{\citenamefont {Krien}\ \emph {et~al.}(2021)\citenamefont {Krien}, \citenamefont {Kauch},\ and\ \citenamefont {Held}}]{Krien2021}%
  \BibitemOpen
  \bibfield  {author} {\bibinfo {author} {\bibfnamefont {F.}~\bibnamefont {Krien}}, \bibinfo {author} {\bibfnamefont {A.}~\bibnamefont {Kauch}},\ and\ \bibinfo {author} {\bibfnamefont {K.}~\bibnamefont {Held}},\ }\bibfield  {title} {\bibinfo {title} {{Tiling with triangles: parquet and $GW\ensuremath{\gamma}$ methods unified}},\ }\href {https://doi.org/10.1103/PhysRevResearch.3.013149} {\bibfield  {journal} {\bibinfo  {journal} {Phys. Rev. Res.}\ }\textbf {\bibinfo {volume} {3}},\ \bibinfo {pages} {013149} (\bibinfo {year} {2021})}\BibitemShut {NoStop}%
\bibitem [{\citenamefont {Katanin}(2004)}]{Katanin2004}%
  \BibitemOpen
  \bibfield  {author} {\bibinfo {author} {\bibfnamefont {A.~A.}\ \bibnamefont {Katanin}},\ }\bibfield  {title} {\bibinfo {title} {{Fulfillment of Ward identities in the functional renormalization group approach}},\ }\href {https://doi.org/10.1103/PhysRevB.70.115109} {\bibfield  {journal} {\bibinfo  {journal} {Phys. Rev. B}\ }\textbf {\bibinfo {volume} {70}},\ \bibinfo {pages} {115109} (\bibinfo {year} {2004})}\BibitemShut {NoStop}%
\bibitem [{\citenamefont {Floerchinger}\ and\ \citenamefont {Wetterich}(2009)}]{Floerchinger2009Exact}%
  \BibitemOpen
  \bibfield  {author} {\bibinfo {author} {\bibfnamefont {S.}~\bibnamefont {Floerchinger}}\ and\ \bibinfo {author} {\bibfnamefont {C.}~\bibnamefont {Wetterich}},\ }\bibfield  {title} {\bibinfo {title} {Exact flow equation for composite operators},\ }\href {https://doi.org/https://doi.org/10.1016/j.physletb.2009.09.014} {\bibfield  {journal} {\bibinfo  {journal} {Physics Letters B}\ }\textbf {\bibinfo {volume} {680}},\ \bibinfo {pages} {371} (\bibinfo {year} {2009})}\BibitemShut {NoStop}%
\bibitem [{\citenamefont {Friederich}\ \emph {et~al.}(2010)\citenamefont {Friederich}, \citenamefont {Krahl},\ and\ \citenamefont {Wetterich}}]{Friederich2010Fourpoint}%
  \BibitemOpen
  \bibfield  {author} {\bibinfo {author} {\bibfnamefont {S.}~\bibnamefont {Friederich}}, \bibinfo {author} {\bibfnamefont {H.~C.}\ \bibnamefont {Krahl}},\ and\ \bibinfo {author} {\bibfnamefont {C.}~\bibnamefont {Wetterich}},\ }\bibfield  {title} {\bibinfo {title} {Four-point vertex in the {H}ubbard model and partial bosonization},\ }\href {https://doi.org/10.1103/PhysRevB.81.235108} {\bibfield  {journal} {\bibinfo  {journal} {Phys. Rev. B}\ }\textbf {\bibinfo {volume} {81}},\ \bibinfo {pages} {235108} (\bibinfo {year} {2010})}\BibitemShut {NoStop}%
\bibitem [{\citenamefont {Homenda}\ \emph {et~al.}(2024)\citenamefont {Homenda}, \citenamefont {Jakubczyk},\ and\ \citenamefont {Yamase}}]{Homenda2024Generalized}%
  \BibitemOpen
  \bibfield  {author} {\bibinfo {author} {\bibfnamefont {M.}~\bibnamefont {Homenda}}, \bibinfo {author} {\bibfnamefont {P.}~\bibnamefont {Jakubczyk}},\ and\ \bibinfo {author} {\bibfnamefont {H.}~\bibnamefont {Yamase}},\ }\bibfield  {title} {\bibinfo {title} {Generalized {H}ertz action and quantum criticality of two-dimensional {F}ermi systems},\ }\href {https://doi.org/10.1103/PhysRevB.110.L121102} {\bibfield  {journal} {\bibinfo  {journal} {Phys. Rev. B}\ }\textbf {\bibinfo {volume} {110}},\ \bibinfo {pages} {L121102} (\bibinfo {year} {2024})}\BibitemShut {NoStop}%
\bibitem [{\citenamefont {Husemann}\ and\ \citenamefont {Salmhofer}(2009)}]{Husemann2009}%
  \BibitemOpen
  \bibfield  {author} {\bibinfo {author} {\bibfnamefont {C.}~\bibnamefont {Husemann}}\ and\ \bibinfo {author} {\bibfnamefont {M.}~\bibnamefont {Salmhofer}},\ }\bibfield  {title} {\bibinfo {title} {{Efficient parametrization of the vertex function, $\ensuremath{\Omega}$ scheme, and the $t,{t}^{\ensuremath{'}}$ Hubbard model at van Hove filling}},\ }\href {https://doi.org/10.1103/PhysRevB.79.195125} {\bibfield  {journal} {\bibinfo  {journal} {Phys. Rev. B}\ }\textbf {\bibinfo {volume} {79}},\ \bibinfo {pages} {195125} (\bibinfo {year} {2009})}\BibitemShut {NoStop}%
\bibitem [{\citenamefont {Wang}\ \emph {et~al.}(2012)\citenamefont {Wang}, \citenamefont {Xiang}, \citenamefont {Wang}, \citenamefont {Wang}, \citenamefont {Yang},\ and\ \citenamefont {Lee}}]{Wang2012}%
  \BibitemOpen
  \bibfield  {author} {\bibinfo {author} {\bibfnamefont {W.-S.}\ \bibnamefont {Wang}}, \bibinfo {author} {\bibfnamefont {Y.-Y.}\ \bibnamefont {Xiang}}, \bibinfo {author} {\bibfnamefont {Q.-H.}\ \bibnamefont {Wang}}, \bibinfo {author} {\bibfnamefont {F.}~\bibnamefont {Wang}}, \bibinfo {author} {\bibfnamefont {F.}~\bibnamefont {Yang}},\ and\ \bibinfo {author} {\bibfnamefont {D.-H.}\ \bibnamefont {Lee}},\ }\bibfield  {title} {\bibinfo {title} {{Functional renormalization group and variational Monte Carlo studies of the electronic instabilities in graphene near $\frac{1}{4}$ doping}},\ }\href {https://doi.org/10.1103/PhysRevB.85.035414} {\bibfield  {journal} {\bibinfo  {journal} {Phys. Rev. B}\ }\textbf {\bibinfo {volume} {85}},\ \bibinfo {pages} {035414} (\bibinfo {year} {2012})}\BibitemShut {NoStop}%
\bibitem [{\citenamefont {Lichtenstein}\ \emph {et~al.}(2017)\citenamefont {Lichtenstein}, \citenamefont {{S{\'{a}}nchez de la Pe{\~{n}}a}}, \citenamefont {Rohe}, \citenamefont {{Di Napoli}}, \citenamefont {Honerkamp},\ and\ \citenamefont {Maier}}]{Lichtenstein2017}%
  \BibitemOpen
  \bibfield  {author} {\bibinfo {author} {\bibfnamefont {J.}~\bibnamefont {Lichtenstein}}, \bibinfo {author} {\bibfnamefont {D.}~\bibnamefont {{S{\'{a}}nchez de la Pe{\~{n}}a}}}, \bibinfo {author} {\bibfnamefont {D.}~\bibnamefont {Rohe}}, \bibinfo {author} {\bibfnamefont {E.}~\bibnamefont {{Di Napoli}}}, \bibinfo {author} {\bibfnamefont {C.}~\bibnamefont {Honerkamp}},\ and\ \bibinfo {author} {\bibfnamefont {S.}~\bibnamefont {Maier}},\ }\bibfield  {title} {\bibinfo {title} {{High-performance functional Renormalization Group calculations for interacting fermions}},\ }\href {https://doi.org/https://doi.org/10.1016/j.cpc.2016.12.013} {\bibfield  {journal} {\bibinfo  {journal} {Computer Physics Communications}\ }\textbf {\bibinfo {volume} {213}},\ \bibinfo {pages} {100} (\bibinfo {year} {2017})}\BibitemShut {NoStop}%
\bibitem [{\citenamefont {Schober}\ \emph {et~al.}(2018)\citenamefont {Schober}, \citenamefont {Ehrlich}, \citenamefont {Reckling},\ and\ \citenamefont {Honerkamp}}]{Schober2018}%
  \BibitemOpen
  \bibfield  {author} {\bibinfo {author} {\bibfnamefont {G.~A.~H.}\ \bibnamefont {Schober}}, \bibinfo {author} {\bibfnamefont {J.}~\bibnamefont {Ehrlich}}, \bibinfo {author} {\bibfnamefont {T.}~\bibnamefont {Reckling}},\ and\ \bibinfo {author} {\bibfnamefont {C.}~\bibnamefont {Honerkamp}},\ }\bibfield  {title} {\bibinfo {title} {{Truncated-Unity Functional Renormalization Group for Multiband Systems With Spin-Orbit Coupling}},\ }\href {https://doi.org/10.3389/fphy.2018.00032} {\bibfield  {journal} {\bibinfo  {journal} {Front. Phys.}\ }\textbf {\bibinfo {volume} {6}},\ \bibinfo {pages} {32} (\bibinfo {year} {2018})}\BibitemShut {NoStop}%
\bibitem [{\citenamefont {Eckhardt}\ \emph {et~al.}(2020)\citenamefont {Eckhardt}, \citenamefont {Honerkamp}, \citenamefont {Held},\ and\ \citenamefont {Kauch}}]{Eckhardt2020}%
  \BibitemOpen
  \bibfield  {author} {\bibinfo {author} {\bibfnamefont {C.~J.}\ \bibnamefont {Eckhardt}}, \bibinfo {author} {\bibfnamefont {C.}~\bibnamefont {Honerkamp}}, \bibinfo {author} {\bibfnamefont {K.}~\bibnamefont {Held}},\ and\ \bibinfo {author} {\bibfnamefont {A.}~\bibnamefont {Kauch}},\ }\bibfield  {title} {\bibinfo {title} {{Truncated unity parquet solver}},\ }\href {https://doi.org/10.1103/PhysRevB.101.155104} {\bibfield  {journal} {\bibinfo  {journal} {Phys. Rev. B}\ }\textbf {\bibinfo {volume} {101}},\ \bibinfo {pages} {155104} (\bibinfo {year} {2020})}\BibitemShut {NoStop}%
\bibitem [{\citenamefont {Profe}\ and\ \citenamefont {Kennes}(2022)}]{Profe2022}%
  \BibitemOpen
  \bibfield  {author} {\bibinfo {author} {\bibfnamefont {J.~B.}\ \bibnamefont {Profe}}\ and\ \bibinfo {author} {\bibfnamefont {D.~M.}\ \bibnamefont {Kennes}},\ }\bibfield  {title} {\bibinfo {title} {{TU$^2$FRG: a scalable approach for truncated unity functional renormalization group in generic fermionic models}},\ }\href {https://doi.org/10.1140/epjb/s10051-022-00316-x} {\bibfield  {journal} {\bibinfo  {journal} {Eur. Phys. J. B}\ }\textbf {\bibinfo {volume} {95}},\ \bibinfo {pages} {60} (\bibinfo {year} {2022})}\BibitemShut {NoStop}%
\bibitem [{\citenamefont {Gneist}\ \emph {et~al.}(2022{\natexlab{a}})\citenamefont {Gneist}, \citenamefont {Classen},\ and\ \citenamefont {Scherer}}]{Gneist2022}%
  \BibitemOpen
  \bibfield  {author} {\bibinfo {author} {\bibfnamefont {N.}~\bibnamefont {Gneist}}, \bibinfo {author} {\bibfnamefont {L.}~\bibnamefont {Classen}},\ and\ \bibinfo {author} {\bibfnamefont {M.~M.}\ \bibnamefont {Scherer}},\ }\bibfield  {title} {\bibinfo {title} {{Competing instabilities of the extended Hubbard model on the triangular lattice: Truncated-unity functional renormalization group and application to moir\'e materials}},\ }\href {https://doi.org/10.1103/PhysRevB.106.125141} {\bibfield  {journal} {\bibinfo  {journal} {Phys. Rev. B}\ }\textbf {\bibinfo {volume} {106}},\ \bibinfo {pages} {125141} (\bibinfo {year} {2022}{\natexlab{a}})}\BibitemShut {NoStop}%
\bibitem [{\citenamefont {Beyer}\ \emph {et~al.}(2023)\citenamefont {Beyer}, \citenamefont {Profe}, \citenamefont {Klebl}, \citenamefont {Schwemmer}, \citenamefont {Kennes}, \citenamefont {Thomale}, \citenamefont {Honerkamp},\ and\ \citenamefont {Rachel}}]{Beyer2023}%
  \BibitemOpen
  \bibfield  {author} {\bibinfo {author} {\bibfnamefont {J.}~\bibnamefont {Beyer}}, \bibinfo {author} {\bibfnamefont {J.~B.}\ \bibnamefont {Profe}}, \bibinfo {author} {\bibfnamefont {L.}~\bibnamefont {Klebl}}, \bibinfo {author} {\bibfnamefont {T.}~\bibnamefont {Schwemmer}}, \bibinfo {author} {\bibfnamefont {D.~M.}\ \bibnamefont {Kennes}}, \bibinfo {author} {\bibfnamefont {R.}~\bibnamefont {Thomale}}, \bibinfo {author} {\bibfnamefont {C.}~\bibnamefont {Honerkamp}},\ and\ \bibinfo {author} {\bibfnamefont {S.}~\bibnamefont {Rachel}},\ }\bibfield  {title} {\bibinfo {title} {{Rashba spin-orbit coupling in the square-lattice Hubbard model: A truncated-unity functional renormalization group study}},\ }\href {https://doi.org/10.1103/PhysRevB.107.125115} {\bibfield  {journal} {\bibinfo  {journal} {Phys. Rev. B}\ }\textbf {\bibinfo {volume} {107}},\ \bibinfo {pages} {125115} (\bibinfo {year} {2023})}\BibitemShut {NoStop}%
\bibitem [{\citenamefont {Profe}\ \emph {et~al.}(2024)\citenamefont {Profe}, \citenamefont {Kennes},\ and\ \citenamefont {Klebl}}]{Profe2024}%
  \BibitemOpen
  \bibfield  {author} {\bibinfo {author} {\bibfnamefont {J.~B.}\ \bibnamefont {Profe}}, \bibinfo {author} {\bibfnamefont {D.~M.}\ \bibnamefont {Kennes}},\ and\ \bibinfo {author} {\bibfnamefont {L.}~\bibnamefont {Klebl}},\ }\bibfield  {title} {\bibinfo {title} {{divERGe implements various Exact Renormalization Group examples}},\ }\href {https://doi.org/10.21468/SciPostPhysCodeb.26} {\bibfield  {journal} {\bibinfo  {journal} {SciPost Phys. Codebases}\ ,\ \bibinfo {pages} {26}} (\bibinfo {year} {2024})}\BibitemShut {NoStop}%
\bibitem [{\citenamefont {Kugler}(2019)}]{KuglerPhDthesis}%
  \BibitemOpen
  \bibfield  {author} {\bibinfo {author} {\bibfnamefont {F.~B.}\ \bibnamefont {Kugler}},\ }\emph {\bibinfo {title} {{Renormalization group approaches to strongly correlated electron systems}}},\ \href {http://nbn-resolving.de/urn:nbn:de:bvb:19-253594} {Ph.D. thesis},\ \bibinfo  {school} {Ludwig-Maximilians-Universit{\"a}t M{\"u}nchen} (\bibinfo {year} {2019})\BibitemShut {NoStop}%
\bibitem [{\citenamefont {Patricolo}\ \emph {et~al.}(2025)\citenamefont {Patricolo}, \citenamefont {Gievers}, \citenamefont {Fraboulet}, \citenamefont {Al-Eryani}, \citenamefont {Heinzelmann}, \citenamefont {Bonetti}, \citenamefont {Toschi}, \citenamefont {Vilardi},\ and\ \citenamefont {Andergassen}}]{Patricolo2025}%
  \BibitemOpen
  \bibfield  {author} {\bibinfo {author} {\bibfnamefont {M.}~\bibnamefont {Patricolo}}, \bibinfo {author} {\bibfnamefont {M.}~\bibnamefont {Gievers}}, \bibinfo {author} {\bibfnamefont {K.}~\bibnamefont {Fraboulet}}, \bibinfo {author} {\bibfnamefont {A.}~\bibnamefont {Al-Eryani}}, \bibinfo {author} {\bibfnamefont {S.}~\bibnamefont {Heinzelmann}}, \bibinfo {author} {\bibfnamefont {P.~M.}\ \bibnamefont {Bonetti}}, \bibinfo {author} {\bibfnamefont {A.}~\bibnamefont {Toschi}}, \bibinfo {author} {\bibfnamefont {D.}~\bibnamefont {Vilardi}},\ and\ \bibinfo {author} {\bibfnamefont {S.}~\bibnamefont {Andergassen}},\ }\bibfield  {title} {\bibinfo {title} {{Single-boson exchange formulation of the Schwinger-Dyson equation and its application to the functional renormalization group}},\ }\href {https://doi.org/10.21468/SciPostPhys.18.3.078} {\bibfield  {journal} {\bibinfo  {journal} {SciPost Phys.}\ }\textbf {\bibinfo {volume} {18}},\ \bibinfo {pages} {078} (\bibinfo {year} {2025})}\BibitemShut {NoStop}%
\bibitem [{\citenamefont {Hubbard}(1963)}]{Hubbard1963}%
  \BibitemOpen
  \bibfield  {author} {\bibinfo {author} {\bibfnamefont {J.}~\bibnamefont {Hubbard}},\ }\bibfield  {title} {\bibinfo {title} {{Electron correlations in narrow energy bands}},\ }\href {https://doi.org/10.1098/rspa.1963.0204} {\bibfield  {journal} {\bibinfo  {journal} {Proceedings of the Royal Society of London. Series A. Mathematical and Physical Sciences}\ }\textbf {\bibinfo {volume} {276}},\ \bibinfo {pages} {238} (\bibinfo {year} {1963})}\BibitemShut {NoStop}%
\bibitem [{\citenamefont {Karrasch}\ \emph {et~al.}(2008)\citenamefont {Karrasch}, \citenamefont {Hedden}, \citenamefont {Peters}, \citenamefont {Pruschke}, \citenamefont {Schönhammer},\ and\ \citenamefont {Meden}}]{Karrasch2008Finite}%
  \BibitemOpen
  \bibfield  {author} {\bibinfo {author} {\bibfnamefont {C.}~\bibnamefont {Karrasch}}, \bibinfo {author} {\bibfnamefont {R.}~\bibnamefont {Hedden}}, \bibinfo {author} {\bibfnamefont {R.}~\bibnamefont {Peters}}, \bibinfo {author} {\bibfnamefont {T.}~\bibnamefont {Pruschke}}, \bibinfo {author} {\bibfnamefont {K.}~\bibnamefont {Schönhammer}},\ and\ \bibinfo {author} {\bibfnamefont {V.}~\bibnamefont {Meden}},\ }\bibfield  {title} {\bibinfo {title} {{A finite-frequency functional renormalization group approach to the single impurity Anderson model}},\ }\href {https://doi.org/10.1088/0953-8984/20/34/345205} {\bibfield  {journal} {\bibinfo  {journal} {Journal of Physics: Condensed Matter}\ }\textbf {\bibinfo {volume} {20}},\ \bibinfo {pages} {345205} (\bibinfo {year} {2008})}\BibitemShut {NoStop}%
\bibitem [{\citenamefont {Jakobs}\ \emph {et~al.}(2010)\citenamefont {Jakobs}, \citenamefont {Pletyukhov},\ and\ \citenamefont {Schoeller}}]{Jakobs2010Nonequilibrium}%
  \BibitemOpen
  \bibfield  {author} {\bibinfo {author} {\bibfnamefont {S.~G.}\ \bibnamefont {Jakobs}}, \bibinfo {author} {\bibfnamefont {M.}~\bibnamefont {Pletyukhov}},\ and\ \bibinfo {author} {\bibfnamefont {H.}~\bibnamefont {Schoeller}},\ }\bibfield  {title} {\bibinfo {title} {{Nonequilibrium functional renormalization group with frequency-dependent vertex function: A study of the single-impurity Anderson model}},\ }\href {https://doi.org/10.1103/PhysRevB.81.195109} {\bibfield  {journal} {\bibinfo  {journal} {Phys. Rev. B}\ }\textbf {\bibinfo {volume} {81}},\ \bibinfo {pages} {195109} (\bibinfo {year} {2010})}\BibitemShut {NoStop}%
\bibitem [{Note8()}]{Note8}%
  \BibitemOpen
  \bibinfo {note} {We remark that although we use the term ``regulator'', this flow scheme is not really regularizing since it does not handle the singularities of the bare propagator in Eq.~\protect \eqref {eq:anderson_impurity_propagator}.}\BibitemShut {Stop}%
\bibitem [{\citenamefont {Honerkamp}\ \emph {et~al.}(2004)\citenamefont {Honerkamp}, \citenamefont {Rohe}, \citenamefont {Andergassen},\ and\ \citenamefont {Enss}}]{Honerkamp2004}%
  \BibitemOpen
  \bibfield  {author} {\bibinfo {author} {\bibfnamefont {C.}~\bibnamefont {Honerkamp}}, \bibinfo {author} {\bibfnamefont {D.}~\bibnamefont {Rohe}}, \bibinfo {author} {\bibfnamefont {S.}~\bibnamefont {Andergassen}},\ and\ \bibinfo {author} {\bibfnamefont {T.}~\bibnamefont {Enss}},\ }\bibfield  {title} {\bibinfo {title} {{Interaction flow method for many-fermion systems}},\ }\href {https://doi.org/10.1103/PhysRevB.70.235115} {\bibfield  {journal} {\bibinfo  {journal} {Phys. Rev. B}\ }\textbf {\bibinfo {volume} {70}},\ \bibinfo {pages} {235115} (\bibinfo {year} {2004})}\BibitemShut {NoStop}%
\bibitem [{\citenamefont {Schneider}\ \emph {et~al.}(2024)\citenamefont {Schneider}, \citenamefont {Reuther}, \citenamefont {Gonzalez}, \citenamefont {Sbierski},\ and\ \citenamefont {Niggemann}}]{Schneider2023}%
  \BibitemOpen
  \bibfield  {author} {\bibinfo {author} {\bibfnamefont {B.}~\bibnamefont {Schneider}}, \bibinfo {author} {\bibfnamefont {J.}~\bibnamefont {Reuther}}, \bibinfo {author} {\bibfnamefont {M.~G.}\ \bibnamefont {Gonzalez}}, \bibinfo {author} {\bibfnamefont {B.}~\bibnamefont {Sbierski}},\ and\ \bibinfo {author} {\bibfnamefont {N.}~\bibnamefont {Niggemann}},\ }\bibfield  {title} {\bibinfo {title} {{Temperature flow in pseudo-Majorana functional renormalization for quantum spins}},\ }\href {https://doi.org/10.1103/PhysRevB.109.195109} {\bibfield  {journal} {\bibinfo  {journal} {Phys. Rev. B}\ }\textbf {\bibinfo {volume} {109}},\ \bibinfo {pages} {195109} (\bibinfo {year} {2024})}\BibitemShut {NoStop}%
\bibitem [{\citenamefont {Braun}\ \emph {et~al.}(2025)\citenamefont {Braun}, \citenamefont {Scherer},\ and\ \citenamefont {Classen}}]{Braun2025}%
  \BibitemOpen
  \bibfield  {author} {\bibinfo {author} {\bibfnamefont {H.}~\bibnamefont {Braun}}, \bibinfo {author} {\bibfnamefont {M.~M.}\ \bibnamefont {Scherer}},\ and\ \bibinfo {author} {\bibfnamefont {L.}~\bibnamefont {Classen}},\ }\bibfield  {title} {\bibinfo {title} {{Kohn-Luttinger-like mechanism for unconventional charge density waves}},\ }\href {https://doi.org/10.1103/PhysRevB.111.195157} {\bibfield  {journal} {\bibinfo  {journal} {Phys. Rev. B}\ }\textbf {\bibinfo {volume} {111}},\ \bibinfo {pages} {195157} (\bibinfo {year} {2025})}\BibitemShut {NoStop}%
\bibitem [{\citenamefont {Gneist}\ \emph {et~al.}(2022{\natexlab{b}})\citenamefont {Gneist}, \citenamefont {Kiese}, \citenamefont {Henkel}, \citenamefont {Thomale}, \citenamefont {Classen},\ and\ \citenamefont {Scherer}}]{Gneist_2022_2}%
  \BibitemOpen
  \bibfield  {author} {\bibinfo {author} {\bibfnamefont {N.}~\bibnamefont {Gneist}}, \bibinfo {author} {\bibfnamefont {D.}~\bibnamefont {Kiese}}, \bibinfo {author} {\bibfnamefont {R.}~\bibnamefont {Henkel}}, \bibinfo {author} {\bibfnamefont {R.}~\bibnamefont {Thomale}}, \bibinfo {author} {\bibfnamefont {L.}~\bibnamefont {Classen}},\ and\ \bibinfo {author} {\bibfnamefont {M.~M.}\ \bibnamefont {Scherer}},\ }\bibfield  {title} {\bibinfo {title} {Functional renormalization of spinless triangular-lattice fermions: N-patch vs. truncated-unity scheme},\ }\bibfield  {journal} {\bibinfo  {journal} {The European Physical Journal B}\ }\textbf {\bibinfo {volume} {95}},\ \href {https://doi.org/10.1140/epjb/s10051-022-00395-w} {10.1140/epjb/s10051-022-00395-w} (\bibinfo {year} {2022}{\natexlab{b}})\BibitemShut {NoStop}%
\bibitem [{\citenamefont {Igoshev}\ and\ \citenamefont {Katanin}(2023)}]{Igoshev2023}%
  \BibitemOpen
  \bibfield  {author} {\bibinfo {author} {\bibfnamefont {P.~A.}\ \bibnamefont {Igoshev}}\ and\ \bibinfo {author} {\bibfnamefont {A.~A.}\ \bibnamefont {Katanin}},\ }\bibfield  {title} {\bibinfo {title} {{Ferromagnetic instability in itinerant fcc lattice electron systems with higher-order van Hove singularities: Functional renormalization group study}},\ }\href {https://doi.org/10.1103/PhysRevB.107.115105} {\bibfield  {journal} {\bibinfo  {journal} {Phys. Rev. B}\ }\textbf {\bibinfo {volume} {107}},\ \bibinfo {pages} {115105} (\bibinfo {year} {2023})}\BibitemShut {NoStop}%
\bibitem [{\citenamefont {Laakso}\ \emph {et~al.}(2014)\citenamefont {Laakso}, \citenamefont {Kennes}, \citenamefont {Jakobs},\ and\ \citenamefont {Meden}}]{Laakso2014Functional}%
  \BibitemOpen
  \bibfield  {author} {\bibinfo {author} {\bibfnamefont {M.~A.}\ \bibnamefont {Laakso}}, \bibinfo {author} {\bibfnamefont {D.~M.}\ \bibnamefont {Kennes}}, \bibinfo {author} {\bibfnamefont {S.~G.}\ \bibnamefont {Jakobs}},\ and\ \bibinfo {author} {\bibfnamefont {V.}~\bibnamefont {Meden}},\ }\bibfield  {title} {\bibinfo {title} {{Functional renormalization group study of the Anderson--Holstein model}},\ }\href {https://iopscience.iop.org/article/10.1088/1367-2630/16/2/023007} {\bibfield  {journal} {\bibinfo  {journal} {New Journal of Physics}\ }\textbf {\bibinfo {volume} {16}},\ \bibinfo {pages} {023007} (\bibinfo {year} {2014})}\BibitemShut {NoStop}%
\bibitem [{\citenamefont {Hauke}\ \emph {et~al.}(2016)\citenamefont {Hauke}, \citenamefont {Heyl}, \citenamefont {Tagliacozzo},\ and\ \citenamefont {Zoller}}]{Hauke2016Measuring}%
  \BibitemOpen
  \bibfield  {author} {\bibinfo {author} {\bibfnamefont {P.}~\bibnamefont {Hauke}}, \bibinfo {author} {\bibfnamefont {M.}~\bibnamefont {Heyl}}, \bibinfo {author} {\bibfnamefont {L.}~\bibnamefont {Tagliacozzo}},\ and\ \bibinfo {author} {\bibfnamefont {P.}~\bibnamefont {Zoller}},\ }\bibfield  {title} {\bibinfo {title} {Measuring multipartite entanglement through dynamic susceptibilities},\ }\href {https://www.nature.com/articles/nphys3700} {\bibfield  {journal} {\bibinfo  {journal} {Nature Physics}\ }\textbf {\bibinfo {volume} {12}},\ \bibinfo {pages} {778} (\bibinfo {year} {2016})}\BibitemShut {NoStop}%
\bibitem [{\citenamefont {Bippus}\ \emph {et~al.}(2025)\citenamefont {Bippus}, \citenamefont {Krsnik}, \citenamefont {Kitatani}, \citenamefont {Akšamović}, \citenamefont {Kauch}, \citenamefont {Barišić},\ and\ \citenamefont {Held}}]{Bippus2025}%
  \BibitemOpen
  \bibfield  {author} {\bibinfo {author} {\bibfnamefont {F.}~\bibnamefont {Bippus}}, \bibinfo {author} {\bibfnamefont {J.}~\bibnamefont {Krsnik}}, \bibinfo {author} {\bibfnamefont {M.}~\bibnamefont {Kitatani}}, \bibinfo {author} {\bibfnamefont {L.}~\bibnamefont {Akšamović}}, \bibinfo {author} {\bibfnamefont {A.}~\bibnamefont {Kauch}}, \bibinfo {author} {\bibfnamefont {N.}~\bibnamefont {Barišić}},\ and\ \bibinfo {author} {\bibfnamefont {K.}~\bibnamefont {Held}},\ }\bibfield  {title} {\bibinfo {title} {{Entanglement in the pseudogap regime of cuprate superconductors}},\ }\bibfield  {journal} {\bibinfo  {journal} {Physical Review B}\ }\textbf {\bibinfo {volume} {112}},\ \href {https://doi.org/10.1103/xk42-b9cx} {10.1103/xk42-b9cx} (\bibinfo {year} {2025})\BibitemShut {NoStop}%
\bibitem [{Note9()}]{Note9}%
  \BibitemOpen
  \bibinfo {note} {Since $\protect \overline {\psi }$ and $\psi $ are Grassmann fields in the present case, this expansion must be performed about $\protect \overline {\psi }=\psi =0$}\BibitemShut {NoStop}%
\bibitem [{Note10()}]{Note10}%
  \BibitemOpen
  \bibinfo {note} {By ``does not constitute an efficient approximation scheme'', we exclude here the interaction flow scheme of Ref.~\cite {Honerkamp2004}, which was originally developed within the conventional fRG framework of the vertex expansion for fermionic models with a quartic interaction. However, as explained in Sec.~\ref {sec:ModelDefinitionsCutoffSchemes}, the dependence of the bare interaction with respect to the flow parameter is entirely reshuffled in the bare propagator within this approach, so that the corresponding flow equations reduce to Eqs.~\protect \eqref {eq:1lfRG} within the level-2 truncation.}\BibitemShut {Stop}%
\bibitem [{\citenamefont {Salmhofer}\ and\ \citenamefont {Honerkamp}(2001)}]{Salmhofer2001}%
  \BibitemOpen
  \bibfield  {author} {\bibinfo {author} {\bibfnamefont {M.}~\bibnamefont {Salmhofer}}\ and\ \bibinfo {author} {\bibfnamefont {C.}~\bibnamefont {Honerkamp}},\ }\bibfield  {title} {\bibinfo {title} {{Fermionic Renormalization Group Flows: Technique and Theory}},\ }\href {https://doi.org/10.1143/PTP.105.1} {\bibfield  {journal} {\bibinfo  {journal} {Progress of Theoretical Physics}\ }\textbf {\bibinfo {volume} {105}},\ \bibinfo {pages} {1} (\bibinfo {year} {2001})},\ \Eprint {https://arxiv.org/abs/https://academic.oup.com/ptp/article-pdf/105/1/1/5164880/105-1-1.pdf} {https://academic.oup.com/ptp/article-pdf/105/1/1/5164880/105-1-1.pdf} \BibitemShut {NoStop}%
\bibitem [{\citenamefont {Gievers}\ \emph {et~al.}(2025)\citenamefont {Gievers}, \citenamefont {Schmidt}, \citenamefont {von Delft},\ and\ \citenamefont {Kugler}}]{Gievers2025Subleading}%
  \BibitemOpen
  \bibfield  {author} {\bibinfo {author} {\bibfnamefont {M.}~\bibnamefont {Gievers}}, \bibinfo {author} {\bibfnamefont {R.}~\bibnamefont {Schmidt}}, \bibinfo {author} {\bibfnamefont {J.}~\bibnamefont {von Delft}},\ and\ \bibinfo {author} {\bibfnamefont {F.~B.}\ \bibnamefont {Kugler}},\ }\bibfield  {title} {\bibinfo {title} {{Subleading logarithmic behavior in the parquet formalism}},\ }\href {https://doi.org/10.1103/PhysRevB.111.085151} {\bibfield  {journal} {\bibinfo  {journal} {Phys. Rev. B}\ }\textbf {\bibinfo {volume} {111}},\ \bibinfo {pages} {085151} (\bibinfo {year} {2025})}\BibitemShut {NoStop}%
\bibitem [{Note11()}]{Note11}%
  \BibitemOpen
  \bibinfo {note} {Like the full vertex $V$~\cite {Salmhofer2001}, the vertex $A_{ph}$ can be expressed as $A_{ph}^{\uparrow \uparrow }=A_{ph}^{\uparrow \downarrow }+A_{ph}^{\setbox \z@ \hbox {\mathsurround \z@ $\textstyle \uparrow \downarrow $}\mathaccent "0362{\uparrow \downarrow }}$ whereas its crossing symmetry implies that $A_{ph}^{\setbox \z@ \hbox {\mathsurround \z@ $\textstyle \uparrow \downarrow $}\mathaccent "0362{\uparrow \downarrow }}=-A_{\protect \overline {ph}}^{\uparrow \downarrow }$. According to these two relations, one can derive the equality $A^{\uparrow \uparrow }_{ph} - A^{\uparrow \downarrow }_{ph} = - A^{\uparrow \downarrow }_{\protect \overline {ph}}$ used in Eq.~\protect \eqref {eq:DefinitionMchannel}.}\BibitemShut {Stop}%
\bibitem [{Note12()}]{Note12}%
  \BibitemOpen
  \bibinfo {note} {We have $\lambda ^{\setbox \z@ \hbox {\mathsurround \z@ $\textstyle \uparrow \downarrow $}\mathaccent "0362{\uparrow \downarrow }}_{ph}=+\lambda ^{\uparrow \downarrow }_{\protect \overline {ph}}$ whereas $A^{\setbox \z@ \hbox {\mathsurround \z@ $\textstyle \uparrow \downarrow $}\mathaccent "0362{\uparrow \downarrow }}_{ph}=-A^{\uparrow \downarrow }_{\protect \overline {ph}}$ for Eq.~\protect \eqref {eq:DefinitionMchannel} (see notably Eq.~(4.31a) and related discussions in Ref.~\cite {Gievers2022}). The definition $\lambda _{\protect \text {M}}=+\lambda ^{\uparrow \downarrow }_{\protect \overline {ph}}$ in Eq.~\protect \eqref {eq:DefinitionlambdaMchannel} is then obtained by combining the relation $\lambda ^{\setbox \z@ \hbox {\mathsurround \z@ $\textstyle \uparrow \downarrow $}\mathaccent "0362{\uparrow \downarrow }}_{ph}=+\lambda ^{\uparrow \downarrow }_{\protect \overline {ph}}$ with $\lambda _{ph}^{\uparrow \uparrow }=\lambda _{ph}^{\uparrow \downarrow }+\lambda _{ph}^{\setbox \z@ \hbox
  {\mathsurround \z@ $\textstyle \uparrow \downarrow $}\mathaccent "0362{\uparrow \downarrow }}$.}\BibitemShut {Stop}%
\bibitem [{Note13()}]{Note13}%
  \BibitemOpen
  \bibinfo {note} {As explained in Refs.~\cite {Patricolo2025,GieversPhDthesis}, it is also possible to reformulate the product~\protect \eqref {eq:Definitionfcircproduct} in such a way that Eqs.~\protect \eqref {eq:ABFormulasfcircproduct} and~\protect \eqref {eq:ABCFormulasfcircproduct} can be directly derived from standard matrix products}\BibitemShut {NoStop}%
\bibitem [{\citenamefont {Heinzelmann}(2023)}]{HeinzelmannPhDThesis}%
  \BibitemOpen
  \bibfield  {author} {\bibinfo {author} {\bibfnamefont {S.}~\bibnamefont {Heinzelmann}},\ }\emph {\bibinfo {title} {{The single-boson exchange formalism and its application to the functional renormalization group}}},\ \href@noop {} {Ph.D. thesis},\ \bibinfo  {school} {Universit\"at T\"ubingen} (\bibinfo {year} {2023})\BibitemShut {NoStop}%
\bibitem [{\citenamefont {De~Polsi}\ and\ \citenamefont {Wschebor}(2022)}]{DePolsi2022}%
  \BibitemOpen
  \bibfield  {author} {\bibinfo {author} {\bibfnamefont {G.}~\bibnamefont {De~Polsi}}\ and\ \bibinfo {author} {\bibfnamefont {N.}~\bibnamefont {Wschebor}},\ }\bibfield  {title} {\bibinfo {title} {{Regulator dependence in the functional renormalization group: A quantitative explanation}},\ }\href {https://doi.org/10.1103/PhysRevE.106.024111} {\bibfield  {journal} {\bibinfo  {journal} {Phys. Rev. E}\ }\textbf {\bibinfo {volume} {106}},\ \bibinfo {pages} {024111} (\bibinfo {year} {2022})}\BibitemShut {NoStop}%
\end{thebibliography}%
\end{document}